\documentclass[traditabstract,longauth]{aa}
\usepackage{epsfig}
\usepackage{graphicx}
\usepackage{url}
\usepackage{color}
\usepackage[nonamebreak]{natbib}
\usepackage[breaklinks,colorlinks,citecolor=blue]{hyperref}
\usepackage{ifthen}
\usepackage{amssymb,,amsmath}
\usepackage{longtable}
\usepackage{multirow}
\usepackage{fixltx2e}
\usepackage{breakurl}
\usepackage{txfonts}
\usepackage[usenames,svgnames]{xcolor}
\bibpunct{(}{)}{;}{a}{}{,}

\newcommand{\cc}{\mathcal{C}} % "cc" stands for "colour correction"
\def\lsim{\mathrel{\raise .4ex\hbox{\rlap{$<$}\lower 1.2ex\hbox{$\sim$}}}}
\def\gsim{\mathrel{\raise .4ex\hbox{\rlap{$>$}\lower 1.2ex\hbox{$\sim$}}}}
\def\arcm{\ifmmode {^{\scriptscriptstyle\prime}}
          \else $^{\scriptscriptstyle\prime}$\fi}

\def\setsymbol#1#2{\expandafter\def\csname #1\endcsname{#2}}
\def\getsymbol#1{\csname #1\endcsname}

\def\Planck{\textit{Planck}}

\def\allearlypapers{\nocite{planck2011-1.1, planck2011-1.3, planck2011-1.4, planck2011-1.5, planck2011-1.6, planck2011-1.7, planck2011-1.10, planck2011-1.10sup, planck2011-5.1a, planck2011-5.1b, planck2011-5.2a, planck2011-5.2b, planck2011-5.2c, planck2011-6.1, planck2011-6.2, planck2011-6.3a, planck2011-6.4a, planck2011-6.4b, planck2011-6.6, planck2011-7.0, planck2011-7.2, planck2011-7.3, planck2011-7.7a, planck2011-7.7b, planck2011-7.12, planck2011-7.13}}

\def\alltwentythirteenresultspapers{\nocite{planck2013-p01, planck2013-p02, planck2013-p02a, planck2013-p02d, planck2013-p02b, planck2013-p03, planck2013-p03c, planck2013-p03f, planck2013-p03d, planck2013-p03e, planck2013-p01a, planck2013-p06, planck2013-p03a, planck2013-pip88, planck2013-p08, planck2013-p11, planck2013-p12, planck2013-p13, planck2013-p14, planck2013-p15, planck2013-p05b, planck2013-p17, planck2013-p09, planck2013-p09a, planck2013-p20, planck2013-p19, planck2013-pipaberration, planck2013-p05, planck2013-p05a, planck2013-pip56, planck2013-p06b, planck2013-p01a}}

\def\alltwentyfifteenresultspapers{\nocite{planck2014-a01, planck2014-a03, planck2014-a04, planck2014-a05, planck2014-a06, planck2014-a07, planck2014-a08, planck2014-a09, planck2014-a11, planck2014-a12, planck2014-a13, planck2014-a14, planck2014-a15, planck2014-a16, planck2014-a17, planck2014-a18, planck2014-a19, planck2014-a20, planck2014-a22, planck2014-a24, planck2014-a26, planck2014-a28, planck2014-a29, planck2014-a30, planck2014-a31, planck2014-a35, planck2014-a36, planck2014-a37, planck2014-ES}}

\newbox\tablebox    \newdimen\tablewidth
\def\leaderfil{\leaders\hbox to 5pt{\hss.\hss}\hfil}
\def\endPlancktable{\tablewidth=\columnwidth 
    $$\hss\copy\tablebox\hss$$
    \vskip-\lastskip\vskip -2pt}
\def\endPlancktablewide{\tablewidth=\textwidth 
    $$\hss\copy\tablebox\hss$$
    \vskip-\lastskip\vskip -2pt}
\def\tablenote#1 #2\par{\begingroup \parindent=0.8em
    \abovedisplayshortskip=0pt\belowdisplayshortskip=0pt
    \noindent
    $$\hss\vbox{\hsize\tablewidth \hangindent=\parindent \hangafter=1 \noindent
    \hbox to \parindent{$^#1$\hss}\strut#2\strut\par}\hss$$
    \endgroup}
\def\doubleline{\vskip 3pt\hrule \vskip 1.5pt \hrule \vskip 5pt}

\def\L2{\ifmmode L_2\else $L_2$\fi}

\def\DeltaT{\ifmmode \Delta T\else $\Delta T$\fi}
\def\deltat{\ifmmode \Delta t\else $\Delta t$\fi}
\def\fknee{\ifmmode f_{\rm knee}\else $f_{\rm knee}$\fi}
\def\Fmax{\ifmmode F_{\rm max}\else $F_{\rm max}$\fi}
\def\solar{\ifmmode{\rm M}_{\mathord\odot}\else${\rm M}_{\mathord\odot}$\fi}
\def\Msolar{\ifmmode{\rm M}_{\mathord\odot}\else${\rm M}_{\mathord\odot}$\fi}
\def\Lsolar{\ifmmode{\rm L}_{\mathord\odot}\else${\rm L}_{\mathord\odot}$\fi}
\def\inv{\ifmmode^{-1}\else$^{-1}$\fi}
\def\mo{\ifmmode^{-1}\else$^{-1}$\fi}
\def\sup#1{\ifmmode ^{\rm #1}\else $^{\rm #1}$\fi}
\def\expo#1{\ifmmode \times 10^{#1}\else $\times 10^{#1}$\fi}
\def\,{\thinspace}
\def\lsim{\mathrel{\raise .4ex\hbox{\rlap{$<$}\lower 1.2ex\hbox{$\sim$}}}}
\def\gsim{\mathrel{\raise .4ex\hbox{\rlap{$>$}\lower 1.2ex\hbox{$\sim$}}}}

\def\simprop{\mathrel{\raise .4ex\hbox{\rlap{$\propto$}\lower 1.2ex\hbox{$\sim$}}}}
\def\deg{\ifmmode^\circ\else$^\circ$\fi}
\def\pdeg{\ifmmode $\setbox0=\hbox{$^{\circ}$}\rlap{\hskip.11\wd0 .}$^{\circ}
          \else \setbox0=\hbox{$^{\circ}$}\rlap{\hskip.11\wd0 .}$^{\circ}$\fi}
\def\arcs{\ifmmode {^{\scriptstyle\prime\prime}}
          \else $^{\scriptstyle\prime\prime}$\fi}
\def\arcm{\ifmmode {^{\scriptstyle\prime}}
          \else $^{\scriptstyle\prime}$\fi}
\newdimen\sa  \newdimen\sb
\def\parcs{\sa=.07em \sb=.03em
     \ifmmode \hbox{\rlap{.}}^{\scriptstyle\prime\kern -\sb\prime}\hbox{\kern -\sa}
     \else \rlap{.}$^{\scriptstyle\prime\kern -\sb\prime}$\kern -\sa\fi}
\def\parcm{\sa=.08em \sb=.03em
     \ifmmode \hbox{\rlap{.}\kern\sa}^{\scriptstyle\prime}\hbox{\kern-\sb}
     \else \rlap{.}\kern\sa$^{\scriptstyle\prime}$\kern-\sb\fi}
\def\ra[#1 #2 #3.#4]{#1\sup{h}#2\sup{m}#3\sup{s}\llap.#4}
\def\dec[#1 #2 #3.#4]{#1\deg#2\arcm#3\arcs\llap.#4}
\def\deco[#1 #2 #3]{#1\deg#2\arcm#3\arcs}
\def\rra[#1 #2]{#1\sup{h}#2\sup{m}}

\def\dots{\relax\ifmmode \ldots\else $\ldots$\fi}
\def\WHzsr{\ifmmode $W\,Hz\mo\,sr\mo$\else W\,Hz\mo\,sr\mo\fi}
\def\mHz{\ifmmode $\,mHz$\else \,mHz\fi}
\def\GHz{\ifmmode $\,GHz$\else \,GHz\fi}
\def\mKs{\ifmmode $\,mK\,s$^{1/2}\else \,mK\,s$^{1/2}$\fi}
\def\muKs{\ifmmode \,\mu$K\,s$^{1/2}\else \,$\mu$K\,s$^{1/2}$\fi}
\def\muKRJs{\ifmmode \,\mu$K$_{\rm RJ}$\,s$^{1/2}\else \,$\mu$K$_{\rm RJ}$\,s$^{1/2}$\fi}
\def\muKHz{\ifmmode \,\mu$K\,Hz$^{-1/2}\else \,$\mu$K\,Hz$^{-1/2}$\fi}
\def\MJysr{\ifmmode \,$MJy\,sr\mo$\else \,MJy\,sr\mo\fi}
\def\MJysrmK{\ifmmode \,$MJy\,sr\mo$\,mK$_{\rm CMB}\mo\else \,MJy\,sr\mo\,mK$_{\rm CMB}\mo$\fi}
\def\microns{\ifmmode \,\mu$m$\else \,$\mu$m\fi}

\def\muK{\ifmmode \,\mu$K$\else \,$\mu$\hbox{K}\fi}
\def\microK{\ifmmode \,\mu$K$\else \,$\mu$\hbox{K}\fi}
\def\muW{\ifmmode \,\mu$W$\else \,$\mu$\hbox{W}\fi}
\def\kms{\ifmmode $\,km\,s$^{-1}\else \,km\,s$^{-1}$\fi}
\def\kmsMpc{\ifmmode $\,\kms\,Mpc\mo$\else \,\kms\,Mpc\mo\fi}
\setsymbol{LFI:center:frequency:70GHz:units}{70.3\,GHz}
\setsymbol{LFI:center:frequency:44GHz:units}{44.1\,GHz}
\setsymbol{LFI:center:frequency:30GHz:units}{28.5\,GHz}

\setsymbol{LFI:center:frequency:70GHz}{70.3}
\setsymbol{LFI:center:frequency:44GHz}{44.1}
\setsymbol{LFI:center:frequency:30GHz}{28.5}

\setsymbol{LFI:center:frequency:LFI18:Rad:M:units}{71.7\GHz}
\setsymbol{LFI:center:frequency:LFI19:Rad:M:units}{67.5\GHz}
\setsymbol{LFI:center:frequency:LFI20:Rad:M:units}{69.2\GHz}
\setsymbol{LFI:center:frequency:LFI21:Rad:M:units}{70.4\GHz}
\setsymbol{LFI:center:frequency:LFI22:Rad:M:units}{71.5\GHz}
\setsymbol{LFI:center:frequency:LFI23:Rad:M:units}{70.8\GHz}
\setsymbol{LFI:center:frequency:LFI24:Rad:M:units}{44.4\GHz}
\setsymbol{LFI:center:frequency:LFI25:Rad:M:units}{44.0\GHz}
\setsymbol{LFI:center:frequency:LFI26:Rad:M:units}{43.9\GHz}
\setsymbol{LFI:center:frequency:LFI27:Rad:M:units}{28.3\GHz}
\setsymbol{LFI:center:frequency:LFI28:Rad:M:units}{28.8\GHz}
\setsymbol{LFI:center:frequency:LFI18:Rad:S:units}{70.1\GHz}
\setsymbol{LFI:center:frequency:LFI19:Rad:S:units}{69.6\GHz}
\setsymbol{LFI:center:frequency:LFI20:Rad:S:units}{69.5\GHz}
\setsymbol{LFI:center:frequency:LFI21:Rad:S:units}{69.5\GHz}
\setsymbol{LFI:center:frequency:LFI22:Rad:S:units}{72.8\GHz}
\setsymbol{LFI:center:frequency:LFI23:Rad:S:units}{71.3\GHz}
\setsymbol{LFI:center:frequency:LFI24:Rad:S:units}{44.1\GHz}
\setsymbol{LFI:center:frequency:LFI25:Rad:S:units}{44.1\GHz}
\setsymbol{LFI:center:frequency:LFI26:Rad:S:units}{44.1\GHz}
\setsymbol{LFI:center:frequency:LFI27:Rad:S:units}{28.5\GHz}
\setsymbol{LFI:center:frequency:LFI28:Rad:S:units}{28.2\GHz}

\setsymbol{LFI:center:frequency:LFI18:Rad:M}{71.7}
\setsymbol{LFI:center:frequency:LFI19:Rad:M}{67.5}
\setsymbol{LFI:center:frequency:LFI20:Rad:M}{69.2}
\setsymbol{LFI:center:frequency:LFI21:Rad:M}{70.4}
\setsymbol{LFI:center:frequency:LFI22:Rad:M}{71.5}
\setsymbol{LFI:center:frequency:LFI23:Rad:M}{70.8}
\setsymbol{LFI:center:frequency:LFI24:Rad:M}{44.4}
\setsymbol{LFI:center:frequency:LFI25:Rad:M}{44.0}
\setsymbol{LFI:center:frequency:LFI26:Rad:M}{43.9}
\setsymbol{LFI:center:frequency:LFI27:Rad:M}{28.3}
\setsymbol{LFI:center:frequency:LFI28:Rad:M}{28.8}
\setsymbol{LFI:center:frequency:LFI18:Rad:S}{70.1}
\setsymbol{LFI:center:frequency:LFI19:Rad:S}{69.6}
\setsymbol{LFI:center:frequency:LFI20:Rad:S}{69.5}
\setsymbol{LFI:center:frequency:LFI21:Rad:S}{69.5}
\setsymbol{LFI:center:frequency:LFI22:Rad:S}{72.8}
\setsymbol{LFI:center:frequency:LFI23:Rad:S}{71.3}
\setsymbol{LFI:center:frequency:LFI24:Rad:S}{44.1}
\setsymbol{LFI:center:frequency:LFI25:Rad:S}{44.1}
\setsymbol{LFI:center:frequency:LFI26:Rad:S}{44.1}
\setsymbol{LFI:center:frequency:LFI27:Rad:S}{28.5}
\setsymbol{LFI:center:frequency:LFI28:Rad:S}{28.2}

\setsymbol{LFI:white:noise:sensitivity:70GHz:units}{134.7\muKs}
\setsymbol{LFI:white:noise:sensitivity:44GHz:units}{164.7\muKs}
\setsymbol{LFI:white:noise:sensitivity:30GHz:units}{143.4\muKs}

\setsymbol{LFI:white:noise:sensitivity:70GHz}{134.7}
\setsymbol{LFI:white:noise:sensitivity:44GHz}{164.7}
\setsymbol{LFI:white:noise:sensitivity:30GHz}{143.4}

\setsymbol{LFI:white:noise:sensitivity:LFI18:Rad:M:units}{512.0\muKs}
\setsymbol{LFI:white:noise:sensitivity:LFI19:Rad:M:units}{581.4\muKs}
\setsymbol{LFI:white:noise:sensitivity:LFI20:Rad:M:units}{590.8\muKs}
\setsymbol{LFI:white:noise:sensitivity:LFI21:Rad:M:units}{455.2\muKs}
\setsymbol{LFI:white:noise:sensitivity:LFI22:Rad:M:units}{492.0\muKs}
\setsymbol{LFI:white:noise:sensitivity:LFI23:Rad:M:units}{507.7\muKs}
\setsymbol{LFI:white:noise:sensitivity:LFI24:Rad:M:units}{462.2\muKs}
\setsymbol{LFI:white:noise:sensitivity:LFI25:Rad:M:units}{413.6\muKs}
\setsymbol{LFI:white:noise:sensitivity:LFI26:Rad:M:units}{478.6\muKs}
\setsymbol{LFI:white:noise:sensitivity:LFI27:Rad:M:units}{277.7\muKs}
\setsymbol{LFI:white:noise:sensitivity:LFI28:Rad:M:units}{312.3\muKs}
\setsymbol{LFI:white:noise:sensitivity:LFI18:Rad:S:units}{465.7\muKs}
\setsymbol{LFI:white:noise:sensitivity:LFI19:Rad:S:units}{555.6\muKs}
\setsymbol{LFI:white:noise:sensitivity:LFI20:Rad:S:units}{623.2\muKs}
\setsymbol{LFI:white:noise:sensitivity:LFI21:Rad:S:units}{564.1\muKs}
\setsymbol{LFI:white:noise:sensitivity:LFI22:Rad:S:units}{534.4\muKs}
\setsymbol{LFI:white:noise:sensitivity:LFI23:Rad:S:units}{542.4\muKs}
\setsymbol{LFI:white:noise:sensitivity:LFI24:Rad:S:units}{399.2\muKs}
\setsymbol{LFI:white:noise:sensitivity:LFI25:Rad:S:units}{392.6\muKs}
\setsymbol{LFI:white:noise:sensitivity:LFI26:Rad:S:units}{418.6\muKs}
\setsymbol{LFI:white:noise:sensitivity:LFI27:Rad:S:units}{302.9\muKs}
\setsymbol{LFI:white:noise:sensitivity:LFI28:Rad:S:units}{285.3\muKs}

\setsymbol{LFI:white:noise:sensitivity:LFI18:Rad:M}{512.0}
\setsymbol{LFI:white:noise:sensitivity:LFI19:Rad:M}{581.4}
\setsymbol{LFI:white:noise:sensitivity:LFI20:Rad:M}{590.8}
\setsymbol{LFI:white:noise:sensitivity:LFI21:Rad:M}{455.2}
\setsymbol{LFI:white:noise:sensitivity:LFI22:Rad:M}{492.0}
\setsymbol{LFI:white:noise:sensitivity:LFI23:Rad:M}{507.7}
\setsymbol{LFI:white:noise:sensitivity:LFI24:Rad:M}{462.2}
\setsymbol{LFI:white:noise:sensitivity:LFI25:Rad:M}{413.6}
\setsymbol{LFI:white:noise:sensitivity:LFI26:Rad:M}{478.6}
\setsymbol{LFI:white:noise:sensitivity:LFI27:Rad:M}{277.7}
\setsymbol{LFI:white:noise:sensitivity:LFI28:Rad:M}{312.3}
\setsymbol{LFI:white:noise:sensitivity:LFI18:Rad:S}{465.7}
\setsymbol{LFI:white:noise:sensitivity:LFI19:Rad:S}{555.6}
\setsymbol{LFI:white:noise:sensitivity:LFI20:Rad:S}{623.2}
\setsymbol{LFI:white:noise:sensitivity:LFI21:Rad:S}{564.1}
\setsymbol{LFI:white:noise:sensitivity:LFI22:Rad:S}{534.4}
\setsymbol{LFI:white:noise:sensitivity:LFI23:Rad:S}{542.4}
\setsymbol{LFI:white:noise:sensitivity:LFI24:Rad:S}{399.2}
\setsymbol{LFI:white:noise:sensitivity:LFI25:Rad:S}{392.6}
\setsymbol{LFI:white:noise:sensitivity:LFI26:Rad:S}{418.6}
\setsymbol{LFI:white:noise:sensitivity:LFI27:Rad:S}{302.9}
\setsymbol{LFI:white:noise:sensitivity:LFI28:Rad:S}{285.3}

\setsymbol{LFI:knee:frequency:70GHz:units}{29.5\mHz}
\setsymbol{LFI:knee:frequency:44GHz:units}{56.2\mHz}
\setsymbol{LFI:knee:frequency:30GHz:units}{113.7\mHz}

\setsymbol{LFI:knee:frequency:70GHz}{29.5}
\setsymbol{LFI:knee:frequency:44GHz}{56.2}
\setsymbol{LFI:knee:frequency:30GHz}{113.7}

\setsymbol{LFI:knee:frequency:LFI18:Rad:M:units}{16.3\mHz}
\setsymbol{LFI:knee:frequency:LFI19:Rad:M:units}{15.1\mHz}
\setsymbol{LFI:knee:frequency:LFI20:Rad:M:units}{18.7\mHz}
\setsymbol{LFI:knee:frequency:LFI21:Rad:M:units}{37.2\mHz}
\setsymbol{LFI:knee:frequency:LFI22:Rad:M:units}{12.7\mHz}
\setsymbol{LFI:knee:frequency:LFI23:Rad:M:units}{34.6\mHz}
\setsymbol{LFI:knee:frequency:LFI24:Rad:M:units}{46.2\mHz}
\setsymbol{LFI:knee:frequency:LFI25:Rad:M:units}{24.9\mHz}
\setsymbol{LFI:knee:frequency:LFI26:Rad:M:units}{67.6\mHz}
\setsymbol{LFI:knee:frequency:LFI27:Rad:M:units}{187.4\mHz}
\setsymbol{LFI:knee:frequency:LFI28:Rad:M:units}{122.2\mHz}
\setsymbol{LFI:knee:frequency:LFI18:Rad:S:units}{17.7\mHz}
\setsymbol{LFI:knee:frequency:LFI19:Rad:S:units}{22.0\mHz}
\setsymbol{LFI:knee:frequency:LFI20:Rad:S:units}{8.7\mHz}
\setsymbol{LFI:knee:frequency:LFI21:Rad:S:units}{25.9\mHz}
\setsymbol{LFI:knee:frequency:LFI22:Rad:S:units}{15.8\mHz}
\setsymbol{LFI:knee:frequency:LFI23:Rad:S:units}{129.8\mHz}
\setsymbol{LFI:knee:frequency:LFI24:Rad:S:units}{100.9\mHz}
\setsymbol{LFI:knee:frequency:LFI25:Rad:S:units}{38.9\mHz}
\setsymbol{LFI:knee:frequency:LFI26:Rad:S:units}{58.9\mHz}
\setsymbol{LFI:knee:frequency:LFI27:Rad:S:units}{104.4\mHz}
\setsymbol{LFI:knee:frequency:LFI28:Rad:S:units}{40.7\mHz}

\setsymbol{LFI:knee:frequency:LFI18:Rad:M}{16.3}
\setsymbol{LFI:knee:frequency:LFI19:Rad:M}{15.1}
\setsymbol{LFI:knee:frequency:LFI20:Rad:M}{18.7}
\setsymbol{LFI:knee:frequency:LFI21:Rad:M}{37.2}
\setsymbol{LFI:knee:frequency:LFI22:Rad:M}{12.7}
\setsymbol{LFI:knee:frequency:LFI23:Rad:M}{34.6}
\setsymbol{LFI:knee:frequency:LFI24:Rad:M}{46.2}
\setsymbol{LFI:knee:frequency:LFI25:Rad:M}{24.9}
\setsymbol{LFI:knee:frequency:LFI26:Rad:M}{67.6}
\setsymbol{LFI:knee:frequency:LFI27:Rad:M}{187.4}
\setsymbol{LFI:knee:frequency:LFI28:Rad:M}{122.2}
\setsymbol{LFI:knee:frequency:LFI18:Rad:S}{17.7}
\setsymbol{LFI:knee:frequency:LFI19:Rad:S}{22.0}
\setsymbol{LFI:knee:frequency:LFI20:Rad:S}{8.7}
\setsymbol{LFI:knee:frequency:LFI21:Rad:S}{25.9}
\setsymbol{LFI:knee:frequency:LFI22:Rad:S}{15.8}
\setsymbol{LFI:knee:frequency:LFI23:Rad:S}{129.8}
\setsymbol{LFI:knee:frequency:LFI24:Rad:S}{100.9}
\setsymbol{LFI:knee:frequency:LFI25:Rad:S}{38.9}
\setsymbol{LFI:knee:frequency:LFI26:Rad:S}{58.9}
\setsymbol{LFI:knee:frequency:LFI27:Rad:S}{104.4}
\setsymbol{LFI:knee:frequency:LFI28:Rad:S}{40.7}

\setsymbol{LFI:slope:70GHz:units}{$-1.03$\mHz}
\setsymbol{LFI:slope:44GHz:units}{$-0.89$\mHz}
\setsymbol{LFI:slope:30GHz:units}{$-0.87$\mHz}

\setsymbol{LFI:slope:70GHz}{$-1.03$}
\setsymbol{LFI:slope:44GHz}{$-0.89$}
\setsymbol{LFI:slope:30GHz}{$-0.87$}

\setsymbol{LFI:slope:LFI18:Rad:M:units}{$-1.04$\mHz}
\setsymbol{LFI:slope:LFI19:Rad:M:units}{$-1.09$\mHz}
\setsymbol{LFI:slope:LFI20:Rad:M:units}{$-0.69$\mHz}
\setsymbol{LFI:slope:LFI21:Rad:M:units}{$-1.56$\mHz}
\setsymbol{LFI:slope:LFI22:Rad:M:units}{$-1.01$\mHz}
\setsymbol{LFI:slope:LFI23:Rad:M:units}{$-0.96$\mHz}
\setsymbol{LFI:slope:LFI24:Rad:M:units}{$-0.83$\mHz}
\setsymbol{LFI:slope:LFI25:Rad:M:units}{$-0.91$\mHz}
\setsymbol{LFI:slope:LFI26:Rad:M:units}{$-0.95$\mHz}
\setsymbol{LFI:slope:LFI27:Rad:M:units}{$-0.87$\mHz}
\setsymbol{LFI:slope:LFI28:Rad:M:units}{$-0.88$\mHz}
\setsymbol{LFI:slope:LFI18:Rad:S:units}{$-1.15$\mHz}
\setsymbol{LFI:slope:LFI19:Rad:S:units}{$-1.00$\mHz}
\setsymbol{LFI:slope:LFI20:Rad:S:units}{$-0.95$\mHz}
\setsymbol{LFI:slope:LFI21:Rad:S:units}{$-0.92$\mHz}
\setsymbol{LFI:slope:LFI22:Rad:S:units}{$-1.01$\mHz}
\setsymbol{LFI:slope:LFI23:Rad:S:units}{$-0.95$\mHz}
\setsymbol{LFI:slope:LFI24:Rad:S:units}{$-0.73$\mHz}
\setsymbol{LFI:slope:LFI25:Rad:S:units}{$-1.16$\mHz}
\setsymbol{LFI:slope:LFI26:Rad:S:units}{$-0.79$\mHz}
\setsymbol{LFI:slope:LFI27:Rad:S:units}{$-0.82$\mHz}
\setsymbol{LFI:slope:LFI28:Rad:S:units}{$-0.91$\mHz}

\setsymbol{LFI:slope:LFI18:Rad:M}{$-1.04$}
\setsymbol{LFI:slope:LFI19:Rad:M}{$-1.09$}
\setsymbol{LFI:slope:LFI20:Rad:M}{$-0.69$}
\setsymbol{LFI:slope:LFI21:Rad:M}{$-1.56$}
\setsymbol{LFI:slope:LFI22:Rad:M}{$-1.01$}
\setsymbol{LFI:slope:LFI23:Rad:M}{$-0.96$}
\setsymbol{LFI:slope:LFI24:Rad:M}{$-0.83$}
\setsymbol{LFI:slope:LFI25:Rad:M}{$-0.91$}
\setsymbol{LFI:slope:LFI26:Rad:M}{$-0.95$}
\setsymbol{LFI:slope:LFI27:Rad:M}{$-0.87$}
\setsymbol{LFI:slope:LFI28:Rad:M}{$-0.88$}
\setsymbol{LFI:slope:LFI18:Rad:S}{$-1.15$}
\setsymbol{LFI:slope:LFI19:Rad:S}{$-1.00$}
\setsymbol{LFI:slope:LFI20:Rad:S}{$-0.95$}
\setsymbol{LFI:slope:LFI21:Rad:S}{$-0.92$}
\setsymbol{LFI:slope:LFI22:Rad:S}{$-1.01$}
\setsymbol{LFI:slope:LFI23:Rad:S}{$-0.95$}
\setsymbol{LFI:slope:LFI24:Rad:S}{$-0.73$}
\setsymbol{LFI:slope:LFI25:Rad:S}{$-1.16$}
\setsymbol{LFI:slope:LFI26:Rad:S}{$-0.79$}
\setsymbol{LFI:slope:LFI27:Rad:S}{$-0.82$}
\setsymbol{LFI:slope:LFI28:Rad:S}{$-0.91$}

\setsymbol{LFI:FWHM:70GHz:units}{13\parcm01}
\setsymbol{LFI:FWHM:44GHz:units}{27\parcm92}
\setsymbol{LFI:FWHM:30GHz:units}{32\parcm65}

\setsymbol{LFI:FWHM:70GHz}{13.01}
\setsymbol{LFI:FWHM:44GHz}{27.92}
\setsymbol{LFI:FWHM:30GHz}{32.65}

\setsymbol{LFI:FWHM:LFI18:units}{13\parcm39}
\setsymbol{LFI:FWHM:LFI19:units}{13\parcm01}
\setsymbol{LFI:FWHM:LFI20:units}{12\parcm75}
\setsymbol{LFI:FWHM:LFI21:units}{12\parcm74}
\setsymbol{LFI:FWHM:LFI22:units}{12\parcm87}
\setsymbol{LFI:FWHM:LFI23:units}{13\parcm27}
\setsymbol{LFI:FWHM:LFI24:units}{22\parcm98}
\setsymbol{LFI:FWHM:LFI25:units}{30\parcm46}
\setsymbol{LFI:FWHM:LFI26:units}{30\parcm31}
\setsymbol{LFI:FWHM:LFI27:units}{32\parcm65}
\setsymbol{LFI:FWHM:LFI28:units}{32\parcm66}

\setsymbol{LFI:FWHM:LFI18}{13.39}
\setsymbol{LFI:FWHM:LFI19}{13.01}
\setsymbol{LFI:FWHM:LFI20}{12.75}
\setsymbol{LFI:FWHM:LFI21}{12.74}
\setsymbol{LFI:FWHM:LFI22}{12.87}
\setsymbol{LFI:FWHM:LFI23}{13.27}
\setsymbol{LFI:FWHM:LFI24}{22.98}
\setsymbol{LFI:FWHM:LFI25}{30.46}
\setsymbol{LFI:FWHM:LFI26}{30.31}
\setsymbol{LFI:FWHM:LFI27}{32.65}
\setsymbol{LFI:FWHM:LFI28}{32.66}

\setsymbol{LFI:FWHM:uncertainty:LFI18:units}{0.170\arcm}
\setsymbol{LFI:FWHM:uncertainty:LFI19:units}{0.174\arcm}
\setsymbol{LFI:FWHM:uncertainty:LFI20:units}{0.170\arcm}
\setsymbol{LFI:FWHM:uncertainty:LFI21:units}{0.156\arcm}
\setsymbol{LFI:FWHM:uncertainty:LFI22:units}{0.164\arcm}
\setsymbol{LFI:FWHM:uncertainty:LFI23:units}{0.171\arcm}
\setsymbol{LFI:FWHM:uncertainty:LFI24:units}{0.652\arcm}
\setsymbol{LFI:FWHM:uncertainty:LFI25:units}{1.075\arcm}
\setsymbol{LFI:FWHM:uncertainty:LFI26:units}{1.131\arcm}
\setsymbol{LFI:FWHM:uncertainty:LFI27:units}{1.266\arcm}
\setsymbol{LFI:FWHM:uncertainty:LFI28:units}{1.287\arcm}

\setsymbol{LFI:FWHM:uncertainty:LFI18}{0.170}
\setsymbol{LFI:FWHM:uncertainty:LFI19}{0.174}
\setsymbol{LFI:FWHM:uncertainty:LFI20}{0.170}
\setsymbol{LFI:FWHM:uncertainty:LFI21}{0.156}
\setsymbol{LFI:FWHM:uncertainty:LFI22}{0.164}
\setsymbol{LFI:FWHM:uncertainty:LFI23}{0.171}
\setsymbol{LFI:FWHM:uncertainty:LFI24}{0.652}
\setsymbol{LFI:FWHM:uncertainty:LFI25}{1.075}
\setsymbol{LFI:FWHM:uncertainty:LFI26}{1.131}
\setsymbol{LFI:FWHM:uncertainty:LFI27}{1.266}
\setsymbol{LFI:FWHM:uncertainty:LFI28}{1.287}

\setsymbol{HFI:center:frequency:100GHz:units}{100\,GHz}
\setsymbol{HFI:center:frequency:143GHz:units}{143\,GHz}
\setsymbol{HFI:center:frequency:217GHz:units}{217\,GHz}
\setsymbol{HFI:center:frequency:353GHz:units}{353\,GHz}
\setsymbol{HFI:center:frequency:545GHz:units}{545\,GHz}
\setsymbol{HFI:center:frequency:857GHz:units}{857\,GHz}

\setsymbol{HFI:center:frequency:100GHz}{100}
\setsymbol{HFI:center:frequency:143GHz}{143}
\setsymbol{HFI:center:frequency:217GHz}{217}
\setsymbol{HFI:center:frequency:353GHz}{353}
\setsymbol{HFI:center:frequency:545GHz}{545}
\setsymbol{HFI:center:frequency:857GHz}{857}

\setsymbol{HFI:Ndetectors:100GHz}{8}
\setsymbol{HFI:Ndetectors:143GHz}{11}
\setsymbol{HFI:Ndetectors:217GHz}{12}
\setsymbol{HFI:Ndetectors:353GHz}{12}
\setsymbol{HFI:Ndetectors:545GHz}{3}
\setsymbol{HFI:Ndetectors:857GHz}{4}

\setsymbol{HFI:FWHM:Maps:100GHz:units}{9\parcm88}
\setsymbol{HFI:FWHM:Maps:143GHz:units}{7\parcm18}
\setsymbol{HFI:FWHM:Maps:217GHz:units}{4\parcm87}
\setsymbol{HFI:FWHM:Maps:353GHz:units}{4\parcm65}
\setsymbol{HFI:FWHM:Maps:545GHz:units}{4\parcm72}
\setsymbol{HFI:FWHM:Maps:857GHz:units}{4\parcm39}
\setsymbol{HFI:FWHM:Maps:100GHz}{9.88}
\setsymbol{HFI:FWHM:Maps:143GHz}{7.18}
\setsymbol{HFI:FWHM:Maps:217GHz}{4.87}
\setsymbol{HFI:FWHM:Maps:353GHz}{4.65}
\setsymbol{HFI:FWHM:Maps:545GHz}{4.72}
\setsymbol{HFI:FWHM:Maps:857GHz}{4.39}

\setsymbol{HFI:beam:ellipticity:Maps:100GHz}{1.15}
\setsymbol{HFI:beam:ellipticity:Maps:143GHz}{1.01}
\setsymbol{HFI:beam:ellipticity:Maps:217GHz}{1.06}
\setsymbol{HFI:beam:ellipticity:Maps:353GHz}{1.05}
\setsymbol{HFI:beam:ellipticity:Maps:545GHz}{1.14}
\setsymbol{HFI:beam:ellipticity:Maps:857GHz}{1.19}

\setsymbol{HFI:FWHM:Mars:100GHz:units}{9\parcm37}
\setsymbol{HFI:FWHM:Mars:143GHz:units}{7\parcm04}
\setsymbol{HFI:FWHM:Mars:217GHz:units}{4\parcm68}
\setsymbol{HFI:FWHM:Mars:353GHz:units}{4\parcm43}
\setsymbol{HFI:FWHM:Mars:545GHz:units}{3\parcm80}
\setsymbol{HFI:FWHM:Mars:857GHz:units}{3\parcm67}

\setsymbol{HFI:FWHM:Mars:100GHz}{9.37}
\setsymbol{HFI:FWHM:Mars:143GHz}{7.04}
\setsymbol{HFI:FWHM:Mars:217GHz}{4.68}
\setsymbol{HFI:FWHM:Mars:353GHz}{4.43}
\setsymbol{HFI:FWHM:Mars:545GHz}{3.80}
\setsymbol{HFI:FWHM:Mars:857GHz}{3.67}

\setsymbol{HFI:beam:ellipticity:Mars:100GHz}{1.18}
\setsymbol{HFI:beam:ellipticity:Mars:143GHz}{1.03}
\setsymbol{HFI:beam:ellipticity:Mars:217GHz}{1.14}
\setsymbol{HFI:beam:ellipticity:Mars:353GHz}{1.09}
\setsymbol{HFI:beam:ellipticity:Mars:545GHz}{1.25}
\setsymbol{HFI:beam:ellipticity:Mars:857GHz}{1.03}

\setsymbol{HFI:CMB:relative:calibration:100GHz}{$\lsim 1\%$}
\setsymbol{HFI:CMB:relative:calibration:143GHz}{$\lsim 1\%$}
\setsymbol{HFI:CMB:relative:calibration:217GHz}{$\lsim 1\%$}
\setsymbol{HFI:CMB:relative:calibration:353GHz}{$\lsim 1\%$}
\setsymbol{HFI:CMB:relative:calibration:545GHz}{}
\setsymbol{HFI:CMB:relative:calibration:857GHz}{}

\setsymbol{HFI:CMB:absolute:calibration:100GHz}{$\lsim 2\%$}
\setsymbol{HFI:CMB:absolute:calibration:143GHz}{$\lsim 2\%$}
\setsymbol{HFI:CMB:absolute:calibration:217GHz}{$\lsim 2\%$}
\setsymbol{HFI:CMB:absolute:calibration:353GHz}{$\lsim 2\%$}
\setsymbol{HFI:CMB:absolute:calibration:545GHz}{}
\setsymbol{HFI:CMB:absolute:calibration:857GHz}{}

\setsymbol{HFI:FIRAS:gain:calibration:accuracy:statistical:100GHz}{}
\setsymbol{HFI:FIRAS:gain:calibration:accuracy:statistical:143GHz}{}
\setsymbol{HFI:FIRAS:gain:calibration:accuracy:statistical:217GHz}{}
\setsymbol{HFI:FIRAS:gain:calibration:accuracy:statistical:353GHz}{2.5\%}
\setsymbol{HFI:FIRAS:gain:calibration:accuracy:statistical:545GHz}{1\%}
\setsymbol{HFI:FIRAS:gain:calibration:accuracy:statistical:857GHz}{0.5\%}

\setsymbol{HFI:FIRAS:gain:calibration:accuracy:systematic:100GHz}{}
\setsymbol{HFI:FIRAS:gain:calibration:accuracy:systematic:143GHz}{}
\setsymbol{HFI:FIRAS:gain:calibration:accuracy:systematic:217GHz}{}
\setsymbol{HFI:FIRAS:gain:calibration:accuracy:systematic:353GHz}{}
\setsymbol{HFI:FIRAS:gain:calibration:accuracy:systematic:545GHz}{7\%}
\setsymbol{HFI:FIRAS:gain:calibration:accuracy:systematic:857GHz}{7\%}

\setsymbol{HFI:FIRAS:zero:point:accuracy:100GHz:units}{0.8\MJysr}
\setsymbol{HFI:FIRAS:zero:point:accuracy:143GHz:units}{}
\setsymbol{HFI:FIRAS:zero:point:accuracy:217GHz:units}{}
\setsymbol{HFI:FIRAS:zero:point:accuracy:353GHz:units}{1.4\MJysr}
\setsymbol{HFI:FIRAS:zero:point:accuracy:545GHz:units}{2.2\MJysr}
\setsymbol{HFI:FIRAS:zero:point:accuracy:857GHz:units}{1.7\MJysr}

\setsymbol{HFI:FIRAS:zero:point:accuracy:100GHz}{0.8}
\setsymbol{HFI:FIRAS:zero:point:accuracy:143GHz}{}
\setsymbol{HFI:FIRAS:zero:point:accuracy:217GHz}{}
\setsymbol{HFI:FIRAS:zero:point:accuracy:353GHz}{1.4}
\setsymbol{HFI:FIRAS:zero:point:accuracy:545GHz}{2.2}
\setsymbol{HFI:FIRAS:zero:point:accuracy:857GHz}{1.7}

\setsymbol{HFI:unit:conversion:100GHz:units}{0.2415\MJysrmK}
\setsymbol{HFI:unit:conversion:143GHz:units}{0.3694\MJysrmK}
\setsymbol{HFI:unit:conversion:217GHz:units}{0.4811\MJysrmK}
\setsymbol{HFI:unit:conversion:353GHz:units}{0.2883\MJysrmK}
\setsymbol{HFI:unit:conversion:545GHz:units}{0.05826\MJysrmK}
\setsymbol{HFI:unit:conversion:857GHz:units}{0.002238\MJysrmK}

\setsymbol{HFI:unit:conversion:100GHz}{0.2415}
\setsymbol{HFI:unit:conversion:143GHz}{0.3694}
\setsymbol{HFI:unit:conversion:217GHz}{0.4811}
\setsymbol{HFI:unit:conversion:353GHz}{0.2883}
\setsymbol{HFI:unit:conversion:545GHz}{0.05826}
\setsymbol{HFI:unit:conversion:857GHz}{0.002238}

\setsymbol{HFI:colour:correction:alpha=-2:V1.01:100GHz}{0.9893}
\setsymbol{HFI:colour:correction:alpha=-2:V1.01:143GHz}{0.9759}
\setsymbol{HFI:colour:correction:alpha=-2:V1.01:217GHz}{1.0007}
\setsymbol{HFI:colour:correction:alpha=-2:V1.01:353GHz}{1.0028}
\setsymbol{HFI:colour:correction:alpha=-2:V1.01:545GHz}{1.0019}
\setsymbol{HFI:colour:correction:alpha=-2:V1.01:857GHz}{0.9889}

\setsymbol{HFI:colour:correction:alpha=0:V1.01:100GHz}{1.0008}
\setsymbol{HFI:colour:correction:alpha=0:V1.01:143GHz}{1.0148}
\setsymbol{HFI:colour:correction:alpha=0:V1.01:217GHz}{0.9909}
\setsymbol{HFI:colour:correction:alpha=0:V1.01:353GHz}{0.9888}
\setsymbol{HFI:colour:correction:alpha=0:V1.01:545GHz}{0.9878}
\setsymbol{HFI:colour:correction:alpha=0:V1.01:857GHz}{1.0014}

\providecommand{\sorthelp}[1]{}

\def\setsymbol#1#2{\expandafter\def\csname #1\endcsname{#2}}
\def\getsymbol#1{\csname #1\endcsname}

\def\Planck{{\it Planck\/}}

\def\allearlypapers{\nocite{planck2011-1.1, planck2011-1.3, planck2011-1.4, planck2011-1.5, planck2011-1.6, planck2011-1.7, planck2011-1.10, planck2011-1.10sup, planck2011-5.1a, planck2011-5.1b, planck2011-5.2a, planck2011-5.2b, planck2011-5.2c, planck2011-6.1, planck2011-6.2, planck2011-6.3a, planck2011-6.4a, planck2011-6.4b, planck2011-6.6, planck2011-7.0, planck2011-7.2, planck2011-7.3, planck2011-7.7a, planck2011-7.7b, planck2011-7.12, planck2011-7.13}}

\newbox\tablebox    \newdimen\tablewidth
\def\leaderfil{\leaders\hbox to 5pt{\hss.\hss}\hfil}
\def\endPlancktable{\tablewidth=\columnwidth 
    $$\hss\copy\tablebox\hss$$
    \vskip-\lastskip\vskip -2pt}
\def\endPlancktablewide{\tablewidth=\textwidth 
    $$\hss\copy\tablebox\hss$$
    \vskip-\lastskip\vskip -2pt}
\def\tablenote#1 #2\par{\begingroup \parindent=0.8em
    \abovedisplayshortskip=0pt\belowdisplayshortskip=0pt
    \noindent
    $$\hss\vbox{\hsize\tablewidth \hangindent=\parindent \hangafter=1 \noindent
    \hbox to \parindent{\sup{\rm #1}\hss}\strut#2\strut\par}\hss$$
    \endgroup}
\def\doubleline{\vskip 3pt\hrule \vskip 1.5pt \hrule \vskip 5pt}

\def\L2{\ifmmode L_2\else $L_2$\fi}

\def\DeltaT{\ifmmode \Delta T\else $\Delta T$\fi}
\def\deltat{\ifmmode \Delta t\else $\Delta t$\fi}
\def\fknee{\ifmmode f_{\rm knee}\else $f_{\rm knee}$\fi}
\def\Fmax{\ifmmode F_{\rm max}\else $F_{\rm max}$\fi}
\def\solar{\ifmmode{\rm M}_{\mathord\odot}\else${\rm M}_{\mathord\odot}$\fi}

\def\inv{\ifmmode^{-1}\else$^{-1}$\fi}
\def\mo{\ifmmode^{-1}\else$^{-1}$\fi}
\def\sup#1{\ifmmode ^{\rm #1}\else $^{\rm #1}$\fi}
\def\expo#1{\ifmmode \times 10^{#1}\else $\times 10^{#1}$\fi}
\def\,{\thinspace}
\def\lsim{\mathrel{\raise .4ex\hbox{\rlap{$<$}\lower 1.2ex\hbox{$\sim$}}}}
\def\gsim{\mathrel{\raise .4ex\hbox{\rlap{$>$}\lower 1.2ex\hbox{$\sim$}}}}

\def\simprop{\mathrel{\raise .4ex\hbox{\rlap{$\propto$}\lower 1.2ex\hbox{$\sim$}}}}
\def\deg{\ifmmode^\circ\else$^\circ$\fi}
\def\pdeg{\ifmmode $\setbox0=\hbox{$^{\circ}$}\rlap{\hskip.11\wd0 .}$^{\circ}
          \else \setbox0=\hbox{$^{\circ}$}\rlap{\hskip.11\wd0 .}$^{\circ}$\fi}
\def\arcs{\ifmmode {^{\scriptstyle\prime\prime}}
          \else $^{\scriptstyle\prime\prime}$\fi}
\def\arcm{\ifmmode {^{\scriptstyle\prime}}
          \else $^{\scriptstyle\prime}$\fi}
\newdimen\sa  \newdimen\sb
\def\parcs{\sa=.07em \sb=.03em
     \ifmmode \hbox{\rlap{.}}^{\scriptstyle\prime\kern -\sb\prime}\hbox{\kern -\sa}
     \else \rlap{.}$^{\scriptstyle\prime\kern -\sb\prime}$\kern -\sa\fi}
\def\parcm{\sa=.08em \sb=.03em
     \ifmmode \hbox{\rlap{.}\kern\sa}^{\scriptstyle\prime}\hbox{\kern-\sb}
     \else \rlap{.}\kern\sa$^{\scriptstyle\prime}$\kern-\sb\fi}
\def\ra[#1 #2 #3.#4]{#1\sup{h}#2\sup{m}#3\sup{s}\llap.#4}
\def\dec[#1 #2 #3.#4]{#1\deg#2\arcm#3\arcs\llap.#4}
\def\deco[#1 #2 #3]{#1\deg#2\arcm#3\arcs}
\def\rra[#1 #2]{#1\sup{h}#2\sup{m}}

\def\dots{\relax\ifmmode \ldots\else $\ldots$\fi}
\def\WHzsr{\ifmmode $W\,Hz\mo\,sr\mo$\else W\,Hz\mo\,sr\mo\fi}
\def\mHz{\ifmmode $\,mHz$\else \,mHz\fi}
\def\GHz{\ifmmode $\,GHz$\else \,GHz\fi}
\def\mKs{\ifmmode $\,mK\,s$^{1/2}\else \,mK\,s$^{1/2}$\fi}
\def\muKs{\ifmmode \,\mu$K\,s$^{1/2}\else \,$\mu$K\,s$^{1/2}$\fi}
\def\muKRJs{\ifmmode \,\mu$K$_{\rm RJ}$\,s$^{1/2}\else \,$\mu$K$_{\rm RJ}$\,s$^{1/2}$\fi}
\def\muKCMBs{\ifmmode \,\mu$K$_{\rm CMB}$\,s$^{1/2}\else \,$\mu$K$_{\rm CMB}$\,s$^{1/2}$\fi} % A. Mennella, Sept 2011
\def\muKHz{\ifmmode \,\mu$K\,Hz$^{-1/2}\else \,$\mu$K\,Hz$^{-1/2}$\fi}
\def\MJysr{\ifmmode \,$MJy\,sr\mo$\else \,MJy\,sr\mo\fi}
\def\MJysrmK{\ifmmode \,$MJy\,sr\mo$\,mK$_{\rm CMB}\mo\else \,MJy\,sr\mo\,mK$_{\rm CMB}\mo$\fi}
\def\microns{\ifmmode \,\mu$m$\else \,$\mu$m\fi}

\def\muK{\ifmmode \,\mu$K$\else \,$\mu$\hbox{K}\fi}
\def\muKRJ{\ifmmode \,\mu$K$_{\rm RJ}$\else \,$\mu$\hbox{K}$_{\rm RJ}$\fi}    % A. Mennella, Sept 2011
\def\muKCMB{\ifmmode \,\mu$K$_{\rm CMB}$\else \,$\mu$\hbox{K}$_{\rm CMB}$\fi} % A. Mennella, Sept 2011
\def\microK{\ifmmode \,\mu$K$\else \,$\mu$\hbox{K}\fi}
\def\muW{\ifmmode \,\mu$W$\else \,$\mu$\hbox{W}\fi}
\def\kms{\ifmmode $\,km\,s$^{-1}\else \,km\,s$^{-1}$\fi}
\def\kmsMpc{\ifmmode $\,\kms\,Mpc\mo$\else \,\kms\,Mpc\mo\fi}
\setsymbol{LFI:center:frequency:70GHz:units}{$70.4$\,GHz}
\setsymbol{LFI:center:frequency:44GHz:units}{$44.1$\,GHz}
\setsymbol{LFI:center:frequency:30GHz:units}{$28.4$\,GHz}

\setsymbol{LFI:center:frequency:70GHz}{$70.4$}
\setsymbol{LFI:center:frequency:44GHz}{$44.1$}
\setsymbol{LFI:center:frequency:30GHz}{$28.4$}

\setsymbol{LFI:white:noise:sensitivity:LFI18:Rad:M:units}{513.0\muKs}
\setsymbol{LFI:white:noise:sensitivity:LFI19:Rad:M:units}{579.6\muKs}
\setsymbol{LFI:white:noise:sensitivity:LFI20:Rad:M:units}{587.3\muKs}
\setsymbol{LFI:white:noise:sensitivity:LFI21:Rad:M:units}{451.0\muKs}
\setsymbol{LFI:white:noise:sensitivity:LFI22:Rad:M:units}{490.8\muKs}
\setsymbol{LFI:white:noise:sensitivity:LFI23:Rad:M:units}{504.3\muKs}
\setsymbol{LFI:white:noise:sensitivity:LFI24:Rad:M:units}{463.0\muKs}
\setsymbol{LFI:white:noise:sensitivity:LFI25:Rad:M:units}{415.3\muKs}
\setsymbol{LFI:white:noise:sensitivity:LFI26:Rad:M:units}{483.0\muKs}
\setsymbol{LFI:white:noise:sensitivity:LFI27:Rad:M:units}{281.5\muKs}
\setsymbol{LFI:white:noise:sensitivity:LFI28:Rad:M:units}{317.7\muKs}
\setsymbol{LFI:white:noise:sensitivity:LFI18:Rad:S:units}{467.2\muKs}
\setsymbol{LFI:white:noise:sensitivity:LFI19:Rad:S:units}{555.0\muKs}
\setsymbol{LFI:white:noise:sensitivity:LFI20:Rad:S:units}{620.5\muKs}
\setsymbol{LFI:white:noise:sensitivity:LFI21:Rad:S:units}{560.1\muKs}
\setsymbol{LFI:white:noise:sensitivity:LFI22:Rad:S:units}{531.3\muKs}
\setsymbol{LFI:white:noise:sensitivity:LFI23:Rad:S:units}{539.7\muKs}
\setsymbol{LFI:white:noise:sensitivity:LFI24:Rad:S:units}{400.7\muKs}
\setsymbol{LFI:white:noise:sensitivity:LFI25:Rad:S:units}{395.4\muKs}
\setsymbol{LFI:white:noise:sensitivity:LFI26:Rad:S:units}{423.2\muKs}
\setsymbol{LFI:white:noise:sensitivity:LFI27:Rad:S:units}{303.2\muKs}
\setsymbol{LFI:white:noise:sensitivity:LFI28:Rad:S:units}{286.5\muKs}

\setsymbol{LFI:white:noise:sensitivity:uncertainty:LFI18:Rad:M:units}{2.1\muKs} 
\setsymbol{LFI:white:noise:sensitivity:uncertainty:LFI19:Rad:M:units}{2.2\muKs} 
\setsymbol{LFI:white:noise:sensitivity:uncertainty:LFI20:Rad:M:units}{2.1\muKs} 
\setsymbol{LFI:white:noise:sensitivity:uncertainty:LFI21:Rad:M:units}{1.7\muKs} 
\setsymbol{LFI:white:noise:sensitivity:uncertainty:LFI22:Rad:M:units}{1.5\muKs} 
\setsymbol{LFI:white:noise:sensitivity:uncertainty:LFI23:Rad:M:units}{1.8\muKs} 
\setsymbol{LFI:white:noise:sensitivity:uncertainty:LFI24:Rad:M:units}{1.4\muKs} 
\setsymbol{LFI:white:noise:sensitivity:uncertainty:LFI25:Rad:M:units}{1.5\muKs} 
\setsymbol{LFI:white:noise:sensitivity:uncertainty:LFI26:Rad:M:units}{1.9\muKs} 
\setsymbol{LFI:white:noise:sensitivity:uncertainty:LFI27:Rad:M:units}{2.1\muKs} 
\setsymbol{LFI:white:noise:sensitivity:uncertainty:LFI28:Rad:M:units}{2.4\muKs} 
\setsymbol{LFI:white:noise:sensitivity:uncertainty:LFI18:Rad:S:units}{2.3\muKs} 
\setsymbol{LFI:white:noise:sensitivity:uncertainty:LFI19:Rad:S:units}{2.2\muKs} 
\setsymbol{LFI:white:noise:sensitivity:uncertainty:LFI20:Rad:S:units}{2.7\muKs} 
\setsymbol{LFI:white:noise:sensitivity:uncertainty:LFI21:Rad:S:units}{2.0\muKs} 
\setsymbol{LFI:white:noise:sensitivity:uncertainty:LFI22:Rad:S:units}{2.3\muKs} 
\setsymbol{LFI:white:noise:sensitivity:uncertainty:LFI23:Rad:S:units}{1.8\muKs} 
\setsymbol{LFI:white:noise:sensitivity:uncertainty:LFI24:Rad:S:units}{1.3\muKs} 
\setsymbol{LFI:white:noise:sensitivity:uncertainty:LFI25:Rad:S:units}{2.9\muKs} 
\setsymbol{LFI:white:noise:sensitivity:uncertainty:LFI26:Rad:S:units}{2.5\muKs} 
\setsymbol{LFI:white:noise:sensitivity:uncertainty:LFI27:Rad:S:units}{1.8\muKs} 
\setsymbol{LFI:white:noise:sensitivity:uncertainty:LFI28:Rad:S:units}{2.3\muKs} 

\setsymbol{LFI:white:noise:sensitivity:LFI18:Rad:M}{513.0}
\setsymbol{LFI:white:noise:sensitivity:LFI19:Rad:M}{579.6}
\setsymbol{LFI:white:noise:sensitivity:LFI20:Rad:M}{587.3}
\setsymbol{LFI:white:noise:sensitivity:LFI21:Rad:M}{451.0}
\setsymbol{LFI:white:noise:sensitivity:LFI22:Rad:M}{490.8}
\setsymbol{LFI:white:noise:sensitivity:LFI23:Rad:M}{504.3}
\setsymbol{LFI:white:noise:sensitivity:LFI24:Rad:M}{463.0}
\setsymbol{LFI:white:noise:sensitivity:LFI25:Rad:M}{415.3}
\setsymbol{LFI:white:noise:sensitivity:LFI26:Rad:M}{483.0}
\setsymbol{LFI:white:noise:sensitivity:LFI27:Rad:M}{281.5}
\setsymbol{LFI:white:noise:sensitivity:LFI28:Rad:M}{317.7}
\setsymbol{LFI:white:noise:sensitivity:LFI18:Rad:S}{467.2}
\setsymbol{LFI:white:noise:sensitivity:LFI19:Rad:S}{555.0}
\setsymbol{LFI:white:noise:sensitivity:LFI20:Rad:S}{620.5}
\setsymbol{LFI:white:noise:sensitivity:LFI21:Rad:S}{560.1}
\setsymbol{LFI:white:noise:sensitivity:LFI22:Rad:S}{531.3}
\setsymbol{LFI:white:noise:sensitivity:LFI23:Rad:S}{539.7}
\setsymbol{LFI:white:noise:sensitivity:LFI24:Rad:S}{400.7}
\setsymbol{LFI:white:noise:sensitivity:LFI25:Rad:S}{395.4}
\setsymbol{LFI:white:noise:sensitivity:LFI26:Rad:S}{423.2}
\setsymbol{LFI:white:noise:sensitivity:LFI27:Rad:S}{303.2}
\setsymbol{LFI:white:noise:sensitivity:LFI28:Rad:S}{286.5}

\setsymbol{LFI:white:noise:sensitivity:uncertainty:LFI18:Rad:M}{2.1} 
\setsymbol{LFI:white:noise:sensitivity:uncertainty:LFI19:Rad:M}{2.2} 
\setsymbol{LFI:white:noise:sensitivity:uncertainty:LFI20:Rad:M}{2.1} 
\setsymbol{LFI:white:noise:sensitivity:uncertainty:LFI21:Rad:M}{1.7} 
\setsymbol{LFI:white:noise:sensitivity:uncertainty:LFI22:Rad:M}{1.5} 
\setsymbol{LFI:white:noise:sensitivity:uncertainty:LFI23:Rad:M}{1.8} 
\setsymbol{LFI:white:noise:sensitivity:uncertainty:LFI24:Rad:M}{1.4} 
\setsymbol{LFI:white:noise:sensitivity:uncertainty:LFI25:Rad:M}{1.5} 
\setsymbol{LFI:white:noise:sensitivity:uncertainty:LFI26:Rad:M}{1.9} 
\setsymbol{LFI:white:noise:sensitivity:uncertainty:LFI27:Rad:M}{2.1} 
\setsymbol{LFI:white:noise:sensitivity:uncertainty:LFI28:Rad:M}{2.4} 
\setsymbol{LFI:white:noise:sensitivity:uncertainty:LFI18:Rad:S}{2.3} 
\setsymbol{LFI:white:noise:sensitivity:uncertainty:LFI19:Rad:S}{2.2} 
\setsymbol{LFI:white:noise:sensitivity:uncertainty:LFI20:Rad:S}{2.7} 
\setsymbol{LFI:white:noise:sensitivity:uncertainty:LFI21:Rad:S}{2.0} 
\setsymbol{LFI:white:noise:sensitivity:uncertainty:LFI22:Rad:S}{2.3} 
\setsymbol{LFI:white:noise:sensitivity:uncertainty:LFI23:Rad:S}{1.8} 
\setsymbol{LFI:white:noise:sensitivity:uncertainty:LFI24:Rad:S}{1.3} 
\setsymbol{LFI:white:noise:sensitivity:uncertainty:LFI25:Rad:S}{2.9} 
\setsymbol{LFI:white:noise:sensitivity:uncertainty:LFI26:Rad:S}{2.5} 
\setsymbol{LFI:white:noise:sensitivity:uncertainty:LFI27:Rad:S}{1.8} 
\setsymbol{LFI:white:noise:sensitivity:uncertainty:LFI28:Rad:S}{2.3}

\setsymbol{LFI:knee:frequency:LFI18:Rad:M:units}{14.8\mHz}
\setsymbol{LFI:knee:frequency:LFI19:Rad:M:units}{11.7\mHz}
\setsymbol{LFI:knee:frequency:LFI20:Rad:M:units}{8.0\mHz}
\setsymbol{LFI:knee:frequency:LFI21:Rad:M:units}{37.9\mHz}
\setsymbol{LFI:knee:frequency:LFI22:Rad:M:units}{9.7\mHz}
\setsymbol{LFI:knee:frequency:LFI23:Rad:M:units}{29.7\mHz}
\setsymbol{LFI:knee:frequency:LFI24:Rad:M:units}{26.8\mHz}
\setsymbol{LFI:knee:frequency:LFI25:Rad:M:units}{20.1\mHz}
\setsymbol{LFI:knee:frequency:LFI26:Rad:M:units}{64.4\mHz}
\setsymbol{LFI:knee:frequency:LFI27:Rad:M:units}{174.5\mHz}
\setsymbol{LFI:knee:frequency:LFI28:Rad:M:units}{130.1\mHz}
\setsymbol{LFI:knee:frequency:LFI18:Rad:S:units}{17.8\mHz}
\setsymbol{LFI:knee:frequency:LFI19:Rad:S:units}{13.7\mHz}
\setsymbol{LFI:knee:frequency:LFI20:Rad:S:units}{5.7\mHz}
\setsymbol{LFI:knee:frequency:LFI21:Rad:S:units}{13.3\mHz}
\setsymbol{LFI:knee:frequency:LFI22:Rad:S:units}{14.8\mHz}
\setsymbol{LFI:knee:frequency:LFI23:Rad:S:units}{59.0\mHz}
\setsymbol{LFI:knee:frequency:LFI24:Rad:S:units}{88.3\mHz}
\setsymbol{LFI:knee:frequency:LFI25:Rad:S:units}{46.4\mHz}
\setsymbol{LFI:knee:frequency:LFI26:Rad:S:units}{68.2\mHz}
\setsymbol{LFI:knee:frequency:LFI27:Rad:S:units}{108.8\mHz}
\setsymbol{LFI:knee:frequency:LFI28:Rad:S:units}{43.1\mHz}

\setsymbol{LFI:knee:frequency:uncertainty:LFI18:Rad:M:units}{2.5\mHz}
\setsymbol{LFI:knee:frequency:uncertainty:LFI19:Rad:M:units}{1.2\mHz}
\setsymbol{LFI:knee:frequency:uncertainty:LFI20:Rad:M:units}{1.9\mHz}
\setsymbol{LFI:knee:frequency:uncertainty:LFI21:Rad:M:units}{5.2\mHz}
\setsymbol{LFI:knee:frequency:uncertainty:LFI22:Rad:M:units}{2.3\mHz}
\setsymbol{LFI:knee:frequency:uncertainty:LFI23:Rad:M:units}{1.1\mHz}
\setsymbol{LFI:knee:frequency:uncertainty:LFI24:Rad:M:units}{1.3\mHz}
\setsymbol{LFI:knee:frequency:uncertainty:LFI25:Rad:M:units}{0.7\mHz}
\setsymbol{LFI:knee:frequency:uncertainty:LFI26:Rad:M:units}{1.9\mHz}
\setsymbol{LFI:knee:frequency:uncertainty:LFI27:Rad:M:units}{2.9\mHz}
\setsymbol{LFI:knee:frequency:uncertainty:LFI28:Rad:M:units}{4.4\mHz}
\setsymbol{LFI:knee:frequency:uncertainty:LFI18:Rad:S:units}{1.5\mHz}
\setsymbol{LFI:knee:frequency:uncertainty:LFI19:Rad:S:units}{1.3\mHz}
\setsymbol{LFI:knee:frequency:uncertainty:LFI20:Rad:S:units}{1.5\mHz}
\setsymbol{LFI:knee:frequency:uncertainty:LFI21:Rad:S:units}{1.5\mHz}
\setsymbol{LFI:knee:frequency:uncertainty:LFI22:Rad:S:units}{6.7\mHz}
\setsymbol{LFI:knee:frequency:uncertainty:LFI23:Rad:S:units}{1.4\mHz}
\setsymbol{LFI:knee:frequency:uncertainty:LFI24:Rad:S:units}{8.9\mHz}
\setsymbol{LFI:knee:frequency:uncertainty:LFI25:Rad:S:units}{1.8\mHz}
\setsymbol{LFI:knee:frequency:uncertainty:LFI26:Rad:S:units}{9.5\mHz}
\setsymbol{LFI:knee:frequency:uncertainty:LFI27:Rad:S:units}{2.5\mHz}
\setsymbol{LFI:knee:frequency:uncertainty:LFI28:Rad:S:units}{2.4\mHz}

\setsymbol{LFI:knee:frequency:LFI18:Rad:M}{14.8}
\setsymbol{LFI:knee:frequency:LFI19:Rad:M}{11.7}
\setsymbol{LFI:knee:frequency:LFI20:Rad:M}{8.0}
\setsymbol{LFI:knee:frequency:LFI21:Rad:M}{37.9}
\setsymbol{LFI:knee:frequency:LFI22:Rad:M}{9.7}
\setsymbol{LFI:knee:frequency:LFI23:Rad:M}{29.7}
\setsymbol{LFI:knee:frequency:LFI24:Rad:M}{26.8}
\setsymbol{LFI:knee:frequency:LFI25:Rad:M}{20.1}
\setsymbol{LFI:knee:frequency:LFI26:Rad:M}{64.4}
\setsymbol{LFI:knee:frequency:LFI27:Rad:M}{174.5}
\setsymbol{LFI:knee:frequency:LFI28:Rad:M}{130.1}
\setsymbol{LFI:knee:frequency:LFI18:Rad:S}{17.8}
\setsymbol{LFI:knee:frequency:LFI19:Rad:S}{13.7}
\setsymbol{LFI:knee:frequency:LFI20:Rad:S}{5.7}
\setsymbol{LFI:knee:frequency:LFI21:Rad:S}{13.3}
\setsymbol{LFI:knee:frequency:LFI22:Rad:S}{14.8}
\setsymbol{LFI:knee:frequency:LFI23:Rad:S}{59.0}
\setsymbol{LFI:knee:frequency:LFI24:Rad:S}{88.3}
\setsymbol{LFI:knee:frequency:LFI25:Rad:S}{46.4}
\setsymbol{LFI:knee:frequency:LFI26:Rad:S}{68.2}
\setsymbol{LFI:knee:frequency:LFI27:Rad:S}{108.8}
\setsymbol{LFI:knee:frequency:LFI28:Rad:S}{43.1}

\setsymbol{LFI:knee:frequency:uncertainty:LFI18:Rad:M}{2.5}
\setsymbol{LFI:knee:frequency:uncertainty:LFI19:Rad:M}{1.2}
\setsymbol{LFI:knee:frequency:uncertainty:LFI20:Rad:M}{1.9}
\setsymbol{LFI:knee:frequency:uncertainty:LFI21:Rad:M}{5.2}
\setsymbol{LFI:knee:frequency:uncertainty:LFI22:Rad:M}{2.3}
\setsymbol{LFI:knee:frequency:uncertainty:LFI23:Rad:M}{1.1}
\setsymbol{LFI:knee:frequency:uncertainty:LFI24:Rad:M}{1.3}
\setsymbol{LFI:knee:frequency:uncertainty:LFI25:Rad:M}{0.7}
\setsymbol{LFI:knee:frequency:uncertainty:LFI26:Rad:M}{1.9}
\setsymbol{LFI:knee:frequency:uncertainty:LFI27:Rad:M}{2.9}
\setsymbol{LFI:knee:frequency:uncertainty:LFI28:Rad:M}{4.4}
\setsymbol{LFI:knee:frequency:uncertainty:LFI18:Rad:S}{1.5}
\setsymbol{LFI:knee:frequency:uncertainty:LFI19:Rad:S}{1.3}
\setsymbol{LFI:knee:frequency:uncertainty:LFI20:Rad:S}{1.5}
\setsymbol{LFI:knee:frequency:uncertainty:LFI21:Rad:S}{1.5}
\setsymbol{LFI:knee:frequency:uncertainty:LFI22:Rad:S}{6.7}
\setsymbol{LFI:knee:frequency:uncertainty:LFI23:Rad:S}{1.4}
\setsymbol{LFI:knee:frequency:uncertainty:LFI24:Rad:S}{8.9}
\setsymbol{LFI:knee:frequency:uncertainty:LFI25:Rad:S}{1.8}
\setsymbol{LFI:knee:frequency:uncertainty:LFI26:Rad:S}{9.5}
\setsymbol{LFI:knee:frequency:uncertainty:LFI27:Rad:S}{2.5}
\setsymbol{LFI:knee:frequency:uncertainty:LFI28:Rad:S}{2.4}

\setsymbol{LFI:slope:LFI18:Rad:M}{$-1.06$}
\setsymbol{LFI:slope:LFI19:Rad:M}{$-1.21$}
\setsymbol{LFI:slope:LFI20:Rad:M}{$-1.20$}
\setsymbol{LFI:slope:LFI21:Rad:M}{$-1.25$}
\setsymbol{LFI:slope:LFI22:Rad:M}{$-1.42$}
\setsymbol{LFI:slope:LFI23:Rad:M}{$-1.07$}
\setsymbol{LFI:slope:LFI24:Rad:M}{$-0.94$}
\setsymbol{LFI:slope:LFI25:Rad:M}{$-0.85$}
\setsymbol{LFI:slope:LFI26:Rad:M}{$-0.92$}
\setsymbol{LFI:slope:LFI27:Rad:M}{$-0.93$}
\setsymbol{LFI:slope:LFI28:Rad:M}{$-0.93$}
\setsymbol{LFI:slope:LFI18:Rad:S}{$-1.18$}
\setsymbol{LFI:slope:LFI19:Rad:S}{$-1.11$}
\setsymbol{LFI:slope:LFI20:Rad:S}{$-1.30$}
\setsymbol{LFI:slope:LFI21:Rad:S}{$-1.21$}
\setsymbol{LFI:slope:LFI22:Rad:S}{$-1.24$}
\setsymbol{LFI:slope:LFI23:Rad:S}{$-1.21$}
\setsymbol{LFI:slope:LFI24:Rad:S}{$-0.91$}
\setsymbol{LFI:slope:LFI25:Rad:S}{$-0.90$}
\setsymbol{LFI:slope:LFI26:Rad:S}{$-0.76$}
\setsymbol{LFI:slope:LFI27:Rad:S}{$-0.91$}
\setsymbol{LFI:slope:LFI28:Rad:S}{$-0.90$}

\setsymbol{LFI:slope:uncertainty:LFI18:Rad:M}{$0.10$}
\setsymbol{LFI:slope:uncertainty:LFI19:Rad:M}{$0.26$}
\setsymbol{LFI:slope:uncertainty:LFI20:Rad:M}{$0.36$}
\setsymbol{LFI:slope:uncertainty:LFI21:Rad:M}{$0.09$}
\setsymbol{LFI:slope:uncertainty:LFI22:Rad:M}{$0.23$}
\setsymbol{LFI:slope:uncertainty:LFI23:Rad:M}{$0.03$}
\setsymbol{LFI:slope:uncertainty:LFI24:Rad:M}{$0.01$}
\setsymbol{LFI:slope:uncertainty:LFI25:Rad:M}{$0.01$}
\setsymbol{LFI:slope:uncertainty:LFI26:Rad:M}{$0.01$}
\setsymbol{LFI:slope:uncertainty:LFI27:Rad:M}{$0.01$}
\setsymbol{LFI:slope:uncertainty:LFI28:Rad:M}{$0.01$}
\setsymbol{LFI:slope:uncertainty:LFI18:Rad:S}{$0.13$}
\setsymbol{LFI:slope:uncertainty:LFI19:Rad:S}{$0.14$}
\setsymbol{LFI:slope:uncertainty:LFI20:Rad:S}{$0.41$}
\setsymbol{LFI:slope:uncertainty:LFI21:Rad:S}{$0.09$}
\setsymbol{LFI:slope:uncertainty:LFI22:Rad:S}{$0.30$}
\setsymbol{LFI:slope:uncertainty:LFI23:Rad:S}{$0.02$}
\setsymbol{LFI:slope:uncertainty:LFI24:Rad:S}{$0.01$}
\setsymbol{LFI:slope:uncertainty:LFI25:Rad:S}{$0.01$}
\setsymbol{LFI:slope:uncertainty:LFI26:Rad:S}{$0.07$}
\setsymbol{LFI:slope:uncertainty:LFI27:Rad:S}{$0.01$}
\setsymbol{LFI:slope:uncertainty:LFI28:Rad:S}{$0.02$}

\setsymbol{LFI:systematic:effects:p2p:uncertainty:70GHz}{7.87}
\setsymbol{LFI:systematic:effects:p2p:uncertainty:44GHz}{5.61}
\setsymbol{LFI:systematic:effects:p2p:uncertainty:30GHz}{21.02}

\begin{document}
\title{\textit{Planck} 2015 results. II. Low Frequency Instrument data
 processing}

%This author list corresponds to \title{Author list for A03\_LFI\_processing}
%Prepared by M. Lopez-Caniego (Marcos.Lopez.Caniego@sciops.esa.int), ESAC/ESA
%This version is from Tue Feb  9 10:24:47 2016 CET
%\subtitle{There are 216 co-authors in this list}
\author{\small
Planck Collaboration: P.~A.~R.~Ade\inst{91}
\and
N.~Aghanim\inst{62}
\and
M.~Ashdown\inst{73, 6}
\and
J.~Aumont\inst{62}
\and
C.~Baccigalupi\inst{89}
\and
M.~Ballardini\inst{50, 52, 31}
\and
A.~J.~Banday\inst{99, 9}
\and
R.~B.~Barreiro\inst{68}
\and
N.~Bartolo\inst{30, 69}
\and
S.~Basak\inst{89}
% ...
}
\institute{\small
APC, AstroParticule et Cosmologie, Universit\'{e} Paris Diderot, CNRS/IN2P3, CEA/lrfu, Observatoire de Paris, Sorbonne Paris Cit\'{e}, 10, rue Alice Domon et L\'{e}onie Duquet, 75205 Paris Cedex 13, France\goodbreak
\and
Aalto University Mets\"{a}hovi Radio Observatory and Dept of Radio Science and Engineering, P.O. Box 13000, FI-00076 AALTO, Finland\goodbreak
\and
African Institute for Mathematical Sciences, 6-8 Melrose Road, Muizenberg, Cape Town, South Africa\goodbreak
\and
Agenzia Spaziale Italiana Science Data Center, Via del Politecnico snc, 00133, Roma, Italy\goodbreak
\and
Aix Marseille Universit\'{e}, CNRS, LAM (Laboratoire d'Astrophysique de Marseille) UMR 7326, 13388, Marseille, France\goodbreak
\and
Astrophysics Group, Cavendish Laboratory, University of Cambridge, J J Thomson Avenue, Cambridge CB3 0HE, U.K.\goodbreak
\and
CGEE, SCS Qd 9, Lote C, Torre C, 4$^{\circ}$ andar, Ed. Parque Cidade Corporate, CEP 70308-200, Bras\'{i}lia, DF, Brazil\goodbreak
\and
CITA, University of Toronto, 60 St. George St., Toronto, ON M5S 3H8, Canada\goodbreak
\and
CNRS, IRAP, 9 Av. colonel Roche, BP 44346, F-31028 Toulouse cedex 4, France\goodbreak
\and
CRANN, Trinity College, Dublin, Ireland\goodbreak
\and
California Institute of Technology, Pasadena, California, U.S.A.\goodbreak
\and
Centre for Theoretical Cosmology, DAMTP, University of Cambridge, Wilberforce Road, Cambridge CB3 0WA, U.K.\goodbreak
\and
Computational Cosmology Center, Lawrence Berkeley National Laboratory, Berkeley, California, U.S.A.\goodbreak
\and
Consejo Superior de Investigaciones Cient\'{\i}ficas (CSIC), Madrid, Spain\goodbreak
\and
DSM/Irfu/SPP, CEA-Saclay, F-91191 Gif-sur-Yvette Cedex, France\goodbreak
\and
DTU Space, National Space Institute, Technical University of Denmark, Elektrovej 327, DK-2800 Kgs. Lyngby, Denmark\goodbreak
\and
D\'{e}partement de Physique Th\'{e}orique, Universit\'{e} de Gen\`{e}ve, 24, Quai E. Ansermet,1211 Gen\`{e}ve 4, Switzerland\goodbreak
\and
Departamento de Astrof\'{i}sica, Universidad de La Laguna (ULL), E-38206 La Laguna, Tenerife, Spain\goodbreak
\and
Departamento de F\'{\i}sica, Universidad de Oviedo, Avda. Calvo Sotelo s/n, Oviedo, Spain\goodbreak
\and
Department of Astronomy and Astrophysics, University of Toronto, 50 Saint George Street, Toronto, Ontario, Canada\goodbreak
\and
Department of Astrophysics/IMAPP, Radboud University Nijmegen, P.O. Box 9010, 6500 GL Nijmegen, The Netherlands\goodbreak
\and
Department of Physics \& Astronomy, University of British Columbia, 6224 Agricultural Road, Vancouver, British Columbia, Canada\goodbreak
\and
Department of Physics and Astronomy, Dana and David Dornsife College of Letter, Arts and Sciences, University of Southern California, Los Angeles, CA 90089, U.S.A.\goodbreak
\and
Department of Physics and Astronomy, University College London, London WC1E 6BT, U.K.\goodbreak
\and
Department of Physics, Florida State University, Keen Physics Building, 77 Chieftan Way, Tallahassee, Florida, U.S.A.\goodbreak
\and
Department of Physics, Gustaf H\"{a}llstr\"{o}min katu 2a, University of Helsinki, Helsinki, Finland\goodbreak
\and
Department of Physics, Princeton University, Princeton, New Jersey, U.S.A.\goodbreak
\and
Department of Physics, University of California, Santa Barbara, California, U.S.A.\goodbreak
\and
Department of Physics, University of Illinois at Urbana-Champaign, 1110 West Green Street, Urbana, Illinois, U.S.A.\goodbreak
\and
Dipartimento di Fisica e Astronomia G. Galilei, Universit\`{a} degli Studi di Padova, via Marzolo 8, 35131 Padova, Italy\goodbreak
\and
Dipartimento di Fisica e Astronomia, ALMA MATER STUDIORUM, Universit\`{a} degli Studi di Bologna, Viale Berti Pichat 6/2, I-40127, Bologna, Italy\goodbreak
\and
Dipartimento di Fisica e Scienze della Terra, Universit\`{a} di Ferrara, Via Saragat 1, 44122 Ferrara, Italy\goodbreak
\and
Dipartimento di Fisica, Universit\`{a} La Sapienza, P. le A. Moro 2, Roma, Italy\goodbreak
\and
Dipartimento di Fisica, Universit\`{a} degli Studi di Milano, Via Celoria, 16, Milano, Italy\goodbreak
\and
Dipartimento di Fisica, Universit\`{a} degli Studi di Trieste, via A. Valerio 2, Trieste, Italy\goodbreak
\and
Dipartimento di Fisica, Universit\`{a} di Roma Tor Vergata, Via della Ricerca Scientifica, 1, Roma, Italy\goodbreak
\and
Dipartimento di Matematica, Universit\`{a} di Roma Tor Vergata, Via della Ricerca Scientifica, 1, Roma, Italy\goodbreak
\and
Discovery Center, Niels Bohr Institute, Blegdamsvej 17, Copenhagen, Denmark\goodbreak
\and
Discovery Center, Niels Bohr Institute, Copenhagen University, Blegdamsvej 17, Copenhagen, Denmark\goodbreak
\and
European Space Agency, ESAC, Planck Science Office, Camino bajo del Castillo, s/n, Urbanizaci\'{o}n Villafranca del Castillo, Villanueva de la Ca\~{n}ada, Madrid, Spain\goodbreak
\and
European Space Agency, ESTEC, Keplerlaan 1, 2201 AZ Noordwijk, The Netherlands\goodbreak
\and
Gran Sasso Science Institute, INFN, viale F. Crispi 7, 67100 L'Aquila, Italy\goodbreak
\and
HGSFP and University of Heidelberg, Theoretical Physics Department, Philosophenweg 16, 69120, Heidelberg, Germany\goodbreak
\and
Haverford College Astronomy Department, 370 Lancaster Avenue, Haverford, Pennsylvania, U.S.A.\goodbreak
\and
Helsinki Institute of Physics, Gustaf H\"{a}llstr\"{o}min katu 2, University of Helsinki, Helsinki, Finland\goodbreak
\and
INAF - Osservatorio Astrofisico di Catania, Via S. Sofia 78, Catania, Italy\goodbreak
\and
INAF - Osservatorio Astronomico di Padova, Vicolo dell'Osservatorio 5, Padova, Italy\goodbreak
\and
INAF - Osservatorio Astronomico di Roma, via di Frascati 33, Monte Porzio Catone, Italy\goodbreak
\and
INAF - Osservatorio Astronomico di Trieste, Via G.B. Tiepolo 11, Trieste, Italy\goodbreak
\and
INAF/IASF Bologna, Via Gobetti 101, Bologna, Italy\goodbreak
\and
INAF/IASF Milano, Via E. Bassini 15, Milano, Italy\goodbreak
\and
INFN, Sezione di Bologna, Via Irnerio 46, I-40126, Bologna, Italy\goodbreak
\and
INFN, Sezione di Roma 1, Universit\`{a} di Roma Sapienza, Piazzale Aldo Moro 2, 00185, Roma, Italy\goodbreak
\and
INFN, Sezione di Roma 2, Universit\`{a} di Roma Tor Vergata, Via della Ricerca Scientifica, 1, Roma, Italy\goodbreak
\and
INFN/National Institute for Nuclear Physics, Via Valerio 2, I-34127 Trieste, Italy\goodbreak
\and
ISDC, Department of Astronomy, University of Geneva, ch. d'Ecogia 16, 1290 Versoix, Switzerland\goodbreak
\and
IUCAA, Post Bag 4, Ganeshkhind, Pune University Campus, Pune 411 007, India\goodbreak
\and
Imperial College London, Astrophysics group, Blackett Laboratory, Prince Consort Road, London, SW7 2AZ, U.K.\goodbreak
\and
Infrared Processing and Analysis Center, California Institute of Technology, Pasadena, CA 91125, U.S.A.\goodbreak
\and
Institut N\'{e}el, CNRS, Universit\'{e} Joseph Fourier Grenoble I, 25 rue des Martyrs, Grenoble, France\goodbreak
\and
Institut Universitaire de France, 103, bd Saint-Michel, 75005, Paris, France\goodbreak
\and
Institut d'Astrophysique Spatiale, CNRS, Univ. Paris-Sud, Universit\'{e} Paris-Saclay, B\^{a}t. 121, 91405 Orsay cedex, France\goodbreak
\and
Institut d'Astrophysique de Paris, CNRS (UMR7095), 98 bis Boulevard Arago, F-75014, Paris, France\goodbreak
\and
Institut f\"ur Theoretische Teilchenphysik und Kosmologie, RWTH Aachen University, D-52056 Aachen, Germany\goodbreak
\and
Institute of Astronomy, University of Cambridge, Madingley Road, Cambridge CB3 0HA, U.K.\goodbreak
\and
Institute of Theoretical Astrophysics, University of Oslo, Blindern, Oslo, Norway\goodbreak
\and
Instituto de Astrof\'{\i}sica de Canarias, C/V\'{\i}a L\'{a}ctea s/n, La Laguna, Tenerife, Spain\goodbreak
\and
Instituto de F\'{\i}sica de Cantabria (CSIC-Universidad de Cantabria), Avda. de los Castros s/n, Santander, Spain\goodbreak
\and
Istituto Nazionale di Fisica Nucleare, Sezione di Padova, via Marzolo 8, I-35131 Padova, Italy\goodbreak
\and
Jet Propulsion Laboratory, California Institute of Technology, 4800 Oak Grove Drive, Pasadena, California, U.S.A.\goodbreak
\and
Jodrell Bank Centre for Astrophysics, Alan Turing Building, School of Physics and Astronomy, The University of Manchester, Oxford Road, Manchester, M13 9PL, U.K.\goodbreak
\and
Kavli Institute for Cosmological Physics, University of Chicago, Chicago, IL 60637, USA\goodbreak
\and
Kavli Institute for Cosmology Cambridge, Madingley Road, Cambridge, CB3 0HA, U.K.\goodbreak
\and
Kazan Federal University, 18 Kremlyovskaya St., Kazan, 420008, Russia\goodbreak
\and
LAL, Universit\'{e} Paris-Sud, CNRS/IN2P3, Orsay, France\goodbreak
\and
LERMA, CNRS, Observatoire de Paris, 61 Avenue de l'Observatoire, Paris, France\goodbreak
\and
Laboratoire AIM, IRFU/Service d'Astrophysique - CEA/DSM - CNRS - Universit\'{e} Paris Diderot, B\^{a}t. 709, CEA-Saclay, F-91191 Gif-sur-Yvette Cedex, France\goodbreak
\and
Laboratoire Traitement et Communication de l'Information, CNRS (UMR 5141) and T\'{e}l\'{e}com ParisTech, 46 rue Barrault F-75634 Paris Cedex 13, France\goodbreak
\and
Laboratoire de Physique Subatomique et Cosmologie, Universit\'{e} Grenoble-Alpes, CNRS/IN2P3, 53, rue des Martyrs, 38026 Grenoble Cedex, France\goodbreak
\and
Laboratoire de Physique Th\'{e}orique, Universit\'{e} Paris-Sud 11 \& CNRS, B\^{a}timent 210, 91405 Orsay, France\goodbreak
\and
Lawrence Berkeley National Laboratory, Berkeley, California, U.S.A.\goodbreak
\and
Lebedev Physical Institute of the Russian Academy of Sciences, Astro Space Centre, 84/32 Profsoyuznaya st., Moscow, GSP-7, 117997, Russia\goodbreak
\and
Max-Planck-Institut f\"{u}r Astrophysik, Karl-Schwarzschild-Str. 1, 85741 Garching, Germany\goodbreak
\and
McGill Physics, Ernest Rutherford Physics Building, McGill University, 3600 rue University, Montr\'{e}al, QC, H3A 2T8, Canada\goodbreak
\and
National University of Ireland, Department of Experimental Physics, Maynooth, Co. Kildare, Ireland\goodbreak
\and
Nicolaus Copernicus Astronomical Center, Bartycka 18, 00-716 Warsaw, Poland\goodbreak
\and
Niels Bohr Institute, Blegdamsvej 17, Copenhagen, Denmark\goodbreak
\and
Niels Bohr Institute, Copenhagen University, Blegdamsvej 17, Copenhagen, Denmark\goodbreak
\and
SISSA, Astrophysics Sector, via Bonomea 265, 34136, Trieste, Italy\goodbreak
\and
SMARTEST Research Centre, Universit\`{a} degli Studi e-Campus, Via Isimbardi 10, Novedrate (CO), 22060, Italy\goodbreak
\and
School of Physics and Astronomy, Cardiff University, Queens Buildings, The Parade, Cardiff, CF24 3AA, U.K.\goodbreak
\and
School of Physics and Astronomy, University of Nottingham, Nottingham NG7 2RD, U.K.\goodbreak
\and
Sorbonne Universit\'{e}-UPMC, UMR7095, Institut d'Astrophysique de Paris, 98 bis Boulevard Arago, F-75014, Paris, France\goodbreak
\and
Space Sciences Laboratory, University of California, Berkeley, California, U.S.A.\goodbreak
\and
Special Astrophysical Observatory, Russian Academy of Sciences, Nizhnij Arkhyz, Zelenchukskiy region, Karachai-Cherkessian Republic, 369167, Russia\goodbreak
\and
Sub-Department of Astrophysics, University of Oxford, Keble Road, Oxford OX1 3RH, U.K.\goodbreak
\and
Theory Division, PH-TH, CERN, CH-1211, Geneva 23, Switzerland\goodbreak
\and
UPMC Univ Paris 06, UMR7095, 98 bis Boulevard Arago, F-75014, Paris, France\goodbreak
\and
Universit\'{e} de Toulouse, UPS-OMP, IRAP, F-31028 Toulouse cedex 4, France\goodbreak
% ...
}

\titlerunning{LFI data processing}
\authorrunning{Planck Collaboration}

\abstract{We present an updated description of the \Planck\ Low Frequency Instrument (LFI)
data processing pipeline, associated with the 2015 data release.  We point out
the places where our results and methods have remained unchanged since
the 2013 paper and we highlight the changes made for the 2015 release,
describing the products (especially timelines) and the ways in which they were
obtained.  We demonstrate that the pipeline is self-consistent (principally
based on simulations) and report all null tests. For the first time, we present LFI maps in Stokes $Q$ and $U$ polarization. We refer to other related
papers where more detailed descriptions of the LFI data processing pipeline
may be found if needed.}

\keywords{Space vehicles: instruments -- Methods:data analysis
 -- cosmic microwave background}
\maketitle
\allearlypapers

\section{Introduction}
\label{sec_introduction}
    This paper, one of a set associated with the 2015 release of data
from the \Planck\footnote{\Planck\ (\url{http://www.esa.int/Planck}) is a project of the European Space Agency  (ESA) with instruments provided by two scientific consortia funded by ESA member states and led by Principal Investigators from France and Italy, telescope reflectors provided through a collaboration between ESA and a scientific consortium led and funded by Denmark, and additional contributions from NASA (USA).} mission
\citep{planck2014-a01}, describes the Low Frequency Instrument
(LFI) data processing that supports the second \Planck\
cosmological release. Following the nominal mission of 15.5 months,
the LFI in-flight operation was extended 
to fully exploit the lifetime of the Planck 20\,K to 4\,K cryogenic system,
leading to a total of 48 months of observation 
(or eight full-sky surveys) with essentially unchanged instrument performance. 
This paper is an updated description of the LFI data
processing \citep{planck2013-p02} that was part of the second wave of
astrophysical results published in 
early 2014 (Planck Collaboration VIII--XXVI 2013), now incorporating the
analysis of the full mission data, both in temperature and in polarization. 
This work describes the overall data flow of the pipeline implemented at the
LFI data processing centre (DPC), including scientific telemetry from the
instrument, housekeeping data, and frequency maps, as well as the tests
applied to validate the data products. Detailed descriptions of
critical aspects of the data analysis and products, including
improvements in some of the algorithms used in the pipeline, are
given in four companion papers.
These discuss, respectively: systematic
effects and the overall error budget \citep{planck2014-a04};
the determination of the LFI main beams and window functions from in-flight
planet-crossing measurements and optical modelling \citep{planck2014-a05};
photometric calibration, including methods adopted and related
uncertainties \citep{planck2014-a06}; and  mapmaking, including the process used to
obtain the low-resolution maps and their
associated full noise covariance matrices \citep{planck2014-a07}.
The main results and reference tables on all these topics are
summarized in this paper.  

The structure of this paper is organized as follows.  We summarize
the overall data processing pipeline
in Sect.~\ref{sec_dataproc_overview}.
Processing of the time ordered information (TOI) is described in 
Sect.~\ref{sec_toiprocessing},
with an emphasis on changes since \citet{planck2013-p02}.
Section~\ref{sec_beamrecovery} describes important changes to our calculations
of LFI beams, which in turn has an effect on calibration, 
described in Sect.~\ref{sec_calibration}. LFI noise properties are described in Sect.~\ref{sec_general_noise}.
Section~\ref{sec_mmaking_intro} and Sect.~\ref{sec_OverviewMaps} present
\Planck\ maps at 30, 44, and 70\,GHz, both in temperature and in $Q$ and $U$
polarization, including the low-multipole maps needed
to construct the \Planck\ likelihood \citep{planck2014-a13}.
Section~\ref{sec_polarization} presents the major new results for this release,
LFI polarization maps, and an analysis of systematic effects peculiar to
polarization.
Validation of the LFI products, especially by means of null
tests, is discussed in Sect.~\ref{sec_dataval_intro}, and the special issue of
data selection for low-$\ell$ analysis is considered in Sect.~\ref{sec_low_l}.
Section~\ref{sec_Product description} summarizes the LFI data products
(for further details, see the Explanatory
Supplement\footnote{\url{http://pla.esac.esa.int/pla/index.html}}
that accompanies the release of products and provides its detailed
description).  We conclude briefly in Sect.~\ref{sec_conclusion}.

\section{In-flight behaviour and operations}
\label{sec_flightbahavior}
    The \Planck\ LFI instrument is described in
\citet{bersanelli2010} and \citet{mennella2010}. It comprises 11
radiometer chain assemblies (RCAs), two at 30\,GHz, three at
44\,GHz, and six at 70\,GHz, each composed of two independent
pseudo-correlation radiometers sensitive to orthogonal linear
polarization modes. Each radiometer has two independent square-law
diodes for detection, integration, and conversion from radio
frequency signals to DC voltages. The focal plane is
cryogenically cooled to 20\,K, while the pseudo-correlation design
uses internal, blackbody reference loads cooled to 4.5\,K.  The
radiometer timelines are produced by taking differences between
the signals from the sky, $V_{\rm sky}$, and from the reference
loads, $V_{\rm ref}$. Radiometer balance is optimized by
introducing a gain modulation factor, typically stable within
0.02\,\% throughout the entire mission, which greatly reduces $1/f$ noise
and improves immunity to a wide class of systematic effects
\citep{planck2011-1.4}. During the full operation period (ignoring
a brief, less stable thermal period due to the sorption cooler switchover), the
behaviour of all 22 LFI radiometers was stable, with $1/f$ knee
frequencies unchanging within 9\,\% and white noise levels within
0.5\,\%. These results are in line with those found for the $15.5$ month nominal
mission period \citep{planck2013-p02}.

    \subsection{Operations}
    \label{sec_operations}
        The data set released together with this paper was acquired from 12 August 2009 to 3 August 2013, roughly four years of observations. 
The first two years of data (from Survey~1 to Survey~4) were acquired scanning
the sky with a phase angle of $340$\deg, whereas for the last two years
(from Survey~5 to Survey~8) the phase angle was shifted to $250$\deg\
\citep[see][for details]{planck2014-a01}.  This shift has allowed for more
thorough investigation of systematic effects, including
better characterization of the beam and the related Galactic straylight
(see Sect.~\ref{sec_strylight_removal}) using null tests based on survey
differences.  During the last three Jupiter crossing, the scanning strategy was optimized to obtain a better beam determination (see Sect.~\ref{sec_beamrecovery}). The period from 03 August 2013 to 03 October 2013
was used to perform deep scanning of the Crab Nebula and of the
regions near the minima of the cosmic microwave background (CMB) dipole,
with the aim of determining the dipole direction with an alternative approach.
These data are not included in this release, since they require specialized
analysis, which is not yet complete.

    \subsection{Instrument performance update} \label{sec_instrupdate}
    \label{sec_instrupdate}
        In Table~\ref{tab_summary_performance} we present a top-level summary of
the instrument performance parameters measured in flight during the four years
of operation of LFI.  Optical properties have been reconstructed from Jupiter
transits \citep{planck2014-a05} and are in agreement with estimations made
for the 2013 release \citep{planck2013-p02d}. White noise sensitivity and
parameters describing the 1/$f$ noise component are in line with the 2013
values \citep{planck2013-p02}, demonstrating that cryogenic operation of the low-noise amplifiers and phase switches do not result in any significant aging effects over a period of four years. Overall calibration uncertainty, determined as the sum
of absolute and relative calibration, is $0.35$\,\%, $0.26$\,\%, and $0.20$\,\% at 30, 44,
and 70\,GHz respectively, improving by more than a factor of 2 over the LFI 2013 calibration \citep{planck2013-p02b}. The residual systematic uncertainty was computed for
both temperature and polarization; it varies between 1 and
3\,$\mu\mathrm{K}_\mathrm{CMB}$ \citep{planck2014-a04} in temperature and
polarization. It should be noted that the uncertainty arising from systematic
effects is lower than in the previous release \citep{planck2013-p02d}; this is
principally due to the straylight removal and the new iterative calibration
algorithm now used.

\begin{table*}
\begingroup
\newdimen\tblskip \tblskip=5pt
\caption{LFI performance parameters.}
\label{tab_summary_performance}
\nointerlineskip
\vskip -3mm
\footnotesize
\setbox\tablebox=\vbox{
   \newdimen\digitwidth
   \setbox0=\hbox{\rm 0}
   \digitwidth=\wd0
   \catcode`*=\active
   \def*{\kern\digitwidth}
   \newdimen\signwidth
   \setbox0=\hbox{+}
   \signwidth=\wd0
   \catcode`!=\active
   \def!{\kern\signwidth}
\halign{\hbox to 2.7in{#\leaderfil}\tabskip=3em&
        \hfil#\hfil&
        \hfil#\hfil&
        \hfil#\hfil\tabskip=0pt\cr
\noalign{\doubleline}
\noalign{\vskip -3pt}
\omit\hfil Parameter\hfil&30\,GHz&44\,GHz&70\,GHz\cr
\noalign{\vskip 3pt\hrule\vskip 5pt}
Centre frequency [GHz]&28.4&44.1&70.4\cr
\noalign{\vskip 3pt}
Scanning beam FWHM$^{\rm a}$ [arcmin]&33.10&27.94&13.08\cr
\noalign{\vskip 3pt}
Scanning beam ellipticity$^{\rm a}$&1.37&1.25&1.27\cr
\noalign{\vskip 3pt}
Effective beam FWHM$^{\rm b}$ [arcmin]&32.29&27.00&13.21\cr
\noalign{\vskip 3pt}
White noise level in timelines$^{\rm c}$ [\muKCMBs]&148.1&174.2&152.0\cr
\noalign{\vskip 3pt}
$f_{\rm knee}$$^{\rm c}$ [mHz]&114&52&19\cr
\noalign{\vskip 3pt}
$1/f$ slope$^{\rm c}$&\llap{$-$}0.92&\llap{$-$}0.88&\llap{$-$}1.20\cr
\noalign{\vskip 3pt}
Overall calibration uncertainty$^{\rm d}$ [\%]&0.35&0.26&0.20\cr
\noalign{\vskip 3pt}
Systematic effects uncertainty in Stokes $I^{\rm e}$ [\muKCMB]&0.88&1.97&1.87\cr
\noalign{\vskip 3pt}
Systematic effects uncertainty in Stokes $Q^{\rm e}$ [\muKCMB]&1.11&1.14&2.25\cr
\noalign{\vskip 3pt}
Systematic effects uncertainty in Stokes $U^{\rm e}$ [\muKCMB]&0.95&1.20&2.22\cr
\noalign{\vskip 5pt\hrule\vskip 3pt}}}
\endPlancktablewide
\tablenote a Determined by fitting Jupiter observations
directly in the timelines.\par 
\tablenote b Calculated from the main beam solid angle of the effective beam (Sect.~\ref{sec_effectivebeam}). These values
are used in the source extraction pipeline~\citep{planck2014-a35}.\par
\tablenote c Typical values derived from fitting noise spectra (Sect.~\ref{sec_noise}).\par
\tablenote d Sum of the error determined from the absolute and relative calibration, see \citet{planck2014-a05}.\par
\tablenote e Peak-to-peak difference between 99\,\% and 1\,\% quantiles in the pixel value
distributions from simulated maps (see \citealt{planck2014-a04}).\par
\endgroup
\end{table*}

\section{Data processing overview}
\label{sec_dataproc_overview}
\begin{figure*} [th]
\centering
\includegraphics[width=18cm]{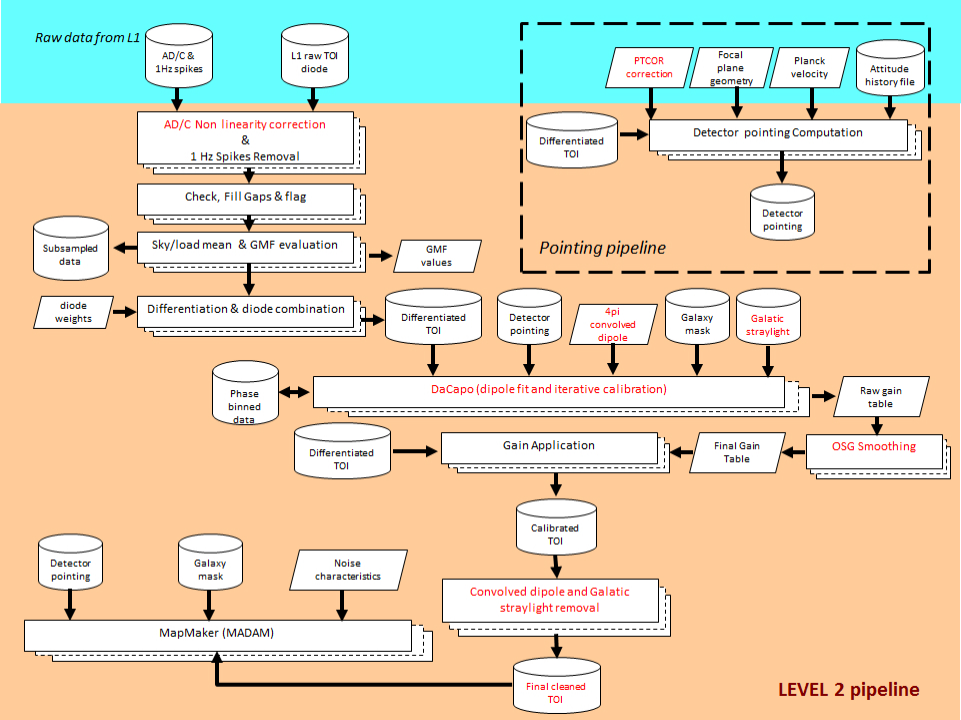}
\caption{Schematic representation of the Level~2 and pointing pipelines of
the LFI DPC; elements in red identify those modified or augmented with
respect to \citet{planck2013-p02}}
\label{dpcpipeline}
\end{figure*}

    As in \cite{planck2013-p02}, the processing of LFI data is divided into three
levels, shown schematically in Fig.~\ref{dpcpipeline}.  The main changes
compared to the earlier release are related to the way in which we
take into account the beam information in the pipeline processing, as well as
an entire overhaul of the iterative algorithm used to calibrate the raw data.
According to the LFI scheme, processing  starts at
Level~1, which retrieves all the necessary information from data packets
and auxiliary data received from the Mission Operation
Centre, and transforms the scientific packets and housekeeping
data into a form manageable by Level~2.  Level~2 uses scientific
and housekeeping information to:

\begin{itemize}

\item build the LFI reduced instrument model (RIMO), which contains the main
characteristics of the instrument;

\item remove analogue-to-digital converter (ADC) non-linearities and 1\,Hz
spikes diode by diode (see Sects.~\ref{sec_adc_nonlinearity} and
\ref{sec_electronic_spikes});

\item compute and apply the gain modulation factor to minimize $1/f$ noise
(see Sect.~\ref{sec_gain_modulation});

\item combine signals from the two diodes of each radiometer (see Sect.~\ref{sec_comb_diodes});

\item compute the appropriate detector pointing for each sample, based on
auxiliary data and beam information corrected by a model (PTCOR) built using
solar distance and radiometer electronics box assembly (REBA) temperature
information (see Sect.~\ref{sec_pointing});

\item calibrate the scientific timelines to physical units
($\mathrm{K}_\mathrm{CMB}$), fitting the total CMB dipole convolved with the
$4\pi$ beam representation (see Sect.~\ref{sec_calibration}), without taking
into account the signature due to Galactic straylight
(see Sect.~\ref{sec_strylight_removal});

\item remove the solar and orbital dipole convolved with the $4\pi$ beam
representation and the Galactic emission convolved with the beam sidelobes
(see Sect.~\ref{sec_strylight_removal}) from the scientific calibrated
timeline;

\item combine the calibrated time-ordered information (TOI) into aggregate
products, such as maps at each frequency (see Sect.~\ref{sec_mmaking_intro}).

\end{itemize}

Level~3 collects Level~2 outputs from both HFI
\citep{planck2014-a09} and LFI and derives various products, such
as component-separated maps of astrophysical foregrounds,
catalogues of different classes of sources, and the likelihood of
cosmological and astrophysical models given in the maps.

\section{Time-ordered information (TOI) processing}
\label{sec_toiprocessing}
    The Level~1 pipeline, which has the responsibility to receive telemetry data
and sort them into a form manageable by the Level~2 pipeline, has not
changed with respect to the $2013$ release; we therefore refer to
\citet{planck2013-p02} for its description.  In this section, we move directly
to a discussion of the Level~2 pipeline.

    \subsection{Input flags}
    \label{sec_flagging}
        The flagging procedure used was exactly the same as described in
\citet{planck2013-p02}.  In Table~\ref{tab_data_flags_percentage} we give the
percentage of usable and unused data for the full mission.  It should be noted
that compared with the same table in
\citet{planck2013-p02} the amount of missing data (where by ``missing'' we mean
packets that were not been received on the ground) is larger due to two
technical problems that were experienced with the spacecraft, resulting in
data not being downloaded for 2 days of observation.  On the other hand the
anomalies were lessened due to better control of the instrument's temperature
stability.  The percentages of time spent on spacecraft manoeuvres are the same
for the three frequencies, and as a consequence the fraction of data used in
the science analysis was similar (at more than 90\,\%) at each frequency.

\begin{table}[tmb]
  \begingroup
  \newdimen\tblskip \tblskip=5pt
  \caption{Percentage of LFI observation time lost due to missing or unusable
  data, and to manoeuvres$^{\rm a}$.}
  \label{tab_data_flags_percentage}
  \nointerlineskip
  \vskip -3mm
  \footnotesize
  \setbox\tablebox=\vbox{
    \newdimen\digitwidth
    \setbox0=\hbox{\rm 0}
    \digitwidth=\wd0
    \catcode`*=\active
    \def*{\kern\digitwidth}
    \newdimen\signwidth
    \setbox0=\hbox{+}
    \signwidth=\wd0
    \catcode`!=\active
    \def!{\kern\signwidth}
    \halign{\hbox to 1.31in{#\leaderfil}\tabskip=1em&
      \hfil#\hfil&
      \hfil#\hfil&
      \hfil#\hfil\tabskip=0pt\cr                            
      \noalign{\doubleline}
      \omit\hfil Category\hfil& 30\,GHz& 44\,GHz& 70\,GHz\cr
      \noalign{\vskip 3pt\hrule\vskip 5pt}
        Missing [\%]& *0.153& *0.154& *0.153\cr
      Anomalies [\%]& *0.375& *0.448& *0.631\cr
     Manoeuvres [\%]& *8.032& *8.032& *8.032\cr
\noalign{\vskip 3pt}
      Usable [\%]& 91.440& 91.366& 91.184\cr
      \noalign{\vskip 5pt\hrule\vskip 3pt}
    }}
  \endPlancktable
  \par \tablenote a The remaining percentage (listed in the last row) is used
 in scientific analysis. \par
  \endgroup
\end{table}

    \subsection{ADC non-linearity correction}
    \label{sec_adc_nonlinearity}
        The ADCs convert the analogue detector voltages to numbers,
their linearity is as important as that of the receivers and
detectors, with any departure appearing as a distortion in the
system power response curve. While the algorithm for determining the ADC corrections remains the same as
described in \citet{planck2013-p02d}, some changes were made in its implementation
and execution. First, the full mission data are now used, so that when
detector voltages are revisited there will be an improvement in signal
to noise (although in the particular case of radiometer 21M, some of
the voltages were too poorly sampled to generate an adequate
solution). Second, instead of determining the white noise amplitude
via a Fourier transform, we now use the difference between the sum of
the variances and twice the covariance of adjacent paired points in
the timestream, such that white noise variance $\sigma^2_{\rm WN}
= \text{Var}[X_{\rm o}]+\text{Var}[X_{\rm e}]
 - 2\text{Cov}[X_{\rm o},X_{\rm e}]$, where $X_{\rm o}$
and $X_{\rm e}$ are data points with odd and even indices, respectively.
This not only increased the speed of calculating the noise amplitude,
but avoided the iteration steps, since these can be done analytically
from the initial variance-covariance estimates. Finally, data
acquisition electronics (DAE) offset changes made on operational day
(OD) 953 to avoid saturation also shifted the apparent ADC voltage
relative to the true detector voltage.  A separate ADC correction had
to be generated and applied to radiometers 22M and 23S using only the
post OD 953 data.

The ability to recover the correct ADC solution and the level of the
residuals was assessed by simulating time-ordered data with the same
noise statistics, voltage drift, gain fluctuations, and sky signal,
with a known ADC error. As the correction in the DPC pipeline is a
lookup table of input to output detector voltages to which a spline
is fitted to interpolate the TOI voltages, we introduced the ADC error
as the spline curve with the input and output voltages swapped and
thereby generate the inverse of the measured ADC effect. Comparing the
spline curves used to the ones recovered proved to be at the
level of a few percent, leading to rms errors on the residual
simulated frequency maps of $\approx 0.1\,\mu{\rm K}_{\rm CMB}$ at 30 and 44\,GHz and $\approx 0.4\,\mu{\rm K}_{\rm CMB}$ at 70\,GHz, both
temperature and polarization. These simulations and results are
summarized in more detail in \citet{planck2014-a04}.

    \subsection{Corrections for electronic spikes}
    \label{sec_electronic_spikes}
        Electronic spikes are caused by the interaction between the 
electronics clock and the scientific data lines. They occur in the data 
acquisition electronics (DAE) after the detector diodes and before
the analogue-to-digital converters
\citep[ADC,][]{meinhold2009,mennella2010,planck2011-1.4}.
The signal is detected in all the LFI radiometers time-domain outputs 
as a 1\,s square wave with a rising edge near $0.5\,\text{s}$ 
and a falling edge near $0.75\,\text{s}$, synchronous with the on-board 
time signal. In the frequency domain it appears as a spike signal at multiples 
of $1\,\text{Hz}$. The 44\,GHz channels are the only one that are significantly 
affected by this effect. Consequently the spike signal is removed from 
the data only in this channel. The procedure consists of the subtraction 
of a fitted square wave template from the time-domain data as described 
in \citet{planck2014-a04}. We are evaluating the possibility of further reducing the residual effect of the spikes signal at the map level, as described in
\citet{planck2014-a04}, for the next \Planck\ data release by adopting one or more of the following approaches:
\begin{itemize}
  \item increasing the resolution of square wave template, at the moment at $80$\,Hz;
  \item using time varying template instead of the fixed one over the whole
   mission;
  \item removing spikes signal from the 30\,GHz and 70\,GHz channels.
\end{itemize}

    \subsection{Demodulation: gain modulation factor estimation and application}
    \label{sec_gain_modulation}
        Each \Planck\ LFI diode switches at 4096\,Hz \citep{mennella2010} between the
sky and the 4\,K reference load.  The data acquired in this way are dominated
by $1/f$ noise that is highly correlated between 
the two streams \citep{bersanelli2010}; differencing those streams results in
a strong reduction of $1/f$ noise.
The procedure applied differs from that discussed in \citet{planck2013-p02} 
in only one way: the Galaxy and point sources are masked from the time-ordered data used in the computation of the gain modulation factor $R$ (GMF in Fig.~\ref{dpcpipeline}). The overall variation of $R$ over the whole mission is less than 0.02\,\% for every LFI channel.
A full description of the theory of this correction can be found in 
\citet{planck2011-1.4}.

    \subsection{Combining diodes}
    \label{sec_comb_diodes}
        Two detector diodes provide the output for each \Planck\ LFI receiver channel. 
To minimize the impact of imperfect isolation between the two diodes, 
we perform a weighted average of the time-ordered data from the two diodes of
each receiver just before the differencing. 
The procedure applied is the same described in \citet{planck2013-p02} and for
the sake of completeness, we report in Table~\ref{tab_diode_weights}
the values of weights used; the receiver channels are indicated either with \emph{M} (main) or \emph{S} (side).
The weights are kept fixed during the whole mission.

\begin{table}[tmb]
  \begingroup
  \newdimen\tblskip \tblskip=5pt
  \caption{Weights used in combining diodes$^{\rm a}$.}
  \label{tab_diode_weights}
  \nointerlineskip
  \vskip -3mm
  \footnotesize
  \setbox\tablebox=\vbox{
    \newdimen\digitwidth
    \setbox0=\hbox{\rm 0}
    \digitwidth=\wd0
    \catcode`*=\active
    \def*{\kern\digitwidth}
    \newdimen\signwidth
    \setbox0=\hbox{+}
    \signwidth=\wd0
    \catcode`!=\active
    \def!{\kern\signwidth}
    %
%    \halign{\hbox to 1.0in{#\leaderfil}\tabskip=2em&
    \halign{\tabskip=0pt#\hfil\tabskip=2em&
      \hfil#\hfil&
      \hfil#\hfil&
      \hfil#\hfil&
      \hfil#\hfil\tabskip=0pt\cr                            % Template goes here.
      \noalign{\doubleline}
      \omit&\multispan4\hfil Diode\hfil\cr
      \noalign{\vskip -3pt}
      \omit&\multispan4\hrulefill\cr
      \noalign{\vskip 2pt}
      \omit\hfil Radiometer\hfil& M-00& M-01& S-10& S-11\cr
      \noalign{\vskip 3pt\hrule\vskip 5pt}
      {\bf 70\,GHz}&&&\cr
      \noalign{\vskip 3pt}
      LFI 18& 0.567& 0.433& 0.387& 0.613\cr
      LFI 19& 0.502& 0.498& 0.551& 0.449\cr
      LFI 20& 0.523& 0.477& 0.477& 0.523\cr
      LFI 21& 0.500& 0.500& 0.564& 0.436\cr
      LFI 22& 0.536& 0.464& 0.554& 0.446\cr
      LFI 23& 0.508& 0.492& 0.362& 0.638\cr
      \noalign{\vskip 5pt\hrule\vskip 3pt}
      {\bf 44\,GHz}&&&\cr
      \noalign{\vskip 3pt}      
      LFI 24& 0.602& 0.398& 0.456& 0.544\cr
      LFI 25& 0.482& 0.518& 0.370& 0.630\cr
      LFI 26& 0.593& 0.407& 0.424& 0.576\cr
      \noalign{\vskip 5pt\hrule\vskip 3pt}
      {\bf 30\,GHz}&&&\cr
      \noalign{\vskip 3pt}      
      LFI 27& 0.520& 0.480& 0.485& 0.515\cr
      LFI 28& 0.553& 0.447& 0.468& 0.532\cr
      \noalign{\vskip 5pt\hrule\vskip 3pt}
    }}
  \endPlancktable
  \par \tablenote a A perfect instrument would have weights of 0.500 for both
  diodes. \par
  \endgroup
\end{table}

\section{Pointing}
\label{sec_pointing}
    The long time scale pointing correction, PTCOR, has been modified, and is now based
on the solar distance and radiometer electronics box assembly (REBA)
thermometry.  Unlike in 2013, the reconstructed satellite attitude is now
uniform across both of the \Planck\ instruments and is discussed in detail
in the mission overview paper, \cite{planck2014-a01}.

\section{Main beams and the geometrical calibration of the focal plane}
\label{sec_beamrecovery}
    The in-flight assessment of the LFI main beams relies on the measurements
performed during seven Jupiter crossings;
the first four transits (``J1'' to ``J4'') occurred in nominal scan mode
(spin shift 2\,arcmin, 1\,deg per day), and the last three scans
(``J5'' to ``J7'') in a deeper coverage mode (spin shift 0.5\,arcmin,
15\,arcmin per day).  The period of time corresponding to each Jupiter
observation is reported in Table~\ref{tab:ods}.
By stacking data from the seven scans, we measure the main beam profiles
down to $-25\,$dB at 30 and 44\,GHz, and down to $-30$\,dB at 70\,GHz.
If we fit the main beam shapes with elliptical Gaussian profiles, the
uncertainties of the measured scanning beams can be expressed in terms of
statistical errors on these Gaussian parameters.
With respect to the 2013 release, the improvement in the signal-to-noise ratio
due to the number of samples and to better sky coverage is about a factor
of 2.  The beam full width half maximum is determined with a typical
uncertainty of 0.2\,\% at 30 and 44\,GHz, and 0.1\,\% at 70\,GHz,
approximately a factor of 2 better than the value achieved in 2013.
The fitting procedure also returns the main beam pointing directions in the
\Planck\ field of view (i.e. the focal plane geometry), centred along the
nominal line of sight as defined in \cite{tauber2010b}.

We determined the focal plane geometry of LFI independently for each Jupiter
crossing \citep{planck2014-a05}, using the same procedure as adopted in the
2013 release.  The solutions for the seven crossings agree within 4\,arcsec
at 70\,GHz, and 7\,arcsec at 30 and 44\,GHz.
The uncertainty in the determination of the main beam pointing directions
evaluated from the single scans is about 4\,arcsec for the nominal scans,
and 2.5\,arcsec for the deep scans at 70\,GHz (27\,arcsec for the nominal
scan and 19\,arcsec for the deep scan, at 30 and 44\,GHz).
Stacking the seven Jupiter transits, the uncertainty in the reconstructed main
beam pointing directions becomes 0.6\,arcsec at 70\,GHz and 2\,arcsec at
30 and 44\,GHz.
With respect to the 2013 release, we have found a difference in the main beam
pointing directions of about 5\,arcsec in the cross-scan direction and
0.6\,arcsec in the in-scan direction.
The beam centres and polarization orientation are defined by four parameters, $\theta_{uv}$ and $\phi_{uv}$, which define the beam pointing reconstructed using the stacked Jupiter transits; and $\psi_{\mathrm  uv}$ and $\psi_{\rm pol}$ defining the polarization orientation of the beam (see \citealt{planck2014-a05} and \citealt{planck2013-p28} for the definitions of these angles); their values for all the LFI radiometers are reported in Table~\ref{tab_fpg}.
Only $\theta_{uv}$ and $\phi_{uv}$, which are the beam pointing in spherical
coordinates referred to the line of sight, can be determined with Jupiter
observations. The polarization orientation of the beams, defined by $\psi_{uv} + \psi_{\rm pol}$, is estimated based on the geometry of the waveguide components in the LFI
focal plane (which for coherent detectors defines the polarization planes to high precision), reprojected in the sky through our GRASP model. As
discussed in \cite{planck2014-a04}, direct measurements of bright polarized sources (such as the Crab Nebula) provide only loose
constraints, and our final uncertainties on the polarization angles have been evaluated through simulations.

Details of the LFI main beam reconstruction and focal plane geometry
evaluation are reported in \cite{planck2014-a05}. 

\begin{table}[tmb]
  \begingroup
  \newdimen\tblskip \tblskip=5pt
  \caption{Approximate dates of the Jupiter observations.}
  \label{tab:ods}
  \nointerlineskip
  \vskip -3mm
  \footnotesize
  \setbox\tablebox=\vbox{
    \newdimen\digitwidth
    \setbox0=\hbox{\rm 0}
    \digitwidth=\wd0
    \catcode`*=\active
    \def*{\kern\digitwidth}
    \newdimen\signwidth
    \setbox0=\hbox{+}
    \signwidth=\wd0
    \catcode`!=\active
    \def!{\kern\signwidth}
    \halign{\tabskip=0pt\hfil#\hfil\tabskip=2em&
      \hfil#\hfil\tabskip=0pt\cr
      \noalign{\doubleline}
Jupiter transit & Date \cr
\noalign{\vskip 5pt\hrule\vskip 3pt}
Scan 1 (J1)& 21 Oct -- 5 Nov, 2009\cr
Scan 2 (J2)& 27 Jun -- 12 Jul, 2010\cr
Scan 3 (J3)& 3 -- 18 Dec, 2010\cr
Scan 4 (J4)& 30 Jul -- 8 Aug, 2011\cr
Scan 5 (J6)& 8 -- 30 Jan, 2012\cr
Scan 6 (J6)& 1 -- 14 Sept, 2012\cr
Scan 7 (J7)& 7 -- 28 Feb, 2013\cr
\noalign{\vskip 5pt\hrule\vskip 3pt}
}}
\endPlancktable
\endgroup
\end{table}

\begin{table}[tmb]
  \begingroup
  \newdimen\tblskip \tblskip=5pt
  \caption{Focal plane geometry.}
  \label{tab_fpg}
  \nointerlineskip
  \vskip -3mm
  \footnotesize
  \setbox\tablebox=\vbox{
    \newdimen\digitwidth
    \setbox0=\hbox{\rm 0}
    \digitwidth=\wd0
    \catcode`*=\active
    \def*{\kern\digitwidth}
    \newdimen\signwidth
    \setbox0=\hbox{+}
    \signwidth=\wd0
    \catcode`!=\active
    \def!{\kern\signwidth}
    \halign{\tabskip=0pt#\hfil\tabskip=2em&
      \hfil#\hfil&
      \hfil#\hfil&
      \hfil#\hfil&
      \hfil#\hfil\tabskip=0pt\cr
      \noalign{\doubleline}
      \noalign{\vskip -3pt}
      \omit\hfil Radiometer\hfil& $\theta_{uv}$\rlap{$^{\rm a}$}&
        $\phi_{uv}$\rlap{$^{\rm a}$}& $\psi_{uv}$\rlap{$^{\rm b}$}&
        $\psi_{\rm pol}$\rlap{$^{\rm b}$}\cr
      \noalign{\vskip 3pt\hrule\vskip 5pt}
      \omit& [deg]& [deg]& [deg]& [deg]\cr
      \noalign{\vskip 3pt\hrule\vskip 5pt}
      {\bf 70\,GHz}&&&\cr
      \noalign{\vskip 3pt}
      LFI 18M& 3.334& $-$131.828& !*22.15& 90.2\cr
      LFI 18S& 3.334& $-$131.820& !*22.15& *0.0\cr
      LFI 19M& 3.209& $-$150.482& !*22.40& 90.0\cr
      LFI 19S& 3.209& $-$150.488& !*22.40& *0.0\cr
      LFI 20M& 3.184& $-$168.182& !*22.38& 89.9\cr
      LFI 20S& 3.185& $-$168.194& !*22.38& *0.0\cr
      LFI 21M& 3.186&   !169.281&  *$-$22.38& 90.1\cr
      LFI 21S& 3.185&   !169.271&  *$-$22.38& *0.0\cr
      LFI 22M& 3.174&   !151.360&  *$-$22.34& 90.1\cr
      LFI 22S& 3.174&   !151.371&  *$-$22.34& *0.1\cr
      LFI 23M& 3.281&   !132.259&  *$-$22.08& 89.7\cr
      LFI 23S& 3.281&   !132.280&  *$-$22.08& *0.0\cr
      \noalign{\vskip 5pt\hrule\vskip 3pt}
      {\bf 44\,GHz}&&&\cr
      \noalign{\vskip 3pt}
      LFI 24M& 4.073& $-$179.540& !**0.01& 90.0\cr
      LFI 24S& 4.071& $-$179.505& !**0.01& *0.0\cr
      LFI 25M& 4.984&   !*61.093& $-$113.23& 89.5\cr
      LFI 25S& 4.983&   !*61.125& $-$113.23& *0.0\cr
      LFI 26M& 5.036& *$-$61.670& !113.23& 90.5\cr
      LFI 26S& 5.036& *$-$61.675& !113.23& *0.0\cr
      \noalign{\vskip 5pt\hrule\vskip 3pt}
      {\bf 30\,GHz}&&&\cr
      \noalign{\vskip 3pt}
      LFI 27M& 4.346&   !153.987& *$-$22.46& 89.7\cr
      LFI 27S& 4.346&   !153.985& *$-$22.46& *0.0\cr
      LFI 28M& 4.376& $-$153.424& !*22.45& 90.3\cr
      LFI 28S& 4.375& $-$153.418& !*22.45& *0.0\cr
      \noalign{\vskip 5pt\hrule\vskip 3pt}
    }}
  \endPlancktable
  \tablenote a Beam pointing reconstructed using the stacked Jupiter
  transits. \par
  \tablenote b Polarization orientation of the beam derived from
  simulations. \par
  \endgroup
\end{table}

    \subsection{Scanning beams}
    \label{sec_scanningbeam}
\begin{figure}[!hb]
\centering
\centering\includegraphics[width=8.8cm]{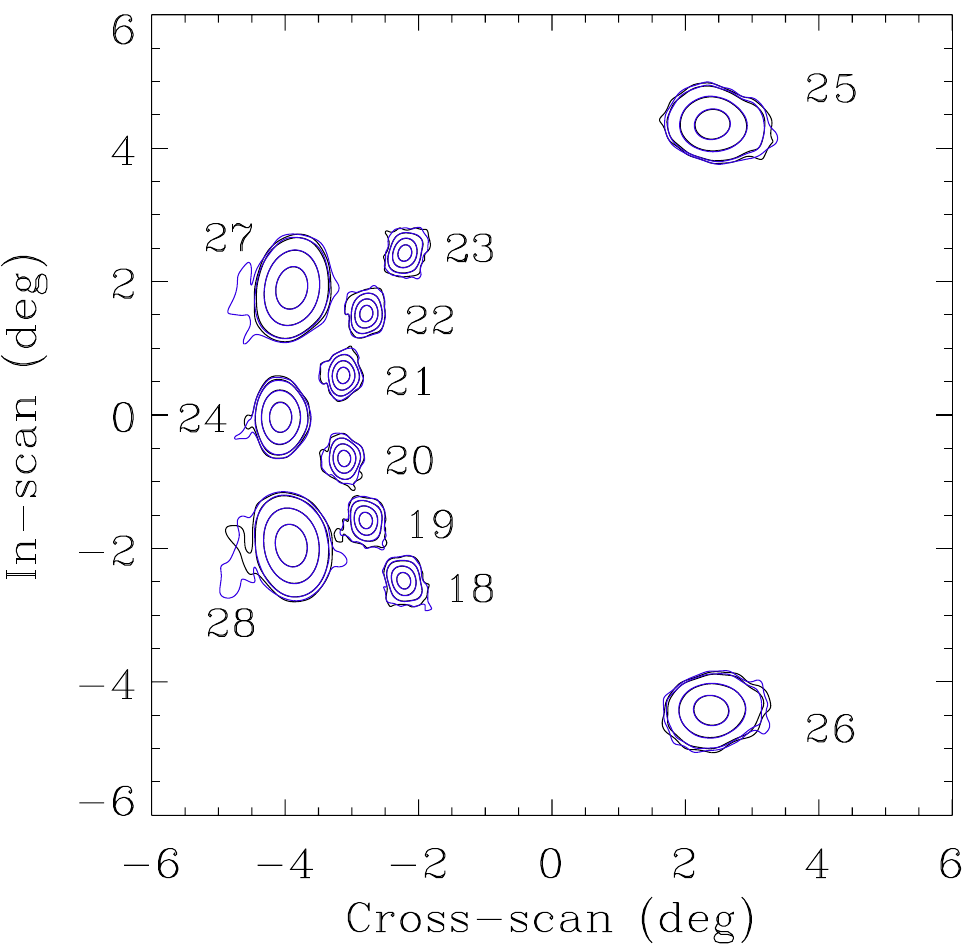}
\caption{Scanning beams reconstructed from Jupiter observations. The beams are
plotted in logarithmic contours of $-3$, $-10$, $-20$, and $-30$\,dB from the
peak, for the 70\,GHz channel (horns 18-23), and $-3$, $-10$, $-20$, and
$-25$\,dB from the peak, for the 30 and 44\,GHz channel (horns 27 and 28, and
24-26, respectively).  The main and side arms are indicated with
black and blue lines, respectively.} 
\label{fig:uvplane}
\end{figure}

        The ``scanning beams'', see Table~\ref{tab:imo} for main beam descriptive parameters, used in the LFI pipeline (affecting calibration,
effective beams, and beam window functions) are very similar to those
presented in \cite{planck2013-p02d}: they are {\tt GRASP} beams properly
smeared to take into account the satellite motion. 
They come from a tuned optical model and represent the most realistic fit to
the available measurements of the LFI main beams. 
These beams have now been validated using seven Jupiter transits.
The Jupiter scans allow us to measure the total field, that is the co- and
cross-polar components combined in quadrature.
The adopted beam model has the added advantage that it allows the co- and
cross-polar pattern to be defined separately; it also permits us to properly
consider the beam cross-polarization in every step of the LFI pipeline.
The scanning beams reconstructed from Jupiter transits are shown in
Fig.~\ref{fig:uvplane}

Unlike in \cite{planck2013-p02d}, where the main beams were full-power main
beams and the resulting beam window functions were normalized to unity
(because the calibration was performed assuming a pencil beam),
a different beam normalization is introduced here to properly take into
account the power entering the main beam (typically about 99\,\% of the total
power).  Indeed, as described in \cite{planck2014-a06}, the current LFI
calibration takes into account the full 4$\pi$ beam (i.e. the main beam,
as well as near and far sidelobes).  
Consequently, in the calculation of the window function, the beams are not
normalized to unity; instead, their normalization takes into account the real
efficiency calculated by considering the variation across the band of the
optical response (coupling between feedhorn pattern and telescope) and the
radiometric response (band shape).  This affects flux densities derived from
the maps (see Sect.~\ref{sec_iterative_calib_wf}).

In addition, ``hybrid beams'' have been created using planet measurements above
20\,dB from the main beam power peak and {\tt GRASP} beams below this
threshold.  The hybrid beams have been normalized to match the {\tt GRASP}
beams (i.e. the main beam efficiency is set to be the same).  
Hybrid beams have been used to perform a further check on the consistency
between the {\tt GRASP} model and the planet data, in terms of window
functions.  Further details are reported in \cite{planck2014-a05}.

\begin{table}[tmb]
\begingroup
\newdimen\tblskip \tblskip=5pt
\caption{Main beam de\-scrip\-tive pa\-ram\-e\-ters of the scanning beams,
with $\pm1\,\sigma$ uncertainties. $\psi_{\rm ell}$ represents the beam orientation as defined in  \cite{planck2013-p02d}}
\label{tab:imo}
\nointerlineskip
\vskip -3mm
\footnotesize
\setbox\tablebox=\vbox{
  \newdimen\digitwidth
  \setbox0=\hbox{\rm 0}
  \digitwidth=\wd0
  \catcode`*=\active
  \def*{\kern\digitwidth}
  \newdimen\signwidth
  \setbox0=\hbox{+}
  \signwidth=\wd0
  \catcode`!=\active
  \def!{\kern\signwidth}
  \halign{\tabskip=0pt#\hfil\tabskip=2em&
      \hfil#\hfil&
      \hfil#\hfil&
      \hfil#\hfil\tabskip=0pt\cr
  \noalign{\doubleline}
  \noalign{\vskip -3pt}
Beam&     FWHM& Ellipticity& $\psi_{\rm ell}$\cr
    & [arcmin]&            & [degrees]\cr
\noalign{\vskip 5pt\hrule\vskip 3pt}
{\bf 70\,GHz}&&&\cr
\noalign{\vskip 4pt}
LFI 18M& $13.40\pm0.02$& $1.235\pm0.004$& $*85.74\pm0.41$\cr
LFI 18S& $13.46\pm0.02$& $1.278\pm0.004$& $*86.41\pm0.33$\cr
LFI 19M& $13.14\pm0.02$& $1.249\pm0.003$& $*78.82\pm0.35$\cr
LFI 19S& $13.09\pm0.02$& $1.281\pm0.002$& $*79.15\pm0.30$\cr
LFI 20M& $12.83\pm0.02$& $1.270\pm0.003$& $*71.59\pm0.32$\cr
LFI 20S& $12.83\pm0.02$& $1.289\pm0.004$& $*72.69\pm0.31$\cr
LFI 21M& $12.75\pm0.02$& $1.280\pm0.003$& $107.99\pm0.27$\cr
LFI 21S& $12.86\pm0.02$& $1.294\pm0.003$& $106.96\pm0.29$\cr
LFI 22M& $12.92\pm0.02$& $1.264\pm0.003$& $101.87\pm0.30$\cr
LFI 22S& $12.99\pm0.02$& $1.279\pm0.003$& $101.61\pm0.30$\cr
LFI 23M& $13.32\pm0.02$& $1.235\pm0.004$& $*93.53\pm0.40$\cr
LFI 23S& $13.33\pm0.02$& $1.279\pm0.004$& $*93.49\pm0.36$\cr
\noalign{\vskip 5pt\hrule\vskip 3pt}
{\bf 44\,GHz}&&&\cr
\noalign{\vskip 4pt} 
LFI 24M& $23.18\pm0.05$& $1.388\pm0.005$& $*89.82\pm0.33$\cr
LFI 24S& $23.03\pm0.04$& $1.344\pm0.003$& $*89.97\pm0.34$\cr
LFI 25M& $30.02\pm0.07$& $1.191\pm0.005$& $115.95\pm0.75$\cr
LFI 25S& $30.79\pm0.07$& $1.188\pm0.005$& $117.70\pm0.74$\cr
LFI 26M& $30.13\pm0.08$& $1.191\pm0.006$& $*61.89\pm0.84$\cr
LFI 26S& $30.52\pm0.08$& $1.189\pm0.006$& $*61.53\pm0.77$\cr
\noalign{\vskip 5pt\hrule\vskip 3pt}
{\bf 30\,GHz}&&&\cr
\noalign{\vskip 4pt}  
LFI 27M& $32.96\pm0.06$& $1.364\pm0.005$& $101.20\pm0.34$\cr
LFI 27S& $33.16\pm0.07$& $1.379\pm0.005$& $101.29\pm0.34$\cr
LFI 28M& $33.17\pm0.07$& $1.366\pm0.006$& $*78.17\pm0.36$\cr
LFI 28S& $33.12\pm0.07$& $1.367\pm0.005$& $*78.47\pm0.33$\cr
\noalign{\vskip 5pt\hrule\vskip 3pt}
  }}
\endPlancktable
\endgroup
\end{table}

    \subsection{Effective beams}
    \label{sec_effectivebeam}
        The {\tt GRASP} combined co- and cross-polar main beams are used to calculate
the ``effective beams'', which take into account the specific scanning strategy
and pointing information in order to include any smearing and orientation
effects on the beams themselves. 
We compute the effective beam at each LFI frequency, using
the scanning beam and scan
history in real space using the {\tt FEBeCoP} \citep{mitra2010} method. 
Effective beams are used to calculate the effective beam window function, as
reported in \citep{planck2014-a05} and in the source detection pipeline used
to generate the PCCS catalogue \citep{planck2014-a35}. 
Table~\ref{tab_eff} lists the mean and rms variation across the sky of the
main parameters computed with {\tt FEBeCoP}. 
Note that the FWHM and ellipticity in Table~\ref{tab_eff} differ slightly from
the values reported in Table~\ref{tab:imo}.  
This results from the different way in which the Gaussian fit was applied. 
The scanning beam fit is determined by fitting the profile of Jupiter to
timelines and limiting the fit to the data with signal-to-noise ratio greater
than 3, while the fit of the effective beam is computed on {\tt GRASP} maps
projected in several positions of the sky ~\citep{planck2014-a05}. 
The latter are less affected by the noise.

\begin{table*}[tmb]
  \begingroup
  \newdimen\tblskip \tblskip=5pt
  \caption{Mean and rms variation across the sky of FWHM, ellipticity,
  orientation, and solid angle of the {\tt FEBeCop} effective beams computed
  with the {\tt GRASP} beam-fitted scanning beams.  Here FWHM$_{\mathrm {eff}}$
  is the effective FWHM estimated from the main beam solid angle of the
  effective beam, $\Omega_{\mathrm {eff}} = {\mathrm{mean}}(\Omega)$.}
  \label{tab_eff}
  \nointerlineskip
  \vskip -3mm
  \footnotesize
  \setbox\tablebox=\vbox{
    \newdimen\digitwidth
    \setbox0=\hbox{\rm 0}
    \digitwidth=\wd0
    \catcode`*=\active
    \def*{\kern\digitwidth}
    \newdimen\signwidth
    \setbox0=\hbox{+}
    \signwidth=\wd0
    \catcode`!=\active
    \def!{\kern\signwidth}
    \newdimen\pointwidth
    \setbox0=\hbox{\rm .}
    \pointwidth=\wd0
    \catcode`?=\active
    \def?{\kern\pointwidth}
    \halign{\hbox to 0.9 in{#\leaderfil}\tabskip=2em&
      \hfil#\hfil&
      \hfil#\hfil&
      \hfil#\hfil&
      \hfil#\hfil&
      \hfil#\hfil\tabskip=0pt\cr                            
      \noalign{\doubleline}
      \noalign{\vskip -3pt}
\omit\hfil Maps\hfil&FWHM & $e$& $\psi$& $\Omega$& FWHM$_{\rm eff}$\cr
\omit& [arcmin]& & [deg]& [arcmin$^2$]& [arcmin]\cr
      \noalign{\vskip 3pt\hrule\vskip 5pt}
{\sc Frequency Maps}&&&&&\cr
\noalign{\vskip 4pt}
LFI 70\,GHz& $13.213\pm0.034$& $1.223\pm0.026$& $*3\pm54$& $*200.90\pm*0.99$&  $13.315\pm0.033$\cr
LFI 44$^{\rm a}$\,GHz& $27.000\pm0.590$& $1.035\pm0.035$& $*0\pm50$& $*832.00\pm34.00$&  $27.100\pm0.57*$\cr
LFI 30\,GHz& $32.293\pm0.024$& $1.318\pm0.037$& $*0\pm54$& $1190.06\pm*0.69$&  $32.408\pm0.009$\cr
\noalign{\vskip 5pt\hrule\vskip 3pt}
{\sc Quadruplet Maps}&&&&&\cr
\noalign{\vskip 4pt}
LFI 18$-$23 & $13.525\pm0.021$& $1.188\pm0.021$& $*3\pm54$& $*210.13\pm*0.63$&  $13.618\pm0.020$\cr
LFI 19$-$22 & $13.154\pm0.037$& $1.230\pm0.027$& $*2\pm54$& $*199.19\pm*0.64$&  $13.259\pm0.021$\cr
LFI 20$-$21 & $12.910\pm0.037$& $1.256\pm0.036$& $*3\pm54$& $*192.58\pm*0.67$&  $13.037\pm0.023$\cr
LFI 25$-$26 & $29.975\pm0.013$& $1.177\pm0.030$& $-2\pm47$& $1019.63\pm*0.65$&  $29.998\pm0.009$\cr
LFI 24 & $23.036\pm0.014$& $1.341\pm0.033$& $*1\pm54$& $*603.61\pm*0.78$&  $23.080\pm0.015$\cr
%\noalign{\vskip 4pt}  
      \noalign{\vskip 5pt\hrule\vskip 3pt}
    }}
  \endPlancktablewide
  \tablenote a Associated errors are artificially large due to the fact that
 the 44\,GHz maps combine the beams of the horns 24, 25, and 26, which
 are very different from each other and are located far out in the focal plane
 (see Fig.~\ref{fig:uvplane}). We suggests using the quadruplet maps of
 LFI 25-26 and LFI 24 separately.\par 
  \endgroup
\end{table*}

    \subsection{Window functions}
    \label{sec_windowfunction}
        Window functions based on the LFI beams are needed for the production
of the LFI likelihoods and power spectra. They are based on the 
revised FEBeCoP (effective) beams discussed earlier in this section, 
and account for the renormalization of the beams described in
Sect.~\ref{sec_iterative_calib_wf}.
The derivation of the 2015 window functions is fully described in
\citet{planck2014-a05}, as are the uncertainties in
the window functions.  The uncertainties are sharply reduced from the
previous release and are: $0.7\,\%$ for the 30\,GHz band 
(evaluated at $\ell=600$); $1.0\,\%$ at 44\,GHz (also evaluated at
$\ell = 600$); and 0.5\,\% in the 70\,GHz window function at $\ell = 1000$.

\section{Photometric calibration}
\label{sec_calibration}
    With the term ``photometric calibration,'' we indicate the process
that converts the raw voltages $V$ measured by the LFI radiometers into a
thermodynamic temperature. The response of an LFI radiometer to a change in
the temperature coming from the sky can be modelled by the following
equation:
\begin{equation}
  \label{eq:CalRadiometerEquation}
  V(t) = G \times \left[ B * \bigl(D + T_\mathrm{CMB} + T_\mathrm{sky}\bigr)
 + T_0 \right],
\end{equation}
where $B$ is the beam response, the temperature $T = D +
T_\mathrm{CMB} + T_\mathrm{sky}$ is decomposed into the sum of three
terms (the dipole induced by the motion of the Solar system plus the
\Planck{} spacecraft, the CMB, and any other foregrounds), and $T_0$ is
a constant offset, which includes both instrumental offsets and the
CMB monopole. The quantity $G$ is the unknown term in the calibration
problem, and its inverse $K = G^{-1}$, the ``calibration
constant,'' is used to convert the timestream of voltages $V(t)$ into
temperatures.

\Planck's calibration source has always been the dipole term, $D$.
However, since the previous \Planck\ data release
\citep{planck2013-p02b} we have implemented a number of important
changes in the pipeline used to calibrate the voltages measured by the
LFI radiometers. In this section we provide an overview of the most
important result; we refer the reader interested in further details
to \citet{planck2014-a06}.

We use as a calibration source the signal $B * D$ in
Eq.~\eqref{eq:CalRadiometerEquation}, which is induced by the combined
motion of the spacecraft and the Solar System with respect to the CMB
rest frame. We have characterized the dipole by means of \Planck\
data and have estimated the amplitude to be
$(3364.5\pm2.0)\,\mu\mathrm{K}_{\rm CMB}$ in the direction
$ l = 264\pdeg00\pm0\pdeg03,\; b=48\pdeg24\pm0\pdeg20$ in Galactic
coordinates \citep{planck2014-a01}.
This represents an approximately 0.3\,\% increase in the amplitude
with respect to the dipole used in the 2013 data release, which was
based on the results of \citet{hinshaw2009}.

    \subsection{4$\pi$ calibration}
    \label{sec_iterative_calib}
        When we apply Eq.~\eqref{eq:CalRadiometerEquation} to solve the
calibration problem, we compute the value of $B*D$ by means of a full $4\pi$ 
convolution over the sphere, between the dipole signal (plus the
relativistic quadrupole component) and the beam response. This is
different from what other experiments have done when using the dipole as a
calibrator, e.g. WMAP and HFI assume the beam to be a Dirac delta function.
Our approach allows us to properly take into account the asymmetric
effect of the sidelobes and the efficiency of the main beam during the
calibration, which is critical for polarization, especially at low multipoles. Indeed, as discussed in \citet{planck2013-p02b}, the introduction of $4\pi$ calibration resulted in a significant improvement in the self-consistency of survey maps as demonstrated by null tests analysis.

 It can be demonstrated
\citep{planck2013-p02b,planck2014-a06} that the average level
$\tilde{C}^{4\pi}_\ell$ of the power spectrum, before convolving it
with the beam window function, changes with respect to the Dirac delta
case $\tilde{C}^{\delta}_\ell$ according to the following formula:

\begin{equation}
  \label{eq:ClFourPiVsClPencilBeam}
  \tilde{C}^{4\pi}_\ell = \tilde{C}^{\delta}_\ell \left(\frac{1 -
    f_\mathrm{sl} - \phi_\mathrm{sky} + \phi_D}{1 - \phi'_\textrm{sky}}\right)^2,
\end{equation}
where $f_\mathrm{sl}$ is the sidelobe fraction of the beam, and
$\phi_D \lesssim 0.2\,\%$,
$\phi_\mathrm{sky} \approx 0.01\,\%$, and $\phi'_\mathrm{sky} \ll 0.01\,\%$ are
quantities defined and discussed in \citet{planck2014-a06}; they depend on
the beam and the scanning strategy, and they are therefore
radiometer-dependent. The typical value of $\tilde{C}^{4\pi}_\ell /
\tilde{C}^{\delta}_\ell$ for the LFI radiometers deviates from unity by less
than 1\,\%.

The solution of the Eq.~\eqref{eq:CalRadiometerEquation} is provided by an
iterative destriper, {\tt DaCapo}, which supersedes our previous dipole-fitting
code used in the 2013 data release.  At each step the iterative procedure
determines the radiometer gains by fitting $D$ to the data, at the same time
extracting the contribution from the sky signal. Because of the degeneracy
between the overall gain level and the signal $D$, it makes sense to constrain
the map dipole to the model.  For this to work the contribution of foregrounds
to the dipole on the sky must be included in the dipole model.

\begin{figure}
\centering
\includegraphics[width=8.8cm]{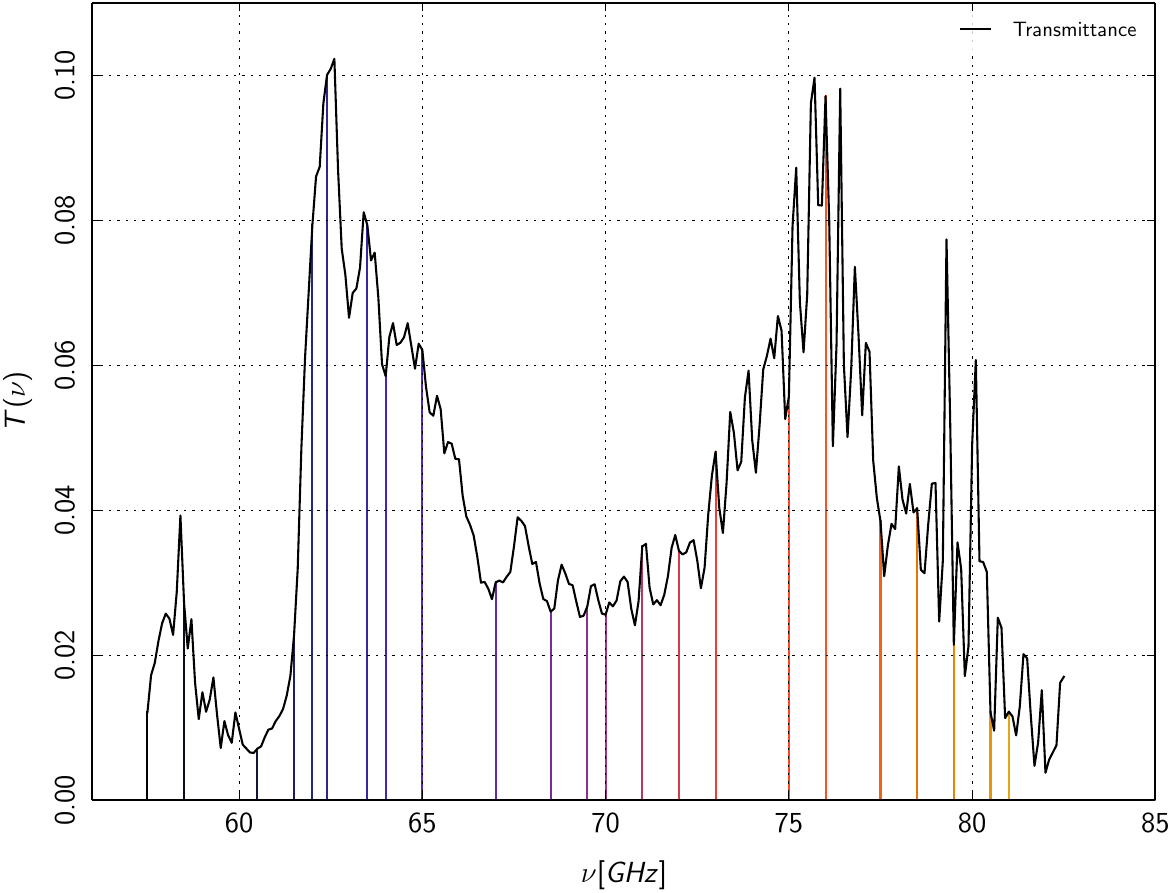}
\caption{Illustration of the method used to produce LFI synthetised beams 
weighted for the radiometer response (in this case LFI 18M). The vertical
lines identify the frequencies at which the beam has been simulated within the
radiometer bandpass $T(\nu)$. The results are then used to construct a 
weigthed $4\pi$ beam for each radiometer. Details on the bandpass measurements can be found in \cite{villa2010}.}
\label{fig:bandpass_18}
\end{figure}

The $4\pi$ beam model used in the calibration has been created taking into
account the radiometer bandpass of each radiometer (measured before flight).
For each radiometer about 25 realizations of the main beam, intermediate beam,
and sidelobe have been produced at fixed frequencies, chosen to fully sample
the shape of the bandpass (as shown for the LFI 18M bandpass and selected
frequencies in Fig.~\ref{fig:bandpass_18}). Those realization were then used
to construct a weighted $4\pi$ beam for each radiometer.

    \subsection{Impact of 4$\pi$ calibration on beam functions and source fluxes}
    \label{sec_iterative_calib_wf}
        The mapping procedure assumes a pencil beam \citep{planck2014-a07}, which, in
the ideal case of a circularly-symmetric beam, would yield a map of the
beam-convolved sky; therefore a fraction of the signal from any source appears
in the far sidelobes, and would be missed by integration of the map over the
main beam alone. 
By the same token, bright resolved features in the map have temperatures
fractionally lower than in the sky, due to signal lost in the sidelobes. 
In essence this description remains true even given the highly asymmetric
sidelobes of the Planck beam: the main difference is that the far sidelobe
contribution to a given pixel varies according to the orientation of the
satellite at the time of observation.
For LFI beams, roughly 1\,\% of the signal is in the sidelobes and this must
be accounted for in any analysis of the maps. 
In particular, the flux densities of compact sources measured from the maps
must be scaled up by the multiplicative factors $f_{{\rm sour}}$ reported in
Table~\ref{tab:fmber}.
These values have been computed from: 
\begin{itemize}
\item the main beam efficiencies \citep{planck2014-a05};
\item a re-normalization factor introduced by the calibration pipeline to
 compensate for the missing power in the 4$\pi$ beam \citep{planck2014-a06}
 (re-normalized beam efficiencies $\eta_{{\rm norm}}$ are also reported in
 Table~\ref{tab:fmber});
\item a factor that takes in account the horn uniform weights applied during
 the mapmaking process \citep{planck2014-a07}.
\end{itemize}

The re-normalization factor was introduced to take in account the
``missing power'' due to the first-order approximation adopted in the
computation carried out with the {\tt GRASP} Multireflector Geometrical Theory
of Diffraction ({\tt MrGTD}) software package \citep{planck2014-a05}.
The missing power was proportionally distributed between main, intermediate,
and sidelobe parts; this procedure has an effect on the previously computed
beam functions, and hence these have now been scaled by the factor 
$f_{{\rm Bl}}$ reported in Table~\ref{tab:fmber}.

\begin{table}[tmb]
\begingroup
\newdimen\tblskip \tblskip=5pt
\caption{Multiplicative factors that should be used to determine the correct
flux densities from compact sources. $\eta_{{\rm renorm}}$ is the re-normalized
main beam efficiency.
$f_{{\rm sour}}$ and $f_{{\rm Bl}}$ are multiplicative factors for
flux densities and beam function (already applied in the delivered LFI beam functions and flux densities of sources in the PCCS2~\citep{planck2014-a35}).}
\label{tab:fmber}
\nointerlineskip
\vskip -3mm
\footnotesize
\setbox\tablebox=\vbox{
  \newdimen\digitwidth
  \setbox0=\hbox{\rm 0}
  \digitwidth=\wd0
  \catcode`*=\active
  \def*{\kern\digitwidth}
  \newdimen\signwidth
  \setbox0=\hbox{+}
  \signwidth=\wd0
  \catcode`!=\active
  \def!{\kern\signwidth}
  \halign{\tabskip=0pt\hbox to 1.5in{#\leaderfil}\tabskip 2em&
      \hfil#\hfil&
      \hfil#\hfil&
      \hfil#\hfil\tabskip=0pt\cr
  \noalign{\doubleline}
  \noalign{\vskip -3pt}
\omit& $\eta_{{\rm renorm}}$& $f_{{\rm sour}}$& $f_{{\rm Bl}}$\cr
\noalign{\vskip 5pt\hrule\vskip 3pt}
\omit{\sc Frequency Maps}\hfil&&&\cr
\noalign{\vskip 4pt}
LFI 70\,GHz& 99.3582& 1.00646& 1.00346\cr
LFI 44\,GHz& 99.8827& 1.00117& 1.00143\cr
LFI 30\,GHz& 99.1983& 1.00808& 1.00258\cr
\noalign{\vskip 5pt\hrule\vskip 3pt}
\omit{\sc Quadruplet Maps}\hfil&&&\cr
\noalign{\vskip 4pt} 
LFI 18$-$23& 99.4556& 1.00547& 1.00333\cr
LFI 19$-$22& 99.3764& 1.00628& 1.00339\cr
LFI 20$-$21& 99.2238& 1.00782& 1.00368\cr
LFI 25$-$26& 99.9119& 1.00088& \dots\cr
LFI 24&      99.8228& 1.00177& \dots\cr
\noalign{\vskip 5pt\hrule\vskip 3pt}
  }}
\endPlancktable
\endgroup
\end{table}

In practice, in order to make consistent comparison with external data, it is essential that:
\begin{itemize}
\item users interested in CMB and diffuse component analysis, should use the official
LFI beam functions (in the LFI RIMO available in the \Planck\ Legacy Archive
interface\footnote{\url{http://archive.esac.esa.int/pla2}}) which already include the 
rescaling factor $f_{\rm Bl}$. Alternatively, users who wish to perform their
own beam deconvolution should multiply their beam functions by the
factor $f_{\rm Bl}$
\item users interested in point sources, the recalibration factors $f_{\rm sour}$ should be
used to obtain proper flux densities for sources extracted directly from LFI maps.
\end{itemize}

    \subsection{Smoothing algorithm}
    \label{sec_smoothing_calib}
\begin{figure}[!hb]
\centering
\includegraphics[width=8.8cm]{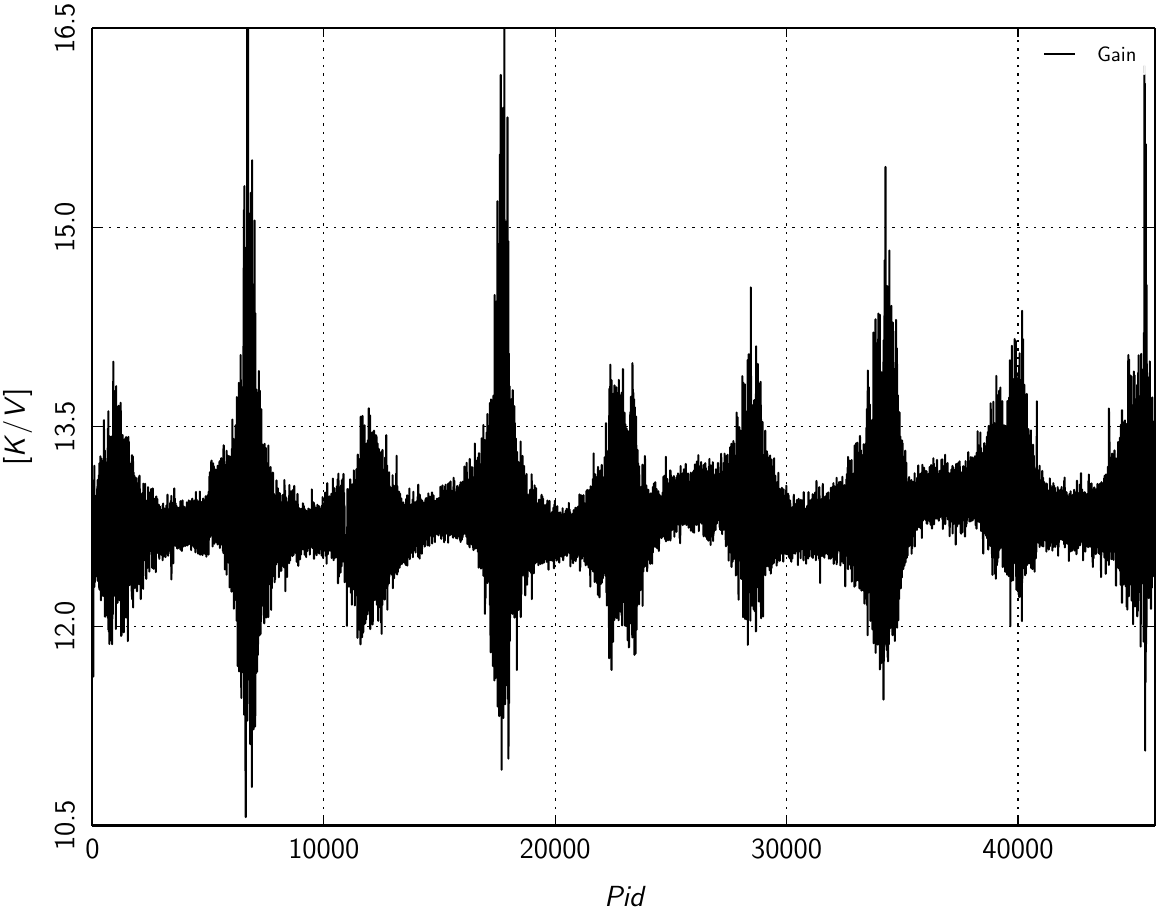}
\caption{Raw gain from radiometer 27M throughout 4 year mission.
$P_{\rm id}$ is a counter for pointings of the spin axis, which had an average duration of about 45 minutes \citet{planck2013-p01}.
The increase of noise corresponding to the periods
of ``minimum dipole'' (see text) are clearly visible for each of the
eight surveys. Survey 2 ($P_{\rm id}$ range approximately $5\,200-10\,000$) and
Survey 4 ($P_{\rm id}$ approximately $15\,700-20\,600$) exhibit a significantly
higher noise, as expected from the unfavourable alignment of the spacecraft spin
axis with the Solar dipole in those two surveys.}
\label{fig:raw_gains}
\end{figure}

        The uncertainty of the calibration constants increases significantly when the \Planck{} spacecraft is aligned such that the observed scan circle measures a low dipole component  (``minimum dipole''). This problem was particularly severe in
the Surveys~2 and 4, as shown in Fig.~\ref{fig:raw_gains}.

To reduce the noise, we apply an adaptive
smoothing algorithm that is also designed to preserve the discontinuities
caused by abrupt changes in the working configuration of the
radiometers (e.g. sudden temperature changes in the focal plane).
Moreover, we apply an additive, zero-mean correction to the
calibration constants derived from measurements of the emission of an
internal load kept at a stable temperature of approximately 4.5\,K,
plus the measurement of a set of temperature sensors mounted on the
focal plane of LFI. The amplitude of this correction is quite small
($\ll 1\,\%$, see Fig.~\ref{fig:fast_gains}), but its purpose is to
account for two phenomena.
\begin{enumerate}
\item During the first survey, the transponder used to download data
  to Earth was repeatedly turned on and off with a 24\,h duty cycle.
  This caused periodic fluctuations in the temperature of the back-end
  amplifiers, which were clearly traceable in the signal of the
  $4.5\,\mathrm{K}$ load \citep{planck2011-1.4}, but are not
  visible in the calibration constants computed using the dipole,
  because of statistical noise (this is particularly true during
  dipole minima).

\begin{figure}
\centering
\includegraphics[width=8.8cm]{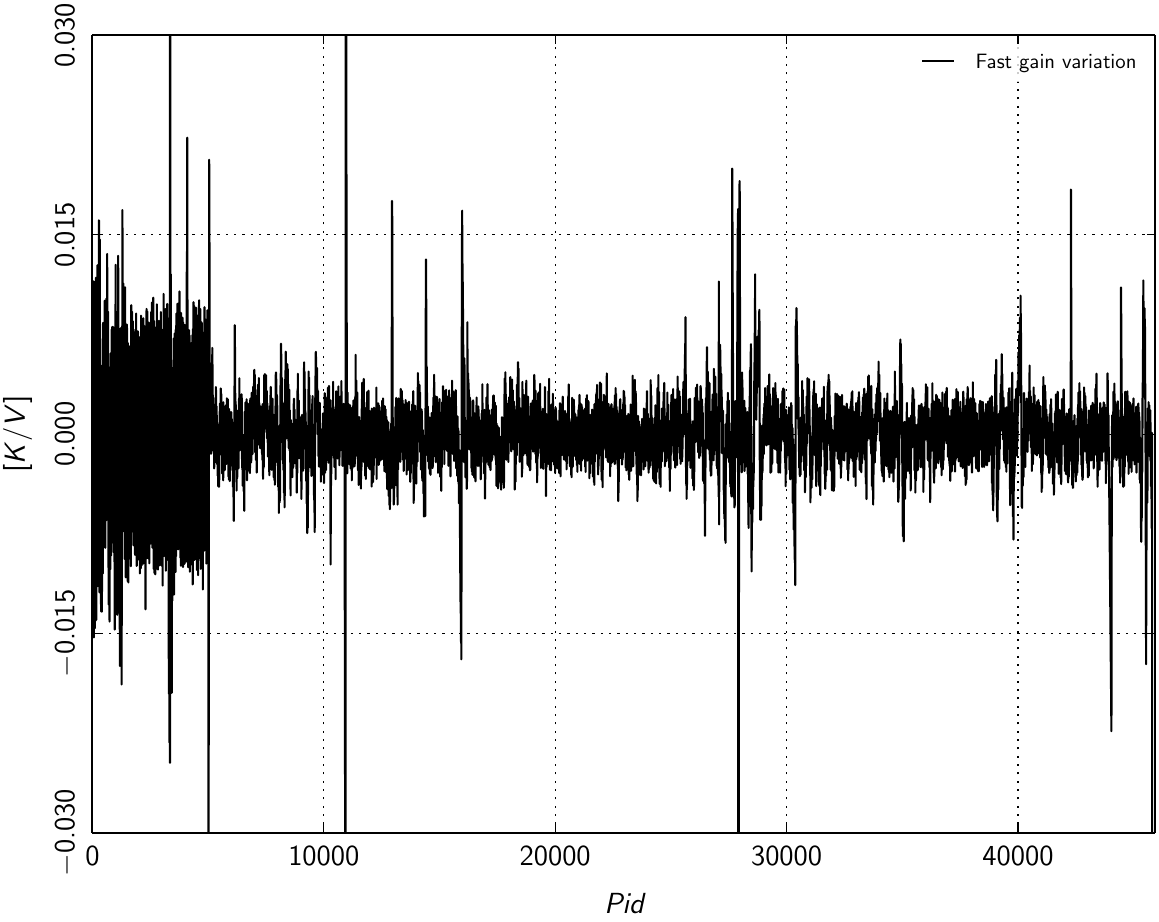}
\caption{High frequency fluctuations of the raw gain from radiometer 27M
throughout the 4 year mission. The major decrease in high frequency
variations occurs after the transponder was left continuously {\it on}
(at $P_{\rm id}$ = 5\,000). Subsequently the high frequency variations are
generally $\ll 1\, \%$.}
\label{fig:fast_gains}
\end{figure}

\item In general, during a dipole minimum, we are not able to keep
  track of variations in the gain of the radiometers. However, the knowledge of
  the internal 4.5\,K signal allows us to estimate an additive
  correction factor that mitigates the problem.
\end{enumerate}

    \subsection{Galactic straylight removal}
    \label{sec_strylight_removal}
        The light incident on the focal plane that does not reflect directly off the
primary mirror (straylight) is a major source of systematic effects,
especially when the Galactic plane intersects the direction of the main
spillover. This effect is now corrected by removing the 
estimated straylight signal from the timelines. To do this the
term $B_\mathrm{sl} * T_\mathrm{sky}$ of Eq.~(\ref{eq:CalRadiometerEquation})
has to be removed from calibrated timelines (here $B_\mathrm{sl}$ represents
the sidelobes contribution to the beam).
This term was computed for each radiometer by convolving both Galactic
and extragalactic emissions with the antenna pattern in the sidelobe region
(with angle $\theta > 5^\circ$ from the main beam pointing direction).
Here $T_\mathrm{sky}$ was estimated using simulated temperature and
polarization maps. These included the main
diffuse Galactic components (synchrotron, free-free, thermal, and anomalous
dust emissions) as well as contribution from faint and
strong radio sources and the thermal and kinetic Sunyaev-Zeldovich
effects (although the last is barely relevant at LFI frequencies), as
described in \cite{planck2014-a11} and \cite{planck2014-a12}.
These maps are weighted across the band using the transmission function
specific to each radiometer and then summed together. For polarization,
the contributions from both synchrotron and thermal dust have been considered. 

\begin{figure*}[htbp]
        \centering
        \begin{tabular}{ccc}
          \includegraphics[width=.32\textwidth]{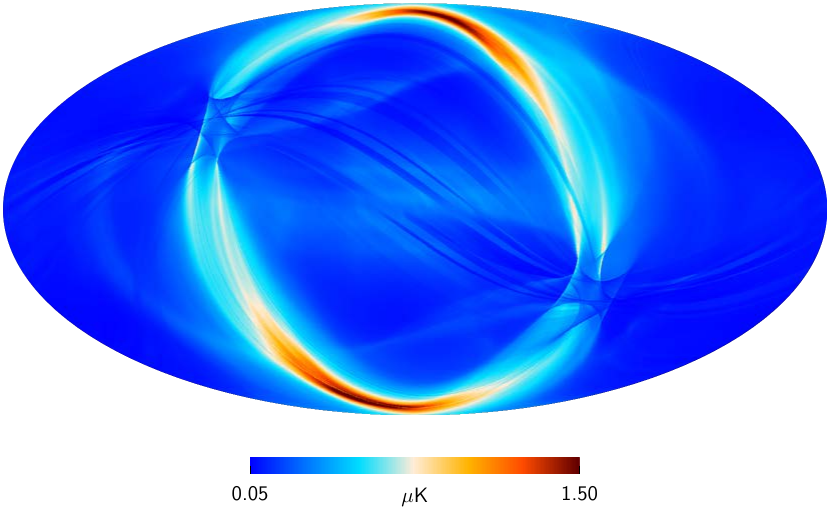} & \includegraphics[width=.32\textwidth]{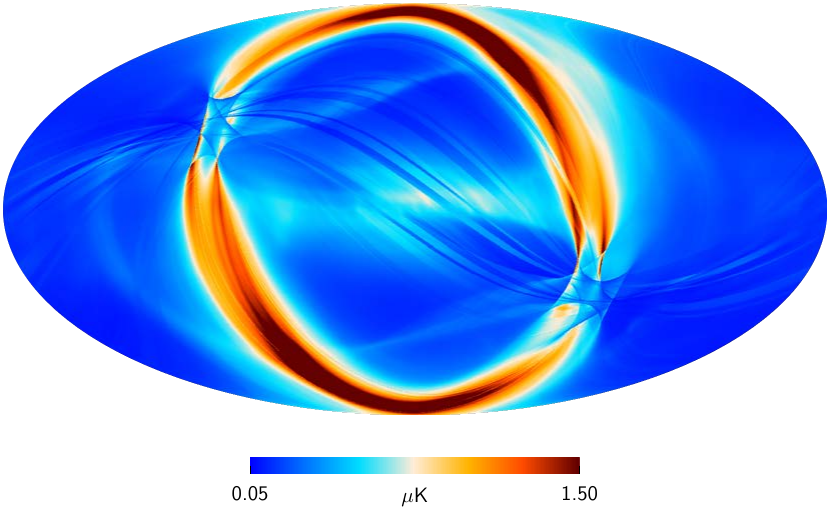}\\
          \includegraphics[width=.32\textwidth]{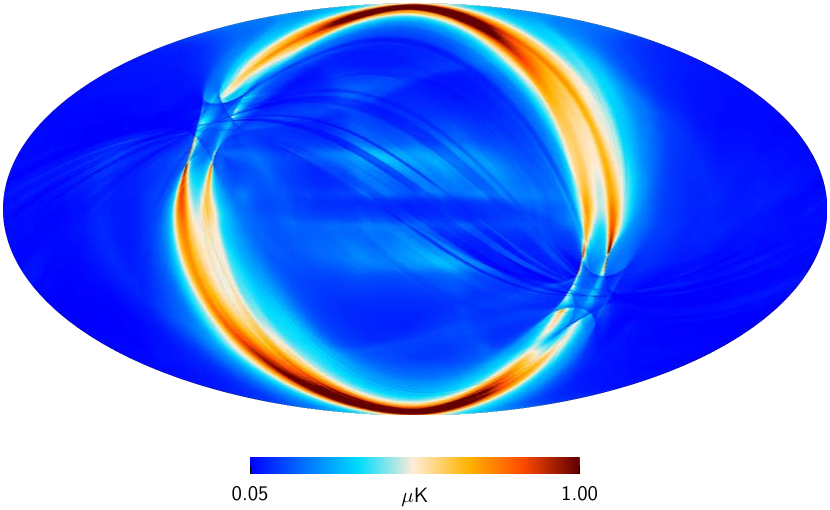} & \includegraphics[width=.32\textwidth]{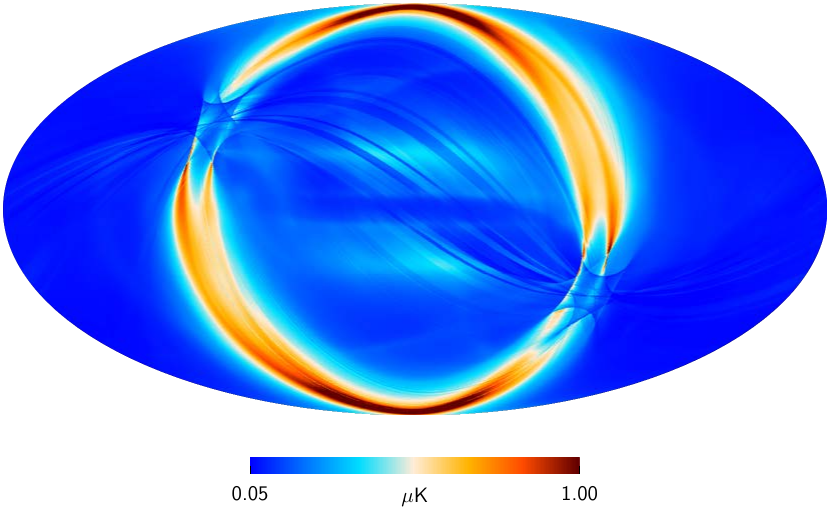}\\
          \includegraphics[width=.32\textwidth]{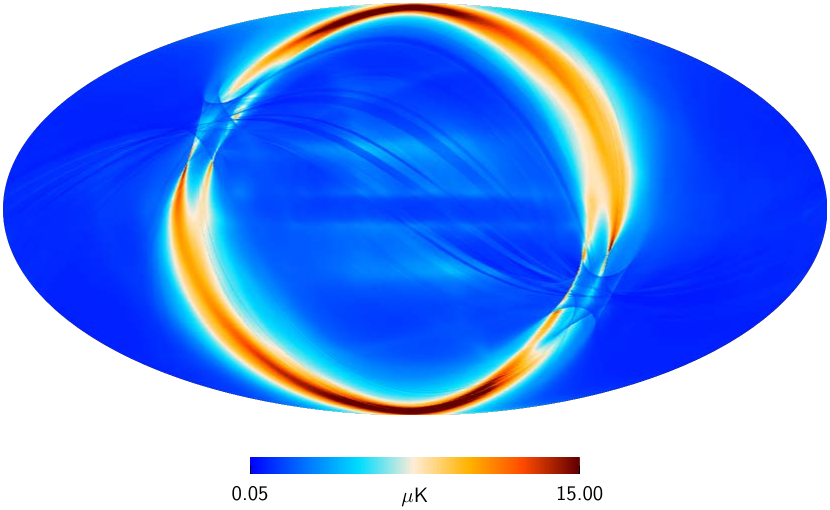} & \includegraphics[width=.32\textwidth]{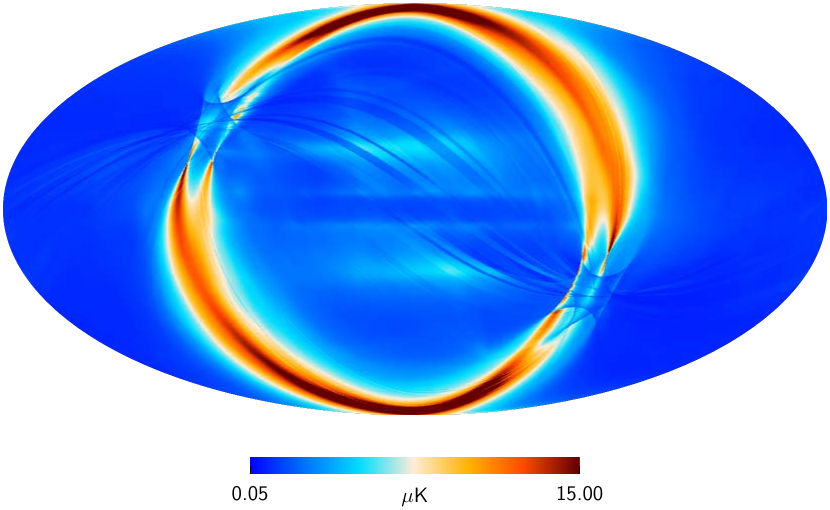}
        \end{tabular}
	\caption{Simulated Galactic straylight in total intensity for representative LFI radiometers for the full mission period.
{\it Top\/}: 70\,GHz radiometer 18M (right) and 18S (left).
{\it Middle\/}: 44\,GHz radiometer 24M (right) and 24S (left).
{\it Bottom\/}: 30\,GHz radiometer 27M and 27S (left).
The faint stripes paralleling the scanning direction are due to the different coverage
of the sky during different surveys.}
	\label{fig:galactic_straylight}
\end{figure*}

The convolution was performed by transforming both the sky and the sidelobe
pattern into spherical harmonics coefficients up to multipole $\ell = 2048$.
These coefficients are then properly multiplied 
to produce an object containing convolution results for each position on the
sky $(\theta,\phi)$ and beam orientation angle $\psi$.  For each sample in
the timeline, the straylight contribution has been evaluated by
performing a polynomial interpolation.
Figure~\ref{fig:galactic_straylight} shows expected Galactic straylight contribution 
in total intensity for a sample of LFI radiometers (both main and side arms), one at each frequency covering
the full mission period. 

    \subsection{Colour correction}
    \label{sec_color_correction}
        Colour corrections are required to adjust LFI measurements for sources or
foregrounds that do not have a thermal spectrum.  Our initial estimates
were listed in \cite{planck2013-p02b} for each LFI
radiometer and frequency band. For power-law spectra they can be well
approximated by a quadratic relationship between flux density and spectral
index $\alpha$
(or equivalently temperature spectral index $\beta = \alpha - 2$),
where the quadratic coefficient is proportional to the square of the fractional 
bandwidth, and the linear term mainly depends on the value of the chosen 
reference frequency \citep{LeahyFoley2006}. The constant component
is constrained by the requirement of zero colour correction for the CMB
spectrum, so there are two free parameters in the model. Accurate quadratic
fits are used in the {\tt fastcc} IDL code included in the \Planck\ unit
conversion and colour correction software package.

The more detailed component separation analysis for the 2015 release
\citep{planck2014-a12} has allowed us to further constrain the colour
corrections, which in the 2013 release were based purely on ground-based
measurement and modelling of the radiometer bandpasses. 
In recent analyses, we used separate
maps from each of the three co-scanning pairs of 70\,GHz horns.  The
analysis uses maps from LFI, HFI, and WMAP,
which includes several pairs of 
channels spaced closely in frequency. Using the nominal colour corrections
for the three instruments, highly significant and systematic residuals were
found to our best-fit models for the strong Galactic emission, which 
resemble gain errors; however, gain errors can be ruled out, because
there were no detectable residuals correlated with the CMB emission.
We thus assume that the previous colour corrections caused the residuals,
and have tried to improve them. 

\begin{figure}
\centering
\includegraphics[width=0.49\textwidth]{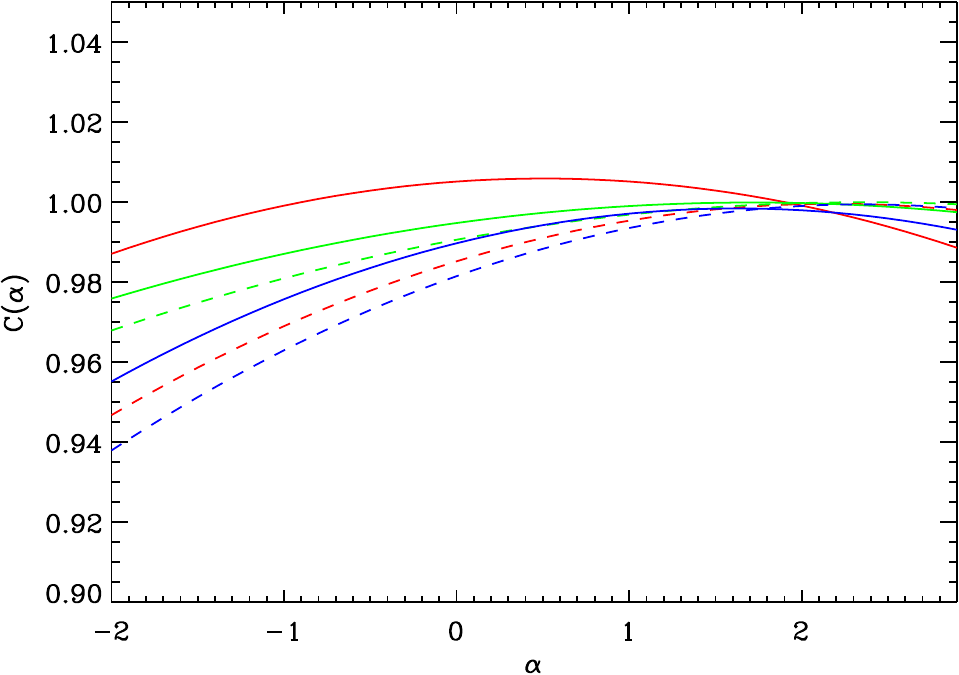}
\caption{Colour corrections $C(\alpha)$ versus intensity spectral index
$\alpha$. Solid lines are the current corrections given by
Table~\protect{\ref{tab:cc}},
while dashed lines are the 2013 values.
Red curves are for the 30\,GHz band,
green for 44\,GHz, and blue for 70\,GHz. Note that the corrections have only
been validated for $\alpha \protect\la 0$ at 30 and 44\,GHz, and for
$0 \protect\la \alpha \protect\la 2.5$ at 70\,GHz.}
\label{fig:cc}
\end{figure}

A first attempt has been made to derive improved colour corrections by fitting
for a frequency shift in the bandpass as part of the component separation 
analysis. This minimal model was adopted to avoid a strong degeneracy 
between the bandpass recalibration and the foreground spectral models; it
is certainly an oversimplification.
The resulting fractional change of frequency is $1.0\pm0.3$\,\%
at 30\,GHz, $0.2\pm 0.2$\,\% at 44\,GHz, and $-0.6$\,\%, $1.6$\,\%
and $0.7$\,\% (all $\pm 1.4$\,\%) for the three 70\,GHz horn pairs 
(18 and 23, 19 and 22, 20 and 21, respectively). 
The uncertainties quoted here are the absolute ones.
For convenience, Table~\ref{tab:cc} lists the
parameters of our parabolic fit to the colour corrections derived from the
shifted bandpasses, where for a map thermodynamic temperature $\tilde{T}$,
the Rayleigh-Jeans brightness temperature at the reference frequency $\nu_0$
is given by
\begin{equation}
T(\nu_0)[{\rm K}_{\rm RJ}] = 
 \tilde{T}[{\rm K}_{\rm CMB}]\,\eta_{\Delta T}(\nu_0)\, \cc(\alpha),
\end{equation}
where $\eta_{\Delta T}(\nu) = \partial T_{\rm RJ}/\partial T|_{T_{\rm CMB}}$,
and the coefficients in Table~\ref{tab:cc} give the colour correction as
$\cc(\alpha) = c_0 + c_1 \alpha + c_2 \alpha^2$.
Because they are based on a simplified analysis, these values
should be treated with some caution; the revised colour corrections
have only been tested for spectral indices near that of the dominant 
foregrounds at each frequency, namely $-1 \la \alpha \la 0$
at 30 and 44\,GHz, and $0 \la \alpha \la 2.5$ at 70\,GHz. 
We plot the old and new corrections in Fig.~\ref{fig:cc}.

\begin{table}
\begingroup
\newdimen\tblskip \tblskip=5pt
\caption{Coefficients for parabolic fits to the LFI colour corrections
$C(\alpha)$, revised from the 2013 values, based on the bandpass shifts
derived by {\tt Commander} component separation code \citep{planck2014-a12}.}
\label{tab:cc}
\nointerlineskip
\vskip -3mm
\footnotesize
\setbox\tablebox=\vbox{
 \newdimen\digitwidth
 \setbox0=\hbox{\rm 0}
 \digitwidth=\wd0
 \catcode`*=\active
 \def*{\kern\digitwidth}
  \newdimen\dpwidth
  \setbox0=\hbox{.}
  \dpwidth=\wd0
  \catcode`!=\active
  \def!{\kern\dpwidth}
\halign{\tabskip=0pt\hbox to 2cm{#\leaderfil}\tabskip 1em&
    \hfil#\hfil\tabskip 1em&
    \hfil#\hfil \tabskip 2em&
    \hfil#\hfil \tabskip 1.3em&
    \hfil#\hfil \tabskip 1em&
    \hfil#\hfil \tabskip 0em\cr
\noalign{\doubleline}
\omit\hfil Horns\hfil& Band& $\nu_0$& $c_0$& $c_1$& $c_2$\cr
\noalign{\vskip 3pt}
\omit& [GHz]& [GHz]& & & \cr
\noalign{\vskip 3pt\hrule\vskip 5pt}
27, 28&     30& 28.4& 1.005& *0.0030& $-0.0030$\cr
24, 25, 26& 44& 44.1& 0.995& *0.0060& $-0.0017$\cr
18, 23&     70& 70.4& 0.983& *0.0142& $-0.0032$\cr
19, 22&     70& 70.4& 1.010& $-0.0007$& $-0.0033$\cr
20, 21&     70& 70.4& 0.977& *0.0176& $-0.0031$\cr
All&        70& 70.4& 0.990& *0.0107& $-0.0033$\cr
\noalign{\vskip 3pt\hrule\vskip 5pt}
}}
\endPlancktable
\endgroup
\end{table}

In \citet{planck2013-p02b} we gave a rough indirect
estimate of the expected uncertainties in the colour corrections, assuming
that the errors for the individual radiometer bandpasses were uncorrelated.
The revised corrections at 30\,GHz differ from our original ones by 
2--3\,\% for for $\alpha \approx -1$, and this change is almost an
order of magnitude larger than our original error estimate. 
In retrospect, our assumption of 
uncorrelated errors was flawed for this particular channel, since our
ground-based estimates of the bandpass shape were particularly sensitive
to modelling assumptions. This arose because the bands still had 
significant response
at the low-frequency end of the the directly-measured range. As a result, 
it seems likely that the actual difference between our 2013 model and 
the true 30\,GHz bandpasses is more in the nature of an upward revision of the 
low-frequency cutoff than a uniform shift to higher frequency; if this is the
case, then our 2015 estimate for $\cc(\alpha)$ will still show too much
curvature in this band.

The estimated bandpass shifts at 44\, and 70\,GHz are not significant, but
they correspond to colour-correction changes of 0.5\,\% and 1\,\%, respectively,
consistent with our original error estimates; moreover, as explained in
\citet{planck2014-a12}, the relative shifts between the 70\,GHz horn pairs 
are known much more accurately than their absolute values, and are
certainly important. We therefore recommend the use of the revised
colour corrections listed here. The uncertainties in the correction should be
taken to be approximately $|\beta|\times 0.3\,\%$ for all channels,
as long as the spectral index
is close to the well-sampled range, $-3 \ga \beta \ga 1$.
We note that, by construction, the colour correction tends to unity for
emission having the colour of the CMB, and so $\cc$ remains accurately 
equal to unity when $\beta = \beta_{\rm CMB}$. 

Since it is possible to make total intensity sky maps from the data for
each individual LFI feed horn (averaging the data from the M and S 
radiometers), it will be possible to improve the colour corrections
individually for each horn, and we plan to do that for the next release.

    \subsection{Summary of changes in LFI calibration}
    \label{sec_summ_changes}
        In this subsection, we summarize the changes in the overall calibration 
of the \Planck LFI channels that have resulted from different 
procedures adopted since \citet{planck2013-p02} and from our deeper 
understanding of instrumental systematics and their effect on calibration.
\begin{itemize}
\item Overall calibration.  Improved accounting for beam effects and other
changes discussed in Sect.~\ref{sec_beamrecovery} and
Sect.~\ref{sec_calibration} produces a small upward shift 
in the calibration for the three LFI channels. In addition, our current use
of the orbital dipole for the determination of the solar dipole used for
calibration has shown that the previous calibration based on the WMAP solar
dipole was 0.28\,\% low for all frequencies.
Combining these effects, we find the following upward shifts in LFI
calibration: 0.83\,\%, 0.72\,\%, and 0.95\,\% for 30\,GHz, 44\,GHz,
and 70\,GHz, respectively.

\item Uncertainties in calibration.  Improved understanding and assessment of
the impact of various systematic effects on calibration have allowed us to
refine our estimates of overall calibration uncertainty.  The uncertainties
are 0.35\,\%, 0.26\,\%, and 0.20\,\% for 30, 44, and 70\,GHz, respectively.

\item Window function. We now use 4$\pi$ beams, rather than a pencil beam
approximation. LFI window functions properly take account 
of the small amount of missing power in the sidelobes
(a roughly $0.4$\,\% effect at most,  see Table.~\ref{tab:fmber}).

\item Flux densities of compact sources. Our current use of a 4$\pi$ beams
also means that flux densities of compact sources need to be boosted by a small
factor if they are derived from the LFI maps (again, see Table~\ref{tab:fmber}). Flux densities in the PCCS2, on the other hand, are already corrected for this factor.

\end{itemize}

\section{LFI noise estimation}
\label{sec_general_noise}

   \subsection{Radiometer noise model}
   \label{sec_noise}
    A detailed knowledge of instrumental noise properties is fundamental for
several stages of the data analysis. First of all evolution in time of basic
noise properties (e.g. white noise variance) throughout the entire mission
lifetime is an important and simple way to track possible variations and
even anomalies in the instrument behaviour. In addition, noise properties
serve as inputs for the Monte Carlo noise simulations (used, e.g. for power
spectrum estimation) and also give correct weights for properly combining
different detectors.

We proceed as already shown in \cite{planck2013-p02} using an
implementation of a Monte Carlo Markov chain (MCMC) approach to estimate basic noise
properties. As before, the noise model is
\begin{equation}
P(f) = \sigma^2 \left[1+\left(\frac{f}{f_{\rm knee}}\right)^\beta \right]\, ,
\label{noise_model}
\end{equation}
where $\sigma^2$ is the white noise level, and $f_{\rm knee}$ and $\beta$
describe the non-white component of the instrumental noise. To evaluate
$\sigma^2$, we take the mean of the noise spectrum in the last few
(typically 10\,\%) of the bins at the highest frequency, which exhibits a flat,
high-frequency tail, as shown in Figure~\ref{noiseps}.
At 30\,GHz the knee-frequency is $f_{\rm knee} \approx 100\, {\rm mHz}$
and therefore a smaller percentage of data has been taken for computing
$\sigma^2$. These values for $\sigma$ are given in
Table~\ref{tab_white_noise_per_radiometer}. Once this is done we can
proceed with the evaluation of the other two parameters. After discarding a
burn-in period from our chains, we obtained the best-fit and variances values
reported in Table~\ref{tab_one_over_f_noise_per_radiometer}.

    \subsection{Updated noise properties}
    \label{sec_update_noise}
        We estimate noise properties at the radiometer level using the MCMC approach.
As already done with the previous data release, we work with calibrated data
and select chunks of data 5~days long and process them with the {\tt roma}
generalized least-squares mapmaking algorithm \citep{degasperis2005}.
The outputs are frequency spectra that are then fitted for the
basic noise parameters. Results are summarized in
Tables~\ref{tab_white_noise_per_radiometer}
and \ref{tab_one_over_f_noise_per_radiometer}, for the white noise
sensitivity and $1/f$ noise parameters, respectively. These numbers are the
medians, computed from the fit results throughout the
whole mission lifetime.

  \begin{table}
  \begingroup
  \newdimen\tblskip \tblskip=5pt
  \caption{White noise sensitivities for the LFI radiometers.}
  \label{tab_white_noise_per_radiometer}
  \nointerlineskip
  \vskip -5mm
  \footnotesize
  \setbox\tablebox=\vbox{
  \newdimen\digitwidth
  \setbox0=\hbox{\rm 0}
  \digitwidth=\wd0
  \catcode`*=\active
  \def*{\kern\digitwidth}
  \newdimen\signwidth
  \setbox0=\hbox{+}
  \signwidth=\wd0
  \catcode`!=\active
  \def!{\kern\signwidth}
  \halign{\hbox to 1.3in{#\leaderfil}\tabskip=2em&
      \hfil#\hfil&
      \hfil#\hfil\tabskip=0pt\cr
  \noalign{\doubleline}
  \omit&\multispan2\hfil W{\sc hite} N{\sc oise} S{\sc ensitivity}\hfil\cr
  \noalign{\vskip -4pt}
  \omit&\multispan2\hrulefill\cr
  \omit&Radiometer M&Radiometer S\cr
  \omit\hfil \hfil&[$\,\mu\mathrm{K}_{\rm CMB}\, \mathrm{s}^{1/2}$]&[$\,\mu\mathrm{K}_{\rm CMB}\, \mathrm{s}^{1/2}$]\cr
  \noalign{\vskip 3pt\hrule\vskip 5pt}
  \omit{\bf 70\,GHz}\hfil\cr
  \noalign{\vskip 4pt}
  \hglue 2em LFI 18& \getsymbol{LFI:white:noise:sensitivity:LFI18:Rad:M}$\,\pm$\getsymbol{LFI:white:noise:sensitivity:uncertainty:LFI18:Rad:M}& \getsymbol{LFI:white:noise:sensitivity:LFI18:Rad:S}$\,\pm$\getsymbol{LFI:white:noise:sensitivity:uncertainty:LFI18:Rad:S}\cr
  \hglue 2em LFI 19& \getsymbol{LFI:white:noise:sensitivity:LFI19:Rad:M}$\,\pm$\getsymbol{LFI:white:noise:sensitivity:uncertainty:LFI19:Rad:M}& \getsymbol{LFI:white:noise:sensitivity:LFI19:Rad:S}$\,\pm$\getsymbol{LFI:white:noise:sensitivity:uncertainty:LFI19:Rad:S}\cr
  \hglue 2em LFI 20& \getsymbol{LFI:white:noise:sensitivity:LFI20:Rad:M}$\,\pm$\getsymbol{LFI:white:noise:sensitivity:uncertainty:LFI20:Rad:M}& \getsymbol{LFI:white:noise:sensitivity:LFI20:Rad:S}$\,\pm$\getsymbol{LFI:white:noise:sensitivity:uncertainty:LFI20:Rad:S}\cr
  \hglue 2em LFI 21& \getsymbol{LFI:white:noise:sensitivity:LFI21:Rad:M}$\,\pm$\getsymbol{LFI:white:noise:sensitivity:uncertainty:LFI21:Rad:M}& \getsymbol{LFI:white:noise:sensitivity:LFI21:Rad:S}$\,\pm$\getsymbol{LFI:white:noise:sensitivity:uncertainty:LFI21:Rad:S}\cr
  \hglue 2em LFI 22& \getsymbol{LFI:white:noise:sensitivity:LFI22:Rad:M}$\,\pm$\getsymbol{LFI:white:noise:sensitivity:uncertainty:LFI22:Rad:M}& \getsymbol{LFI:white:noise:sensitivity:LFI22:Rad:S}$\,\pm$\getsymbol{LFI:white:noise:sensitivity:uncertainty:LFI22:Rad:S}\cr
  \hglue 2em LFI 23& \getsymbol{LFI:white:noise:sensitivity:LFI23:Rad:M}$\,\pm$\getsymbol{LFI:white:noise:sensitivity:uncertainty:LFI23:Rad:M}& \getsymbol{LFI:white:noise:sensitivity:LFI23:Rad:S}$\,\pm$\getsymbol{LFI:white:noise:sensitivity:uncertainty:LFI23:Rad:S}\cr
  \noalign{\vskip 5pt}
  \omit{\bf 44\,GHz}\hfil\cr
  \noalign{\vskip 4pt}
  \hglue 2em LFI 24& \getsymbol{LFI:white:noise:sensitivity:LFI24:Rad:M}$\,\pm$\getsymbol{LFI:white:noise:sensitivity:uncertainty:LFI24:Rad:M}& \getsymbol{LFI:white:noise:sensitivity:LFI24:Rad:S}$\,\pm$\getsymbol{LFI:white:noise:sensitivity:uncertainty:LFI24:Rad:S}\cr
  \hglue 2em LFI 25& \getsymbol{LFI:white:noise:sensitivity:LFI25:Rad:M}$\,\pm$\getsymbol{LFI:white:noise:sensitivity:uncertainty:LFI25:Rad:M}& \getsymbol{LFI:white:noise:sensitivity:LFI25:Rad:S}$\,\pm$\getsymbol{LFI:white:noise:sensitivity:uncertainty:LFI25:Rad:S}\cr
  \hglue 2em LFI 26& \getsymbol{LFI:white:noise:sensitivity:LFI26:Rad:M}$\,\pm$\getsymbol{LFI:white:noise:sensitivity:uncertainty:LFI26:Rad:M}& \getsymbol{LFI:white:noise:sensitivity:LFI26:Rad:S}$\,\pm$\getsymbol{LFI:white:noise:sensitivity:uncertainty:LFI26:Rad:S}\cr
  \noalign{\vskip 5pt}
  \omit{\bf 30\,GHz}\hfil\cr
  \noalign{\vskip 4pt}
  \hglue 2em LFI 27& \getsymbol{LFI:white:noise:sensitivity:LFI27:Rad:M}$\,\pm$\getsymbol{LFI:white:noise:sensitivity:uncertainty:LFI27:Rad:M}& \getsymbol{LFI:white:noise:sensitivity:LFI27:Rad:S}$\,\pm$\getsymbol{LFI:white:noise:sensitivity:uncertainty:LFI27:Rad:S}\cr
  \hglue 2em LFI 28& \getsymbol{LFI:white:noise:sensitivity:LFI28:Rad:M}$\,\pm$\getsymbol{LFI:white:noise:sensitivity:uncertainty:LFI28:Rad:M}& \getsymbol{LFI:white:noise:sensitivity:LFI28:Rad:S}$\,\pm$\getsymbol{LFI:white:noise:sensitivity:uncertainty:LFI28:Rad:S}\cr
  \noalign{\vskip 5pt\hrule\vskip 3pt}}}
  \endPlancktable
  \endgroup
  \end{table}

  \begin{table*}
  \begingroup
  \newdimen\tblskip \tblskip=5pt
  \caption{Knee frequencies and slopes for the LFI radiometers.}
  \label{tab_one_over_f_noise_per_radiometer}
  \nointerlineskip
  \vskip -3mm
  \footnotesize
  \setbox\tablebox=\vbox{
  \newdimen\digitwidth
  \setbox0=\hbox{\rm 0}
  \digitwidth=\wd0
  \catcode`*=\active
  \def*{\kern\digitwidth}
  \newdimen\signwidth
  \setbox0=\hbox{+}
  \signwidth=\wd0
  \catcode`!=\active
  \def!{\kern\signwidth}
  \halign{\hbox to 1.3in{#\leaderfil}\tabskip=2em&
      \hfil#\hfil&
      \hfil#\hfil&
      \hfil#\hfil&
      \hfil#\hfil\tabskip=0pt\cr
  \noalign{\doubleline}
  \omit&\multispan2\hfil K{\sc nee} F{\sc requency} $f_{\rm knee}$ [mHz]\hfil&\multispan2\hfil S{\sc lope} $\beta$\hfil\cr
  \noalign{\vskip -4pt}
  \omit&\multispan2\hrulefill&\multispan2\hrulefill\cr
  \omit& Radiometer M&Radiometer S&Radiometer M&Radiometer S\cr
  \noalign{\vskip 3pt\hrule\vskip 5pt}
  \omit{\bf 70\,GHz}\hfil\cr
  \noalign{\vskip 4pt}
  \hglue 2em LFI 18& *\getsymbol{LFI:knee:frequency:LFI18:Rad:M}$\,\pm\,$\getsymbol{LFI:knee:frequency:uncertainty:LFI18:Rad:M}&
              *\getsymbol{LFI:knee:frequency:LFI18:Rad:S}$\,\pm\,$\getsymbol{LFI:knee:frequency:uncertainty:LFI18:Rad:S}&
              \getsymbol{LFI:slope:LFI18:Rad:M}$\,\pm\,$\getsymbol{LFI:slope:uncertainty:LFI18:Rad:M}&
              \getsymbol{LFI:slope:LFI18:Rad:S}$\,\pm\,$\getsymbol{LFI:slope:uncertainty:LFI18:Rad:S}\cr
  \hglue 2em LFI 19& *\getsymbol{LFI:knee:frequency:LFI19:Rad:M}$\,\pm\,$\getsymbol{LFI:knee:frequency:uncertainty:LFI19:Rad:M}&
              *\getsymbol{LFI:knee:frequency:LFI19:Rad:S}$\,\pm\,$\getsymbol{LFI:knee:frequency:uncertainty:LFI19:Rad:S}&
              \getsymbol{LFI:slope:LFI19:Rad:M}$\,\pm\,$\getsymbol{LFI:slope:uncertainty:LFI19:Rad:M}&
              \getsymbol{LFI:slope:LFI19:Rad:S}$\,\pm\,$\getsymbol{LFI:slope:uncertainty:LFI19:Rad:S}\cr
  \hglue 2em LFI 20& **\getsymbol{LFI:knee:frequency:LFI20:Rad:M}$\,\pm\,$\getsymbol{LFI:knee:frequency:uncertainty:LFI20:Rad:M}&
              **\getsymbol{LFI:knee:frequency:LFI20:Rad:S}$\,\pm\,$\getsymbol{LFI:knee:frequency:uncertainty:LFI20:Rad:S}&
              \getsymbol{LFI:slope:LFI20:Rad:M}$\,\pm\,$\getsymbol{LFI:slope:uncertainty:LFI20:Rad:M}&
              \getsymbol{LFI:slope:LFI20:Rad:S}$\,\pm\,$\getsymbol{LFI:slope:uncertainty:LFI20:Rad:S}\cr
  \hglue 2em LFI 21& *\getsymbol{LFI:knee:frequency:LFI21:Rad:M}$\,\pm\,$\getsymbol{LFI:knee:frequency:uncertainty:LFI21:Rad:M}&
              *\getsymbol{LFI:knee:frequency:LFI21:Rad:S}$\,\pm\,$\getsymbol{LFI:knee:frequency:uncertainty:LFI21:Rad:S}&
              \getsymbol{LFI:slope:LFI21:Rad:M}$\,\pm\,$\getsymbol{LFI:slope:uncertainty:LFI21:Rad:M}&
              \getsymbol{LFI:slope:LFI21:Rad:S}$\,\pm\,$\getsymbol{LFI:slope:uncertainty:LFI21:Rad:S}\cr
  \hglue 2em LFI 22& *\getsymbol{LFI:knee:frequency:LFI22:Rad:M}$\,\pm\,$\getsymbol{LFI:knee:frequency:uncertainty:LFI22:Rad:M}&
              *\getsymbol{LFI:knee:frequency:LFI22:Rad:S}$\,\pm\,$\getsymbol{LFI:knee:frequency:uncertainty:LFI22:Rad:S}&
              \getsymbol{LFI:slope:LFI22:Rad:M}$\,\pm\,$\getsymbol{LFI:slope:uncertainty:LFI22:Rad:M}&
              \getsymbol{LFI:slope:LFI22:Rad:S}$\,\pm\,$\getsymbol{LFI:slope:uncertainty:LFI22:Rad:S}\cr
  \hglue 2em LFI 23& *\getsymbol{LFI:knee:frequency:LFI23:Rad:M}$\,\pm\,$\getsymbol{LFI:knee:frequency:uncertainty:LFI23:Rad:M}&
              *\getsymbol{LFI:knee:frequency:LFI23:Rad:S}$\,\pm\,$\getsymbol{LFI:knee:frequency:uncertainty:LFI23:Rad:S}&
              \getsymbol{LFI:slope:LFI23:Rad:M}$\,\pm\,$\getsymbol{LFI:slope:uncertainty:LFI23:Rad:M}&
              \getsymbol{LFI:slope:LFI23:Rad:S}$\,\pm\,$\getsymbol{LFI:slope:uncertainty:LFI23:Rad:S}\cr
  \noalign{\vskip 5pt}
  \omit{\bf 44\,GHz}\hfil\cr
  \noalign{\vskip 4pt}
  \hglue 2em LFI 24& *\getsymbol{LFI:knee:frequency:LFI24:Rad:M}$\,\pm\,$\getsymbol{LFI:knee:frequency:uncertainty:LFI24:Rad:M}&
              *\getsymbol{LFI:knee:frequency:LFI24:Rad:S}$\,\pm\,$\getsymbol{LFI:knee:frequency:uncertainty:LFI24:Rad:S}&
              \getsymbol{LFI:slope:LFI24:Rad:M}$\,\pm\,$\getsymbol{LFI:slope:uncertainty:LFI24:Rad:M}&
              \getsymbol{LFI:slope:LFI24:Rad:S}$\,\pm\,$\getsymbol{LFI:slope:uncertainty:LFI24:Rad:S}\cr
  \hglue 2em LFI 25& *\getsymbol{LFI:knee:frequency:LFI25:Rad:M}$\,\pm\,$\getsymbol{LFI:knee:frequency:uncertainty:LFI25:Rad:M}&
              *\getsymbol{LFI:knee:frequency:LFI25:Rad:S}$\,\pm\,$\getsymbol{LFI:knee:frequency:uncertainty:LFI25:Rad:S}&
              \getsymbol{LFI:slope:LFI25:Rad:M}$\,\pm\,$\getsymbol{LFI:slope:uncertainty:LFI25:Rad:M}&
              \getsymbol{LFI:slope:LFI25:Rad:S}$\,\pm\,$\getsymbol{LFI:slope:uncertainty:LFI25:Rad:S}\cr
  \hglue 2em LFI 26& *\getsymbol{LFI:knee:frequency:LFI26:Rad:M}$\,\pm\,$\getsymbol{LFI:knee:frequency:uncertainty:LFI26:Rad:M}&
              *\getsymbol{LFI:knee:frequency:LFI26:Rad:S}$\,\pm\,$\getsymbol{LFI:knee:frequency:uncertainty:LFI26:Rad:S}&
              \getsymbol{LFI:slope:LFI26:Rad:M}$\,\pm\,$\getsymbol{LFI:slope:uncertainty:LFI26:Rad:M}&
              \getsymbol{LFI:slope:LFI26:Rad:S}$\,\pm\,$\getsymbol{LFI:slope:uncertainty:LFI26:Rad:S}\cr
  \noalign{\vskip 5pt}
  \omit{\bf 30\,GHz}\hfil\cr
  \noalign{\vskip 4pt}
  \hglue 2em LFI 27& \getsymbol{LFI:knee:frequency:LFI27:Rad:M}$\,\pm\,$\getsymbol{LFI:knee:frequency:uncertainty:LFI27:Rad:M}&
              \getsymbol{LFI:knee:frequency:LFI27:Rad:S}$\,\pm\,$\getsymbol{LFI:knee:frequency:uncertainty:LFI27:Rad:S}&
              \getsymbol{LFI:slope:LFI27:Rad:M}$\,\pm\,$\getsymbol{LFI:slope:uncertainty:LFI27:Rad:M}&
              \getsymbol{LFI:slope:LFI27:Rad:S}$\,\pm\,$\getsymbol{LFI:slope:uncertainty:LFI27:Rad:S}\cr
  \hglue 2em LFI 28& \getsymbol{LFI:knee:frequency:LFI28:Rad:M}$\,\pm\,$\getsymbol{LFI:knee:frequency:uncertainty:LFI28:Rad:M}&
              *\getsymbol{LFI:knee:frequency:LFI28:Rad:S}$\,\pm\,$\getsymbol{LFI:knee:frequency:uncertainty:LFI28:Rad:S}&
              \getsymbol{LFI:slope:LFI28:Rad:M}$\,\pm\,$\getsymbol{LFI:slope:uncertainty:LFI28:Rad:M}&
              \getsymbol{LFI:slope:LFI28:Rad:S}$\,\pm\,$\getsymbol{LFI:slope:uncertainty:LFI28:Rad:S}\cr
  \noalign{\vskip 5pt\hrule\vskip 3pt}}}
  \endPlancktable
  \endgroup
  \end{table*}

\begin{figure*}
\centering
\includegraphics[width=6cm]{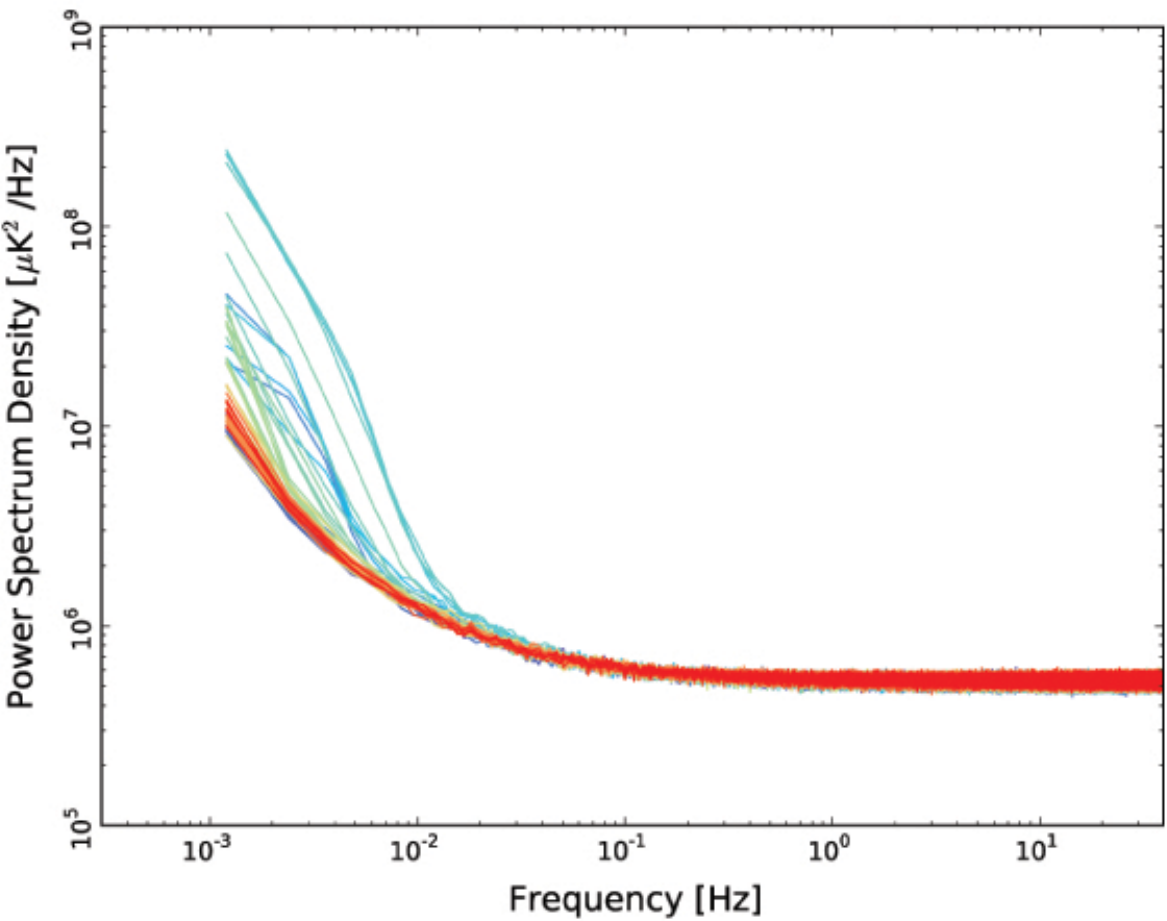}
\includegraphics[width=6cm]{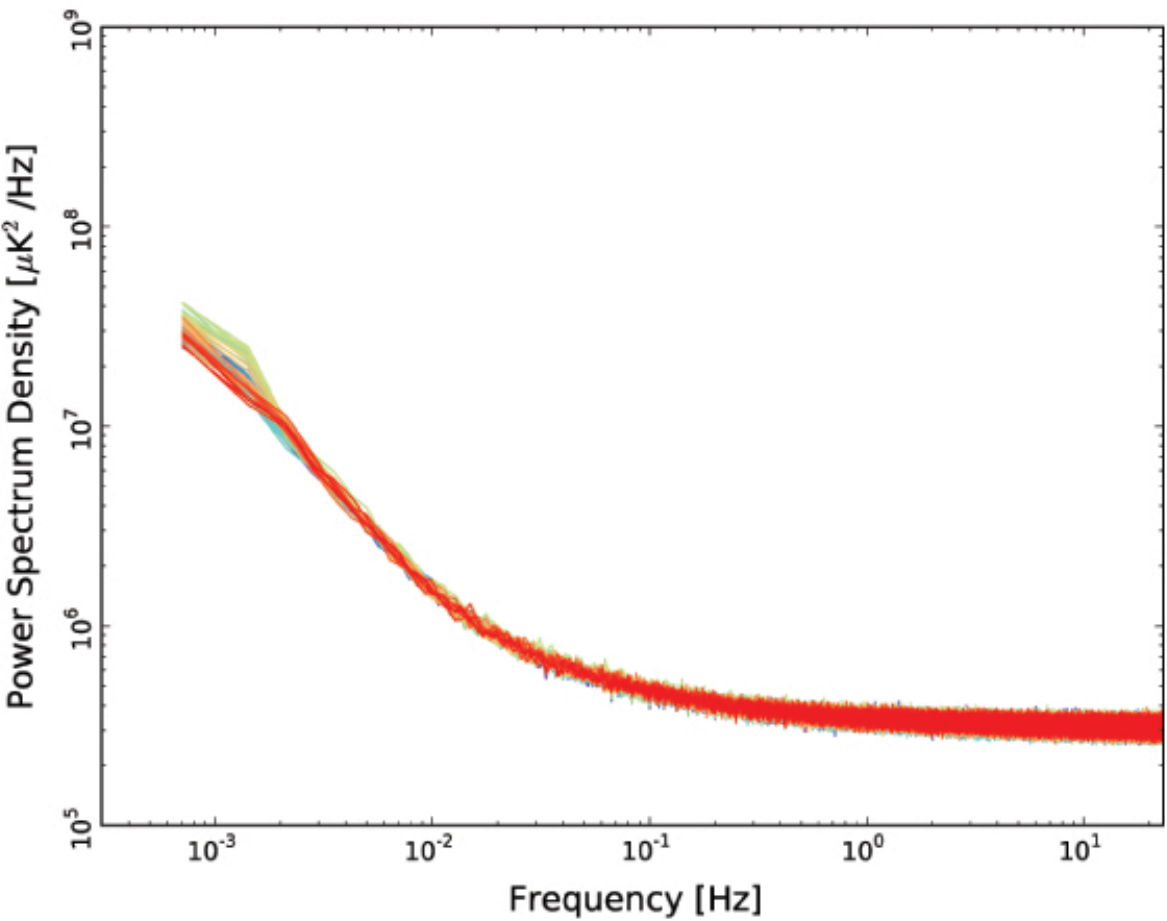}
\includegraphics[width=6cm]{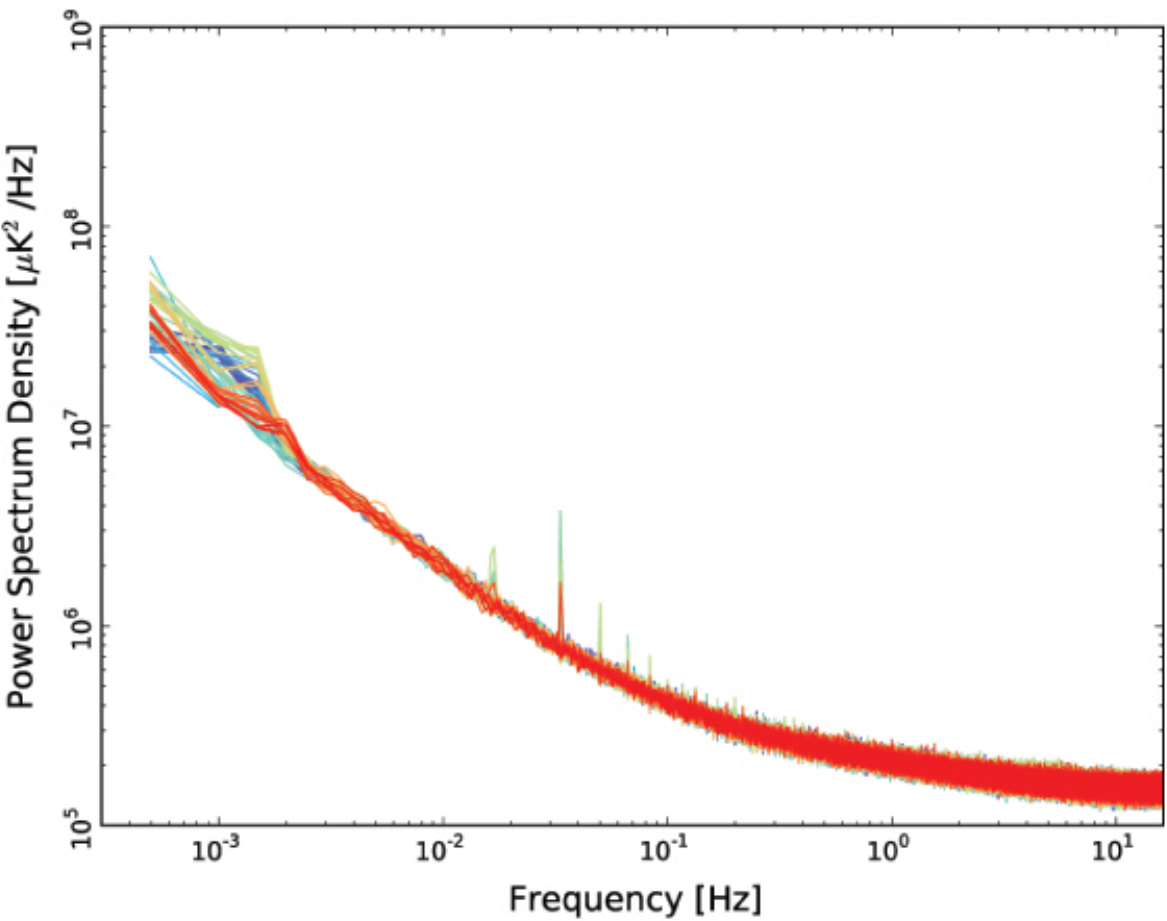}
\caption{Noise spectra throughout the mission lifetime for a 70\,GHz  radiometer 18M
(\emph{left}), 25S (44\,GHz; \emph{middle}), and 27M (30\,GHz; \emph{right}).
Spectra are shown for the ranges from OD 100 (blue) to OD 1526 (red),
spaced about 20 ODs apart. White noise is stable at the level of $0.3\%$, while low-frequency
noise shows variations both in slope and knee-frequency, with different
amplitude for different radiometers.}
\label{noiseps}
\end{figure*}

Time variations of the noise properties are a good indicator of possible
changes in instrument behaviour. There are known events that caused such
variations, such as the sorption cooler switchover at OD 460 \citep{planck2013-p01}. 
Indeed, variations in
noise properties due to changes in temperature are 
expected as the performance of the first cooler degraded, as well as when the
second cooler came in and took time to stabilize the temperature.
Figure~\ref{noiseps} shows a sample of noise spectra for radiometers LFI27M,
LFI25S, and LFI18S, spanning the whole
mission lifetime.  The white noise is stable at the level of $0.3$\%.
As already noted in the previous release,
knee-frequencies and slopes are stable until OD 326 and show significant
variations afterwards, altering the simple ``one slope, one knee'' model.
This is due to the progressive degradation of the first sorption cooler 
and the insertion of the second one. Once the environment became thermally
stable, the spectra moved back towards their initial shape. Of course this is
evident at different levels in the individual radiometers, depending
on their frequency, position on the focal-plane, and susceptibility to thermal
instabilities.

\section{Mapmaking}
\label{sec_mmaking_intro}
    Mapmaking is the last step in the LFI pipeline, after calibration and dipole
removal, and before bandpass correction and component separation.
Mapmaking takes as its input the calibrated timelines, from which the
$4\pi$ convolved dipole and Galactic straylight signal has been removed.
Output consists of sky maps of temperature and $Q$ and $U$ polarization, 
and a description of the residual noise in them.

An important part of the mapmaking step is the removal of correlated $1/f$
noise.  An optimal mapmaking method will remove the noise as accurately as
possible, while simultaneously keeping systematics at an acceptable level.

LFI maps were produced by the {\tt Madam} mapmaking code \cite{keihanen2005}.
The code is the same as used in the 2013 release. 
In the following we give a short overview, and point out aspects relevant to polarization (see \citet{planck2014-a07} for details).

{\tt Madam} removes the correlated noise using a destriping technique.
A noise prior is used to improve the map quality further.
The correlated noise component is modelled by a sequence of baseline offsets.
The choice of the baseline length is a trade-off between computational burden
and optimal noise removal. We have chosen to use 1\,s long baselines for 44
and 70\,GHz, and 0.25\,s for 30\,GHz where the typical knee frequencies are
higher.

The full time-ordered data stream is modelled as
\begin{equation}
y = \tens{P}m+\tens{F}a+n.
\end{equation}
Here vector $a$ represents the baselines, and $\tens{F}$ is formally a matrix
that spreads the baselines into time-ordered data.  Vector $n$ represents
white noise, and $\tens{P}$ is a pointing matrix that picks a time-ordered data
stream from the sky map $m$.
Map $m$ has three columns, corresponding to the three Stokes components $I$,
$Q$, and $U$.

The {\em noise prior\/} describes the expected correlation between baseline
amplitudes,
\begin{equation}
\tens{C}_a=\langle a a^{\rm T}\rangle.
\end{equation}
The prior is constructed from the known noise parameters presented in the
previous section (knee frequency, white noise sigma, and spectral slope).
The noise prior provides an extra constraint which makes it possible to extend
the destriping technique to very short baseline lengths,
allowing for more accurate removal of noise.

With the assumptions above, the baseline vector $a$ can be solved from the
linear system of equations
\begin{equation}
(\tens{F}^{\rm T} \tens{C}_{\rm w}^{-1}\tens{ZF} + \tens{C}_{a}^{-1})a
 = \tens{F}^{\rm T} \tens{C}_{\rm w}^{-1}\tens{Z} y,  \label{baseeq}
\end{equation}
where
\begin{equation}
\tens{Z} = \tens{I}
 -\tens{P}(\tens{P}^{\rm T}\tens{C}_{\rm w}^{-1}\tens{P})^{-1}
  \tens{P}^{\rm T}\tens{C}_{\rm w}^{-1}.   \label{Zmatrix}
\end{equation}
Here $\tens{C}_{\rm w}$ is a diagonal weighting matrix.
The final map is constructed as 
\begin{equation}
  m = (\tens{P}^{\rm T}C_{w}^{-1}\tens{P})^{-1}\tens{P}^{\rm T}
   \tens{C}_{\rm w}^{-1}(y-\tens{F}a).   \label{map-binning}
\end{equation}

The destriping technique constructs the final map through a procedure in which
one first solves for the baselines, and then bins the map from the data
stream from which the baselines have been removed. 
This two-step procedure provides a way of reducing systematics.
We control the ``signal error'' by applying a mask in the destriping phase,
while still binning the final map to cover the whole sky.
Signal error is the uncertainty in baseline determination that arises from
deviations of the actual sky signal from the model $\tens{P}m$.
The main sources of signal error are signal variations within a pixel,
differences in radiometer frequency responses (bandpass mismatch), and beam
shape mismatch.  The error arises mainly at low Galactic latitudes, where
signal gradients are strong.

The choice of the destriping mask is a trade-off between acceptable signal
error level and noise removal.  A mask that is too wide may lead to a situation
where there are not enough crossing points between scanning rings to reliably
determine the noise baselines.

It can be shown that residual noise is minimized when $\tens{C}_{\rm w}$
equals the variance of white noise in time-ordered data.
In order to reduce leakage from temperature to polarization, however,
we apply horn-uniform weighting, which differs from this ideal case.
We replace the white noise variance by the average of the variances of the
two radiometers of the same horn.
This has the effect that the systematic error related to beam shape mismatch,
which is strongly correlated between the radiometers,
largely cancels out in polarization analysis.  Thus we are reducing the
leakage from temperature to polarization.

Along with maps of the sky, {\tt Madam} provides a covariance matrix for
residual white noise in the maps.
This consists of a $3\times3$ matrix for each pixel, describing the
correlations between $I$, $Q$, and $U$ components in the pixel.
White noise is uncorrelated between pixels.
Correlated noise residuals are captured by the low-resolution noise covariance
matrix describe in Sect.~\ref{sec_low_ncvm} below.

{\tt Madam} produces its output maps in {\tt HEALPix} format
\citep{gorski2005}.
For the bulk of the products we used resolution $N_{\rm side}=1024$,
and the same resolution was used when solving the destriping equation.  Maps at
70\,GHz were also produced with $N_{\rm side}=2048$.

To accurately decompose the map into $I$, $Q$, and $U$ components it is
necessary to have several measurements from the same sky pixel, with different
parallactic angles.
If this is not the case, the pixel in question is eliminated from analysis.
{\tt Madam} uses as rejection criterion the reciprocal condition number of
the matrix $\tens{P}^{\rm T}\tens{C}_{\rm w}^{-1}\tens{P}$.

    \subsection{Low-resolution data set}
    \label{sec_low_data}
        Low-resolution products are an integral part of the low-$\ell$ likelihood.
To fully exploit the information contained in the largest structures of the
microwave sky, a full statistical description of the residual noise present
in the maps is required.  This information is provided in the form of
pixel-pixel noise covariance matrices (NCVMs).  However, due to resource
limitations they are impossible to employ at native map resolution.  Therefore
a low-resolution data set is needed for the low-$\ell$ analysis; this data set
consists of low-resolution maps and corresponding noise covariance matrices.
At present, the low-resolution data set can be efficiently used only at
resolution $N_\textrm{side} = 16$, or lower.  All the low-resolution products
are produced at this target resolution.

    \subsection{Low-resolution maps}
    \label{sec_low_res}
\begin{figure*}[htbp]
        \centering
        \begin{tabular}{ccc}
	\includegraphics[width=.25\textwidth]{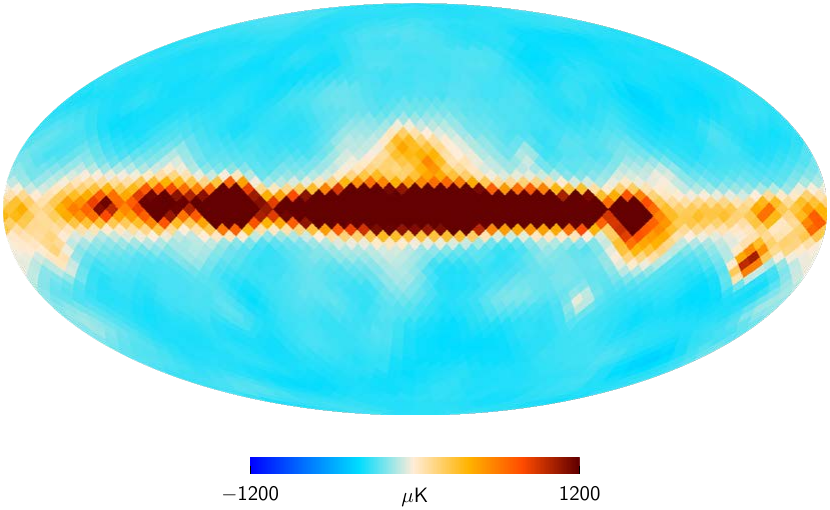} & 
	\includegraphics[width=.25\textwidth]{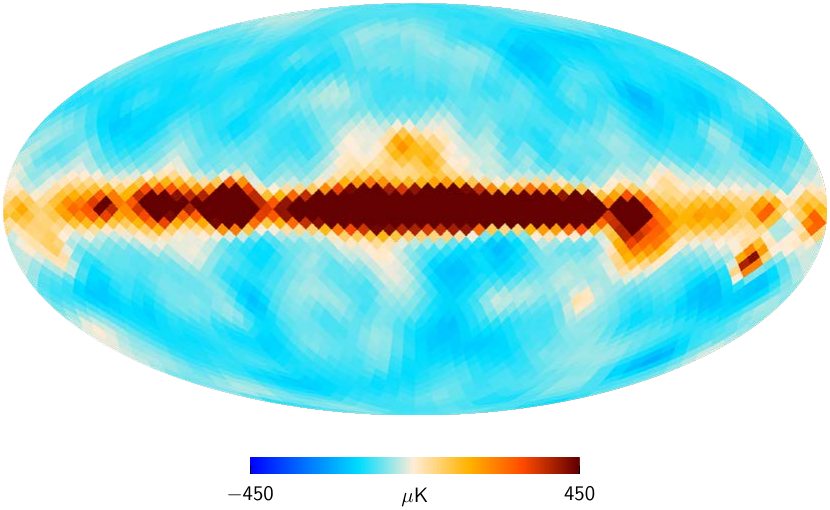} &
	\includegraphics[width=.25\textwidth]{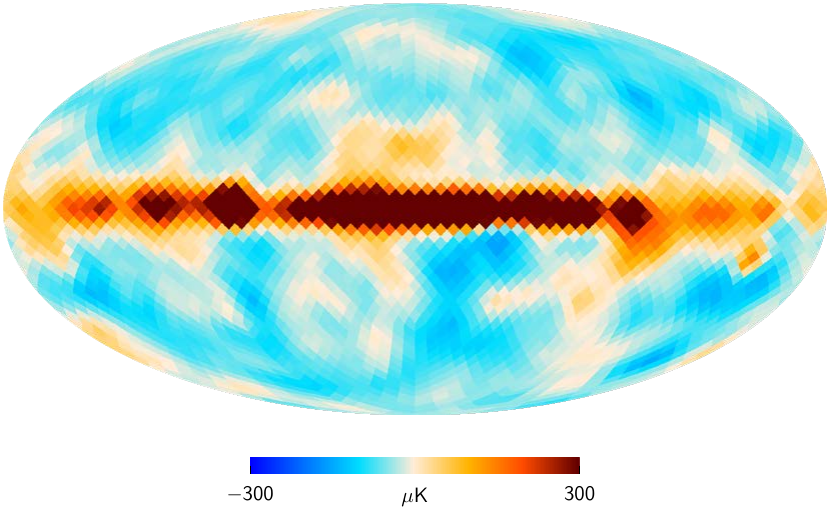} \\
	\includegraphics[width=.25\textwidth]{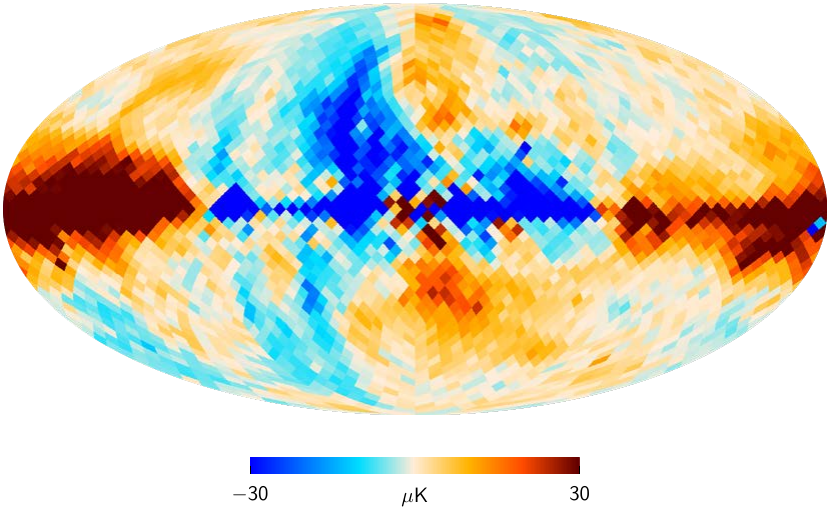} & 
	\includegraphics[width=.25\textwidth]{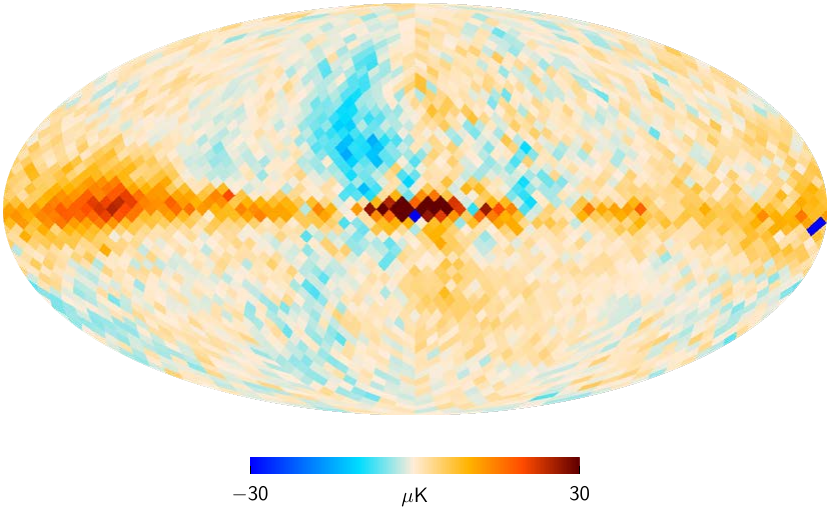} &
	\includegraphics[width=.25\textwidth]{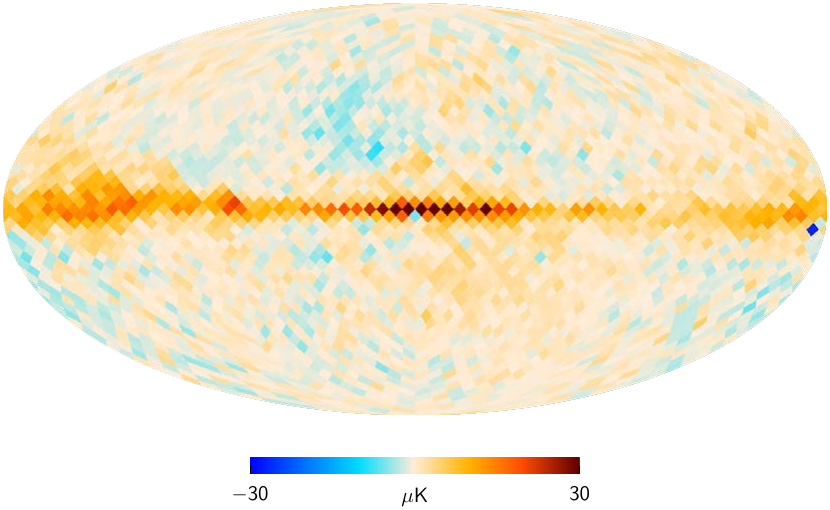} \\
	\includegraphics[width=.25\textwidth]{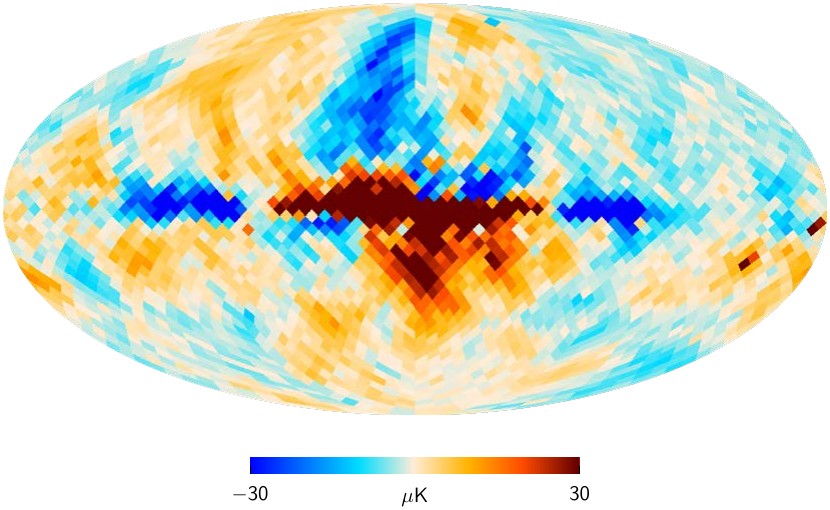} & 
	\includegraphics[width=.25\textwidth]{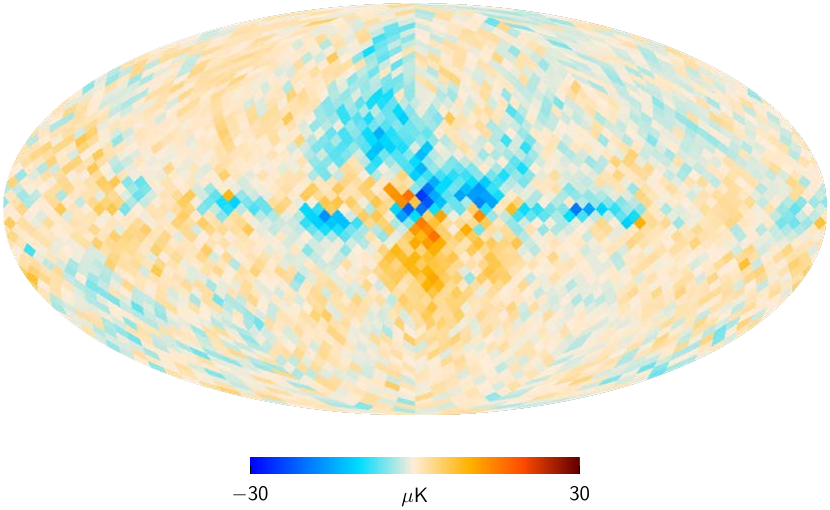} &
	\includegraphics[width=.25\textwidth]{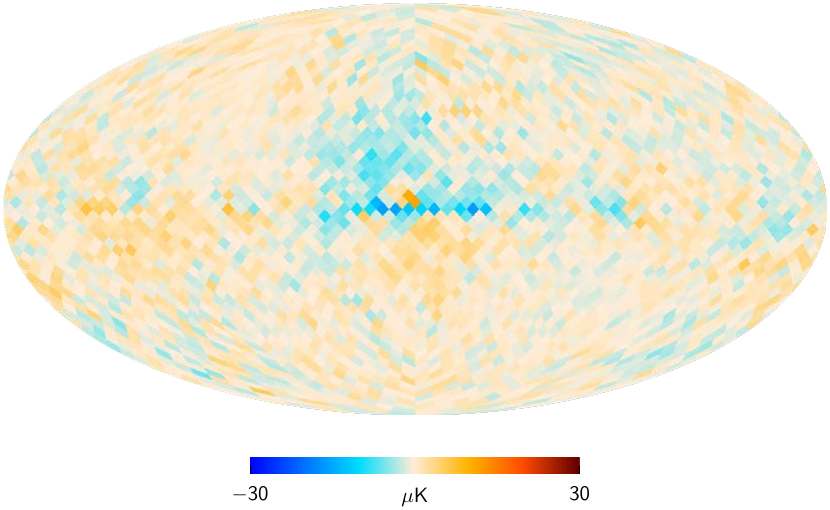} \\
        \end{tabular}
    \caption{LFI full mission low-resolution maps, $N_{\rm side}=16$.
From left to right 30\,GHz, 44\,GHz, and then 70\,GHz: {\it top\/} intensity
$I$; {\it middle\/} polarization $Q$ component; and {\it bottom\/}
polarization $U$ component. Units are $\mu{\rm K}_{CMB}$ .}
        \label{fig:low_maps}
\end{figure*}

        The low-resolution maps, shown in Figure ~\ref{fig:low_maps}, are constructed
by downgrading the high-resolution maps (described in the previous section)
to the target resolution. 
We chose to downgrade the maps using a ``noise-weighted'' scheme.

The noise-weighted scheme has also been used in previous studies
\citep[see, e.g.][]{planck2013-p02}. The noise-weighted map corresponds to a
map that is first destriped at the high resolution, and the destriped TOI is
directly binned onto the low target resolution.
This approach gives adequate control over signal and noise in the resulting
maps. However, concerns have been raised that the noise-weighted scheme
transfers signal from one pixel to another. As a consequence we employ Gaussian smoothing 
to minimize this effect, at the cost of some increase in noise. After downgrading, the temperature component is
smoothed with a Gaussian window function with ${\rm FWHM} = 440\arcm$. We will re-examine this choice in the next release.

In practice the high-resolution maps are noise-weighted to an intermediate
resolution of $N_{\mathrm{side}} = 32$. The Stokes $I$ part of the map is
expanded in spherical harmonics, the expansion is treated with the smoothing
beam, and the final map is then synthesized at the target resolution. The last
step of resolution downgrading for Stokes $Q$ and $U$ maps, however, is
performed by carrying out naive averaging of higher resolution pixels.

Due to the chosen downgrading scheme the resulting NCVM will be singular.
We regularize the problem by adding some white noise both to the maps and
matrices.  Specifically we add 2$\,\mu$K for $I$, and 0.02$\,\mu$K for $Q$ and
$U$ at $N_{\mathrm{side}}= 16$ resolution.
 
    \subsection{Noise covariance matrices}
    \label{sec_low_ncvm}
        The statistical description of the residual noise present in a low-resolution
map is given in the form of a pixel-pixel noise covariance matrix, as described
in \cite{keskitalo2009}. The NCVM formalism describes the noise correlations of
a map produced at the same resolution as the noise covariance matrix.
Therefore, for an exact description we should construct the matrices at
resolution $N_{\textrm{side}} = 1024$ and subsequently downgrade to the target
resolution. This is computationally impractical. Therefore the matrices are
computed at the highest possible initial resolution, and then downgraded to
the target resolution.  For consistency the noise covariance matrices must go
through the same processing steps as applied to the low-resolution maps.

The {\tt Madam/TOAST} code, a Time Ordered Astrophysics Scalable Tools (TOAST)
port of {\tt Madam }, was used to produce the pixel-pixel noise covariance
matrices (\citet{keihanen2010} and \url{http://tskisner.github.io/TOAST/}.
The {\tt TOAST} interface was chosen on the basis of added
flexibility and speed; see \citep{planck2014-a07}.  

The outputs of {\tt Madam/TOAST} software are inverse-noise covariance
matrices, specifically one inverse matrix per radiometer for a given time
period. Because inverse NCVMs are additive, the individual inverse matrices
are merged together to form the actual inverse NCVM. To obtain the noise
covariance matrix from its inverse, the matrices are inverted using the
eigen-decomposition of a matrix. These intermediate-resolution matrices are
then downgraded using the same downgrading scheme as applied to the maps.
The matrices are regularized by adding the same level of white noise to the
diagonal elements of the covariance matrix as to the low-resolution maps.

The noise covariance matrix computation takes two inputs: the detector
pointing; and noise estimates. Since the matrices are calculated with
{\tt Madam/TOAST}, we use the pointing solution provided by {\tt TOAST}.
For more details see \cite{planck2014-a14}. We also use the most
representative noise model available, namely the FFP8 (full focal plane 8 simulations) noise estimates
\citep{planck2014-a14}. The noise model comprises daily $1/f$
model parameters.

The key parameter in the NCVM production is the baseline length. We have
demonstrated in an earlier study \citep{planck2013-p02} that using shorter
baseline lengths when producing the noise covariance matrix production better
models the residual noise. Therefore we chose to use 0.25\,s baselines for the
30\,GHz LFI frequency channel; we show in \citep{planck2014-a07} that 1.0\,s
is adequate for the 44\,GHz and 70\,GHz channels. Reducing the baseline length
still further gives only a
marginal improvement, while the resource requirements increase rapidly.

Previous studies \citep{planck2013-p02} have also shown that matrices should
be calculated at the highest computationally feasible resolution. For the
current release the initial resolution is $N_{\textrm{side}} = 64$.
Increasing the initial resolution beyond $N_{\textrm{side}} = 64$ is likely
to improve results, but the matrix size will be 16 times larger, i.e.
2.5\,TB. Inverting such a matrix is a formidable task.

The noise covariance computation makes two further deviations from the
high-resolution mapmaking: it does not take into account the destriping mask;
and the horns are not uniformly weighted. The effect of these differences
is much smaller than is obtained by either decreasing the baseline length or
increasing the destriping resolution in the production. For more details see
\cite{planck2014-a07}.

\section{Overview of LFI map properties}
\label{sec_OverviewMaps}
\begin{figure*}
\begin{centering}
\includegraphics[width=0.7\textwidth]{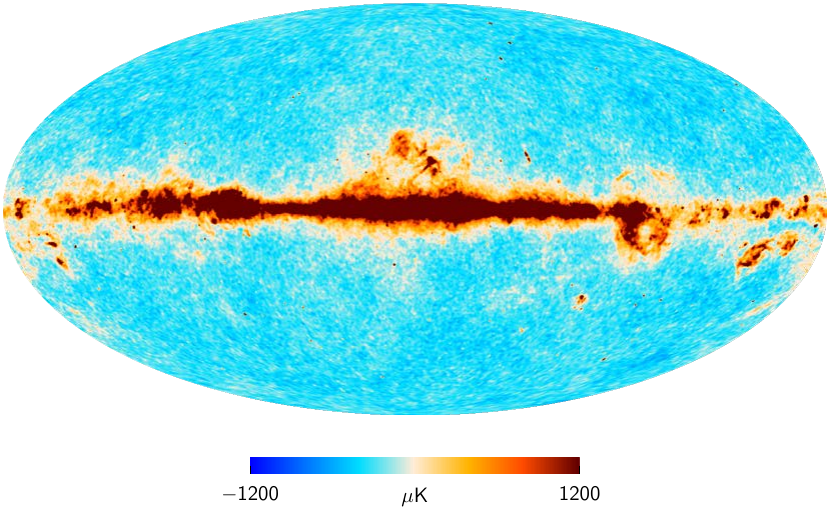}
\includegraphics[width=0.7\textwidth]{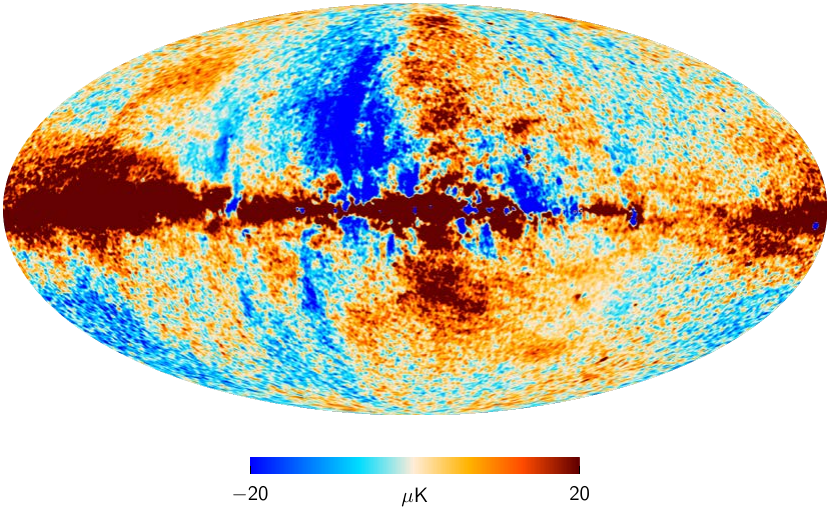}
\includegraphics[width=0.7\textwidth]{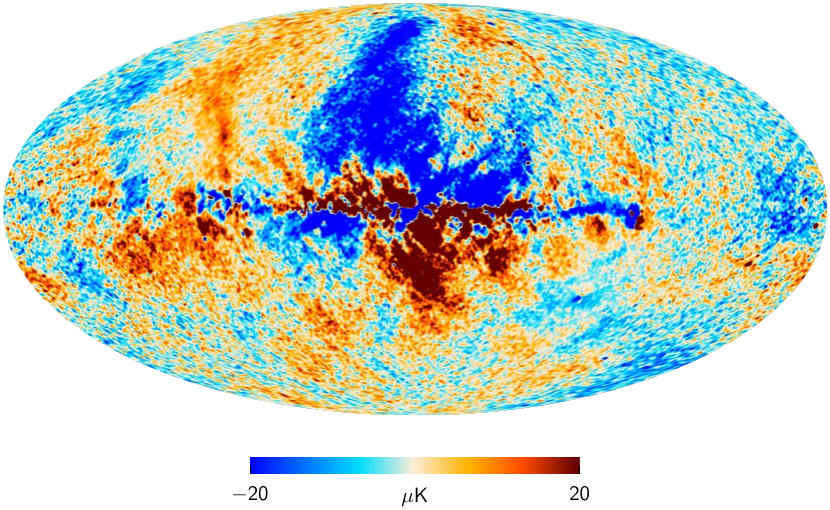}
\caption{LFI maps at 30\,GHz.  {\it Top\/}: intensity $I$.  {\it Middle\/}:
polarization $Q$ component.  {\it Bottom\/}: polarization $U$ component.
Polarization components are at $N_{side}=256$ and smoothed at 1\deg, the
intensity is left at the native $N_{side}=1024$. Units are
$\mu{\rm K}_{\rm CMB}$.
The polarization components have been corrected for the bandpass leakage
effect (see Sect.~\ref{sec_polarization}).}
\label{fig:Imaps-030}
\end{centering}
\end{figure*}

\begin{figure*}
\begin{centering}
\includegraphics[width=0.7\textwidth]{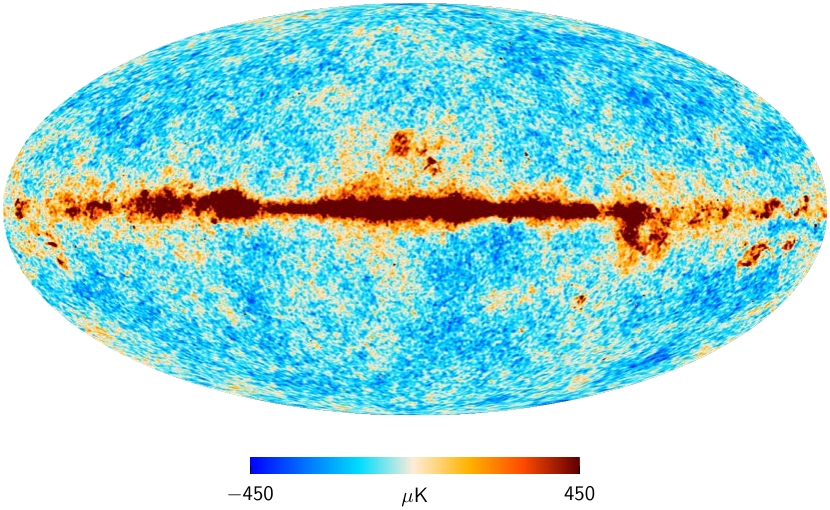}
\includegraphics[width=0.7\textwidth]{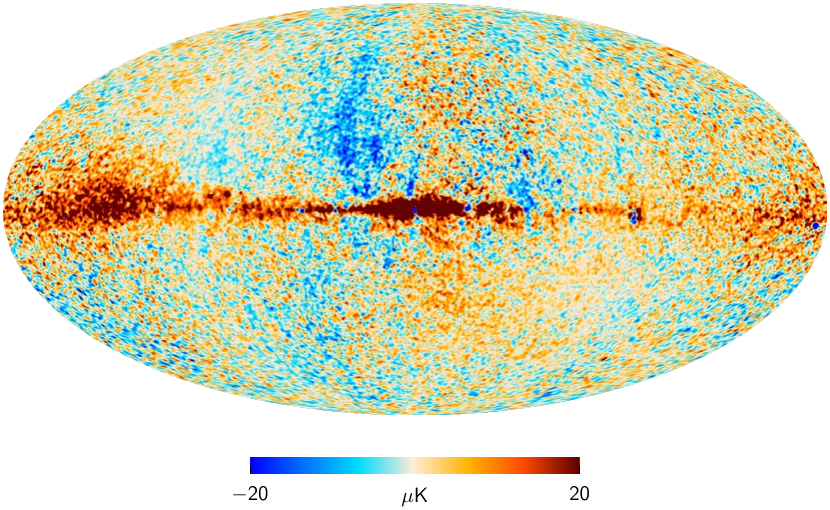}
\includegraphics[width=0.7\textwidth]{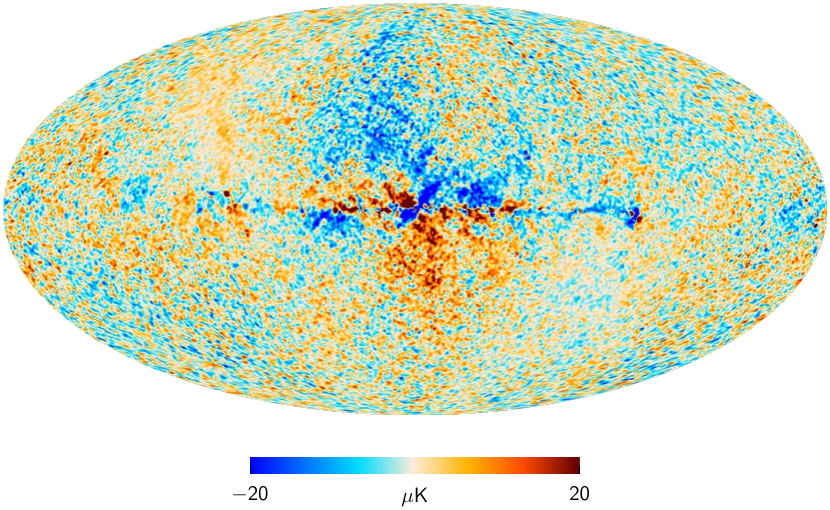}
\caption{Same as Fig.~\ref{fig:Imaps-030} for 44\,GHz.}
\label{fig:Imaps-044}
\end{centering}
\end{figure*}

\begin{figure*}
\begin{centering}
\includegraphics[width=0.7\textwidth]{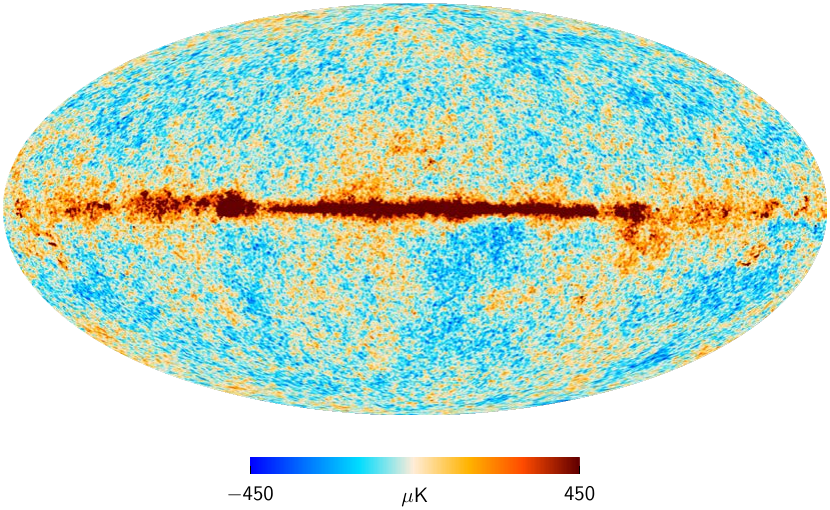}
\includegraphics[width=0.7\textwidth]{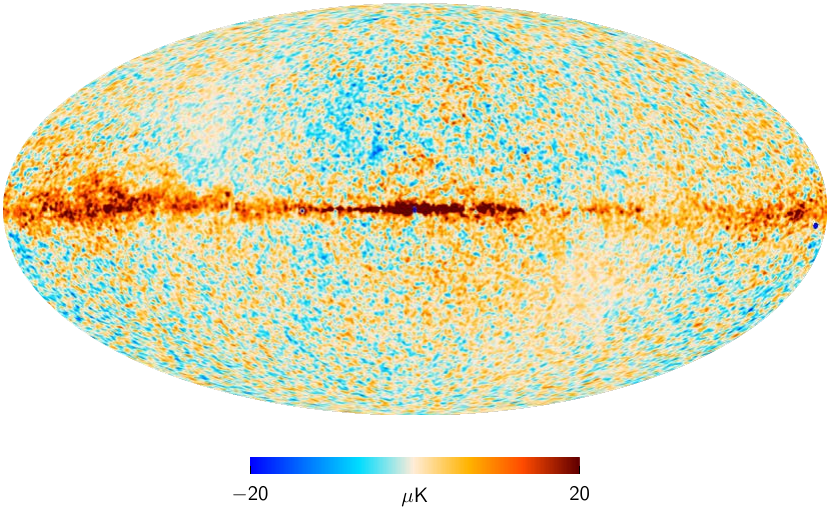}
\includegraphics[width=0.7\textwidth]{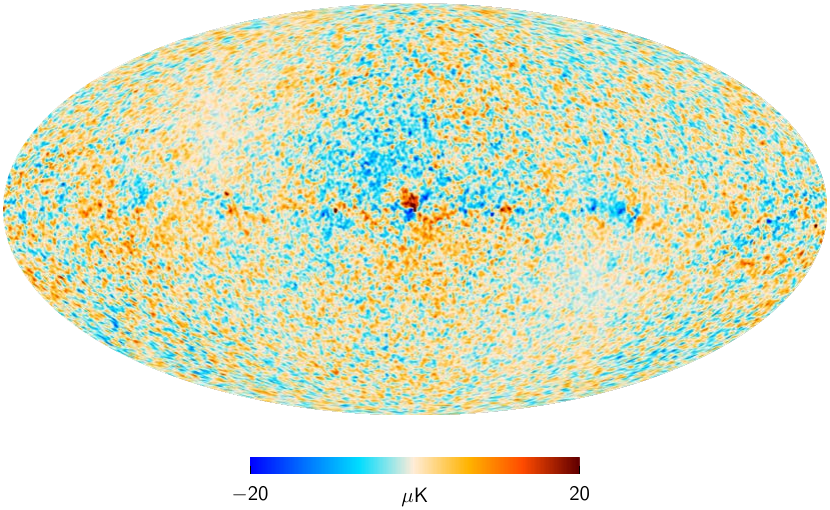}
\caption{Same as Fig.~\ref{fig:Imaps-030} for 70\,GHz.}
\label{fig:Imaps-070}
\end{centering}
\end{figure*}

\begin{figure*}[htbp]
        \centering
        \begin{tabular}{ccc}
	\includegraphics[width=.25\textwidth]{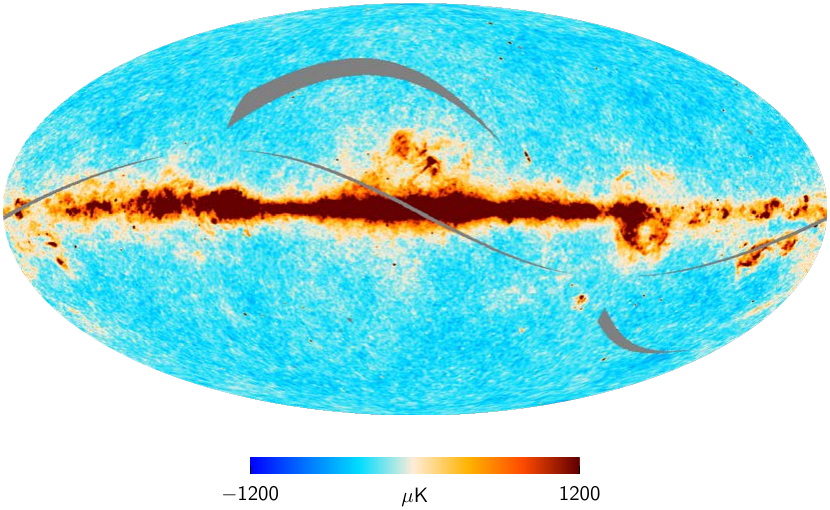} & 
	\includegraphics[width=.25\textwidth]{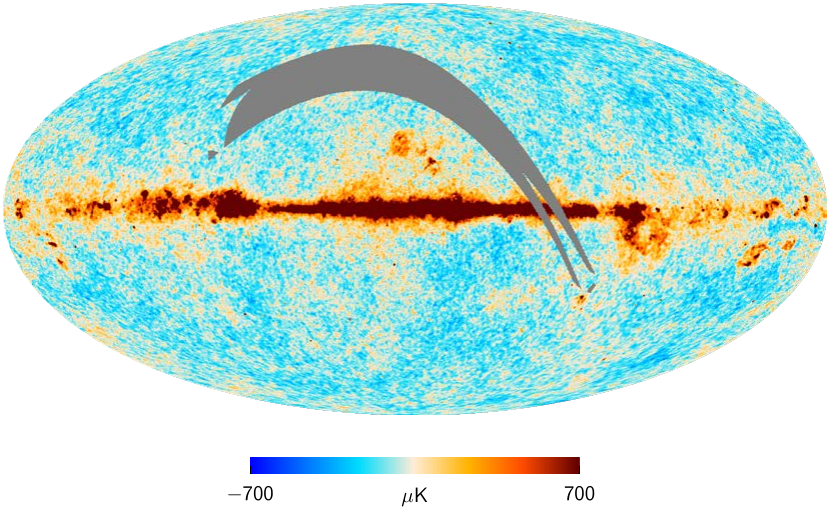} &
	\includegraphics[width=.25\textwidth]{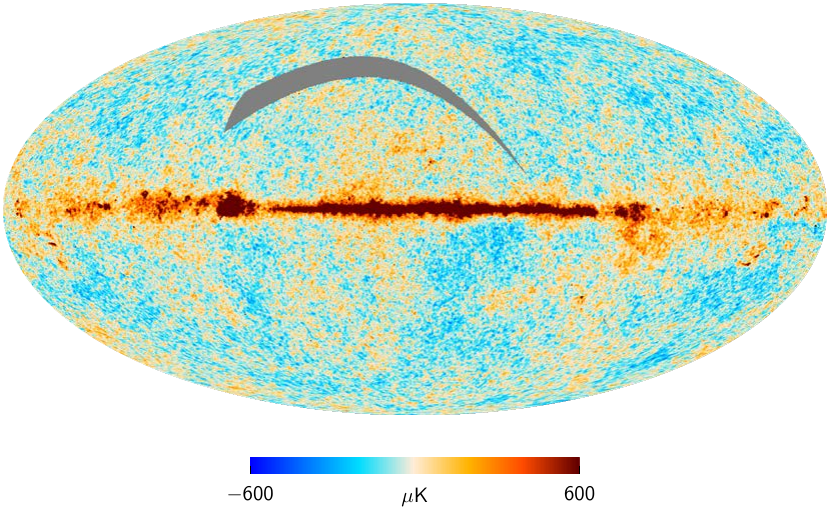} \\
	\includegraphics[width=.25\textwidth]{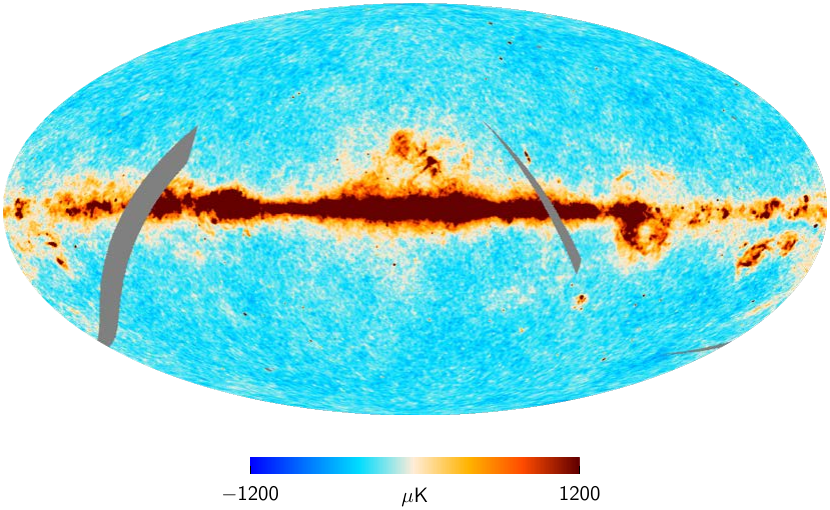} & 
	\includegraphics[width=.25\textwidth]{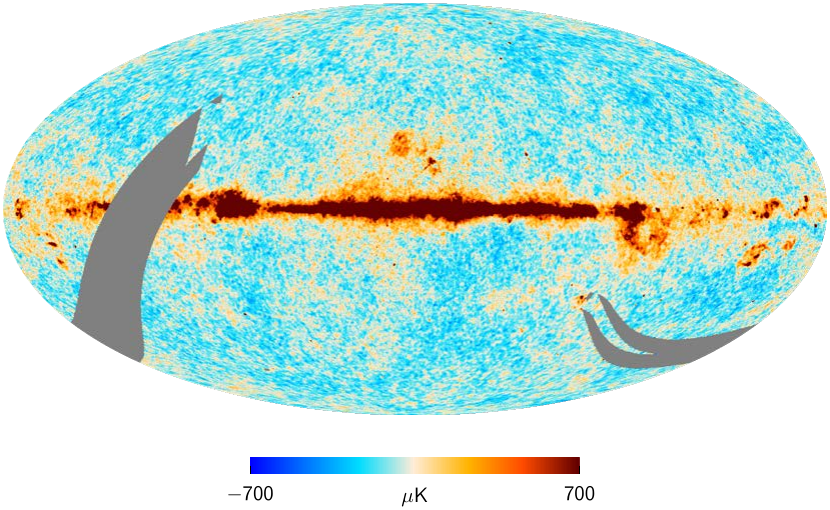} &
	\includegraphics[width=.25\textwidth]{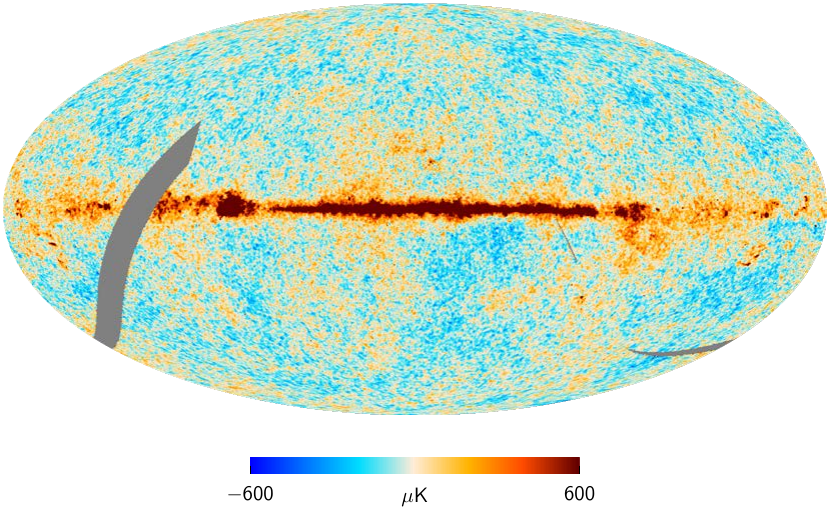} \\
	\includegraphics[width=.25\textwidth]{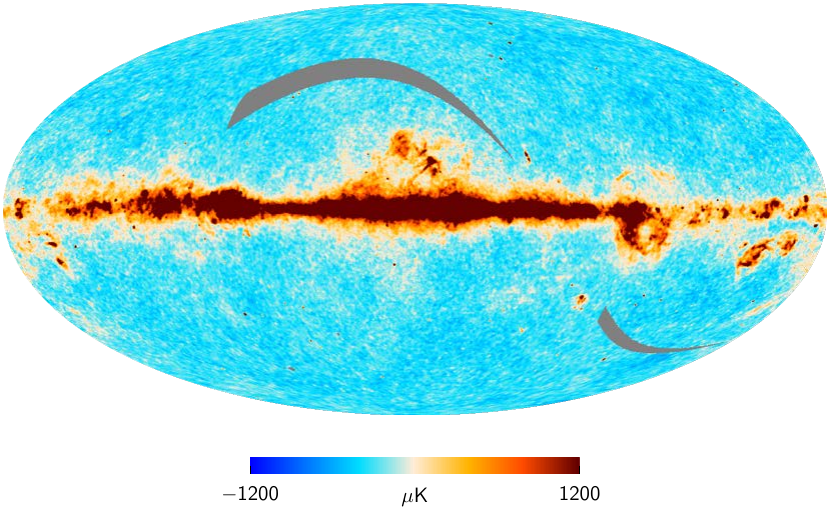} & 
	\includegraphics[width=.25\textwidth]{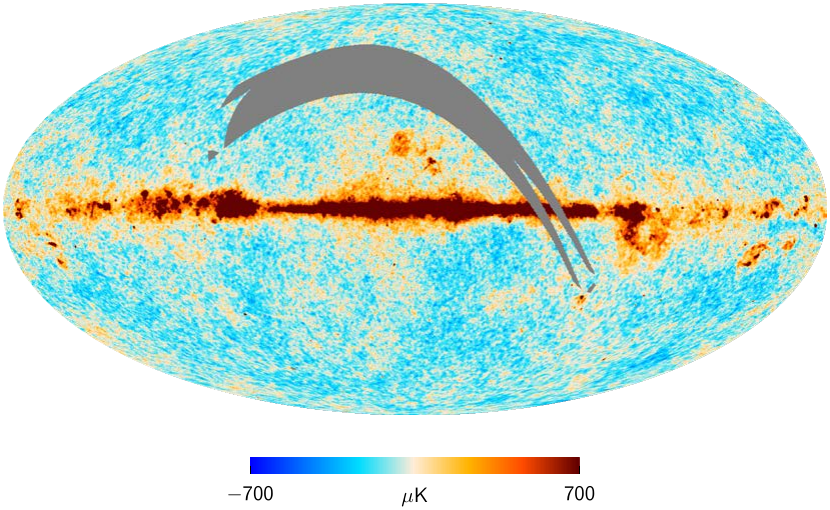} &
	\includegraphics[width=.25\textwidth]{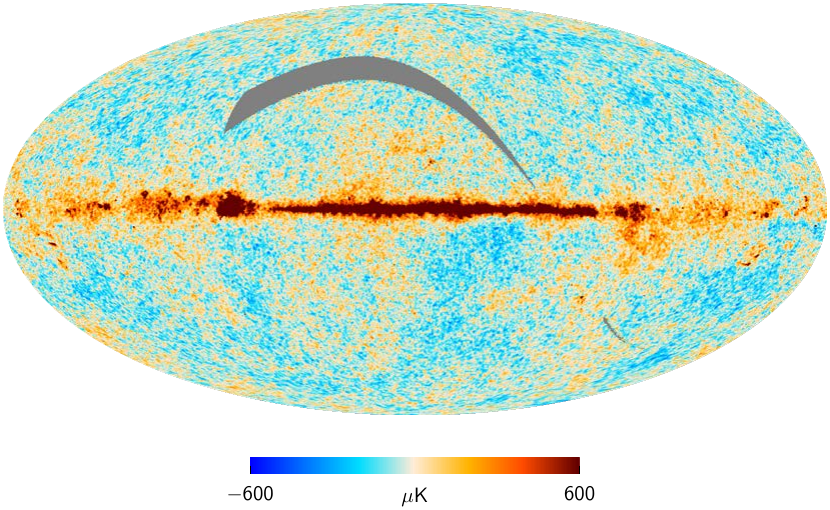} \\
	\includegraphics[width=.25\textwidth]{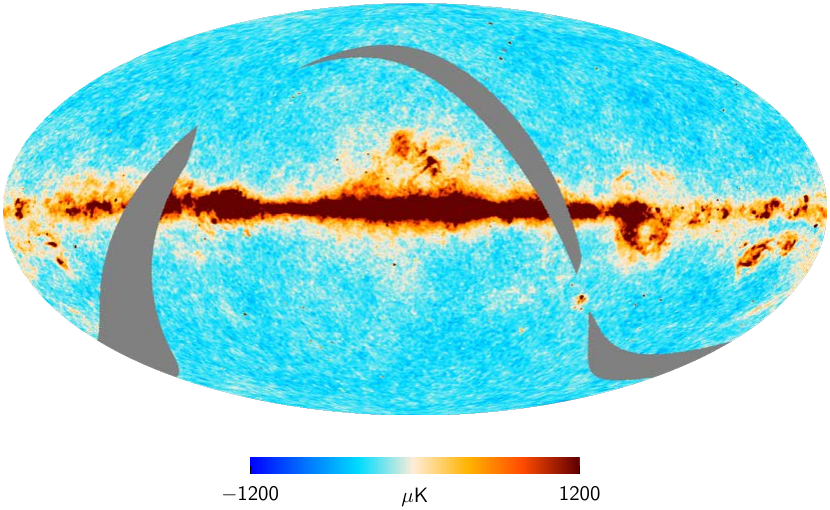} & 
	\includegraphics[width=.25\textwidth]{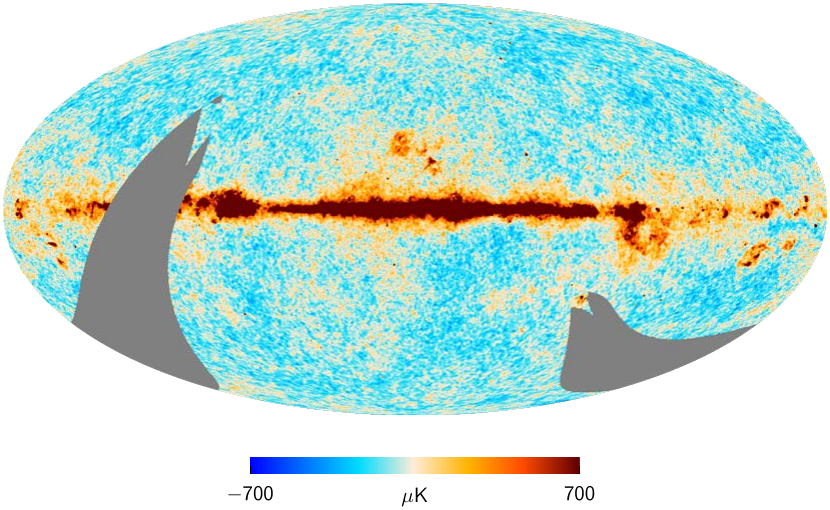} &
	\includegraphics[width=.25\textwidth]{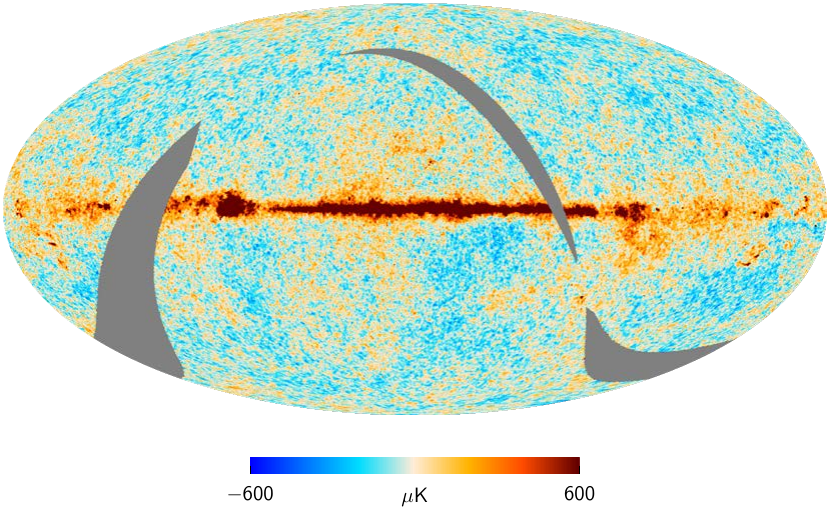} \\
	\includegraphics[width=.25\textwidth]{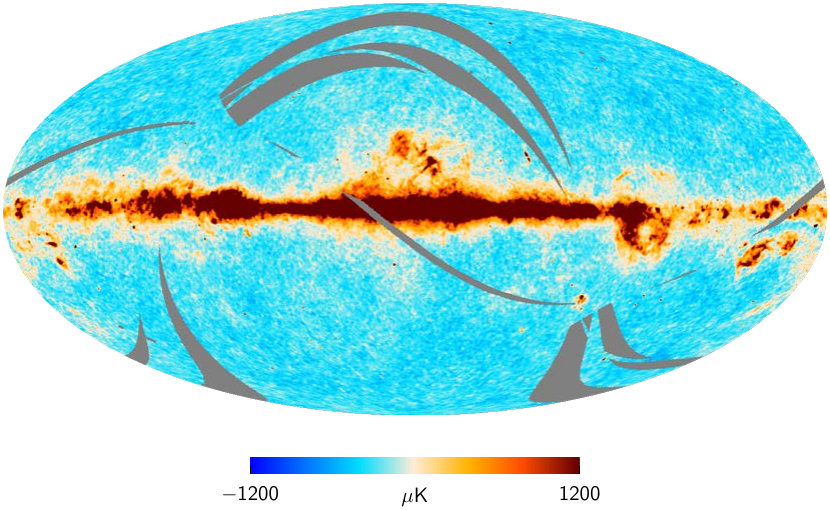} & 
	\includegraphics[width=.25\textwidth]{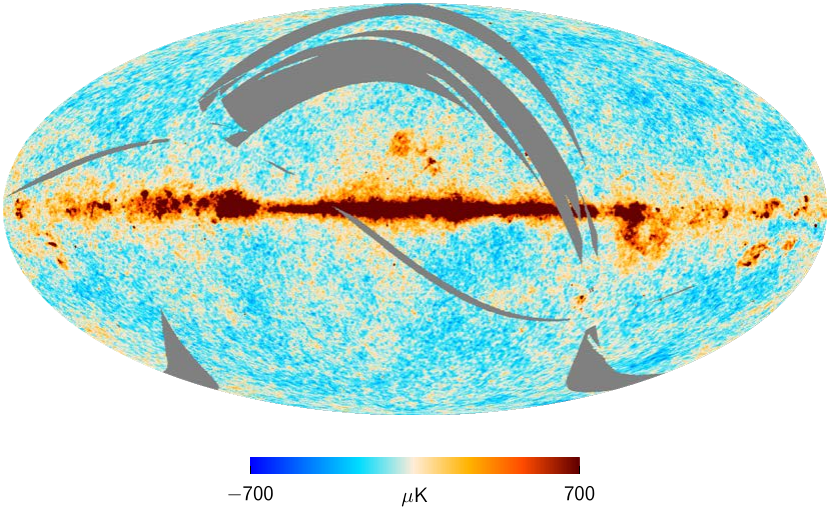} &
	\includegraphics[width=.25\textwidth]{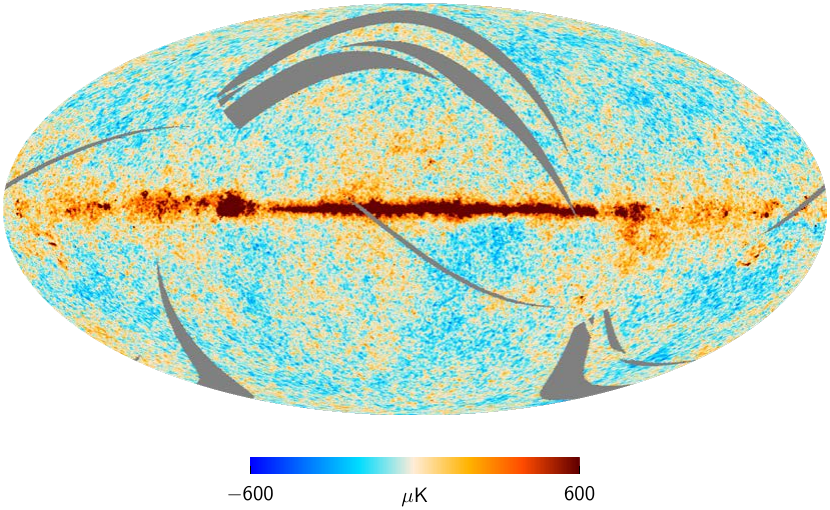} \\
	\includegraphics[width=.25\textwidth]{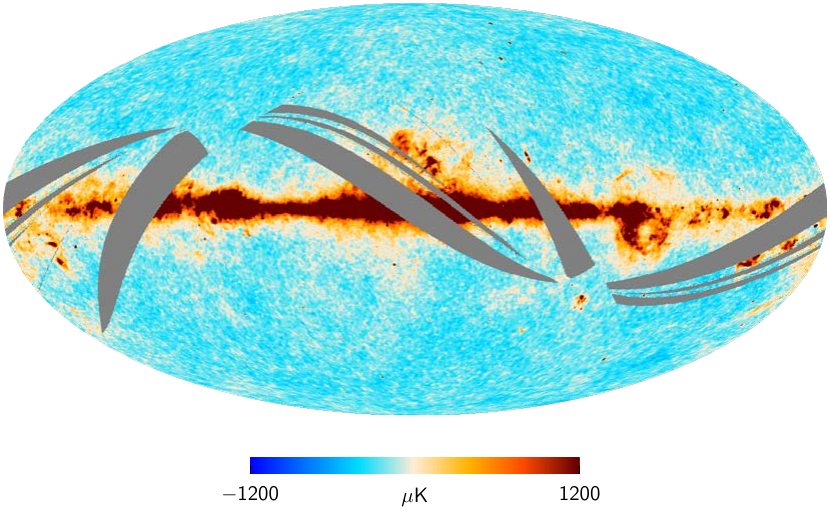} & 
	\includegraphics[width=.25\textwidth]{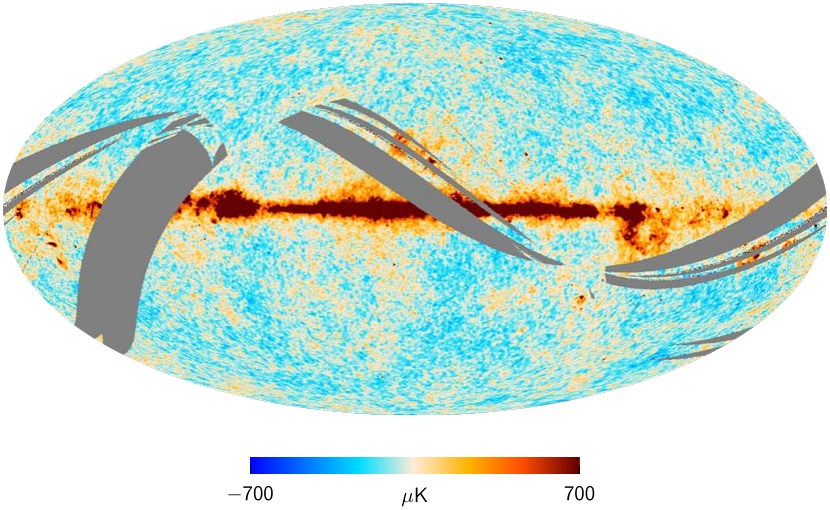} &
	\includegraphics[width=.25\textwidth]{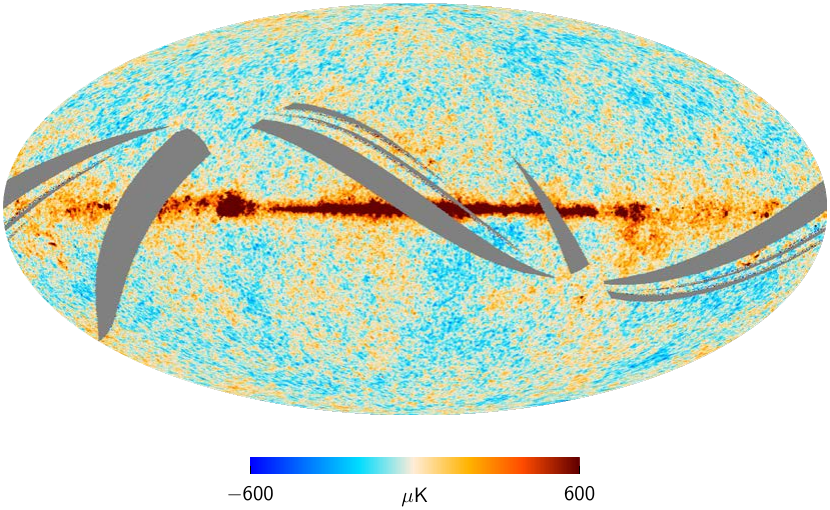} \\
	\includegraphics[width=.25\textwidth]{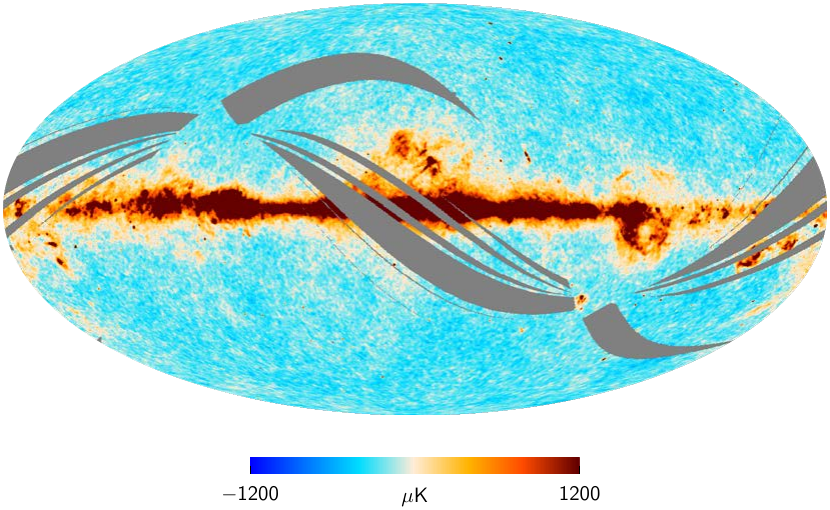} & 
	\includegraphics[width=.25\textwidth]{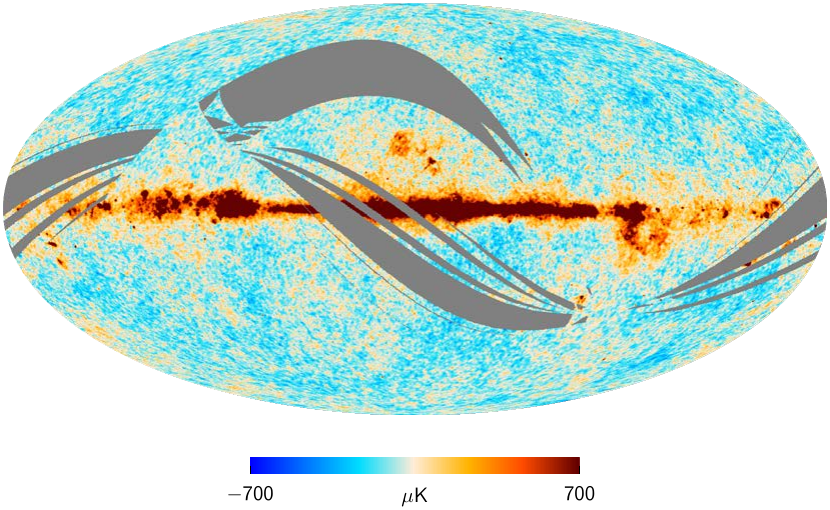} &
	\includegraphics[width=.25\textwidth]{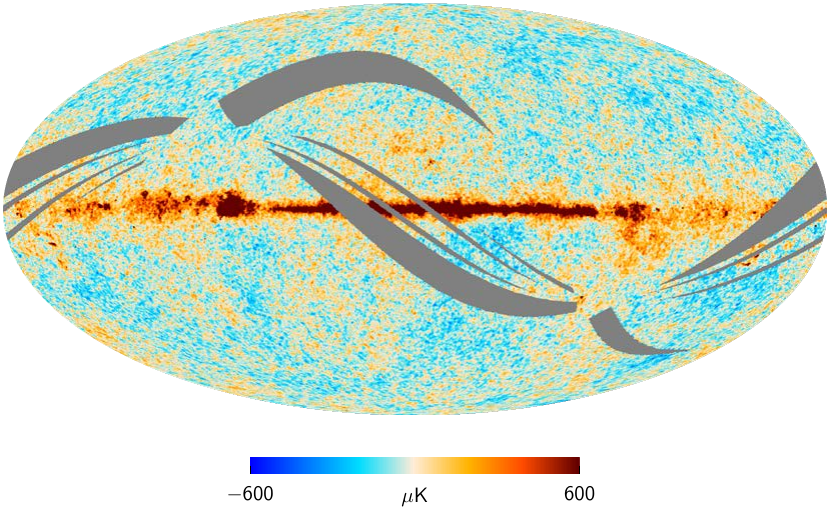} \\
	\includegraphics[width=.25\textwidth]{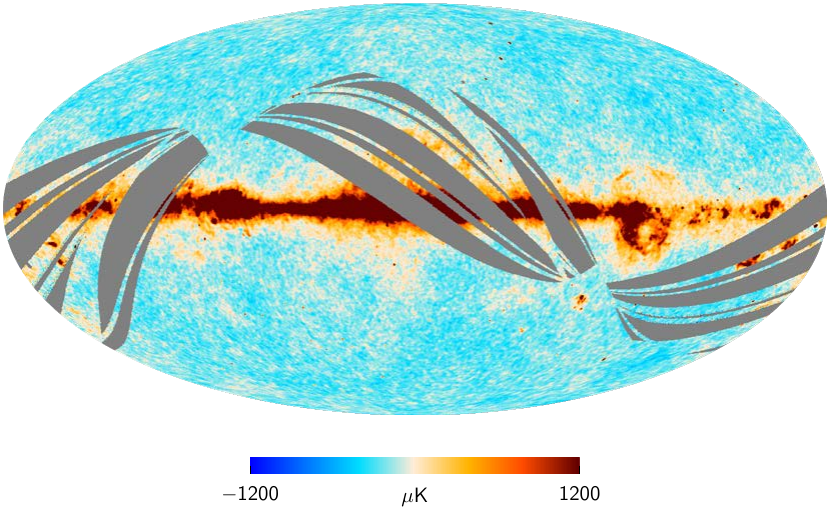} & 
	\includegraphics[width=.25\textwidth]{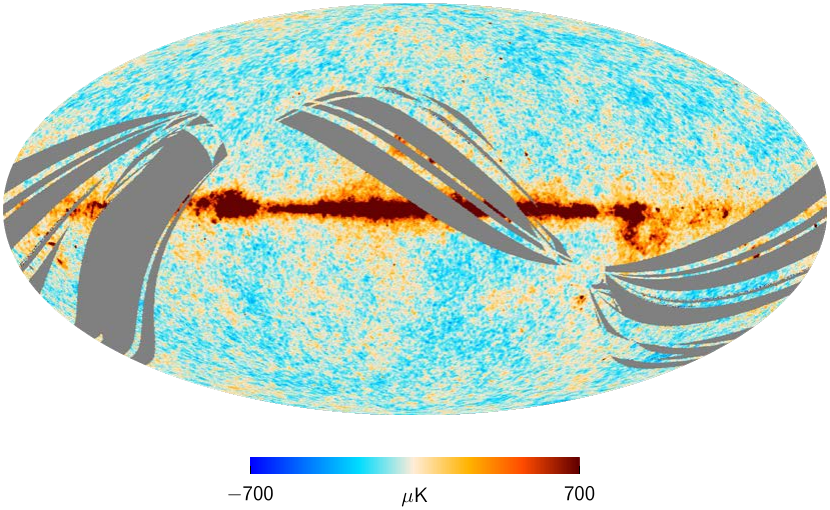} &
	\includegraphics[width=.25\textwidth]{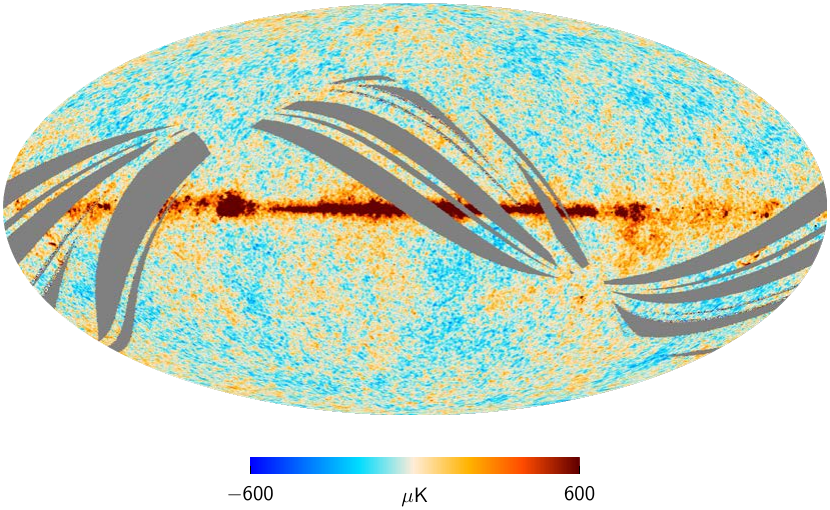} \\
        \end{tabular}
	\caption{Individual survey temperature maps.  {\it Left\/}: 30\,GHz.
        {\it Middle\/}: 44\,GHz.  {\it Right\/}: 70\,GHz. From top to bottom
        are Surveys~1 to 8. the gray area identify the regions of the sky not observed in each survey that depends from the spin axis orientation.}
	\label{fig:surveys}

\end{figure*}

Figures~\ref{fig:Imaps-030} to \ref{fig:Imaps-070} show the 30, 44,
and 70\,GHz frequency maps created from LFI data. 
The top panel in each figure is the temperature ($I$) map, based on the full
observation period at native resolution and {\tt HEALPix} $N_{\rm side}=1024$. 
The middle panel is the $Q$ polarization component,
while the bottom panel is the $U$ polarization component at $N_{\rm side}=256$
smoothed at $1^\circ$ resolution. In
Fig.~\ref{fig:surveys} the eight surveys at 30, 44, and 70\,GHz are shown;
the grey areas identify the regions of the sky not observed in each survey.
Table~\ref{tab:frequency_param} reports the main parameters used in the
mapmaking process.

The delivered maps have been processed in order to remove any spurious
zero-level (or monopole term).
To do this we implemented the following procedure. We derived from LFI data
only an estimation of the CMB signal by processing 1$^\circ$ smoothed maps
with an ILC (Internal Linear Combination) method, as described in
\citet{eriksen2004}. We then smoothed the single frequency LFI maps at the
same resolution and subtracted the CMB estimate. For each map we used the
variation with Galactic latitude of the remaining Galactic emission 
signal to estimate the zero-level. We assumed a simple plane-parallel model
for Galaxy emission and fit the data
with a functional form as $T = A{\rm csc} b + B$ in the range
$-90^\circ<b<-15^\circ$, using the same mask as employed in the mapmaking
procedure.  The value of $B$ is the zero-level we are looking for, which has to
be subtracted from the maps in order to obtain an overall ``null'' zero-level.
This value is reported in Table~\ref{tab:frequency_param}.

Finally Table~\ref{tab:maps_released} lists the delivered maps along with the
data period used to create them.  All have {\tt HEALPix}
resolution $N_{\rm side}=1024$; in the case of 70\,GHz we also provide maps at
a higher resolution, $N_{\rm side}=2048$.

 \begin{table*}
  \begingroup
  \newdimen\tblskip \tblskip=5pt
  \caption{Frequency-specific mapmaking parameters and related information.
 Details are reported in \citet{planck2014-a07}.}
  \label{tab:frequency_param}
  \nointerlineskip
  \vskip -3mm
  \footnotesize
  \setbox\tablebox=\vbox{
 % \vbox{
  \newdimen\digitwidth
  \setbox0=\hbox{\rm 0}
  \digitwidth=\wd0
  \catcode`*=\active
  \def*{\kern\digitwidth}
  \newdimen\signwidth
  \setbox0=\hbox{+}
  \signwidth=\wd0
  \catcode`!=\active
  \def!{\kern\signwidth}
  \halign{\tabskip=0pt\hbox {#}\tabskip=2em&
      \hfil#\hfil&
      \hfil#\hfil&
      \hfil#\hfil&
      \hfil#\hfil&
      \hfil#\hfil&
      \hfil#\hfil\tabskip=0pt\cr
  \noalign{\doubleline}
  \omit&\omit 
  &\multispan2\hfil B{\sc aseline length}$^{\rm b}$\hfil
  &\multispan2\hfil R{\sc esolution}$^{\rm c}$\hfil
  &\hfil M{\sc onopole, B}$^{\rm d}$\hfil
  \cr
  \noalign{\vskip -4pt}
  \omit&\omit &\multispan2\hrulefill &\multispan2\hrulefill &\hrulefill\cr
Channel& $f_{\rm samp}$ [Hz]$^{\rm a}$& [s]& Samples& $N_{\rm side}$& [arcmin]&$\,[\mu\mathrm{K}_{\rm CMB}$]\cr
  \noalign{\vskip 3pt\hrule\vskip 5pt}
30\,GHz& 32.508&  0.246& *8& 1024& 3.44&\llap{$-$}83.5$\,\pm\,$0.7\cr
44\,GHz& 46.545&  0.988& 46&  1024& 3.44&\llap{$-$}40.6$\,\pm\,$0.7\cr
70\,GHz& 78.769& 1.000& 79&  1024/2048& 3.44/1.72&\llap{$-$}35.7$\,\pm\,$0.6\cr
  \noalign{\vskip 5pt\hrule\vskip 3pt}}}
  \endPlancktablewide
  \tablenote a Sampling frequency.\par 
  \tablenote b Baseline length in seconds and as a number of samples.\par
  \tablenote c {\tt HEALPix} $N_{\rm side}$ resolution and pixel averaged
   size.\par
  \tablenote d Monopole removed from the maps; the value is included in the
   header fits.\par
    \endgroup
  \end{table*}

\begin{table*}
  \begingroup
  \newdimen\tblskip \tblskip=5pt
  \caption{Periods covered by the released maps.}
  \label{tab:maps_released}
  \nointerlineskip
  \vskip -3mm
  \footnotesize
  \setbox\tablebox=\vbox{
  \newdimen\digitwidth
  \setbox0=\hbox{\rm 0}
  \digitwidth=\wd0
  \catcode`*=\active
  \def*{\kern\digitwidth}
  \newdimen\signwidth
  \setbox0=\hbox{+}
  \signwidth=\wd0
  \catcode`!=\active
  \def!{\kern\signwidth}
  \halign{\tabskip=0pt\hbox to 1.25in{#\leaderfil}\hfil\tabskip=1em&
      \hfil#\hfil\tabskip=2em& \hfil#\hfil\tabskip=1em&
      \hfil#\hfil\tabskip=2em& \hfil#\hfil\tabskip=1em&
      \hfil#\hfil\tabskip=2em& \hfil#\hfil\tabskip=1em&
      \hfil#\hfil\tabskip=0pt\cr
      \noalign{\doubleline}
\omit& \omit& \multispan2\hfil 30\,GHz\hfil& \multispan2 \hfil 44\,GHz\hfil& \multispan2 \hfil 70\,GHz\hfil\cr
  \noalign{\vskip -4pt}
  \omit& \omit& \multispan2 \hrulefill& \multispan2 \hrulefill& \multispan2 \hrulefill\cr
  \noalign{\vskip 0pt}
\omit Period\hfil& OD range$^{\rm a}$& Sky cov. [\%]&  Horns$^{\rm b}$& Sky cov. [\%]& Horns$^{\rm b}$& Sky cov. [\%]&  Horns$^{\rm b}$\cr
  \noalign{\vskip 3pt\hrule\vskip 5pt}
Full& [**91--1543]& 100.00& 27,28& 100.00& 24,25,26& 100.00& 18,19,20,21,22,23\cr
    \noalign{\vskip 4pt}
Survey~1& [**91--*270]& *97.20& 27,28& *93.93& 24,25,26& *97.94& 18,19,20,21,22,23\cr
Survey~2& [*270--*456]& *97.48& 27,28& *93.31& 24,25,26& *97.47& 18,19,20,21,22,23\cr
Survey~3& [*456--*636]& *97.62& 27,28& *93.65& 24,25,26& *97.61& 18,19,20,21,22,23\cr
Survey~4& [*636--*807]& *91.88& 27,28& *89.53& 24,25,26& *92.40& 18,19,20,21,22,23\cr
Survey~5& [*807--*993]& *90.89& 27,28& *88.43& 24,25,26& *92.44& 18,19,20,21,22,23\cr
Survey~6& [*993--1177]& *87.79& 27,28& *86.10& 24,25,26& *89.95& 18,19,20,21,22,23\cr
Survey~7& [1177--1358]& *85.40& 27,28& *83.70& 24,25,26& *88.43& 18,19,20,21,22,23\cr
Survey~8& [1358--1543]& *80.01& 27,28& *78.92& 24,25,26& *83.83& 18,19,20,21,22,23\cr
    \noalign{\vskip 4pt}
Special survey$^{\rm c}$& S1+S3+S[5--8]& 100.00& 27,28& 100.00& 24,25,26& 100.00& 18,19,20,21,22,23\cr
    \noalign{\vskip 4pt}
Year~1& [**91--*456]& 100.00& 27,28& 100.00& 24,25,26& 100.00& 18,19,20,21,22,23\cr
Year~2& [*456--*807]& *99.98& 27,28& 100.00& 24,25,26& 100.00& 18,19,20,21,22,23\cr
Year~3& [*807--1177]& *99.72& 27,28& *99.91& 24,25,26& *99.65& 18,19,20,21,22,23\cr
Year~4& [1177--1543]& *95.67& 27,28& *96.87& 24,25,26& *97.38& 18,19,20,21,22,23\cr
    \noalign{\vskip 4pt}
Year 1\,$-$\,Year 2$^{\rm d}$& Y1+Y2& 100.00& 27,28& 100.00& 24,25,26& 100.00& 18,19,20,21,22,23\cr
Year 1\,$-$\,Year 3$^{\rm d}$& Y1+Y3& 100.00& 27,28& 100.00& 24,25,26& 100.00& 18,19,20,21,22,23\cr
Year 2\,$-$\,Year 4$^{\rm d}$& Y2+Y4& 100.00& 27,28& 100.00& 24,25,26& 100.00& 18,19,20,21,22,23\cr
Year 3\,$-$\,Year 4$^{\rm d}$& Y3+Y4& *99.77& 27,28& *99.90& 24,25,26& *99.66& 18,19,20,21,22,23\cr
            \noalign{\vskip 5pt\hrule\vskip 3pt}}} 
 \endPlancktablewide
  \tablenote a OD (operational day) is defined as the time period between one
  daily telecommand and the succeeding one; it corresponds to about
  $24$\,h.\par 
  \tablenote b Full period maps have been delivered for frequency, pairs of
  horn (18--23, 19--22, 20--21, 25--26), single radiometers, and in the case
  of horn 24, for a single horn.\par
  \tablenote c This special survey has been created excluding Surveys~2 and
  4 for low multipole analysis, see Sect.~\ref{sec_low_l}.\par
  \tablenote d Year differences.\par
  \endgroup
  \end{table*}

\section{Polarization}
\label{sec_polarization}
The most important new results in this release are polarization measurements.
The maps of Stokes $Q$ and $U$ at each LFI frequency are shown in
Figs.~\ref{fig:Imaps-030}, \ref{fig:Imaps-044}, and \ref{fig:Imaps-070} at
30, 44, and 70\,GHz, respectively.  The 70\,GHz polarized data play a critical
role in the construction of the \Planck\ low-$\ell$ likelihood, as described
in \citet{planck2014-a13}. Given the small amplitude of CMB polarization, we
have paid careful attention to systematic effects that could bias our
polarization results.  The dominant effect is leakage of unpolarized emission
into polarization \citep{leahy2010}, which we describe in detail in section ~\ref{sec_overview_leakage_maps}.
An overview of systematics impacting both temperature and polarization data is provided in section ~\ref{sec_systematics}, 
while a full account of the $2015$ LFI systematic error budget is given in \citet{planck2014-a04}.

\subsection{Bandpass mismatch leakage}
\label{bandpass_mismatch}
Any difference in gain between the two arms of an LFI radiometer will
result in leakage of unpolarized emission into the polarization
signal.  Since gains are calibrated by observations of the CMB dipole,
exact gain calibration would ensure that unpolarized, well resolved,
CMB emission perfectly cancels in the polarization signal.

However, because the bandpasses of the two arms are not identical,
unpolarized foreground emission, if it has a different spectrum from the
CMB, will still appear with different amplitudes in the two arms and
therefore leak into polarization.  This is ``bandpass mismatch'' leakage,
which was discussed extensively in \citet{leahy2010}.

In principle, two approaches can be used to correct for it. The first
exploits the fact that the bandpass leakage is independent of the
polarizer orientation, and performs a ``blind'' separation using
observations of a given pixel with multiple orientations of the same
radiometer.  With the second method, we can characterize both the
instrumental bandpass mismatch, and the foreground spectrum and
intensity, and hence predict the leakage explicitly.

The blind approach was used by \textit{WMAP} \citep{page2007}, but for
most sky pixels it is not effective for \Planck, because only a
relatively small range of detector orientations are available; this
causes very large covariances between the leakage and the true $Q$ and
$U$ values, effectively increasing the $Q$ and $U$ noise by a large
factor. Hence we use the predictive method to calculate the leakage in
our $Q$ and $U$ maps, and subtract it. 
We discuss in turn the determination of the
foreground model, the derivation of the instrumental term, and the
algorithm for making the correction.

\subsection{Leakage maps}
\label{sec_overview_leakage_maps}

The spectra of all important LFI foregrounds are very smooth continua,
and so to a good approximation can be modelled as a power law
within the bandpass at each LFI frequency band. As described
by \citet{leahy2010}, the leakage into the
polarization signal recorded by radiometer $k$ can be written as

\begin{equation}
S_k = a_k (\beta - \beta_{\rm CMB}) T^{\rm F}_{\nu_0}\, ,
\end{equation}
where the $a$-factor characterizes the bandpass mismatch (see next
subsection), $\beta = d\ln T^{\rm F}/d\ln\nu|_{\nu_0}$ is the spectral index
of the foreground within the band and $T^{\rm F}_{\nu_0}$ is the foreground
Rayleigh-Jeans brightness temperature at the band fiducial frequency
$\nu_0$.  We separate this into the instrumental $a$-factor, and an
astrophysical leakage term $L = (\beta - \beta_{\rm CMB})
T^{\rm F}_{\nu_0}$.  We derive $L$ from our Bayesian component separation
analysis, as described in \citet{planck2014-a12}. The analysis
incorporates the \Planck\ full-mission data, along with
the WMAP\ 9-year maps and the \citet{Haslam1982} 408\,MHz map,
to give 15 data points at each pixel.  This was an earlier run than
the one described in \citet{planck2014-a12}: the \Planck\ maps were from
a slightly earlier version of the calibration pipeline; the 
original bandpass models were used to make colour corrections; only a
single spinning dust component was included in the model, not two; and 
the synchrotron template from the {\tt Galprop} code was scaled only
in amplitude, not in frequency. 

This analysis produces numerous Gibbs-sampled realizations of the
astrophysical component parameters, from which $T^{\rm F}$ and $\beta$ can
be reconstructed at any given frequency, for each pixel in each
realization, $j$. In practice, we evaluate these individually for each
component $i$, to find a leakage map for each component, $L_j^i$, and
then sum the components to give $L_j = \sum_i L_j^i$. This is not only
more straightforward to evaluate, but also automatically corrects for
any in-band spectral curvature caused by the superposition of
foregrounds with similar amplitudes but very different spectral
indices. The final leakage map is then simply the average over the
realizations, $L = (1/N) \sum_{j=1}^N L_j$. In practice we use 1000 
realizations taken after the sampling chains have successfully
burnt in.

The uncertainty in $L$ at each pixel is based on the variance over
the Gibbs realizations, $\sigma_L$. However, we also have a measure of
goodness-of-fit of the model $\chi^2_j$, measured per pixel for each
realization. Because our MCMC chains are well burned in, we work with
the average $\chi^2 = (1/N) \sum \chi^2_j$ over all realizations (but
still separate for each pixel).  In regions of strong foreground emission the
component separation is limited not by noise but by a mismatch between
the assumed algorithmic form of the model and the actual spectrum,
signalled by high $\chi^2$.  Because the model is non-linear and many
of the model parameters are subject to strong prior constraints, the
$\chi^2$ statistic is not expected to follow a $\chi^2$ distribution
with the number of degrees of freedom equal to the number of data
points. We therefore define a fiducial $\chi^2$ equal to the median $\chi^2$
over the whole sky, which of course is dominated by the high-latitude
regions, where the foregrounds are weak, and therefore the component
separation residuals are dominated by noise. We adopt an empirical
correction to the uncertainty by multiplying $\sigma_L$ by the square
root of the ratio of the mean $\chi^2$ to our fiducial value wherever
this ratio exceeds unity.

The component separation analysis must be done at identical resolution
for all frequency channels, and this was chosen as 1\deg\ FWHM to allow use
of the 408-MHz survey. Consequently, polarization maps corrected for
bandpass mismatch leakage are only available at this or lower resolution.
Since the full-resolution polarization maps have a signal-to-noise ratio
of much less than unity for nearly all pixels, most scientific analysis
must in any case be done with smoothed maps, or equivalently with only
the low multipoles in harmonic space, so the low resolution of the
leakage maps is not a problem for most purposes. However, full resolution
data are needed to give the most accurate polarimetry of point sources. 
A special procedure was therefore used to correct the
polarization of sources, as described in \citet{planck2014-a35}.

\subsection{$a$-factors}
\label{sec_overview_afactors}

\begin{figure*}[htbp]
	\centering
	\begin{tabular}{cc}
	   \includegraphics[width=.5\textwidth]{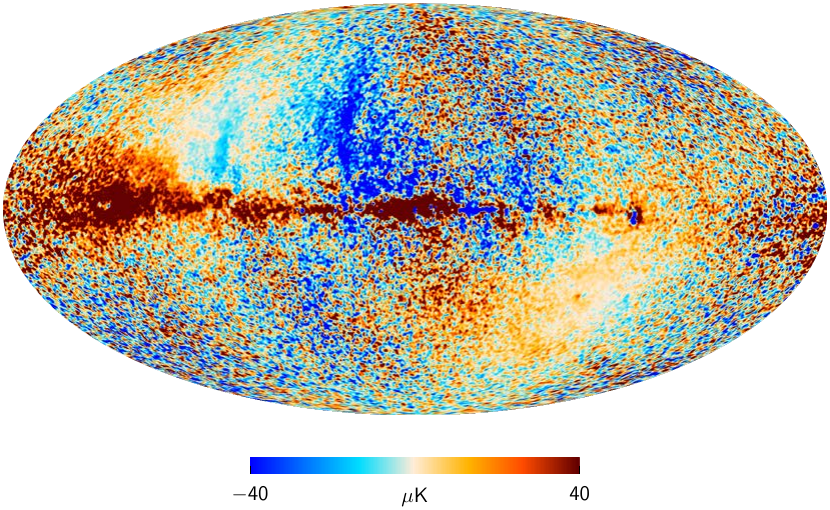} &
	   \includegraphics[width=.5\textwidth]{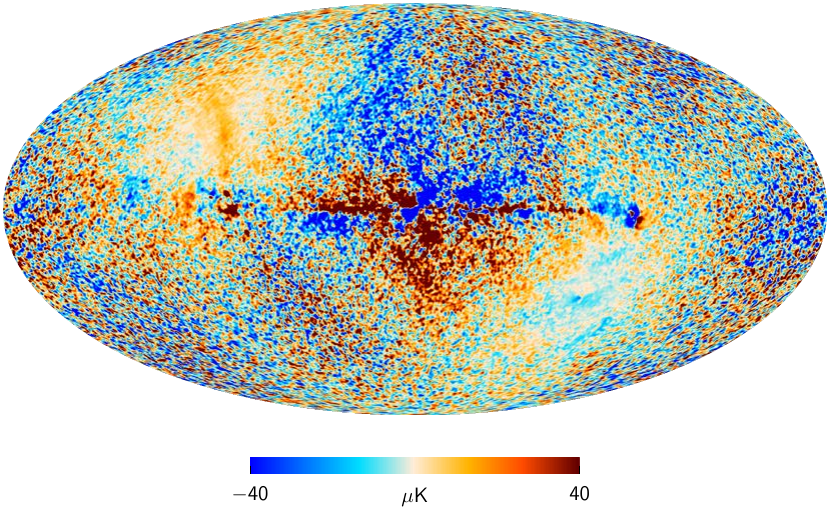} \\
	   \includegraphics[width=.5\textwidth]{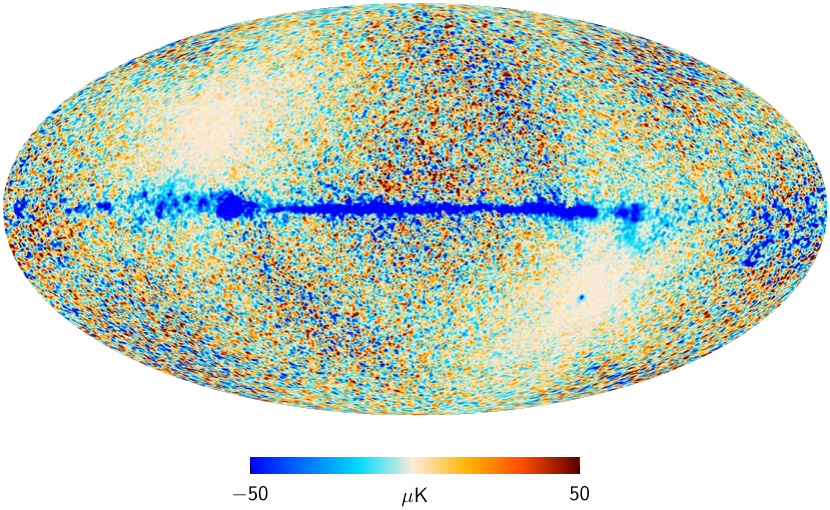} &
	   \includegraphics[width=.5\textwidth]{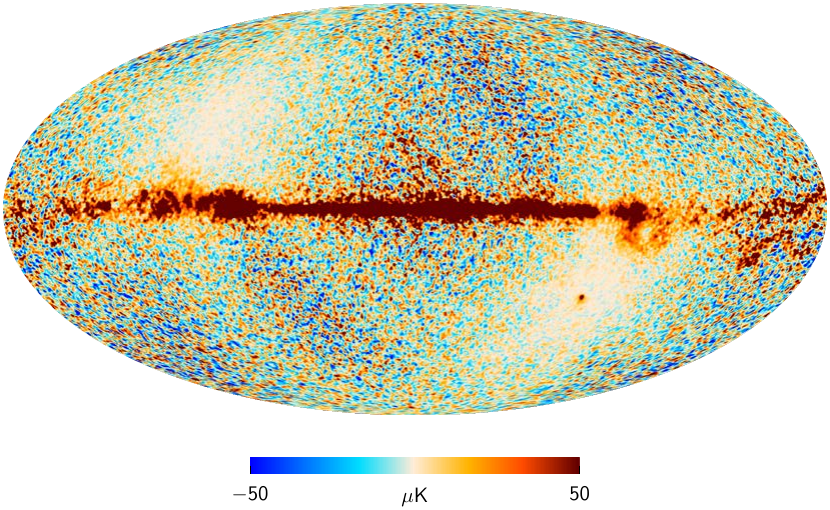} \\
	\end{tabular}
	\caption{$IQUSS$ solution maps at 30\,GHz. \emph{Top left\/}:
 Stokes $Q$. \emph{Top right\/}: Stokes $U$. \emph{Bottom left\/}: spurious
signal from the first RCA, $S_1$.  \emph{Bottom right\/}: spurious signal from
the second RCA, $S_2$. Polarization maps are noisier than the usual mapmaking
solution, since $S_1$ and $S_2$ have to be extracted from the same data.}
	\label{fig_IQUSS_solution}
\end{figure*}

Ground-based measurements of the LFI instrumental bandpasses \citep{zonca2009}
are not accurate enough for our purpose. 
Fortunately, to a good approximation, the bandpass mismatch can be
characterized by a single parameter, the $a$-factor, which 
quantifies the difference in effective frequency between the two bandpasses:
\begin{equation}
  a = {\nu_{\rm eff,s} - \nu_{\rm eff, m} \over 2 \nu_0}\, ,
\end{equation} 
where ``s'' and ``m'' refer to the side and main arms of the radiometer,
respectively, and $\nu_0 = (\nu_{\rm eff,s} + \nu_{\rm eff, m})/ 2$. 
We determine the $a$-factors from flight data by first using the blind
approach at each frequency to estimate $(I, Q, U, S_1, S_2 \ldots)$ 
(hereafter $IQUSS$) at each pixel, where $S_k$ is the spurious signal from
each RCA. Taking the 30\,GHz data as an example, the $S_k$ maps
are defined as
\begin{eqnarray}
       \text{LFI 27} \left\{
	\begin{array}{l l}
	d_{s1} = I + Q{\mathrm{cos}}(2\psi_{\mathrm{s1}})
        + U{\mathrm{sin}}(2\psi_{\mathrm{s1}}) + S_1\, , \nonumber \\
	d_{m1} = I + Q{\mathrm{cos}}(2\psi_{\mathrm{m1}})
        + U{\mathrm{sin}}(2\psi_{\mathrm{m1}}) - S_1\, ,
	\end{array}\right.\\
        \text{LFI 28} \left\{
	\begin{array}{l l}
	d_{s2} = I + Q{\mathrm{cos}}(2\psi_{\mathrm{s2}}) + U{\mathrm{sin}}(2\psi_{\mathrm{s2}}) + S_2\, , \\
	d_{m2} = I + Q{\mathrm{cos}}(2\psi_{\mathrm{m2}}) + U{\mathrm{sin}}(2\psi_{\mathrm{m2}}) - S_2\, . 
	\end{array}\right.
\label{iquss_model}
\end{eqnarray}
This can also be written in a more compact form
\begin{equation}
d_i = I + Q{\mathrm{cos}}(2\psi_i) + U{\mathrm{sin}}(2\psi_i)
 + \alpha_1 S_1 + \alpha_2 S_2\, ,
\end{equation}
where $\alpha_1$ and $\alpha_2$ take the values $-1,0,+1$, depending on the
radiometer.  To estimate $m = [I$, $Q$, $U$, $S_1$, $S_2]$ we need to
solve a problem similar to mapmaking, where the noise covariance
matrix per pixel $\tens{M}_p$ is given by the usual $3\times3$ matrix block
from {\tt Madam}, with two more columns (and rows) in the form
\begin{eqnarray}
\tens{M}_p = \sum_{i \in p} w_i  \nonumber
\end{eqnarray}
\begin{eqnarray}
\left( \begin{array}{ccccc}
\dots & \dots & \dots & \alpha_1 & \alpha_2 \\
\dots & \dots & \dots & \alpha_1\cos(2\psi_i) & \alpha_2\cos(2\psi_i) \\
\dots & \dots & \dots & \alpha_1\sin(2\psi_i) & \alpha_2\sin(2\psi_i) \\
\alpha_1      & \alpha_1\cos(2\psi_i) & \alpha_1\sin(2\psi_i)
  & \alpha_1^2        & 0 \\
\alpha_2      & \alpha_2\cos(2\psi_i) & \alpha_2\sin(2\psi_i) & 0 &
\alpha_2^2
\end{array}\right).
\label{eq:ncvm}
\end{eqnarray}

To ameliorate the limited range of orientations, we perform a joint
solution for all the RCAs at each frequency, in contrast to
the WMAP approach of solving for each radiometer
independently.  In Fig.~\ref{fig_IQUSS_solution} we show
output maps from the $IQUSS$ approach at 30\,GHz: $Q$ and $U$ maps
(top row); and $S_1$ and $S_2$ maps (bottom row). Note that 
$Q$ and $U$ maps are noisier than for the nominal mapmaking
solution. Over most of the sky the resulting maps of spurious
signals are still noisy and therefore we chose a conservative approach to
estimating the $a$-factor for each RCA. This is done with a weighted
least-squares fit of the leakage map $L$ to the spurious signal $S_k$
($S_k = a_k L$ in the absence of errors) using only those pixels
with $|b| < 15^\circ$, since at higher Galactic latitudes the
foregrounds and hence the spurious signals are weak and mainly
contribute noise to the solutions. Our code removes
pixels where the condition number for the noise covariance matrix
$\tens{M}_p$ is less than a given threshold. This now has a negligible effect,
since thanks to the
modification of the \Planck\ scanning strategy after Survey~5, the matrix
$\tens{M}_p$ is very well-behaved; even with our conservative limit of
$8\times 10^{-5}$ for the condition number, fewer than 200 pixels are
excluded at 44\,GHz and none at 30 and 70\,GHz. Our derived values for
the $a$-factors are listed in Table~\ref{tab:afactors}.

\begin{table}
\begingroup
\newdimen\tblskip \tblskip=5pt
\caption{Bandpass mismatch $a$-factors from fitting the leakage model map
 to the spurious maps.}
\label{tab:afactors}
\nointerlineskip
\vskip -3mm
\footnotesize
\setbox\tablebox=\vbox{
\newdimen\digitwidth
\setbox0=\hbox{\rm 0}
\signwidth=\wd0
\catcode`*=\active
\def*{\kern\digitwidth}
\newdimen\signwidth
\setbox0=\hbox{+}
\signwidth=\wd0
\catcode`!=\active
\def!{\kern\signwidth}
\halign{\hbox to 1.3in{#\leaderfil}\tabskip=2em&
	\hfil#\hfil\tabskip=0pt\cr
\noalign{\doubleline}
\noalign{\vskip -3pt}
\omit\hfil Horn\hfil & $a$-factor\cr
\noalign{\vskip 3pt\hrule\vskip 5pt}
\omit{\bf 70\,GHz}\hfil\cr
\noalign{\vskip 4pt}
\hglue 2em LFI 18& $-0.0018\pm0.0022$\cr
\hglue 2em LFI 19& $!0.0124\pm0.0024$\cr
\hglue 2em LFI 20& $!0.0034\pm0.0024$\cr
\hglue 2em LFI 21& $-0.0115\pm0.0024$\cr
\hglue 2em LFI 22& $!0.0039\pm0.0024$\cr
\hglue 2em LFI 23& $!0.0057\pm0.0024$\cr
\noalign{\vskip 5pt}
\omit{\bf 44\,GHz}\hfil\cr
\hglue 2em LFI 24& $!0.0033\pm0.0005$\cr
\hglue 2em LFI 25& $!0.0004\pm0.0004$\cr
\hglue 2em LFI 26& $!0.0014\pm0.0004$\cr
\noalign{\vskip 5pt}
\omit{\bf 30\,GHz}\hfil\cr
\hglue 2em LFI 27& $!0.0046\pm0.0002$\cr
\hglue 2em LFI 28& $-0.0089\pm0.0002$\cr
\noalign{\vskip 5pt\hrule\vskip 3pt}}}
\endPlancktable
\endgroup
\end{table}

Beam-shape mismatch between radiometer arms
can also lead to polarization leakage when there are strong intensity
gradients. We therefore examined the effect of excluding compact sources using
the WMAP 7-year point source mask.  However, this made no
significant changes, apart from a dramatically increased uncertainty, so
our final values do not use such masking.

We compared these values with an independent derivation based on
aperture photometry of bright sources in the $IQUSS$ maps, including
the Tarantula nebula in the LMC, which lies in the ``deep'' region
around the Ecliptic pole, which is scanned with multiple different
polarimeter orientations across a wide range of angles, and hence
allows a particularly accurate blind separation of spurious
signal. Other calibrators were bright \ion{H}{ii} regions at relatively
high Ecliptic latitude, since the range of polarization orientations
observed increases towards the Ecliptic poles.  \ion{H}{ii} regions were
chosen because they have minimal intrinsic polarization, but we did
not force $Q$ and $U$ to zero in the analysis. The $a$-factors derived
from the calibrators were consistent with our preferred values derived
from the large area fit, but somewhat less precise.

\subsection{Production of correction maps}
\label{sec_overview_correction_maps}

The polarization data from a given radiometer constrains one Stokes
parameter (say $Q_H$) in a frame of reference tied to the specific
feed horn (or RCA). This is projected onto the sky according to the
sky orientation of the horn frame. Hence the contribution of the
spurious signal from each radiometer is modulated into the $Q$ and $U$
sky pixels by geometric projection factors. This modulation can be
derived by re-scanning, in a mapmaking fashion, the estimated spurious
map $\hat{S} = a_k L$.  Instead of an actual re-scanning, which is time
consuming, we create projecting maps $A_{Q[U]}$ by solving the
mapmaking system
\begin{equation}
A_{Q[U]}(p) = \tens{M}_p^{-1} m_p,
\end{equation}
where $m_p$ are the maps obtained by binning a stream of $-1$ for the ``side''
and of $+1$ for the ``main'' arms separately. Finally the correction
maps are
\begin{equation}
\Delta Q [U] = L \times \sum_k a_k A_{k,Q[U]}\, .
\end{equation}

One of the main drawbacks of deriving our $L$ maps at 1\deg\ resolution
emerges at this stage. The correction must be applied to $Q$ and $U$
maps matched in resolution, and so the raw $Q$ and $U$ maps are smoothed
to give the required 1\deg\ FWHM Gaussian beam. 
We can regard the raw maps as the true $Q$ and
$U$ sky, smoothed with the instrumental beam, plus the leakage term, 
plus noise. The leakage term in the raw
maps can be thought of as an infinite-resolution leakage sky convolved 
with the instrumental beam, and then multiplied by the leakage projection
maps $P_{Q[U]} = \sum_k a_k A_{k,Q[U]}$, which are defined at the pixel
level according to Eq.~(\ref{eq:ncvm}).  When we smooth this raw map, we
smooth the {\it product\/} $P L$, but when we construct our correction map,
we have only the smoothed $L$ map. The smoothed product is {\it not\/} equal to
the smoothed $L$ map multiplied by the full-resolution $P$ map,
which contains fine-scale structure induced by caustics, lost data,
and abrupt changes in the survey strategy; nor is it equal to the smoothed
$L$-maps multiplied by the smoothed $P$ map, which is over-smoothed in
regions where both $L$ and $P$ vary rapidly. In practice we used the
smoothed $P$ maps, since $P$ only varies rapidly near a small subset
of pixels, and only a few of these will also have rapidly varying $L$.
The issue is most significant for compact sources, for which we recommend
analysis of the raw maps, followed by a leakage correction using the
derived $IQU$ fluxes, as described in \citet{planck2014-a35}.

Although our $a$-factor estimates are relatively stable, our fit of the 
leakage maps to the spurious maps showed significant residuals at the level 
of 18\,$\mu$K, 24\,$\mu$K, and 16\,$\mu$K, respectively at 30, 44, and 70\,GHz.
Contributing factors may include: errors in the leakage maps caused by errors
in the component
separation; residual beam ellipticity after smoothing to 1\deg; and
any variation with time of the bandpass, which would cause corresponding 
changes in the $a$-factors. As a check on our results, we used the $IQUSP$
procedure in which we create a prior for each $S$ map using our component-separation $L$ map and our best $a$-factor 
estimates. This process returns the prior $S$-map essentially unchanged 
over most of the sky, and hence gives $Q$ and $U$
maps indistinguishable from our corrected versions.  However, it 
prefers the $IQUSS$ solution when it differs 
significantly  from the prior (essentially in regions of the brightest 
foreground emission, where the limitations of our simplified emission 
models become apparent). An advantage of the method is that the maps are
returned at full resolution, wherever the data can constrain the resolution
to be higher than in the prior. These maps confirm that most details of the 
structure along the Galactic plane in our corrected LFI polarization maps 
are consistent with the data, i.e. are reproduced in the $IQUSP$ images,
including the most significant discrepancies with WMAP.
Although not fully validated and therefore not included in the current
release, the $IQUSP$ images are likely to form the baseline for our
final-release polarization maps.

\section{Data validation}
\label{sec_dataval_intro}
We verify the quality of the LFI data with a suite of null tests, as well
as with a set of simulations reproducing the main instrumental systematic
effects and the calibration process. In this section we summarize the main
results of our analysis, and refer to \citet{planck2014-a04} and
\citet{planck2014-a06} for more details.

Null tests are performed on blocks of data covering different time scales
(from the pointing period to surveys and years) and considering
different instrument combinations (radiometer, horn, horn-pairs, and frequency)
both in total intensity and polarization (when applicable).

Such null tests can probe different systematic effects depending on the time
and instrument selection considered. Differences at horn level between odd and
even surveys may show effects due to the sidelobe contribution,
since the relative orientation of the horns with respect to the sky is
changed. Furthermore, the comparison of
power spectra at the frequency level may reveal the impact of
calibration uncertainties related to the relative orientation of the scans and
the CMB dipole, our main calibration source as discussed in
Sect.~\ref{sec_calibration}.

    \subsection{Null test results}
    \label{sec_null_test}
\begin{figure*}[th]
\centering
\includegraphics[width=18cm]{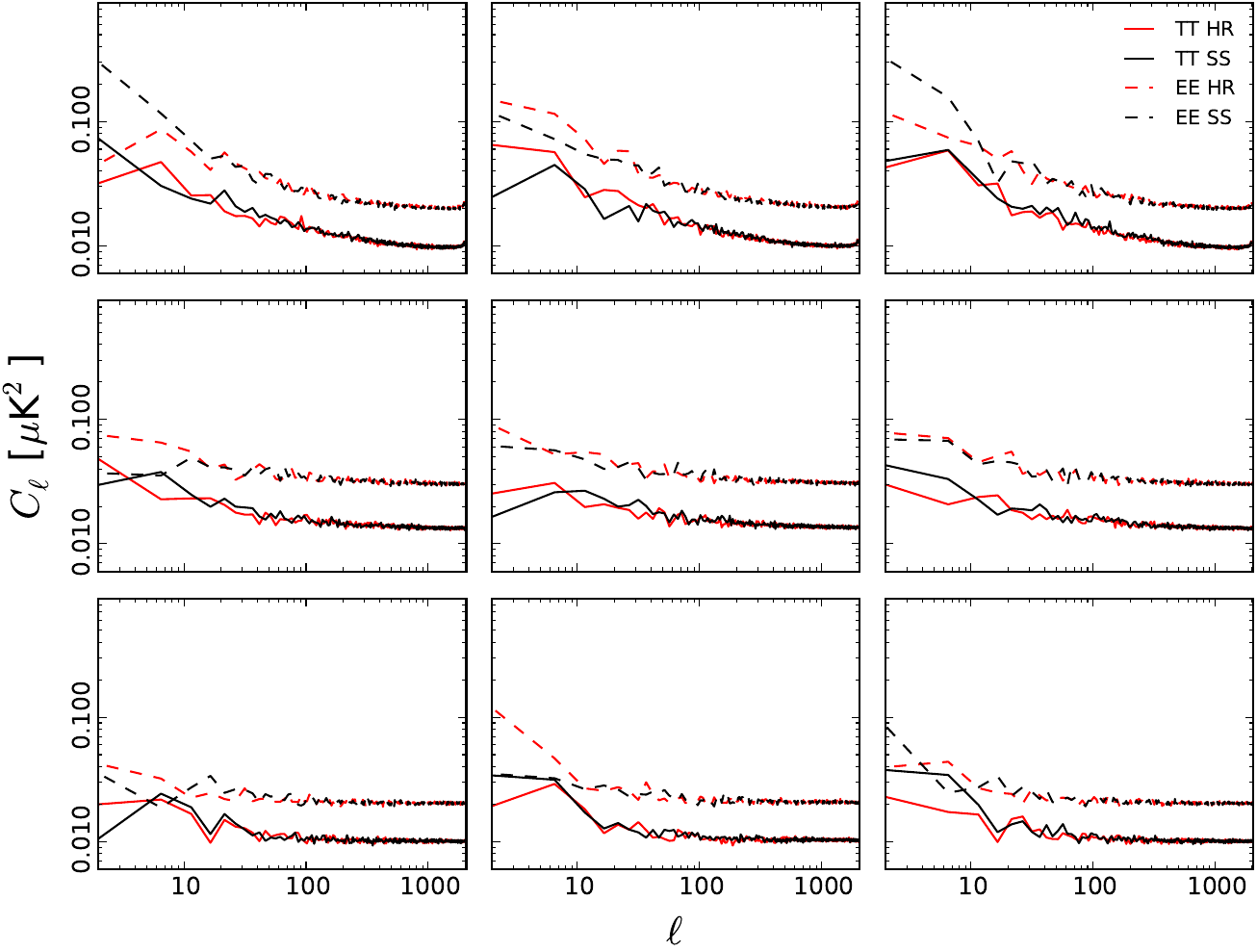}
\caption{Null test results comparing power spectra from survey differences
to those from the half-ring maps. Differences are:
\emph{left\/} Survey~1 $-$ Survey~2;
\emph{middle\/} Survey~1 $-$ Survey~3;
and \emph{right\/} Survey~1 $-$ Survey~4.
These are shown for 30\,GHz (\emph{top\/}), 44\,GHz (\emph{middle\/}),
and 70\,GHz (\emph{bottom\/})
for both $TT$ and $EE$ power spectra.}
\label{nulltest_figures}
\end{figure*}

        In order to assess null test results, it is fundamental to define a clear
figure of merit as a pass-fail criterion. Failure of a specific test is an
indication of a data problem and/or issues
in data processing that should be studied further.
As we already did for the previous release,
we take the noise level as derived from ``half-ring'' difference maps, made of the first and second
half of each stable pointing period (half-ring maps) weighted by
the hit count, as the figure of merit. This quantity traces the actual properties of the data,
including white noise, as well as un-modelled and un-corrected effects.
Figure~\ref{nulltest_figures} shows results at the frequency level for
both $TT$ and $EE$ power spectra when we compare survey differences to the
noise level derived from the corresponding half-ring maps. For simplicity we show here only a subset of
survey differences that are illustrative of the general trend.

When interpreting these results it is important to note that 
we have substantially improved the quality of data
at 30\,GHz by using the new $4\pi$ calibration ~\citep{planck2014-a06}, which accounts for the
impact of the full beam during calibration, and by removing at the TOD
level the modelled sidelobe signals of both the
CMB dipole and Galactic emission (derived from the FFP8 simulation runs).
This is particularly evident from $TT$ spectra, where the null test data
match the level of the half-ring differences. However,
there is still an issue in polarization when considering differences
involving Surveys~2 and 4.  At 44\,GHz, which has the lowest sidelobes among LFI
channels, the agreement with the half-ring noise
is extremely good.  We have almost the same situation at 70\,GHz,
although at very low multipoles ($\ell < 10$) there are discrepancies
between survey difference and half-ring noise; again this is particularly
evident when Surveys~2 and 4 are considered. 

To be more quantitative about these results at 70\,GHz we compute deviations
from the half-ring noise in terms of 
\begin{equation}
\chi^2_\ell = \frac{\sqrt{2\ell +1}}{2}
 \left(\frac{\mathcal{C}_\ell^{\rm SS} 
 - \mathcal{C}_\ell^{\rm hr}}{\mathcal{C}_\ell^{\rm hr}}\right)\, .
\label{chiell}
\end{equation}
We sum up single $\chi_\ell^2$ values in the range 2--50 and from
$\chi2$ and $N_{\rm dof}$ we derive $p-$values of the distribution.
While in principle proper noise simulations should be used, for our purposes it is sufficient to 
consider simple half-ring noise, which is already able to reveal interesting features in the data.
Table~\ref{pvalue_chi2} reports $\chi^2$ and $p-$values for the
three survey differences shown in Fig.~\ref{nulltest_figures}, which suggests that 
Surveys~2 and 4 clearly yield poor $\chi^2$ and problematic $p-$values.

\begin{table}[tmb]
\begingroup
\newdimen\tblskip \tblskip=5pt
\caption{Survey difference $\chi^2$ and $p-$values.}
\label{pvalue_chi2}
\nointerlineskip
\vskip -3mm
\footnotesize
\setbox\tablebox=\vbox{
        \newdimen\digitwidth
        \setbox0=\hbox{\rm 0}
        \digitwidth=\wd0
        \catcode`*=\active
        \def*{\kern\digitwidth}
        \newdimen\signwidth
        \setbox0=\hbox{+}
        \signwidth=\wd0
        \catcode`!=\active
        \def!{\kern\signwidth}
        \newdimen\signwidth
        \setbox0=\hbox{+}
        \signwidth=\wd0
        \catcode`!=\active
        \def!{\kern\signwidth}
        \halign{\tabskip=0pt\hfil#\hfil\tabskip=2em&
           \hfil#\hfil&
           \hfil#\hfil\tabskip=0pt\cr
           \noalign{\doubleline}
           \noalign{\vskip -3pt}
           Survey differences& $\chi^2$& $p-$value\cr
           \noalign{\vskip 3pt\hrule\vskip 5pt}
	  Survey~1 $-$ Survey~2& 83.3& 0.0016\cr
	  Survey~1 $-$ Survey~3& 70.8& 0.0221\cr
	  Survey~1 $-$ Survey~4& 98.3& $3.7\times10^{-5}$\cr
           \noalign{\vskip 3pt\hrule\vskip 3pt}}}
           \endPlancktable
           \endgroup
\end{table}

As discussed further in Sect.~\ref{sec_low_l}, on the basis of these and
other results, we have discarded these two surveys from the released
likelihood.

    \subsection{Half-ring test}
    \label{sec_halfring_test}
	As already pointed out, the half-ring difference maps are the best
direct information about the actual noise in the LFI data. A proper
characterization of the noise is fundamental for the creation of realistic
NCVMs and noise MC required for the following steps in the data analysis.
In this respect such noise modelling has to be validated against the half-ring
maps.  For the current analysis we followed the same procedure exploited in the
previous data release. We computed auto-spectra in temperature and polarization
with {\tt anafast} of both the half-ring difference maps and 10\,000 noise
Monte Carlo simulated maps taken from FFP8. We compared the half-ring spectra
with the distribution of the noise MC simulations and with the white noise
derived from the white noise covariance matrices (WNCVM) calculated by
{\tt Madam} map-making.

\begin{figure*}[th]
\centering
\includegraphics[width=18cm]{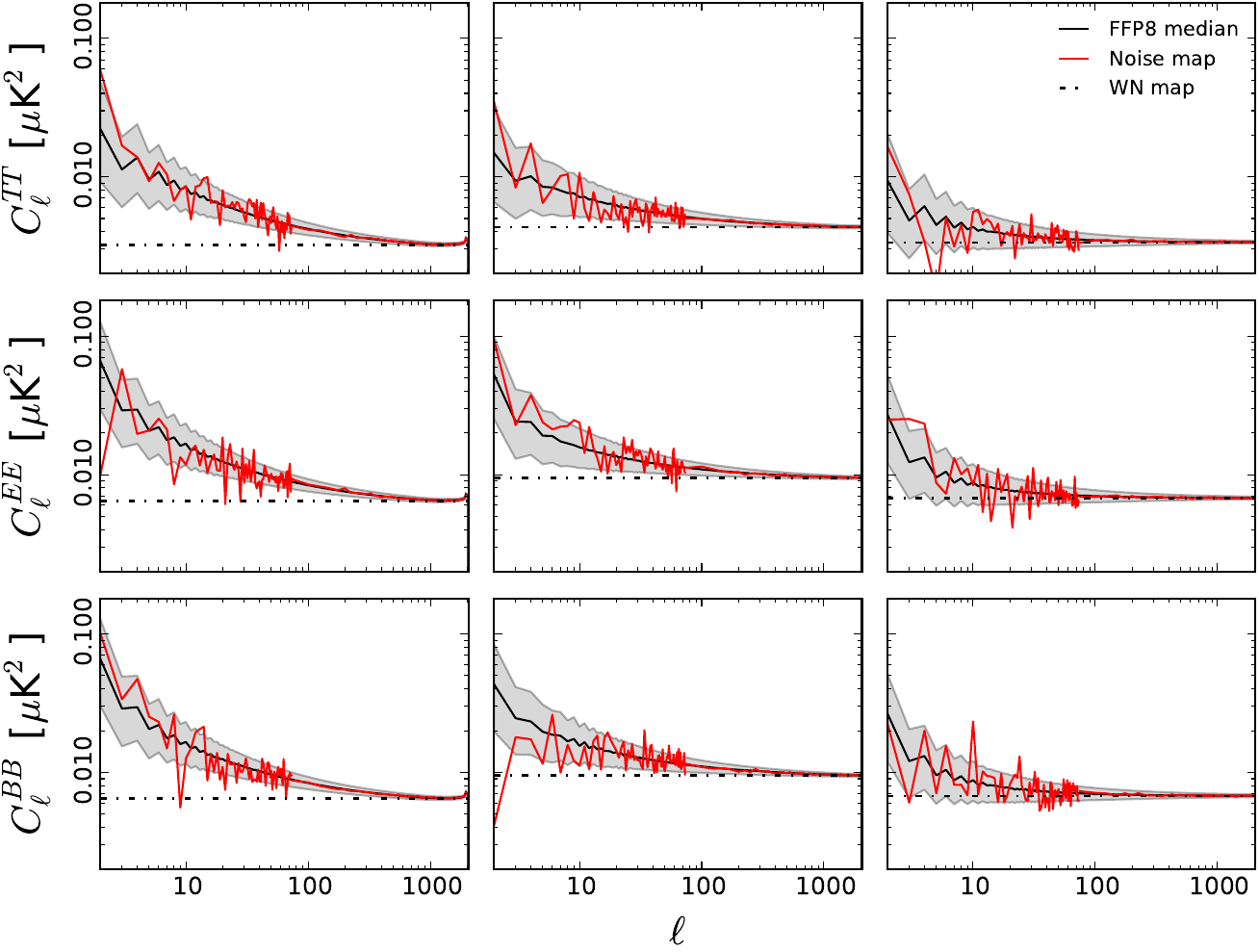}
\caption{Consistency of the noise angular power spectra from the half-ring
difference maps (red), white noise covariance matrix (black dash-dotted line),
and 10\,000 full-noise MCs (grey band showing 50\,\% quantiles, black solid line,
and limits at 16\,\% and 84\,\% quantiles).  From top to bottom we have
$TT$, $EE$, and $BB$ spectra for 30 (left), 44 (centre)
and 70\,GHz (right). Half-ring spectra are binned with $\Delta \ell=75$ for
$\ell \ge 75$.}
\label{fig_halfring_MC}
\end{figure*}

Figure~\ref{fig_halfring_MC} gives a flavour of this comparison for the three
LFI frequencies and for both total intensity $TT$ and polarization
($EE$ and $BB$) power spectra. Note that
the half-ring noise spectra are binned over a range of
$\Delta \ell = 25$ for $\ell \ge 75$. 
The agreement between half-ring noise spectra
and noise MC distribution
is remarkable, and gives us confidence that the LFI noise properties are accurately characterized.

\begin{figure}[th]
\centering
\includegraphics[width=9cm]{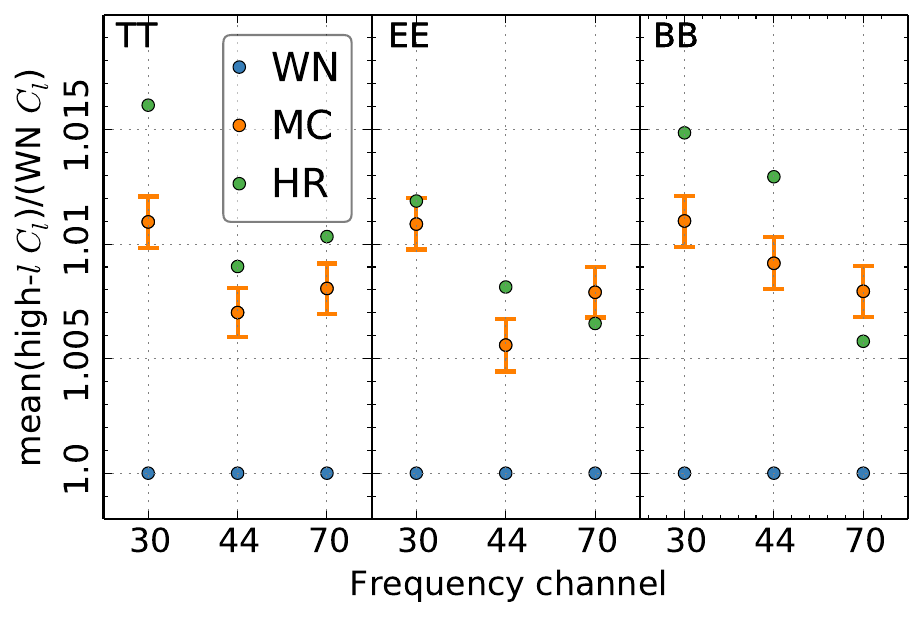}
\caption{Ratio of the mean noise angular power spectrum in the
high-$\ell$ ($1150\le\ell\le1800$) tail to the white noise
as derived from the white noise covariance matrices from {\tt Madam}.}
\label{fig_wncvm}
\end{figure}

We further inspect this comparison computing the mean $C_\ell$ for the
high-$\ell$ tail of the spectrum ($1150\le\ell\le 1800$) and comparing it
with the WNCVM (white noise covariance matrix) estimate (Fig.~\ref{fig_wncvm}). It is clear that there is
some residual $1/f$ noise also at high-$\ell$ as has been already pointed
out in the $2013$ release \citep{planck2013-p02}. This means that both data and noise MCs predict
a slightly higher noise than the WNCVM.  The residual is of the order of
1.6\,\% ($TT$) maximum at 30\,GHz, 1.3\,\% ($BB$) at 44\,GHz,
and 1.0\,\% ($EE$) at 70\,GHz. On the other hand, the agreement between the
actual data and the full noise MCs is extremely good being of the order of
0.5\,\% at 30\,GHz, 0.4\,\% at 44\,GHz, and 0.2\,\% at 70\,GHz.

    \subsection{End-to-end test results}
    \label{sec_null_test}
\begin{figure}[!hb]
\centering
\includegraphics[width=8.8cm]{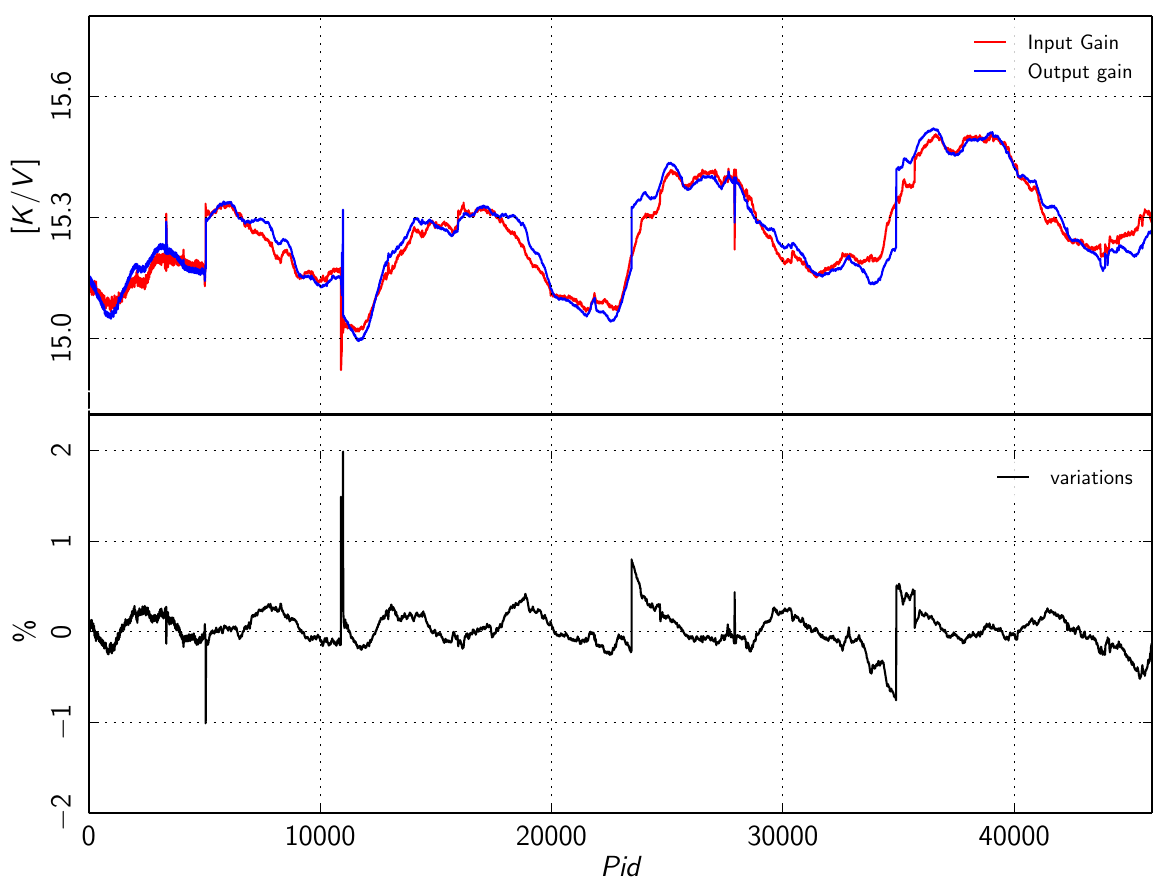}
\caption{Relative variations between input and output of the end-to-end
test for radiometer 27S at 30\,GHz.  In general, we recoverthe input gain to
better than $0.1\,\%$, except for some larger excursions introduced by sudden
changes in the instrument configuration, to which the 30\,GHz radiometers are
particularly sensitive.} 
\label{fig:gains_recon_30}
\end{figure}

\begin{figure}[!hb]
\centering
\includegraphics[width=8.8cm]{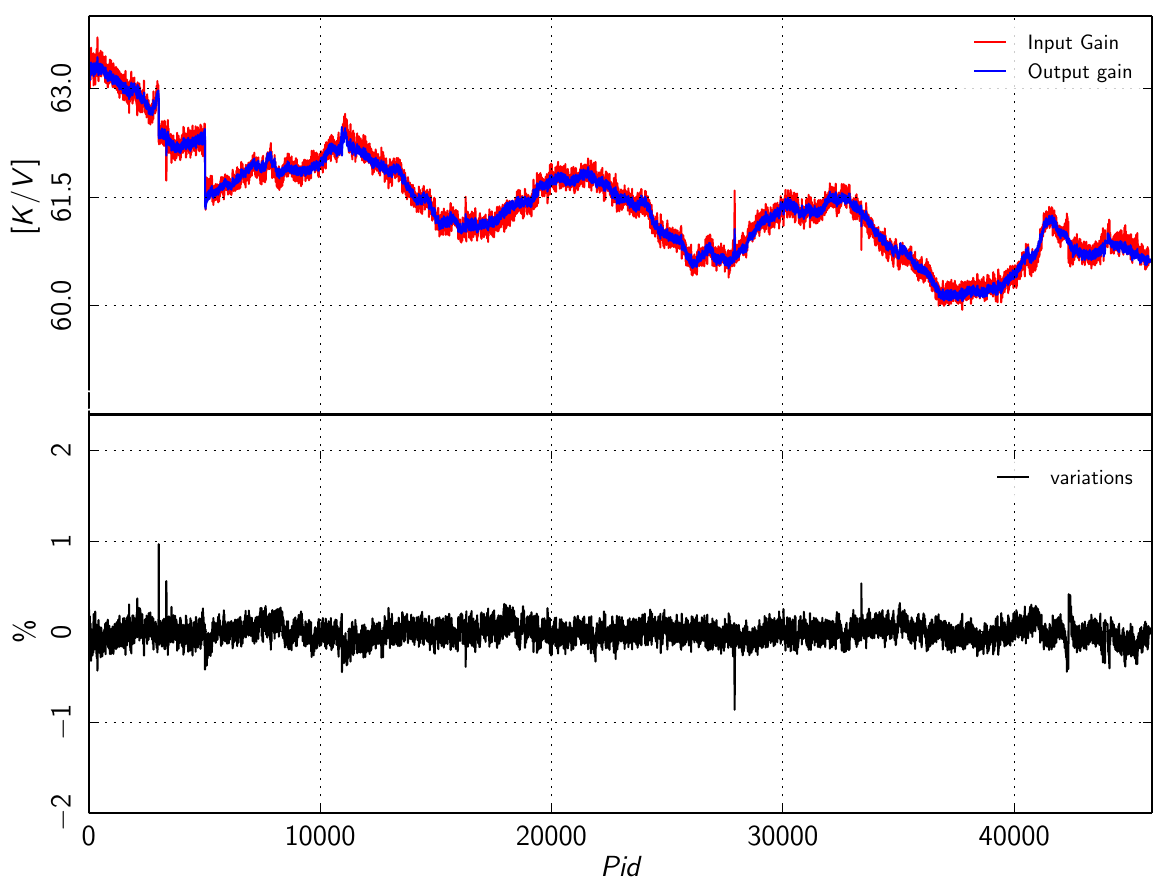}
\caption{Relative variations between input and output of the end-to-end
test for radiometer 22S at 70\,GHz.  The overall recovery is under
$0.1\,\%$, with some spikes in the longest pointing periods.}
\label{fig:gains_recon_70}
\end{figure}

        The LFI calibration pipeline is necessarily quite complex, since it includes
iterative mapmaking, sidelobe removal, Galactic masking, map domain fits to the
$4\pi$ beam-convolved dipole, and filtering.  While the accuracy of the mean
calibration constant is important, particularly for inter-frequency validation
and foreground modelling, we are mostly concerned with quantifying the level of 
systematic errors in our estimation of the calibration over time. 
The gain of the LFI radiometers typically varied by a few percent over the
four years mission lifetime, with changes at time scales from single pointing periods to the full
mission.  Null tests on survey and year time scales set useful limits on
systematic effects, including incorrect calibration estimation, but it is still
important to develop a ``bottom up'' estimation of possible errors.
Consequently we have carried out several parallel efforts to simulate our calibration procedure, each using different software and detailed choices for inputs,
but following the same general approach, which we now summarize.
\begin{itemize}
\item Start with a fiducial sky map (in kelvins), either from the frequency
maps of the data, FFP8, or some other simulation.  This map includes CMB
anisotropies, foregrounds, and possibly some systematics, but no dipole
signals, and can be either temperature only or $Q$ and $U$.
\item ``Unwrap'' or rescan the map to a time ordered signal data set, in
``ring'' basis (still in Kelvin).  This is done using actual flight pointing
data.
\item Add dipole signals, including the solar dipole and the
orbital dipole.  We can choose here whether to use a ``pencil beam'' model,
where the dipole signal that is added has been sampled from the sky model
with a Dirac delta function, or a $4\pi$ model consisting of an all-sky
convolution of the detector beam model with the dipole model. 
\item Add instrument noise, either white noise or full $1/f$ noise
(only white noise turns out to be relevant).   
\item All the steps described so far assume a timeline in kelvins.  Next we
``decalibrate'' these simulated data streams using a fiducial model for the
actual detector gain, and produce timelines in volts.  A standard choice here
is to use the so-called ``Delta $V$'' gain, which is a radiometer gain
estimated directly from the DC-coupled detector data.  While we know from
detailed tests that this gain does not track the actual gain fluctuations
better than about 0.5\,\%, it has the advantage of being a gain estimate with
no smoothing applied, and should reflect closely the true statistics of the
radiometer gain.
\item From this simulated timestream, we proceed with our nominal calibration
pipeline to recover the input gain.  In this way, we can compare the
recovered time domain gain estimate to the fiducial input, as well as the
final calibrated maps to the fiducial input maps. The results of such
comparisons are shown for two radiometers in Figs.~\ref{fig:gains_recon_30}
and \ref{fig:gains_recon_70}.
\end{itemize}

These simulations are designed to test the impact of our procedures on the
results.  They also provide a mechanism for quantitatively determining the
impact that errors in the \textit{inputs}, such as beam shape or far sidelobe
contribution, have on our \textit{output\/} maps and other scientific
products.  They do not provide a way to estimate \textit{what\/} those input
errors are; these must be determined by dedicated investigations on the
optical model or instrument-specific simulations.  Starting from reasonable
estimates of the systematics affecting our instrument, however, we can
introduce changes in the input within the expected range and then test for
deviations in the recovered calibration.  We thus obtain both the sensitivity
to that effect and an estimate of the probable error causing it,
assuming either extreme
values (conservative) or the expected $1\,\sigma$ (typical).  Similarly, we
can use this approach to determine the sensitivity of the calibration process
to Galactic masking.

The basic results of such end-to-end tests of the effects of systematics are
summarized in Table~\ref{e2e_summary_performance}.
Comparison of the difference between input and output gains shows a typical
bias of order $0.2$\,\%.

Figures~\ref{fig:gains_recon_30} and \ref{fig:gains_recon_70} show
input and output gain constants and
the relative variations between the two, at 30\,GHz and 70\,GHz, respectively.
The 30\,GHz channels are the most difficult to calibrate, because they are
more sensitive to changes in instrument configuration, causing
a bigger number of jumps, while the 70\,GHz are more sensitive to the
instrument noise.  Table~\ref{e2e_summary_performance} shows mean and
standard deviations of the 
relative variations between the input gain constants and the output ones,
for all the radiometers. The resulting precision
of the photometric calibration is up to $0.2\,\%$, thus validating the
calibration algorithm.

\begin{table}
\begingroup
\newdimen\tblskip \tblskip=5pt
\caption{Mean and associated error of the percentage variation between input
and output of the end to end tests }
\label{e2e_summary_performance}
\nointerlineskip
\vskip -3mm
\footnotesize
\setbox\tablebox=\vbox{
   \newdimen\digitwidth
   \setbox0=\hbox{\rm 0}
   \digitwidth=\wd0
   \catcode`*=\active
   \def*{\kern\digitwidth}
   \newdimen\signwidth
   \setbox0=\hbox{+}
   \signwidth=\wd0
   \catcode`!=\active
   \def!{\kern\signwidth}
%
%\halign{\hbox to 1.0in{#\leaderfil}\tabskip=3em&
\halign{\tabskip=0pt#\hfil\tabskip=3em&
                \hfil#\hfil\tabskip=0pt\cr
\noalign{\doubleline}
\noalign{\vskip -3pt}
\omit\hfil Radiometer\hfil& Mean difference [\%]\cr
\noalign{\vskip 3pt\hrule\vskip 5pt}
{\bf 70\,GHz}&\cr
\noalign{\vskip 3pt}
LFI 18M& $-0.002\,\pm\,0.057$\cr
\noalign{\vskip 3pt}
LFI 18S& $!0.040\,\pm\,0.140$\cr
\noalign{\vskip 3pt}
LFI 19M& $!0.007\,\pm\,0.169$\cr
\noalign{\vskip 3pt}
LFI 19S& $!0.058\,\pm\,0.090$\cr
\noalign{\vskip 3pt}
LFI 20M& $!0.009\,\pm\,0.081$\cr
\noalign{\vskip 3pt}
LFI 20S& $-0.002\,\pm\,0.071$\cr
\noalign{\vskip 3pt}
LFI 21M& $-0.005\,\pm\,0.070$\cr
\noalign{\vskip 3pt}
LFI 21S& $-0.031\,\pm\,0.066$\cr
\noalign{\vskip 3pt}
LFI 22M& $!0.012\,\pm\,0.093$\cr
\noalign{\vskip 3pt}
LFI 22S& $!0.007\,\pm\,0.062$\cr
\noalign{\vskip 3pt}
LFI 23M& $!0.016\,\pm\,0.087$\cr
\noalign{\vskip 3pt}
LFI23S& $!0.029\,\pm\,0.083$\cr
\noalign{\vskip 5pt\hrule\vskip 3pt}
{\bf 44\,GHz}&\cr
\noalign{\vskip 3pt}
LFI 24M& $-0.001\,\pm\,0.098$\cr
\noalign{\vskip 3pt}
LFI 24S& $-0.023\,\pm\,0.079$\cr
\noalign{\vskip 3pt}
LFI 25M& $-0.009\,\pm\,0.074$\cr
\noalign{\vskip 3pt}
LFI 25S& $!0.015\,\pm\,0.098$\cr
\noalign{\vskip 3pt}
LFI 26M& $!0.034\,\pm\,0.084$\cr
\noalign{\vskip 3pt}
LFI 26S& $-0.027\,\pm\,0.082$\cr
\noalign{\vskip 5pt\hrule\vskip 3pt}
{\bf 30\,GHz}&\cr
\noalign{\vskip 3pt}
LFI 27M& $!0.040\,\pm\,0.094$\cr
\noalign{\vskip 3pt}
LFI 27S& $-0.001\,\pm\,0.127$\cr
\noalign{\vskip 3pt}
LFI 28M& $-0.034\,\pm\,0.092$\cr
\noalign{\vskip 3pt}
LFI 28S& $!0.052\,\pm\,0.136$\cr
\noalign{\vskip 5pt\hrule\vskip 3pt}}}
\endPlancktable
\endgroup
\end{table}

    \subsection{Intra-frequency consistency check}
    \label{sec_consistency_scatter}
        We tested consistency between 30, 44, and 70\,GHz maps by means of power
spectra, as already done in the previous release \citep{planck2013-p02a}. 
In order to avoid the need to estimate
the noise bias, we simply took the cross-spectra
between half-ring maps at the three LFI frequencies.
As in the 2013 data release, we used the {\tt cROMAster} code which 
extends the pseudo-$C_{\ell}$ approach of  \citet{master} to 
cross-power spectrum estimation \citep{polenta_CrossSpectra}.
Although suboptimal with respect to the maximum likelihood approach, 
this method provides accurate results, and is at the same time computationally quick and light.
Consequently, this method is widely used within the CMB community 
(see e.g. \citet{bolpolvscromaster} and references therein 
for a comparison between different power spectrum estimators). 

\begin{figure}[th]
\centering
\includegraphics[width=9cm]{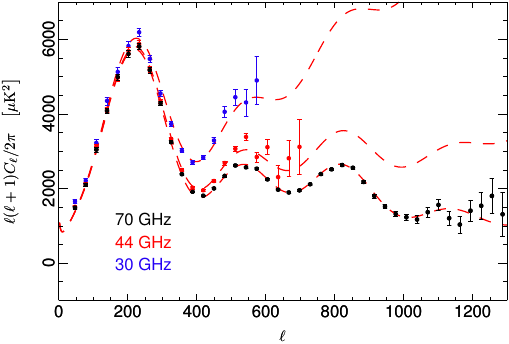}
\caption{Temperature cross-power spectra (from half-ring maps) at 30,
44, and 70\,GHz, binned in multipole space.
Foreground emission is excluded only by means
of a Galactic sky mask, without further component separation.
Best-fit \Planck\ temperature spectra plus contributions from un-resolved point
sources are shown as dashed lines for each LFI band.}
\label{fig_TT_lfi_spectrum}
\end{figure}

Those spectra are computed using a mask that is the 
combination of the G040, G060 and G070 \Planck\ masks,
respectively at 30, 44, and 70\,GHz, together with the
proper frequency-dependent point source mask. 
In Fig.~\ref{fig_TT_lfi_spectrum} cross-spectra
from 30, 44, and 70\,GHz half-ring maps are presented, showing
very good agreement among these maps (especially as 
we did not apply any component separation to 
the maps). All three data sets show strong consistency
with the \Planck\ best-fit $TT$ spectrum (black points)
to which a contribution from unmasked point sources
has been properly added.

\begin{figure*}[th]
\centering
\includegraphics[width=6cm]{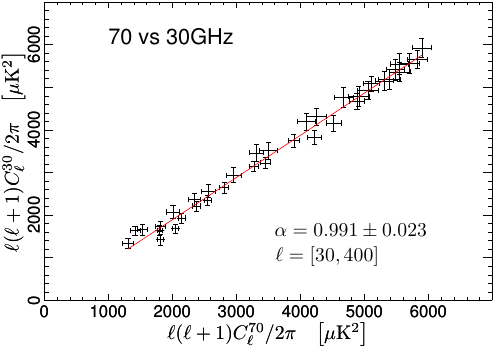}
\includegraphics[width=6cm]{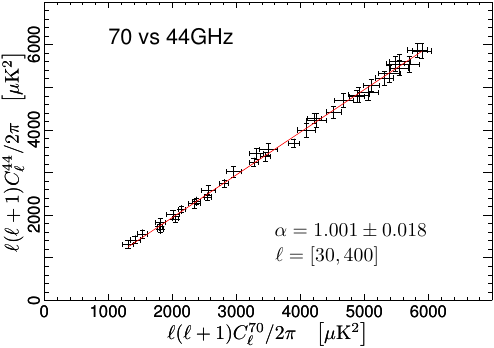}
\includegraphics[width=6cm]{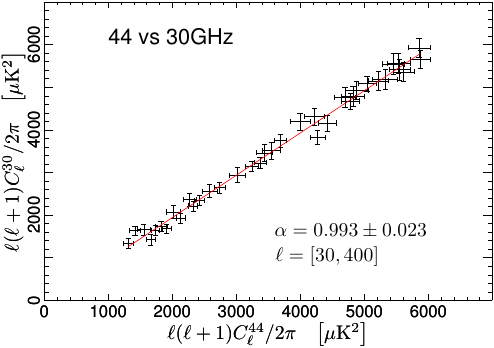}
\caption{Consistency between cross-power spectra at LFI frequencies:
 \emph{left\/} 70\,GHz versus 30\,GHz;
 \emph{middle\/} 70\,GHz versus 44\,GHz;
 and \emph{right\/} 44\,GHz versus 30\,GHz.
The solid red line is the linear regression, accounting for error on both
axes.  Slope values are found to be consistent within the uncertainties.}
\label{fig_tt_plot_lfi}
\end{figure*}

Another more quantitative way for assessing data consistency is
to build scatter, or $TT$-, plots for the three frequency pairs. In order
to do this we have to subtract the contribution of point sources below
the mask threshold at each individual frequency.  
After that we perform a linear
fit, accounting for errors in both $x$- and $y$-axes, to quantify 
the level of agreement between pairs.
Results are presented in Fig.~\ref{fig_tt_plot_lfi}, where we compare
spectra in the multipole range around the first acoustic peak.  The agreement
is extremely good and spectra are consistent with unity within the errors 
(deviations are between 0.9 and 0.1\,\%).  That in turn means a calibration
accuracy in the map at the sub-percent level.  This is very significant
considering
that we did not take into account foreground removal or uncertainties 
on the window function and calibration; therefore we may expect the agreement
to improve when these issues are taken into account.

    \subsection{Internal consistency check}
    \label{sec_int_consis}
\begin{figure}[th]
\centering
\includegraphics[width=9cm]{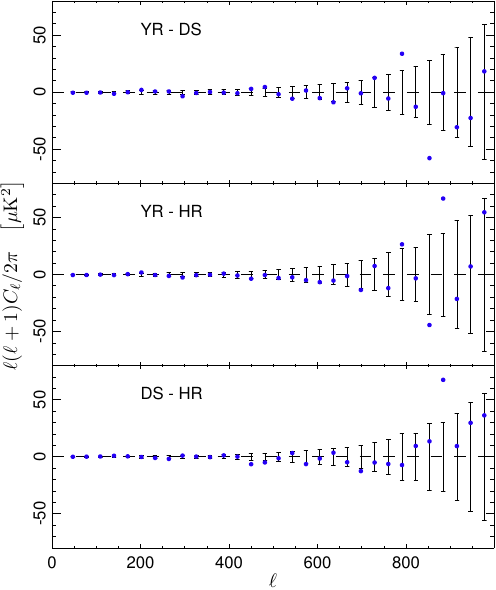}
\caption{Residuals between three different cross-power spectra computed 
from 70\,GHz data: half-ring (HR) maps, quadruplet (detector set, DS) maps,
and year 1-3 / year 2-4 (YR) maps. Error bars are derived from the realistic FFP8 simulations.}
\label{fig_TT_crossps_diff}
\end{figure}

        In order to assess the internal consistency of 70\,GHz data, we build three flavours of 
cross-power spectra that use different kind of data splits,
namely the half-ring maps, the detector set (quadruplet) maps, 
and the year 1-3 and year 2-4 maps. In Fig.~\ref{fig_TT_crossps_diff} we show residuals 
of the three estimates compared to the expected deviations
computed by running the same procedure on the realistic FFP8 Monte Carlo simulations. 
A simple $\chi^2$ analysis shows that
residuals are compatible with the null hypothesis.

\begin{figure*}[th]
\centering
\includegraphics[width=6cm]{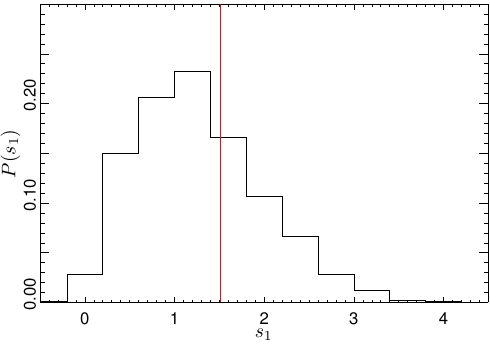}
\includegraphics[width=6cm]{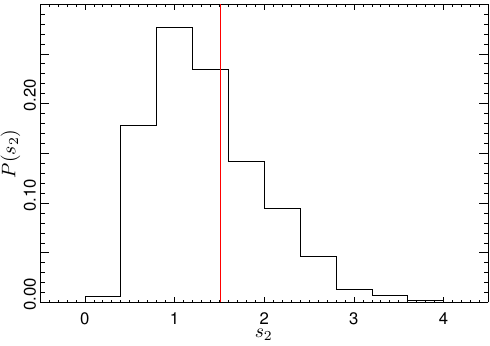}
\includegraphics[width=6cm]{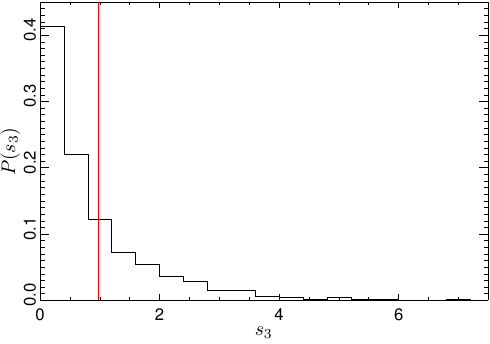}
\caption{From \emph{Left} to \emph{Right}, the empirical
distribution (estimated via FFP8 simulations) of the $s_{1},s_{2},s_{3}$
statistics of the Hausman test (see text). Vertical lines represent the values obtained from \Planck\ 70\,GHz data.}
\label{fig_TT_hausman}
\end{figure*}

We then apply the Hausman test \citep{polenta_CrossSpectra} to further verify the consistency of the 
three power-spectrum estimates. We define the statistic
\begin{equation}
H_{\ell}=\left(\hat{C_{\ell}}-\tilde{C_{\ell}}\right)/\sqrt{Var\left\{ \hat{C_{\ell}}-\tilde{C_{\ell}}\right\} },
\end{equation}
where $\hat{C_{\ell}}$ and $\tilde{C_{\ell}}$ represent two different cross-spectra, and we combine the information from different multipoles
through the quantity
\begin{equation}
B_{L}(r)=\frac{1}{\sqrt{L}}\sum_{\ell=2}^{[Lr]}H_{\ell},r\in\left[0,1\right]
\end{equation}
where $[.]$ denotes integer part. It can be shown that the distribution of $B_L(r)$ converges 
to a Brownian motion process, which can be studied using
three test statistics defined as $s_{1}=\textrm{sup}_{r}B_{L}(r)$,
$s_{2}=\textrm{sup}_{r}|B_{L}(r)|$ and
$s_{3}=\int_{0}^{1}B_{L}^{2}(r)dr$.
Results for the comparison of detector set (DS) and year based (YR) cross-spectra 
are shown in Fig.~\ref{fig_TT_hausman}. Vertical lines represent the values
of the test statistics computed from \Planck\ maps as compared to the empirical 
distribution of the test statistics derived from FFP8 simulations.
The application of the Hausman test to the other cross-spectra combinations produces 
similar results, thus supporting the strong internal consistency of the
LFI 70\,GHz data.

In this second \Planck{} data release, the calibration pipeline
considers the full convolution between the beam response $B$ and the
calibration signal $D$. This is a novel approach, which allows us to
better control the impact of optical systematic effects on the
calibration and to improves the self-consistency of the data.  Note that in the
first data release, the dipole fitting routines used to measure the
calibration constants assumed a pencil-like beam, and the mismatch in
power was fixed by applying a beam window function to the power
spectra.

As \citet{planck2013-p02b} has shown, the convolution $B * D$ retains
the same dipole shape as $D$, but there are two effects of particular
relevance for this discussion:
\begin{enumerate}
\item the finite width of the main beam and the presence of lobes
  reduces the peak-to-peak amplitude of the dipole itself (i.e. the
  peak-to-peak variation in $B*D$ is smaller than the variation in
  $D$);
\item the lack of perfect axial symmetry (particularly in the region
  which is far from the main beam) induces a tilt in the dipole axis.
\end{enumerate}
The first point implies that using the $B * D$ signal as a calibration
source reduces the average value of the calibration constant $K$
($\left[K\right] = \mathrm{K\,V^{-1}}$).
\citet{planck2013-p02b,planck2014-a06} quantify the amount of such
variation in terms of the measured power spectra $\tilde C_\ell$:
\begin{equation}
\label{eq:fourPiPencilPowerSpectrumLevel}
\frac{\tilde C_\ell^{(4\pi)}}{\tilde T^\mathrm{(pencil)}} = \left(\frac{1 -
f_\mathrm{sl} - \phi_\mathrm{sky} + \phi_D}{1 - \phi'_\mathrm{sky}}\right)^2,
\end{equation}
where $f_\mathrm{sl}$ is the fraction of $B$ that falls outside
$5^\circ$ of the main beam (the ``sidelobes''), $\phi_D
= \partial_t B_\mathrm{sl} * D / \partial_t B_\mathrm{main} * D$ is
the ratio between the variation of the dipole signal entering the
sidelobes and the variation of the same signal entering the main beam,
and $\phi_\mathrm{sky}$ and $\phi'_\mathrm{sky}$ are defined similarly
to $\phi_D$ but in terms of the amount of Galactic signal plus CMB
($\phi_\mathrm{sky}$), and of the CMB alone
($\phi'_\mathrm{sky}$).\footnote{Refer to \citet{planck2014-a06} for a
mathematical derivation and a discussion of the formula.}

We have verified the consistency of this approach by producing a set
of maps using data from the current release, but calibrated using
the pencil-beam approximation. By comparing the raw power spectra of
these maps with the official LFI power spectra of the second release,
we have measured excellent agreement (better than 0.03\,\%) with
the estimate provided by
Eq.~\eqref{eq:fourPiPencilPowerSpectrumLevel}, apart from four out of
six 44\,GHz radiometers. In the 44\,GHz case, however, because of the small
level of the sidelobes, the resulting change in the $\tilde C_\ell$ is
(at $< 0.4\,\%$) still smaller than for the other two LFI bands.

    \subsection{Updated systematic effects assessment}
    \label{sec_systematics}
        Known instrumental systematics affecting LFI maps are discussed in detail in
\citet{planck2014-a04} and are listed in
Table~\ref{tab_list_systematic_effects}, along with short descriptions of
their causes and strategies for their removal.  In
Tables~\ref{tab_summary_systematic_effects_maps_30},
\ref{tab_summary_systematic_effects_maps_44},
and \ref{tab_summary_systematic_effects_maps_70} we list both the rms and the
difference between the $99\,\%$ and the $1\,\%$ quantiles in the pixel value
distribution for the $I$, $Q$, and $U$ maps, at 30, 44 and 70\,\GHz\, respectively. 
We refer to the latter as the peak-to-peak (p-p) difference, even though it neglects outliers,
since it effectively approximates the peak-to-peak variation of the effect on
the map. 

Detailed analysis reported in \citet{planck2014-a04} shows that systematic
uncertainties are at least two orders of magnitude below the CMB $TT$ power
spectrum and are not significantly contaminating the $EE$ and $BB$ spectra.

  \begin{table*}[tmb]
  \begingroup
  \newdimen\tblskip \tblskip=5pt
  \caption{List of known instrumental systematic effects in \Planck-LFI.}
  \label{tab_list_systematic_effects}
  \nointerlineskip
  \vskip -3mm
  \footnotesize
  \setbox\tablebox=\vbox{
    \newdimen\digitwidth
    \setbox0=\hbox{\rm 0}
    \digitwidth=\wd0
    \catcode`*=\active
    \def*{\kern\digitwidth}
    \newdimen\signwidth
    \setbox0=\hbox{+}
    \signwidth=\wd0
    \catcode`!=\active
    \def!{\kern\signwidth}
  {
  \halign{
  #\hfil\tabskip=1em&
  #\hfil&
  #\hfil\tabskip=0pt\cr
  \noalign{\doubleline}
  \noalign{\vskip -3pt}
  \omit\hfil Effect\hfil&
  \omit\hfil Source\hfil&
  \hfil Control/removal\cr
  \noalign{\vskip -3pt}
  \noalign{\vskip 5pt\hrule\vskip 3pt}
  \noalign{\vskip 6pt}
  \multispan3\hskip 4cm{\bf Effects independent of sky signal ($T$ and $P$)}\hfil\cr
  \noalign{\vskip 6pt}
  White noise correlation&
  Phase switch imbalance&
  Diode {weighting}\cr
  \noalign{\vskip 6pt}
  1/$f$ noise&
  RF amplifiers&
  Pseudo-correlation and destriping\cr
  \noalign{\vskip 6pt}
  Bias fluctuations&
  RF amplifiers, back-end electronics&
  Pseudo-correlation and destriping\cr
  \noalign{\vskip 6pt}
  Thermal fluctuations&
  4-K, 20-K and 300-K thermal stages&
  Calibration, destriping\cr
  \noalign{\vskip 6pt}
  1-Hz spikes&
  Back-end electronics&
  Template fitting and removal\cr
  \noalign{\vskip 6pt}
  \multispan3\hskip 3.8cm{\bf Effects dependent on the sky signal ($T$ and $P$)}\hfil\cr
  \noalign{\vskip 6pt}
  {Main beam ellipticity}& {Main beams}&
  {Accounted for in window function}\cr
  \noalign{\vskip 6pt}
  {Near sidelobe}& {Optical response at angles}&
  {Masking of Galaxy and point} \cr
  \omit{pickup}\hfil&\omit $<5^\circ$ {from the main beam}\hfil&\omit{sources}\hfil\cr
  \noalign{\vskip 6pt}
  {Far} sidelobe pickup& Main and sub-reflector spillover&
  Model sidelobes removed from timelines\cr
  \omit&\omit&\omit\hfil\cr
  \noalign{\vskip 6pt}
  Analogue-to-digital&
  Back-end analogue-to-digital&
  Template fitting and removal\cr
  \omit converter non-linearity\hfil&\omit  converter\hfill&\omit\cr
  \noalign{\vskip 6pt}
  Imperfect photometric&
  Sidelobe pickup, radiometer noise&Adaptive smoothing algorithm using $4\pi$\cr
  \omit calibration\hfil&\omit temperature changes and other\hfil&\omit beam, 4-K reference load voltage output\hfil\cr
  \omit&\omit non-idealities\hfil&\omit temperature sensor data\hfil\cr
  \noalign{\vskip 6pt}
  Pointing&
  Uncertainties in pointing reconstru-&
  Negligible impact on anisotropy\cr
  \omit&\omit ction, thermal changes affecting\hfil &\omit  measurements\hfil\cr
  \omit&\omit focal plane geometry\hfil&\omit\cr
  \noalign{\vskip 6pt}
  \multispan3\hskip 3.8cm{\bf Effects specifically impacting polarization}\hfil\cr
  \noalign{\vskip 6pt}
  Bandpass asymmetries&
  Differential orthomode transducer&
  Spurious polarization removal\cr
  \omit&\omit and receiver bandpass response\hfil&\omit\cr
  \noalign{\vskip 6pt}
  Polarization angle&
  Uncertainty in the  polarization&
  Negligible impact\cr
  \omit uncertainty&\omit angle in-flight measurement\hfil&\omit\cr
  \noalign{\vskip 6pt}
  Orthomode transducer&
  Imperfect polarization separation&
  Negligible impact\cr
  \omit cross-polarization&\omit \hfil&\omit\cr
  \noalign{\vskip 5pt\hrule\vskip 3pt}
  }
  }}
  \endPlancktablewide
  \endgroup
  \end{table*}

%%%%%%%%%%%%%%%%%%%%%%%%%%%%%%%%%%%%%%%%%%%%%%%%%%%%%%%%%%%%%%%%%%%%%%%%%%%%%%%%%%%%%%%%%%%%%%%%

 \begin{table}[tmb]               
  \begingroup
  \newdimen\tblskip \tblskip=5pt
  \caption{Summary of systematic effect uncertainties on 30\,GHz maps$^{\rm a}$
  in \muKCMB.  Columns give the peak-to-peak (``p-p'') and rms levels
  for Stokes $I$, $Q$, and $U$ maps.}
  \label{tab_summary_systematic_effects_maps_30}                     
  \nointerlineskip
%  \vskip -3mm
  \footnotesize
  \setbox\tablebox=\vbox{
    \newdimen\digitwidth 
    \setbox0=\hbox{\rm 0} 
    \digitwidth=\wd0 
    \catcode`*=\active 
    \def*{\kern\digitwidth}
    \newdimen\signwidth 
    \setbox0=\hbox{+} 
    \signwidth=\wd0 
    \catcode`!=\active 
    \def!{\kern\signwidth}
  {\tabskip=0pt
  \halign{ 
  \hbox to 1.3in{#\leaderfil}\tabskip=0em& 
  \hfil#\hfil\tabskip=1.0em& 
  \hfil#\hfil& 
  \hfil#\hfil& 
  \hfil#\hfil& 
  \hfil#\hfil& 
  \hfil#\hfil\tabskip=0pt\cr                       
  \noalign{\doubleline}
  \omit&
  \multispan2\hfil $I$\hfil& 
  \multispan2\hfil $Q$\hfil&
  \multispan2\hfil $U$\hfil\cr   
  \noalign{\vskip -3pt}
  \omit&\multispan6\hrulefill\cr
  \noalign{\vskip 2pt}
  \omit& p-p& rms& p-p& rms& p-p& rms\cr
    \noalign{\vskip 4pt}
    Near sidelobes&          0.72& 0.13& 0.05& 0.01& 0.05& 0.01\cr            
    \noalign{\vskip 4pt}                                                      
    Pointing&                0.37& 0.07& 0.02& 0.01& 0.02& 0.00\cr            
    \noalign{\vskip 4pt}                                                      
    Polarization angle&      0.02& 0.00& 0.53& 0.11& 0.64& 0.15\cr            
    \noalign{\vskip 4pt}                                                      
    1-Hz spikes&             0.54& 0.11& 0.11& 0.02& 0.09& 0.02\cr            
    \noalign{\vskip 4pt}                                                      
    Bias fluctuations&       0.07& 0.01& 0.07& 0.01& 0.06& 0.01\cr            
    \noalign{\vskip 4pt}                                                      
    ADC non-linearity&        0.42& 0.09& 0.54& 0.11& 0.56& 0.11\cr            
    \noalign{\vskip 4pt}                                                      
    Calibration&             2.43& 0.55& 2.53& 0.46& 2.34& 0.43\cr            
    \noalign{\vskip 4pt}                                                      
    Thermal fluct. (300\,K)& 0.00& 0.00& 0.00& 0.00& 0.00& 0.00\cr            
    \noalign{\vskip 4pt}                                                      
    Thermal fluct. (20\,K)&  0.12& 0.03& 0.06& 0.02& 0.06& 0.02\cr            
    \noalign{\vskip 4pt}                                                      
    Thermal fluct. (4\,K)&   0.29& 0.06& 0.06& 0.01& 0.05& 0.01\cr            
%     \noalign{\vskip 2pt}                                                    
  \omit&\multispan6\hrulefill\cr                                              
  \noalign{\vskip 2pt}
    Total$^{\rm b}$&	            2.72& 0.61& 2.79& 0.52& 2.42& 0.49\cr
  \noalign{\vskip 5pt\hrule\vskip 3pt}
  }
  }}
  \endPlancktable          
  \tablenote {{\rm a}} Calculated for a pixel size approximately equal to the
  average beam FWHM. A null value indicates a residual $<10^{-2}$\,\muKCMB.\par
  \tablenote {{\rm b}} The total has been computed on maps resulting from the
  sum of individual systematic effect maps.\par
  \endgroup
\end{table}   

%%%%%%%%%%%%%%%%%%%%%%%%%%%%%%%%%%%%%%%%%%%%%%%%%%%%%%%%%%%%%%%%%%%%%%%%%%%%%%%%%%%%%%%%%%%%%%%%

\begin{table}[tmb]               
  \begingroup
  \newdimen\tblskip \tblskip=5pt
  \caption{Summary of systematic effect uncertainties on 44\,GHz maps in
  \muKCMB.  Columns give the peak-to-peak (``p-p'') and rms levels
  for Stokes $I$, $Q$, and $U$ maps.}            
  \label{tab_summary_systematic_effects_maps_44}                     
  \nointerlineskip
%  \vskip -3mm
  \footnotesize
  \setbox\tablebox=\vbox{
    \newdimen\digitwidth 
    \setbox0=\hbox{\rm 0} 
    \digitwidth=\wd0 
    \catcode`*=\active 
    \def*{\kern\digitwidth}
    \newdimen\signwidth 
    \setbox0=\hbox{+} 
    \signwidth=\wd0 
    \catcode`!=\active 
    \def!{\kern\signwidth}
  {\tabskip=0pt
  \halign{ 
  \hbox to 1.3in{#\leaderfil}\tabskip=0em& 
  \hfil#\hfil\tabskip=1.0em& 
  \hfil#\hfil& 
  \hfil#\hfil& 
  \hfil#\hfil& 
  \hfil#\hfil& 
  \hfil#\hfil\tabskip=0pt\cr                       
  \noalign{\doubleline}
  \omit&
  \multispan2\hfil $I$\hfil& 
  \multispan2\hfil $Q$\hfil&
  \multispan2\hfil $U$\hfil\cr   
  \noalign{\vskip -3pt}
  \omit&\multispan6\hrulefill\cr
  \noalign{\vskip 2pt}
  \omit& p-p& rms& p-p& rms& p-p& rms\cr
    \noalign{\vskip 4pt}
    Near sidelobes&          0.09& 0.02& 0.00& 0.00& 0.00& 0.00\cr            
    \noalign{\vskip 4pt}                                                      
    Pointing&                0.30& 0.06& 0.01& 0.00& 0.01& 0.00\cr            
    \noalign{\vskip 4pt}                                                      
    Polarization angle&      0.04& 0.01& 0.35& 0.07& 0.38& 0.10\cr            
    \noalign{\vskip 4pt}                                                      
    1-Hz spikes&             1.99& 0.40& 0.88& 0.18& 1.04& 0.21\cr            
    \noalign{\vskip 4pt}                                                      
    Bias fluctuations&       0.04& 0.01& 0.05& 0.01& 0.05& 0.01\cr            
    \noalign{\vskip 4pt}                                                      
    ADC non-linearity&        0.30& 0.06& 0.36& 0.07& 0.34& 0.07\cr            
    \noalign{\vskip 4pt}                                                      
    Calibration&             1.05& 0.18& 1.57& 0.29& 1.31& 0.26\cr            
    \noalign{\vskip 4pt}                                                      
    Thermal fluct. (300\,K)& 0.00& 0.00& 0.00& 0.00& 0.00& 0.00\cr            
    \noalign{\vskip 4pt}                                                      
    Thermal fluct. (20\,K)&  0.04& 0.02& 0.06& 0.01& 0.05& 0.01\cr            
    \noalign{\vskip 4pt}                                                      
    Thermal fluct. (4\,K)&   0.23& 0.05& 0.05& 0.01& 0.06& 0.01\cr            
%     \noalign{\vskip 2pt}                                                    
  \omit&\multispan6\hrulefill\cr                                              
  \noalign{\vskip 2pt}
    Total&                    2.29& 0.45& 1.95& 0.37& 1.76& 0.37\cr
    \noalign{\vskip 4pt}
  \noalign{\vskip 5pt\hrule\vskip 3pt}
  }
  }}
  \endPlancktable          

  \endgroup
\end{table}   

%%%%%%%%%%%%%%%%%%%%%%%%%%%%%%%%%%%%%%%%%%%%%%%%%%%%%%%%%%%%%%%%%%%%%%%%%%%%%%%%%%%%%%%%%%%%%%%%

\begin{table}[tmb]               
  \begingroup
  \newdimen\tblskip \tblskip=5pt
  \caption{Summary of systematic effect uncertainties on 70\,GHz maps in
  \muKCMB.  Columns give the peak-to-peak (``p-p'') and rms levels
  for Stokes $I$, $Q$, and $U$ maps.}
  \label{tab_summary_systematic_effects_maps_70}                     
  \nointerlineskip
%  \vskip -3mm
  \footnotesize
  \setbox\tablebox=\vbox{
    \newdimen\digitwidth 
    \setbox0=\hbox{\rm 0} 
    \digitwidth=\wd0 
    \catcode`*=\active 
    \def*{\kern\digitwidth}
    \newdimen\signwidth 
    \setbox0=\hbox{+} 
    \signwidth=\wd0 
    \catcode`!=\active 
    \def!{\kern\signwidth}
  {\tabskip=0pt
  \halign{ 
  \hbox to 1.3in{#\leaderfil}\tabskip=0em& 
  \hfil#\hfil\tabskip=1.0em& 
  \hfil#\hfil& 
  \hfil#\hfil& 
  \hfil#\hfil& 
  \hfil#\hfil& 
  \hfil#\hfil\tabskip=0pt\cr                       
  \noalign{\doubleline}
  \omit&
  \multispan2\hfil $I$\hfil& 
  \multispan2\hfil $Q$\hfil&
  \multispan2\hfil $U$\hfil\cr   
  \noalign{\vskip -3pt}
  \omit&\multispan6\hrulefill\cr
  \noalign{\vskip 2pt}
  \omit& p-p& rms& p-p& rms& p-p& rms\cr
    \noalign{\vskip 4pt}
    Near sidelobes&          0.30& 0.07& 0.01& 0.00& 0.01& 0.00\cr          
    \noalign{\vskip 4pt}                                                    
    Pointing&                0.60& 0.11& 0.03& 0.01& 0.03& 0.01\cr          
    \noalign{\vskip 4pt}                                                    
    Polarization angle&      0.02& 0.00& 0.08& 0.02& 0.08& 0.02\cr          
    \noalign{\vskip 4pt}                                                    
    1-Hz spikes&             0.39& 0.08& 0.17& 0.03& 0.15& 0.03\cr          
    \noalign{\vskip 4pt}                                                    
    Bias fluctuations&       0.68& 0.14& 0.84& 0.17& 0.95& 0.18\cr          
    \noalign{\vskip 4pt}                                                    
    ADC non-linearity&        1.56& 0.33& 1.92& 0.39& 2.05& 0.41\cr          
    \noalign{\vskip 4pt}                                                    
    Calibration&             1.06& 0.23& 0.98& 0.18& 0.77& 0.16\cr          
    \noalign{\vskip 4pt}                                                    
    Thermal fluct. (300\,K)& 0.00& 0.00& 0.00& 0.00& 0.00& 0.00\cr          
    \noalign{\vskip 4pt}                                                    
    Thermal fluct. (20\,K)&  0.44& 0.08& 0.07& 0.01& 0.08& 0.02\cr          
    \noalign{\vskip 4pt}                                                    
    Thermal fluct. (4\,K)&   0.38& 0.08& 0.04& 0.01& 0.05& 0.01\cr          
%     \noalign{\vskip 2pt}                                                  
  \omit&\multispan6\hrulefill\cr                                            
  \noalign{\vskip 2pt}
    Total&                   2.24& 0.47& 2.27& 0.46& 2.38& 0.48\cr
    \noalign{\vskip 4pt}
  \noalign{\vskip 5pt\hrule\vskip 3pt}
  }
  }}
  \endPlancktable          

  \endgroup
\end{table}

%%%%%%%%%%%%%%%%%%%%%%%%%%%%%%%%%%%%%%%%%%%%%%%%%%%%%%%%%%%%%%%%%%%%%%%%%%%%%%%%%%%%%%%%%%%%%%%%

\section{Low-$\ell$ data selection}
\label{sec_low_l}
The 70\,GHz polarization data are of special importance
since the \Planck\ low-$\ell$ likelihood \citep{planck2014-a13} used to
determine cosmological parameters is based on them.  In order to provide the
best data possible for the construction of the likelihood, we perform
several tests at survey level in order to choose the most reliable data
combination. For this purpose we focus on the very low multipoles,
especially $\ell = 2$--4, which are the most susceptible to systematic errors. 

We compare results from actual data and from noise-only Monte Carlo
realizations made for the FFP8 simulations. 
Specifically, we take differences between the full data set (over the entire
mission lifetime) and some
specific combinations of surveys, for both noise simulations and real data.
We then compute the angular power spectra of these differences to look for
anomalies.

\begin{figure*}
\centering
\includegraphics[width=6cm]{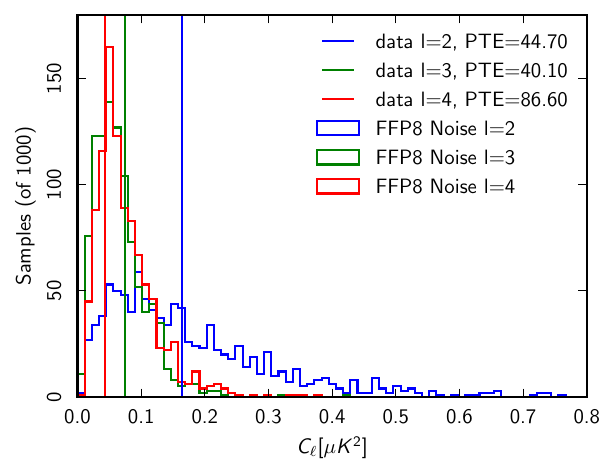}
\includegraphics[width=6cm]{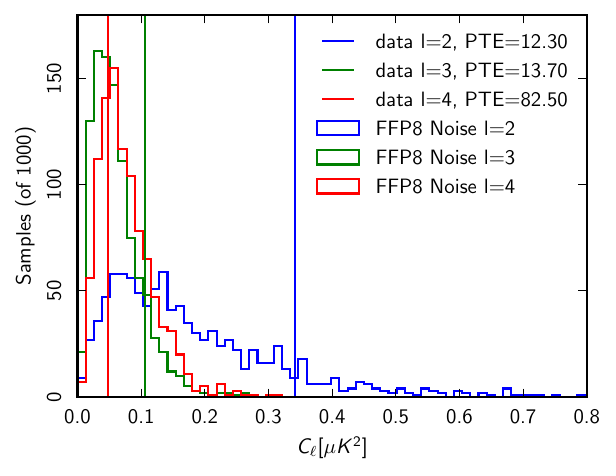}
\includegraphics[width=6cm]{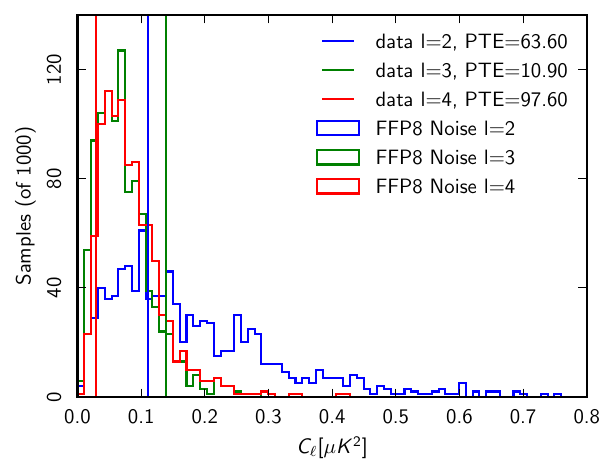}
\includegraphics[width=6cm]{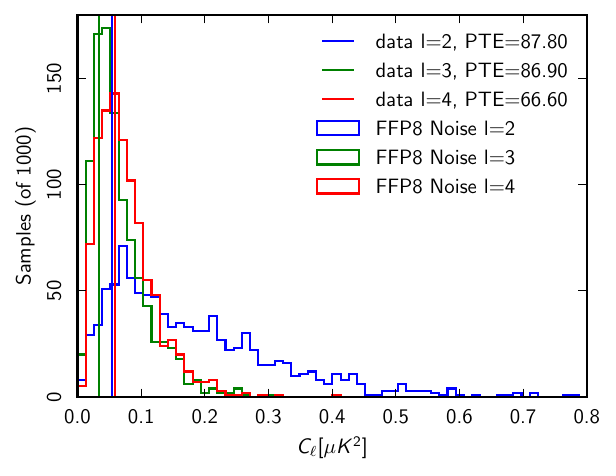}
\includegraphics[width=6cm]{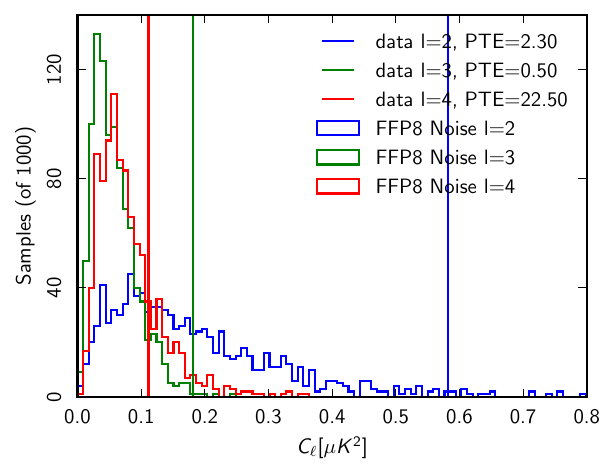}
\includegraphics[width=6cm]{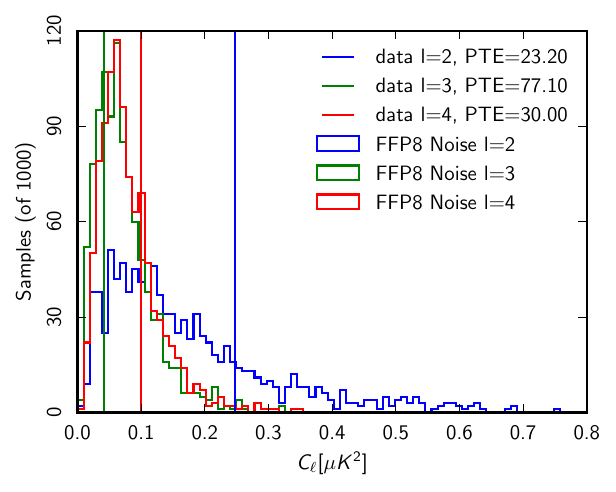}
\caption{Measured LFI 70\,GHz $EE$ (\emph{top}) and $BB$ (\emph{bottom})
null power spectra for $\ell=2$, 3, and 4 (vertical lines), compared to the
distribution derived from noise-only Monte Carlo simulations. Null spectra
from the difference between full data and
specific surveys combinations: \emph{left} Survey~1 and Survey~3;
(\emph{middle}) Survey~2 and Survey~4; and (\emph{right}) Survey~5 and Survey~7.
It is clear that Survey~2/Survey~4 stands out with respect to the others. }
\label{null_lowell}
\end{figure*}

The analysis at the level of surveys is very informative: as a consequence
of the scanning strategy and payload geometry, Survey~1 and Survey~3
share the same beam orientation with respect to the sky. The same is true
for Surveys 2/4, 5/7, and 6/8.  For this reason 
we consider these combinations jointly for the null tests,
thus maximizing signal-to-noise. 
Figure~\ref{null_lowell} shows the distribution of angular power for $E$- and
$B$-modes for each survey pair, as derived from the Monte Carlo simulations,
with results from the actual LFI data indicated by vertical lines. 
Evidently, Survey~2 and Survey~4 are quite anomalous with respect to the rest
of the surveys.  We will offer some possible explanations below.
First, however, 
we can be more quantitative and compute the probability to exceed (PTE) of our
data, based on simulations.  Results are reported in Table~\ref{pt_null}.
\begin{table*}[tmb]
\begingroup
\newdimen\tblskip \tblskip=5pt
\caption{PTE for $EE$ and $BB$ low multipoles, for the differences between full mission and individual surveys.}
\label{pt_null}
\nointerlineskip
\vskip -3mm
\footnotesize
\setbox\tablebox=\vbox{
	\newdimen\digitwidth
	\setbox0=\hbox{\rm 0}
	\digitwidth=\wd0
	\catcode`*=\active
	\def*{\kern\digitwidth}
	\newdimen\signwidth
	\setbox0=\hbox{+}
	\signwidth=\wd0
	\catcode`!=\active
	\def!{\kern\signwidth}
	\halign{\hbox to 1.5in{#\leaderfil}\tabskip=2em&
	   \hfil#\hfil&
	   \hfil#\hfil&
	   \hfil#\hfil&
	   \hfil#\hfil&
	   \hfil#\hfil&
	   \hfil#\hfil&
	   \hfil#\hfil&
	   \hfil#\hfil\tabskip=0pt\cr
	   \noalign{\doubleline}
	   \omit\hfil Multipole \hfil& $Full-S_{1}$& $Full-S_{3}$& $Full-S_{2}$& $Full-S_{4}$& $Full-S_{5}$& $Full-S_{7}$& $Full-S_{6}$& $Full-S_{8}$\cr
%	   \omit\hfil Multipole \hfil& SS1& SS3& SS2& SS4& SS5& SS7& SS6& SS8\cr
	   \noalign{\vskip 3pt\hrule\vskip 5pt}
	   \omit{$\vec{EE}$}\hfil\cr
	   \noalign{\vskip 4pt}
	   \hglue 2em $\ell = 2$& 0.885& 0.307& 0.328& 0.015& 0.241& 0.975& 0.837& 0.090\cr
	   \hglue 2em $\ell = 3$& 0.740& 0.730& 0.137& 0.223& 0.206& 0.566& 0.377& 0.064\cr
	   \hglue 2em $\ell = 4$& 0.807& 0.828& 0.890& 0.535& 0.290& 0.998& 0.932& 0.476\cr
	   \noalign{\vskip 5pt}
	   \omit{$\vec{BB}$}\hfil\cr
	   \noalign{\vskip 4pt}
	   \hglue 2em $\ell = 2$& 0.998& 0.214& 0.030& 0.482& 0.098& 0.680& 0.986& 0.092\cr
	   \hglue 2em $\ell = 3$& 0.833& 0.796& 0.843& 0.002& 0.414& 0.823& 0.516& 0.255\cr
	   \hglue 2em $\ell = 4$& 0.903& 0.105& 0.092& 0.399& 0.052& 0.524& 0.862& 0.950\cr
	   \noalign{\vskip 5pt\hrule\vskip 3pt}}}
	   \endPlancktable
	   \endgroup
\end{table*}	
These probability values seem to indicate that Surveys~2 and 4 show
systematic effects. Guided by these findings, we report the PTE values for
the differences between the full mission and the survey combinations in
Table~\ref{pt_null_comb}.

\begin{table*}[tmb]
\begingroup
\newdimen\tblskip \tblskip=5pt
\caption{PTE for $EE$ and $BB$ low multipoles, for the differences between full mission and survey combinations.}
\label{pt_null_comb}
\nointerlineskip
\vskip -3mm
\footnotesize
\setbox\tablebox=\vbox{
	\newdimen\digitwidth
	\setbox0=\hbox{\rm 0}
	\digitwidth=\wd0
	\catcode`*=\active
	\def*{\kern\digitwidth}
	\newdimen\signwidth
	\setbox0=\hbox{+}
	\signwidth=\wd0
	\catcode`!=\active
	\def!{\kern\signwidth}
	\halign{\hbox to 1.5in{#\leaderfil}\tabskip=2em&
	   \hfil#\hfil&
	   \hfil#\hfil&
	   \hfil#\hfil&
	   \hfil#\hfil\tabskip=0pt\cr
	   \noalign{\doubleline}
	   \omit\hfil Multipole \hfil& $Full-(S_{1}+ S_{3})$& $Full-(S_{2}+ S_{4})$& $Full-(S_{5}+ S_{7})$& $Full-(S_{6}+ S_{8})$\cr
%	   \omit\hfil Multipole \hfil& FULL - SS1 + SS3& FULL - SS2 + SS4& FULL - SS5 + SS7& FULL - SS6 + SS8\cr
	   	   \noalign{\vskip 3pt\hrule\vskip 5pt}
	   \omit{$\vec{EE}$}\hfil\cr
	   \noalign{\vskip 4pt}
	   \hglue 2em $\ell = 2$& 0.491& 0.114& 0.526& 0.578\cr
	   \hglue 2em $\ell = 3$& 0.528& 0.137& 0.109& 0.598\cr
	   \hglue 2em $\ell = 4$& 0.750& 0.825& 0.976& 0.978\cr
	   \noalign{\vskip 5pt}
	   \omit{$\vec{BB}$}\hfil\cr
	   \noalign{\vskip 4pt}
	   \hglue 2em $\ell = 2$& 0.482& 0.023& 0.156& 0.544\cr
	   \hglue 2em $\ell = 3$& 0.698& 0.010& 0.866& 0.320\cr
	   \hglue 2em $\ell = 4$& 0.218& 0.190& 0.152& 0.995\cr
	   \noalign{\vskip 5pt\hrule\vskip 3pt}}}
	   \endPlancktable
	   \endgroup
\end{table*}

We can also combine the PTE results from the survey null tests across these
multipoles. In Table~\ref{pte_unif_test} we report results from a test of
uniformity of the PTEs, simply counting
how many entries are lower than a given threshold.  The $p-$values for these
tests are computed assuming binomial statistics. We report results for
different values of the threshold to show their robustness and stability
with respect to the thresholds.
\begin{table}[tmb]
\begingroup
\newdimen\tblskip \tblskip=5pt
\caption{Uniformity of the PTEs for survey null tests based on the number of
entries lower than a given threshold ($p-$values are from the binomial
distribution).}
\label{pte_unif_test}
\nointerlineskip
\vskip -3mm
\footnotesize
\setbox\tablebox=\vbox{
	\newdimen\digitwidth
	\setbox0=\hbox{\rm 0}
	\digitwidth=\wd0
	\catcode`*=\active
	\def*{\kern\digitwidth}
	\newdimen\signwidth
	\setbox0=\hbox{+}
	\signwidth=\wd0
	\catcode`!=\active
	\def!{\kern\signwidth}
	\halign{\hbox to 1.5in{#\leaderfil}\tabskip=2em&
	   \hfil#\hfil&
	   \hfil#\hfil&
	   \hfil#\hfil&
	   \hfil#\hfil\tabskip=0pt\cr
	   \noalign{\doubleline}
	   \omit\hfil Threshold \hfil& SS1/SS3& SS2/SS4& SS5/SS7& SS6/SS8\cr
	   \noalign{\vskip 3pt\hrule\vskip 5pt}
	   \omit{$N<0.02$}& 0 (0.215)& 2 (0.002)& 0 (0.215)& 0 (0.215)\cr
           \omit{$N<0.05$}& 0 (0.456)& 3 (0.002)& 0 (0.456)& 0 (0.456)\cr
           \omit{$N<0.10$}& 0 (0.716)& 4 (0.004)& 2 (0.111)& 3 (0.026)\cr
           \omit{$N<0.25$}& 2 (0.609)& 6 (0.014)& 4 (0.158)& 3 (0.351)\cr
	   \noalign{\vskip 5pt\hrule\vskip 3pt}}}
	   \endPlancktable
	   \endgroup
\end{table}	

Quantitatively Surveys~2 and 4 again stand out as anomalous at roughly the
$3\,\sigma$ level.  Currently the reason for this is not fully understood,
but we note that this particular survey pair has a scanning strategy
that produces larger uncertainties in gain, as demonstrated in
Fig.~\ref{fig:raw_gains}.  The geometry for these two surveys also increases
the sensitivity of the very low-$\ell$  results to small errors in estimates
of Galactic contamination of the far sidelobes.  
These issues are under investigation and will be addressed further in the
next data release, but for the moment we
choose to be conservative and remove Surveys~2 and 4 from the default
likelihood developed in \citet{planck2014-a13}.  The default likelihood is
used in \citet{planck2014-a15} to derive cosmological parameters.
The optical depth to reionization, $\tau$, is the parameter most affected by
this choice: removing Surveys~2 and 4 changes the value of this parameter by
about $0.5\,\sigma$.

\section{The low-$\ell$ likelihood}
\label{sec_low_likelihood}
The baseline 2015 \Planck\ low-$\ell$ likelihood is described in depth 
in \citet{planck2014-a13}. Here we briefly discuss its polarization content,  
largely based on data from the \Planck\ 70\,GHz channel. As noted in the previous 
sections, Survey 2 and 4 are excerpted from the data set to reduce the chance of 
systematic contamination. In this section, we do not focus on the low-$\ell$ temperature block 
of the likelihood developed in \citet{planck2014-a13} that is based on a CMB map derived using the {\tt Commander} 
algorithm which employs all \Planck\ channels from 30 to 353\,GHz \citep{planck2014-a11}. 

At multipoles $\ell < 30$, we model the likelihood assuming that the maps are Gaussian 
distributed with known covariance \citep{planck2013-p08}:

\begin{equation}
\mathcal{L}(C_{\ell}) = P(m|C_{\ell})=\frac{1}{2\pi^{n/2}|\tens{M}|^{1/2}}
\exp\left(-\frac{1}{2}m^{\rm T}\,\tens{M}^{-1}m\right),
\label{pbLike}
\end{equation}
where $n$ is the total number of observed pixels, 
$\tens{M}(C_{\ell})$ is the covariance matrix of ${m} = [T, Q, U]$, 
being $T$, $Q$, and $U$ the pixel space intensity and linear polarization 
Stokes parameter maps. Note that the covariance matrix depends on the CMB 
model angular power spectra, $C_\ell$, only through the CMB signal covariance 
matrix:
\begin{equation}
\tens{M}(C_{\ell}) = \tens{S}(C_{\ell})+\tens{N}\,. 
\label{pbLike_1}
\end{equation}

In order to clean the 70\,GHz $Q$ and $U$ maps, we perform a template fitting 
procedure using the \Planck\ 30\,GHz channel as a tracer of polarized synchrotron 
emission and the \Planck\ 353\,GHz channel as a tracer of polarized dust emission. 
Restricting from now onwards $\mathbf{m}$ to the $Q$ and $U$ maps (i.e.\  $m \equiv [Q, U]$) we write:
\begin{equation}
{m} = \frac{1}{1-\alpha-\beta}\left ( {m}_{70} - \alpha {m}_{30} - \beta {m}_{353} \right)\,,
\end{equation}
where ${m}_{70}$, ${m}_{30}$ and ${m}_{353}$ are bandpass corrected versions of 
the 70, 30, 353 maps \citep{planck2014-a04,planck2014-a08}, whereas $\alpha$ 
and $\beta$ are the scaling coefficients for  synchrotron and dust emission, respectively. 
The latter are best fitted by minimizing the quantity 
\begin{equation}
\chi^2 = (1-\alpha-\beta)^2\,{m}^{\rm T} \left[\tens{S}(C_\ell)+ \tens{N}_{70}\right]^{-1} {m}\,,
\end{equation} 
where $ \tens{N}_{70}$ is the pure polarization part of the 70\,GHz noise covariance 
matrix\footnote{We assume that the noise induced $TQ$ and $TU$ correlations are negligible.}  
\citep{planck2014-a07}, and $C_\ell$ is taken as the \Planck\ 2013 fiducial model \citep{planck2013-p11}. 
We have verified that changing this model does not impact the results significantly. We find $\alpha = 0.063 $, 
$\beta = 0.0077$, with three sigma uncertainties $\sigma_\alpha = 0.025$ and $\sigma_\beta = 0.0022$. 
The best fit values quoted correspond to a polarization mask that allows 47\% of the sky to pass through. In fact, we 
have repeated this procedure for a set of 24 masks, allowing sky fractions from 80\% to 29\%. Such masks 
have been constructed by rescaling the templates ${m}_{30}$ and   ${m}_{353}$ to 70 GHz assuming fiducial 
spectral indexes, computing the polarized intensity $P=\sqrt{Q^2+U^2}$ and thresholding the latter. 
For each mask, we evaluate the probability to exceed $\mathcal{P}(\chi^2 > \chi^2_0)$. The 47\% analysis 
mask is chosen as the tightest mask satisfying the requirement $\mathcal{P}>5\%$.  

We define the final polarization noise covariance matrix used in Eq.~(\ref{pbLike_1}) as:
\begin{equation}
\tens{N} = \frac{1}{(1-\alpha-\beta)^2}\left ( \tens{N}_{70} + \sigma^2_\alpha {m}_{30} {m}^{\rm T}_{30} +  
\sigma^2_\beta {m}_{353} {m}^{\rm T}_{353}\right)\,. 
\label{lowl_covmat}
\end{equation}
We have verified that the external (column to row) products involving the foreground 
templates are subdominant corrections. We do not include further correction terms associated 
with the band pass leakage error budget since they are completely negligible.

\section{Product description}
\label{sec_Product description}
We now give a list and brief description of \Planck\ LFI released
products, which can be freely accessed via the \Planck\ Legacy Archive
interface\footnote{\url{http://archive.esac.esa.int/pla2}},
based on all the data acquired during routine operation from 12 August 2009
to 23 October 2014; the full format is reported in the Explanatory
Supplement\footnote{\url{ http://pla.esac.esa.int/pla/index.html}}.

\begin{itemize}

\item Pointing timelines:
one FITS file for each OD for each frequency, each FITS file contains the
OBT (onboard time) and the three angles, $\theta$, $\phi$, and $\psi$,
which identify each sample on the sky.

\item Time timelines:
one FITS file for each OD for each frequency, each FITS files containing
the OBT and its corresponding TAI (International Atomic Time)
value (with no leap second) in modified Julian
day format. This will allow the user to cross-correlate OBT with UTC.

\item Housekeeping timelines:
all the housekeeping parameters with their raw and calibrated values
are provided, separated by the housekeeping sources and for each OD.

\item Timelines in volts:
raw scientific data in engineering units for each detector at 30, 44, and
70\,GHz and each OD, before its calibration from which instrumental systematic
effects have been removed;

\item Cleaned and calibrated timelines: provided
in ${\rm K}_{\rm CMB}$ for each detector
at 30, 44, and 70\,GHz and each OD, after scientific calibration from which
the convolved dipole and convolved Galactic straylight have been removed. 

\item Scanning beam: $4\pi$ beam representation used in the calibration
pipeline.

\item Effective beam: sky beam representation as a projection of the scanning
beam on the maps.  

\item Full sky maps at each frequency: maps of the sky at 30, 44, and
70\,GHz in temperature and polarization at $N_{\rm side} = 1024$ and, in the
case of 70\,GHz at $N_{\rm side} = 2048$. Maps are provided for different
data periods, as detailed in Table~\ref{tab:maps_released}. Note that the
polarization convention used for the \Planck\ maps is referred to as
``COSMO'' instead of the ``IAU''; see the Explanatory Supplement for details.

\item Baseline timelines for the full and half-ring periods: these timelines
have the baseline offset removed (the length is specified in
Table~\ref{tab:frequency_param}) during the mapmaking process. 

\item Low-resolution maps: maps provided at $N_{\rm side} = 16$ and their
associated full noise covariance matrices. 

\item RIMO (reduced instrument model): model provided with all parameters that
identify the main instrument characteristics from noise to bandpass and beam
function.

\end{itemize}

\section{Discussion and conclusions}
\label{sec_conclusion}
We have summarized in this paper all the steps taken to assemble, calibrate,
and map the data gathered by the \Planck\ LFI instrument.  
While the focus is on the changes in data and methods since our previous
release in 2013 \citep{planck2013-p02}, this paper provides a complete, 
if brief, description of LFI data processing, and of the resulting temperature
and polarization maps at 30, 44, and 70\,GHz. Many supporting details are provided in four additional papers accompanying
this release, \citet{planck2014-a04}, \citet{planck2014-a05},
\citet{planck2014-a06}, and \citet{planck2014-a07}, 
which treat systematic effects, beams, calibration and mapmaking, respectively.
We note that \citet{planck2014-a08} and \citet{planck2014-a09} cover the same
set of topics for the \Planck\ HFI instrument.

\subsection{Operations, TOI, and beams}
\label{sec_concl_op}
LFI operated stably for all four years of observations (eight sky surveys).  
The last four surveys were performed with a different phase angle
(see Sect.~\ref{sec_flightbahavior}), allowing us to investigate some
systematic effects (and also reducing Galactic straylight).  
The most significant change in LFI operations was the gradual degradation of 
the sorption cooler, and its replacement by a second cooler (on OD 460).
For the current release, construction of the satellite attitude and pointing
takes account of two additional variables, 
the distance to the Sun and the temperature of the REBA \citep{planck2014-a01}.

Routine spacecraft manoeuvers made approximately 8\,\% of the data unusable;
other losses of TOI data were $<1\,\%$ for all three LFI bands.  

The TOI required several small corrections described in
Sect.~\ref{sec_toiprocessing}. These include corrections for ADC non-linearity
and for electronic spikes. Residual effects in the LFI maps are at the
$\muK$ level or below (see \citet{planck2014-a05} for a fuller discussion).

Measurements of LFI beam properties (Sect.~\ref{sec_beamrecovery}) have
substantially improved since the earlier release, based on repeated scans of
Jupiter and better modelling of sidelobes.  The effective beam solid angles at
30, 44, and 70\,GHz are $1190.06$, $832.00$,
and $200.90$ [arcmin$^2$], respectively see Tab.~\ref{tab_eff} for details. 
The remaining sidelobe power outside the main beam is very small,
0.808\,\%, 0.117\,\%, and 0.646\,\% for the three LFI bands.

\subsection{Noise and calibration}
\label{sec_concl_calib}
Calibration of the TOI (to convert to units of $\muKCMBs$) has improved in
several ways since the previous release \citep{planck2013-p02}.  Firstly,
\Planck\ calibration is now based on the dipole signal induced by the annual
motion of the satellite around the Sun (the orbital dipole).  
The calibration thus does not depend on WMAP measurements of the 
larger solar dipole, and it is also absolute, in the sense that it depends
only on well-measured properties of the solar system and fundamental constants.
 
Secondly, LFI calibration is now based on full $4\pi$ convolution of the beam
with the dipole (see Sect.~\ref{sec_iterative_calib}).  
While the calibration is based on the dipole, the dipole signal is removed
from the TOI before mapmaking.

A major source of potential systematic error in calibration is Galactic
straylight (Galactic emission leaking into the LFI horns).  
We model this effect, and correct the TOI accordingly
(Sect.~\ref{sec_strylight_removal}).  Straylight (if not corrected) produces 
evident rings centred on the Galactic centre
(see Fig.~\ref{fig:galactic_straylight}). 

As noted, \Planck\ calibration is carried out on a large-scale source, namely
the orbital dipole, which has a thermal spectrum.  
When assessing the brightness temperature or flux density of other
astronomical objects with non-thermal spectra, 
small colour corrections are necessary; these are provided in
Sect.~\ref{sec_color_correction}.  For compact sources, 
the small amount of power missing from the main beams, listed above, must be
taken into account.  As an example, the flux density of a compact source with
spectral index $-0.5$ extracted from the 30\,GHz map 
requires a $1.00808$ multiplicative correction for missing power and a
multiplicative colour correction of $0.997$.

The noise properties (white noise levels and knee frequencies) of the LFI
receivers are discussed in Sect.~\ref{sec_general_noise}.  The white noise
was stable over the four-year mission for all receivers.

\subsection{Maps}
\label{sec_concl_map}
LFI produces full-sky maps in Stokes parameters $I$, $Q$, and $U$ at all
three frequencies; the map properties are listed in
Sect.~\ref{sec_OverviewMaps}.
Calibrated TOI data are destriped using the {\tt Madam} mapmaking code and
maps are constructed using the same package (see Sect.~\ref{sec_mmaking_intro}
for a description and \citet{planck2014-a07} for full details).  
In destriping, a mask is employed to limit noise introduced by Galactic
emission; the final maps, however, cover the entire sky
(at $N_{\mathrm{side}} = 1024 $ resolution).  {\tt Madam} also produces the
noise covariance matrix (NCVM) for each pixel of the maps.  

We also provide maps at lower resolution ($N_{\mathrm{side}} = 16$;
Fig.~\ref{fig:low_maps}) for use in the construction of the 
low-$\ell$ likelihood (fully described in \citet{planck2014-a13}).
The downgrading scheme to smooth the maps from $N_{\mathrm{side}}=1024$ to
$32$ and then to $16$ is described in Sect.~\ref{sec_low_data} and \ref{sec_low_res}.
Sect.~\ref{sec_low_ncvm} describes the NCVM for these low-resolution products.  

\subsection{Polarization}
\label{sec_concl_pol}
The major new feature of this release is the set of polarized maps and
products.  The low-resolution polarization maps at 70\,GHz, in particular,
play a crucial role in the construction of the \Planck\ low-$\ell$ likelihood
\citep{planck2014-a13} and consequently on \Planck\ values for cosmological
parameters \citep{planck2014-a15}.  We therefore devote considerable attention
to investigating potential systematic errors in these maps (detailed in
Sect.~\ref{sec_polarization}).  The largest source of uncertainty in LFI
polarization measurements is leakage from temperature to polarization.
This leakage is largely caused by differences in 
the frequency responses or bandpasses between the two arms of a given LFI
radiometer (``bandpass mismatch'').  
This mismatch can be quantified by a single parameter; for the 70\,GHz radiometers, it varies between 0.18 and 1.24\,\%.  
The bandpass mismatch correction maps are provided in these release at $N_{\rm side}=256$, those should be applied to LFI $Q$ and $U$ maps.

\subsection{Validation}
\label{sec_concl_val}
We employ suites of both null tests and simulations to assess the quality of
LFI maps and other products derived from them (see
Sect.~\ref{sec_dataval_intro}).  The null tests exploit the many ways in which
the data can be divided: survey by survey, year by year, and on the much
shorter time scale of half-ring differences.  The results of some of these
null tests are shown in Fig.~\ref{nulltest_figures}; 
further details appear in \citep{planck2014-a04} and \citep{planck2014-a06}.
We call attention to the substantially lower residuals (and cleaner maps)
resulting mainly from better calibration. The null tests do, however, 
reveal larger than average residual signals in polarized maps made from
Survey~2 and Survey~4 data (we return to this issue below).  

Another type of null test is to compare the CMB power spectra derived from
different frequencies.  This topic is discussed for the entire mission in
\citep{planck2014-a01}.  Here, we point out that Fig.~\ref{fig_wncvm}, shows good agreement among the three LFI bands.  In the $\ell$ range
40--300 (which covers the first peak of the CMB power spectrum), 
the three LFI power spectra agree to better than $1\,\%$
This agreement extends to measurements of compact sources 
(which involve both a wider $\ell$ range and values for the beam solid angles;
see \citealt{planck2014-a35}).

We validated LFI polarization maps by comparing our bandpass-mismatch-corrected
maps to maps constructed using the $IQUSP$ procedure
(Sect.~\ref{sec_overview_correction_maps}), and found that the Stokes $Q$
and $U$ maps were indistinguishable.  
In particular, the polarized structure along the Galactic plane is reproduced,
including the most significant discrepancies with WMAP maps.  

Simulations based on FFP8 \citep{planck2014-a14} are also used to validate
LFI results.  We perform end-to-end simulations primarily to test the impact
of systematic errors and various 
steps in our calibration and mapmaking procedures on the final results.
Sect.~\ref{sec_systematics} and
Tables~\ref{tab_summary_systematic_effects_maps_30},
\ref{tab_summary_systematic_effects_maps_44}, and
\ref{tab_summary_systematic_effects_maps_70} summarize 
the sources of systematic error and their effects on LFI maps.
The far sidelobes of the LFI beams are the dominant source of uncertainty
in the 30\,GHz maps.  At 44 and 70\,GHz, other instrumental 
effects dominate, particularly 1\,Hz electronic spikes and ADC non-linearity,
respectively. The overall systematic effects uncertainty was estimated to be $0.88$, $1.97$, and $1.87$ \muKCMB in the $I$ component; 
$1.11$, $1.14$, and $2.25$ \muKCMB in the $Q$ component; $0.95$, $1.20$, and $2.22$ in the $U$ component at 30, 44 and 70\,GHz, respectively.

As mentioned above, null tests show that the polarized data from Surveys~2
and 4 contain residual signals (possibly due to contamination from Galactic
emission).  As a consequence, we choose to be conservative and omit these two
surveys (approximately 1/4 of the data) from the low-$\ell$ likelihood.
The studies supporting this decision are described in Sect.~\ref{sec_low_l}.
The optical depth, $\tau$, is the cosmological parameter most affected;
including or omitting Surveys~2 and 4 changes $\tau$ by about $0.5\,\sigma$.

\begin{acknowledgements}

The Planck Collaboration acknowledges the support of: ESA; CNES and CNRS/INSU-IN2P3-INP (France); ASI, CNR, and INAF (Italy); NASA and DoE (USA); STFC and UKSA (UK); CSIC, MINECO, JA, and RES (Spain); Tekes, AoF, and CSC (Finland); DLR and MPG (Germany); CSA (Canada); DTU Space (Denmark); SER/SSO (Switzerland); RCN (Norway); SFI (Ireland); FCT/MCTES (Portugal); ERC and PRACE (EU). A description of the Planck Collaboration and a list of its members, indicating which technical or scientific activities they have been involved in, can be found at \url{http://www.cosmos.esa.int/web/planck/}. 
Finally, we thank Benjamin Walter for a careful reading of the manuscript. 

\end{acknowledgements}

\allearlypapers
\alltwentyfifteenresultspapers
\alltwentythirteenresultspapers

\bibliographystyle{aat}

\bibliography{Planck_bib,LFI_DPC_bib}

\raggedright
\end{document}